\documentclass{article}
\usepackage{epsf}
\usepackage{epubtk}
\usepackage{amsfonts}
\usepackage{amsmath}
\usepackage{bm}

\newcounter{theorem}  \newenvironment{theorem}
{\refstepcounter{theorem}
   \vspace{1 em}
   \noindent{\bf Theorem~\thetheorem}
   \begin{em}}
  {\end{em}
   \newline}

\allowdisplaybreaks

\newcommand{\pz}{\phantom{.0\quad}}
\newcommand{\ws}{\phantom{\quad}}

\DeclareMathOperator{\pf}{Pf}

\begin{document}

\title{Gravitational Radiation from Post-Newtonian Sources \\
       and Inspiralling Compact Binaries}

\author{\epubtkAuthorData{Luc Blanchet}
        {Institut d'Astrophysique de Paris \\
         98${}^\mathrm{bis}$ Boulevard Arago \\
         75014 Paris, France}
        {blanchet@iap.fr}
        {http://www.iap.fr/users/blanchet}}

\date{}
\maketitle


\begin{abstract}
  The article reviews the current status of a theoretical approach to
  the problem of the emission of gravitational waves by isolated systems
  in the context of general relativity. Part~\ref{part:a} of the article
  deals with general post-Newtonian sources. The exterior field of the
  source is investigated by means of a combination of analytic
  post-Minkowskian and multipolar approximations. The physical
  observables in the far-zone of the source are described by a specific
  set of radiative multipole moments. By matching the exterior solution
  to the metric of the post-Newtonian source in the near-zone we obtain
  the explicit expressions of the source multipole moments. The
  relationships between the radiative and source moments involve many
  non-linear multipole interactions, among them those associated with
  the tails (and tails-of-tails) of gravitational waves.
  Part~\ref{part:b} of the article is devoted to the application to
  compact binary systems. We present the equations of binary motion, and
  the associated Lagrangian and Hamiltonian, at the third post-Newtonian
  (3PN) order beyond the Newtonian acceleration. The gravitational-wave
  energy flux, taking consistently into account the relativistic
  corrections in the binary moments as well as the various tail effects,
  is derived through 3.5PN order with respect to the quadrupole
  formalism. The binary's orbital phase, whose prior knowledge is
  crucial for searching and analyzing the signals from inspiralling
  compact binaries, is deduced from an energy balance argument.
\end{abstract}

\epubtkKeywords{gravitational radiation, post-Newtonian approximation,
  multipolar expansion, inspiralling compact binary}

\newpage


\section{Introduction}
\label{sec:1}

The theory of gravitational radiation from isolated sources, in the
context of general relativity, is a fascinating science that can be
explored by means of what was referred to in the French XVIIIth
century as \emph{l'analyse sublime}: the analytical (i.e.\
mathematical) method, and more specifically the resolution of partial
differential equations. Indeed, the field equations of general
relativity, when use is made of the harmonic-coordinate conditions,
take the form of a quasi-linear hyperbolic differential system of
equations, involving the famous wave operator or d'Alembertian
(denoted $\Box$), invented by d'Alembert in his Trait\'e de dynamique
of 1743.

Nowadays, the importance of the field lies in the exciting possibility
of comparing the theory with contemporary astrophysical observations,
made by a new generation of detectors -- large-scale optical
interferometers LIGO, VIRGO, GEO and TAMA -- that should routinely
observe the gravitational waves produced by massive and rapidly evolving
systems such as inspiralling compact binaries. To prepare these
experiments, the required theoretical work consists of carrying out a
sufficiently general solution of the Einstein field equations, valid for
a large class of matter systems, and describing the physical processes
of the emission and propagation of the waves from the source to the
distant detector, as well as their back-reaction onto the source.


\subsection{Gravitational-wave generation formalisms}
\label{subsec:1.1}

The basic problem we face is to relate the asymptotic gravitational-wave
form $h_{ij}$ generated by some isolated source, at the location of some
detector in the wave zone of the source, to the stress-energy tensor
$T^{\alpha\beta}$ of the matter fields\epubtkFootnote{In this article
Greek indices take the values $0, 1, 2, 3 $ and Latin $1, 2, 3$. Our
signature is +2. $G$ and $c$ are Newton's constant and the speed of
light.}. For general sources it is hopeless to solve the problem
\emph{via} a rigorous deduction within the exact theory of general
relativity, and we have to resort to approximation methods, keeping in mind that,
sadly, such methods are often not related in a very precise mathematical
way to the first principles of the theory. Therefore, a general
wave-generation formalism must solve the field equations, and the
non-linearity therein, by imposing some suitable approximation series in
one or several small physical parameters. Of ourse the ultimate aim of
approximation methods is to extract from the theory some firm
predictions for the outcome of experiments such as VIRGO and LIGO. Some
important approximations that we shall use in this article are the
post-Newtonian method (or non-linear $1/c$-expansion), the
post-Minkowskian method or non-linear iteration ($G$-expansion), the
multipole decomposition in irreducible representations of the rotation
group (or equivalently $a$-expansion in the source radius), and the
far-zone expansion ($1/R$-expansion in the distance). In particular, the
post-Newtonian expansion has provided us in the past with our best
insights into the problems of motion and radiation in general
relativity. The most successful wave-generation formalisms make a
gourmet cocktail of all these approximation methods. For reviews on
analytic approximations and applications to the motion and the
gravitational wave-generation see Refs.~\cite{Thhouches, Dcargese, D300,
Th300, W94, Bhouches, Behlers}.

The post-Newtonian approximation is valid under the assumptions of a
weak gravitational field inside the source (we shall see later how to
model neutron stars and black holes), and of slow internal motions. The
main problem with this approximation is its domain of validity, which is
limited to the near zone of the source -- the region surrounding the
source that is of small extent with respect to the wavelength of waves.
A serious consequence is the \emph{a priori\/} inability of the
post-Newtonian expansion to incorporate the boundary conditions at
infinity, which determine the radiation reaction force in the source's
local equations of motion. The post-Minkowskian expansion, by contrast,
is uniformly valid, as soon as the source is weakly self-gravitating,
over all space-time. In a sense, the post-Minkowskian method is more
fundamental than the post-Newtonian one; it can be regarded as an
``upstream'' approximation with respect to the post-Newtonian expansion,
because each coefficient of the post-Minkowskian series can in turn be
re-expanded in a post-Newtonian fashion. Therefore, a way to take into
account the boundary conditions at infinity in the post-Newtonian series
is \emph{first} to perform the post-Minkowskian expansion. Notice that
the post-Minkowskian method is also upstream (in the previous sense)
with respect to the multipole expansion, when considered outside the
source, and with respect to the far-zone expansion, when considered far
from the source.

The most ``downstream'' approximation that we shall use in this article
is the post-Newtonian one; therefore this is the approximation that
dictates the allowed physical properties of our matter source. We assume
mainly that the source is at once \emph{slowly moving} and \emph{weakly
stressed}, and we abbreviate this by saying that the source is
\emph{post-Newtonian}. For post-Newtonian sources, the parameter
defined from the components of the matter stress-energy tensor
$T^{\alpha\beta}$ and the source's Newtonian potential $U$ by
\begin{equation}
  \epsilon = \max \left\{\left|\frac{T^{0i}}{T^{00}}\right|,
  \left|\frac{T^{ij}}{T^{00}}\right|^{1/2}\!\!,
  \left|\frac{U}{c^2}\right|^{1/2}\right\},
  \label{1}
\end{equation}
is much less than one. This parameter represents essentially a slow
motion estimate $\epsilon \sim v/c$, where $v$ denotes a typical
internal velocity. By a slight abuse of notation, following
Chandrasekhar et al.~\cite{C65, CN69, CE70}, we shall henceforth
write $\epsilon\equiv 1/c$, even though $\epsilon$ is dimensionless
whereas $c$ has the dimension of a velocity. The small post-Newtonian
remainders will be denoted ${\cal O}(1/c^n)$. Thus, $1/c \ll 1$ in the
case of post-Newtonian sources. We have $|U/c^2|^{1/2} \ll 1/c$ for
sources with negligible self-gravity, and whose dynamics are therefore
driven by non-gravitational forces. However, we shall generally assume
that the source is self-gravitating; in that case we see that it is
necessarily \emph{weakly} (but not negligibly) self-gravitating,
i.e.\ $|U/c^2|^{1/2}={\cal O}(1/c)$. Note that the adjective
``slow-motion'' is a bit clumsy because we shall in fact consider
\emph{very} relativistic sources such as inspiralling compact binaries, for
which $1/c$ can be as large as $30\%$ in the last rotations, and whose
description necessitates the control of high post-Newtonian
approximations.

The lowest-order wave generation formalism, in the Newtonian limit
$1/c\to 0$, is the famous quadrupole formalism of Einstein~\cite{E18}
and Landau and Lifchitz~\cite{LL}. This formalism can also be referred
to as Newtonian because the evolution of the quadrupole moment of the
source is computed using Newton's laws of gravity. It expresses the
gravitational field $h^\mathrm{TT}_{ij}$ in a transverse and traceless
(TT) coordinate system, covering the far zone of the
source\epubtkFootnote{The TT coordinate system can be extended to the
near zone of the source as well; see for instance Ref.~\cite{KSGE}.}, as
\begin{equation}
  h^\mathrm{TT}_{ij}=\frac{2G}{c^4R} {\cal P}_{ijab}
  ({\bf N})\left\{\frac{d^2\mathrm{Q}_{ab}}{dT^2}(T-R/c)+
  {\cal O}\left(\frac{1}{c}\right)\right\}+
  {\cal O}\left(\frac{1}{R^2}\right),
  \label{2}
\end{equation}
where $R=|{\bf X}|$ is the distance to the source, ${\bf N}={\bf X}/R$
is the unit direction from the source to the observer, and ${\cal
P}_{ijab}= {\cal P}_{ia}{\cal P}_{jb}-\frac{1}{2}\delta_{ij}{\cal
P}_{ij}{\cal P}_{ab}$ is the TT projection operator, with ${\cal
P}_{ij}=\delta_{ij} -N_iN_j$ being the projector onto the plane
orthogonal to ${\bf N}$. The source's quadrupole moment takes the
familiar Newtonian form
\begin{equation}
  \mathrm{Q}_{ij}(t) = \int_\mathrm{source} \!\!\!\!\!\!\! d^3{\bf x} \,
  \rho ({\bf x}, t)\left(x_ix_j-\frac{1}{3}\delta_{ij}{\bf x}^2\right),
  \label{3}
\end{equation}
where $\rho$ is the Newtonian mass density. The total gravitational
power emitted by the source in all directions is given by the Einstein
quadrupole formula
\begin{equation}
  {\cal L} = \frac{G}{5c^5}\left\{\frac{d^3\mathrm{Q}_{ab}}
  {dT^3}\frac{d^3\mathrm{Q}_{ab}}{dT^3}+
  {\cal O}\left(\frac{1}{c^2}\right)\right\}.
  \label{4}
\end{equation}
Our notation ${\cal L}$ stands for the total gravitational
``luminosity'' of the source. The cardinal virtues of the
Einstein--Landau--Lifchitz quadrupole formalism are its generality --
the only restrictions are that the source be Newtonian and bounded --
its simplicity, as it necessitates only the computation of the time
derivatives of the Newtonian quadrupole moment (using the Newtonian laws
of motion), and, most importantly, its agreement with the observation of
the dynamics of the Hulse-Taylor binary pulsar PSR~1913+16~\cite{TFMc79,
TW82, T93}. Indeed the prediction of the quadrupole formalism for the
waves emitted by the binary pulsar system comes from applying
Equation~(\ref{4}) to a system of two point masses moving on an eccentric
orbit (the classic reference is Peters and Mathews~\cite{PM63}; see also
Refs.~\cite{EH75, Wag75}). Then, relying on the energy equation
\begin{equation}
  \frac{dE}{dt} = - {\cal L},
  \label{5}
\end{equation}
where $E$ is the Newtonian binary's center-of-mass energy, we deduce
from Kepler's third law the expression of the ``observable'', that is,
the change in the orbital period $P$ of the pulsar, or ${\dot P}$, as a
function of $P$ itself. From the binary pulsar test, we can say that the
post-Newtonian corrections to the quadrupole formalism, which we shall
compute in this article, have already received, in the case of compact
binaries, strong observational support (in addition to having, as we
shall demonstrate, a sound theoretical basis).

The multipole expansion is one of the most useful tools of physics, but
its use in general relativity is difficult because of the non-linearity
of the theory and the tensorial character of the gravitational
interaction. In the stationary case, the multipole moments are
determined by the expansion of the metric at spatial infinity~\cite{G70,
H74, SB83}, while, in the case of non-stationary fields, the moments,
starting with the quadrupole, are defined at future null infinity. The
multipole moments have been extensively studied in the linearized
theory, which ignores the gravitational forces inside the source. Early
studies have extended the formula~(\ref{4}) to include the
current-quadrupole and mass-octupole moments~\cite{Papa62, Papa71}, and
obtained the corresponding formulas for linear momentum~\cite{Papa62,
Papa71, Bek73, Press77} and angular momentum~\cite{Pe64, CB69}. The
general structure of the infinite multipole series in the linearized
theory was investigated by several works~\cite{SB58, S61, Pi64, Th80},
from which it emerged that the expansion is characterized by two and
only two sets of moments: mass-type and current-type moments. Below we
shall use a particular multipole decomposition of the linearized
(vacuum) metric, parametrized by symmetric and trace-free (STF) mass and
current moments, as given by Thorne~\cite{Th80}. The explicit
expressions of the multipole moments (for instance in STF guise) as
integrals over the source, valid in the linearized theory but
irrespective of a slow motion hypothesis, are completely
known~\cite{M62, CM71, CMM77, DI91b}.

In the full non-linear theory, the (radiative) multipole moments can be
read off the coefficient of $1/R$ in the expansion of the metric when
$R\to +\infty$, with a null coordinate $ T - R / c = \mathrm{const} $.
The solutions of the field equations in the form of a far-field
expansion (power series in $1/R$) have been constructed, and their
properties elucidated, by Bondi et al.~\cite{BBM62} and
Sachs~\cite{Sachs62}. The precise way under which such radiative
space-times fall off asymptotically has been formulated geometrically by
Penrose~\cite{P63, P65} in the concept of an asymptotically simple
space-time (see also Ref.~\cite{GH78}). The resulting
Bondi--Sachs--Penrose approach is very powerful, but it can answer
\emph{a priori} only a part of the problem, because it gives information on
the field only in the limit where $R\to +\infty$, which cannot be
connected in a direct way to the actual behaviour of the source. In
particular the multipole moments that one considers in this approach are
those measured at infinity -- we call them the \emph{radiative} multipole
moments. These moments are distinct, because of non-linearities, from
some more natural \emph{source} multipole moments, which are defined
operationally by means of explicit integrals extending over the matter
and gravitational fields.

An alternative way of defining the multipole expansion within the
complete non-linear theory is that of Blanchet and Damour~\cite{BD86,
B87}, following pioneering work by Bonnor and collaborators~\cite{Bo59,
BoR61, BoR66, HR69} and Thorne~\cite{Th80}. In this approach the basic
multipole moments are the \emph{source} moments, rather than the
radiative ones. In a first stage, the moments are left unspecified, as
being some arbitrary functions of time, supposed to describe an actual
physical source. They are iterated by means of a post-Minkowskian
expansion of the vacuum field equations (valid in the source's
exterior). Technically, the post-Minkowskian approximation scheme is
greatly simplified by the assumption of a multipolar expansion, as one
can consider separately the iteration of the different multipole pieces
composing the exterior field (whereas, the direct attack of the
post-Minkowskian expansion, valid at once inside and outside the source,
faces some calculational difficulties~\cite{ThK75, CTh77}). In this
``multipolar-post-Minkowskian'' formalism, which is physically valid
over the entire weak-field region outside the source, and in particular
in the wave zone (up to future null infinity), the radiative multipole
moments are obtained in the form of some non-linear functionals of the
more basic source moments. \emph{A priori}, the method is not limited to
post-Newtonian sources, however we shall see that, in the current
situation, the \emph{closed-form} expressions of the source multipole
moments can be established only in the case where the source is
post-Newtonian~\cite{B95, B98mult}. The reason is that in this case the
domain of validity of the post-Newtonian iteration (viz.\ the near
zone) overlaps the exterior weak-field region, so that there exists an
intermediate zone in which the post-Newtonian and multipolar expansions
can be matched together. This is a standard application of the method of
matched asymptotic expansions in general relativity~\cite{BuTh70, Bu71}.

To be more precise, we shall show how a systematic multipolar and
post-Minkowskian iteration scheme for the vacuum Einstein field
equations yields the most general physically admissible solution of
these equations~\cite{BD86}. The solution is specified once we give two
and only two sets of time-varying (source) multipole moments. Some
general theorems about the near-zone and far-zone expansions of that
general solution will be stated. Notably, we find~\cite{B87} that the
asymptotic behaviour of the solution at future null infinity is in
agreement with the findings of the Bondi--Sachs--Penrose~\cite{BBM62,
Sachs62, P63, P65, GH78} approach to gravitational radiation. However,
checking that the asymptotic structure of the radiative field is correct
is not sufficient by itself, because the ultimate aim is to relate the
far field to the properties of the source, and we are now obliged to
ask: What are the multipole moments corresponding to a given
stress-energy tensor $T^{\alpha\beta}$ describing the source? Only in
the case of post-Newtonian sources has it been possible to answer this
question. The general expression of the moments was obtained at the
level of the second post-Newtonian (2PN) order in Ref.~\cite{B95}, and
was subsequently proved to be in fact valid up to any post-Newtonian
order in Ref.~\cite{B98mult}. The source moments are given by some
integrals extending over the post-Newtonian expansion of the total
(pseudo) stress-energy tensor $\tau^{\alpha\beta}$, which is made of a
matter part described by $T^{\alpha\beta}$ and a crucial non-linear
gravitational source term $\Lambda^{\alpha\beta}$. These moments carry
in front a particular operation of taking the finite part (${\cal FP}$
as we call it below), which makes them mathematically well-defined
despite the fact that the gravitational part $\Lambda^{\alpha\beta}$ has
a spatially infinite support, which would have made the bound of the
integral at spatial infinity singular (of course the finite part is not
added \emph{a posteriori} to restore the well-definiteness of the
integral, but is \emph{proved} to be actually present in this formalism).
The expressions of the moments had been obtained earlier at the 1PN
level, albeit in different forms, in Ref.~\cite{BD89} for the mass-type
moments (strangely enough, the mass moments admit a compact-support
expression at 1PN order), and in Ref.~\cite{DI91a} for the current-type
ones.

The wave-generation formalism resulting from matching the exterior
multipolar and post-Minkowskian field~\cite{BD86, B87} to the
post-Newtonian source~\cite{B95, B98mult} is able to take into account,
in principle, any post-Newtonian correction to both the source and
radiative multipole moments (for any multipolarity of the moments). The
relationships between the radiative and source moments include many
non-linear multipole interactions, because the source moments mix with
each other as they ``propagate'' from the source to the detector. Such
multipole interactions include the famous effects of wave tails,
corresponding to the coupling between the non-static moments with the
total mass $\mathrm{M}$ of the source. The non-linear multipole
interactions have been computed within the present wave-generation
formalism up to the 3PN order in Refs.~\cite{BD92, B98quad, B98tail}.
Furthermore, the back-reaction of the gravitational-wave emission onto
the source, up to the 1.5PN order relative to the leading order of
radiation reaction, has also been studied within this
formalism~\cite{BD88, B93, B97}. Now, recall that the leading radiation
reaction force, which is quadrupolar, occurs already at the 2.5PN order
in the source's equations of motion. Therefore the 1.5PN ``relative''
order in the radiation reaction corresponds in fact to the 4PN order in
the equations of motion, beyond the Newtonian acceleration. It has been
shown that the gravitational wave tails enter the radiation reaction at
precisely the 1.5PN \emph{relative} order, which means 4PN ``absolute''
order~\cite{BD88}. A systematic post-Newtonian iteration scheme for the
near-zone field, formally taking into account all radiation reaction
effects, has been recently proposed, consistent with the present
formalism~\cite{PB02, BFN05}.

A different wave-generation formalism has been devised by Will and
Wiseman~\cite{WW96} (see also Refs.~\cite{W99, PW00, PW02}), after
earlier attempts by Epstein and Wagoner~\cite{EW75} and
Thorne~\cite{Th80}. This formalism has exactly the same scope as ours,
i.e.\ it applies to any isolated post-Newtonian sources, but it
differs in the definition of the source multipole moments and in many
technical details when properly implemented~\cite{WW96}. In both
formalisms, the moments are generated by the post-Newtonian expansion of
the pseudo-tensor $\tau^{\alpha\beta}$, but in the Will--Wiseman
formalism they are defined by some \emph{compact-support} integrals
terminating at some finite radius ${\cal R}$ enclosing the source,
e.g., the radius of the near zone). By contrast, in our case~\cite{B95,
B98mult}, the moments are given by some integrals covering the whole
space and regularized by means of the finite part ${\cal FP}$. We shall
prove the complete equivalence, at the most general level, between the
two formalisms. What is interesting about both formalisms is that the
source multipole moments, which involve a whole series of relativistic
corrections, are coupled together, in the true non-linear solution, in a
very complicated way. These multipole couplings give rise to the many
tail and related non-linear effects, which form an integral part of the
radiative moments at infinity and thereby of the observed signal.

Part~\ref{part:a} of this article is devoted to a presentation of the
post-Newtonian wave generation formalism. We try to state the main
results in a form that is simple enough to be understood without the
full details, but at the same time we outline some of the proofs when
they present some interest on their own. To emphasize the importance of
some key results, we present them in the form of mathematical theorems.


\subsection{Problem posed by compact binary systems}
\label{subsec:1.2}

Inspiralling compact binaries, containing neutron stars and/or black
holes, are promising sources of gravitational waves detectable by the
detectors LIGO, VIRGO, GEO and TAMA. The two compact objects steadily
lose their orbital binding energy by emission of gravitational
radiation; as a result, the orbital separation between them decreases,
and the orbital frequency increases. Thus, the frequency of the
gravitational-wave signal, which equals twice the orbital frequency for
the dominant harmonics, ``chirps'' in time (i.e.\ the signal
becomes higher and higher pitched) until the two objects collide and
merge.

The orbit of most inspiralling compact binaries can be considered to be
circular, apart from the gradual inspiral, because the gravitational
radiation reaction forces tend to circularize the motion rapidly. For
instance, the eccentricity of the orbit of the Hulse--Taylor binary
pulsar is presently $e_0=0.617$. At the time when the gravitational
waves emitted by the binary system will become visible by the detectors,
i.e.\ when the signal frequency reaches about $10 \mathrm{\ Hz}$ (in a few
hundred million years from now), the eccentricity will be $e=5.3 \times
10^{-6}$ -- a value calculated from the Peters~\cite{Pe64} law, which is
itself based on the quadrupole formula~(\ref{2}).

The main point about modelling the inspiralling compact binary is that a
model made of two structureless point particles, characterized solely by
two mass parameters $m_1$ and $m_2$ (and possibly two spins), is
sufficient. Indeed, most of the non-gravitational effects usually
plaguing the dynamics of binary star systems, such as the effects of a
magnetic field, of an interstellar medium, and so on, are dominated by
gravitational effects. However, the real justification for a model of
point particles is that the effects due to the finite size of the
compact bodies are small. Consider for instance the influence of the
Newtonian quadrupole moments $\mathrm{Q}_1$ and $\mathrm{Q}_2$ induced
by tidal interaction between two neutron stars. Let $a_1$ and $a_2$ be
the radius of the stars, and $L$ the distance between the two centers of
mass. We have, for tidal moments,
\begin{equation}
  \mathrm{Q}_1 = k_1 m_2 \frac{a_1^5}{L^3},
  \qquad
  \mathrm{Q}_2 = k_2 m_1 \frac{a_2^5}{L^3},
  \label{6}
\end{equation}
where $k_1$ and $k_2$ are the star's dimensionless (second) Love
numbers~\cite{Moritz}, which depend on their internal structure, and
are, typically, of the order unity. On the other hand, for compact
objects, we can introduce their ``compactness'', defined by the
dimensionless ratios
\begin{equation}
  K_1 = \frac{G m_1}{a_1 c^2},\qquad K_2 = \frac{G m_2}{a_2 c^2},
  \label{7}
\end{equation}
which equal $\sim 0.2$ for neutron stars (depending on their equation of
state). The quadrupoles $\mathrm{Q}_1$ and $\mathrm{Q}_2$ will affect
both sides of Equation~(\ref{5}), i.e.\ the Newtonian binding energy $E$ of
the two bodies, and the emitted total gravitational flux ${\cal L}$ as
computed using the Newtonian quadrupole formula~(\ref{4}). It is known
that for inspiralling compact binaries the neutron stars are not
co-rotating because the tidal synchronization time is much larger than
the time left till the coalescence. As shown by Kochanek~\cite{Kochanek}
the best models for the fluid motion inside the two neutron stars are
the so-called Roche--Riemann ellipsoids, which have tidally locked
figures (the quadrupole moments face each other at any instant during
the inspiral), but for which the fluid motion has zero circulation in
the inertial frame. In the Newtonian approximation we find that within
such a model (in the case of two identical neutron stars) the orbital
phase, deduced from Equation~(\ref{5}), reads
\begin{equation}
  \phi^\mathrm{finite\ size} - \phi_0 = -\frac{1}{8x^{5/2}}
  \left\{1+\mathrm{const\,} k \left(\frac{x}{K}\right)^5\right\},
  \label{8}
\end{equation}
where $x=(G m\omega/c^3)^{2/3}$ is a standard dimensionless
post-Newtonian parameter $\sim 1/c^2$ ($\omega$ is the orbital
frequency), and where $k$ is the Love number and $K$ is the compactness
of the neutron star. The first term in the right-hand side of Equation~(\ref{8})
corresponds to the gravitational-wave damping of two point masses; the
second term is the finite-size effect, which appears as a relative
correction, proportional to $(x/K)^5$, to the latter radiation damping
effect. Because the finite-size effect is purely Newtonian, its relative
correction $\sim (x/K)^5$ should not depend on $c$; and indeed the
factors $1/c^2$ cancel out in the ratio $x/K$. However, the compactness
$K$ of compact objects is by Equation~(\ref{7}) of the order unity (or, say,
one half), therefore the $1/c^2$ it contains should not be taken into
account numerically in this case, and so the real order of magnitude of
the relative contribution of the finite-size effect in Equation~(\ref{8}) is
given by $x^5$ alone. This means that for compact objects the
finite-size effect should be comparable, numerically, to a
post-Newtonian correction of magnitude $x^5 \sim 1/c^{10}$ namely 5PN
order\epubtkFootnote{See Ref.~\cite{D83houches} for the proof of such an
``effacement'' principle in the context of relativistic equations of
motion.}. This is a much higher post-Newtonian order than the one at
which we shall investigate the gravitational effects on the phasing
formula. Using $k'\equiv \mathrm{const\,} k\sim 1$ and $K\sim 0.2$ for
neutron stars (and the bandwidth of a VIRGO detector between
$10 \mathrm{\ Hz}$ and $1000 \mathrm{\ Hz}$), we find that the
cumulative phase error due to the finite-size
effect amounts to less that one orbital rotation over a total of $\sim
16,000$ produced by the gravitational-wave damping of point masses. The
conclusion is that the finite-size effect can in general be neglected in
comparison with purely gravitational-wave damping effects. But note that
for non-compact or moderately compact objects (such as white dwarfs for
instance) the Newtonian tidal interaction dominates over the radiation
damping.

The inspiralling compact binaries are ideally suited for application of
a high-order post-Newtonian wave generation formalism. The main reason
is that these systems are very relativistic, with orbital velocities as
high as $0.5 c$ in the last rotations (as compared to $\sim 10^{-3}c$ for
the binary pulsar), and it is not surprising that the quadrupole-moment
formalism~(\ref{2}, \ref{3}, \ref{4}, \ref{5}) constitutes a poor
description of the emitted gravitational waves, since many
post-Newtonian corrections play a substantial role. This expectation has
been confirmed in recent years by several
measurement-analyses~\cite{3mn, CFPS93, FCh93, CF94, TNaka94, P95, PW95,
KKS95, DIS98}, which have demonstrated that the post-Newtonian precision
needed to implement successively the optimal filtering technique in the
LIGO/VIRGO detectors corresponds grossly, in the case of neutron-star
binaries, to the 3PN approximation, or $1/c^6$ beyond the quadrupole
moment approximation. Such a high precision is necessary because of the
large number of orbital rotations that will be monitored in the
detector's frequency bandwidth ($\sim 16,000$ in the case of neutron
stars), giving the possibility of measuring very accurately the orbital
phase of the binary. Thus, the 3PN order is required mostly to compute
the time evolution of the orbital phase, which depends, \emph{via} the
energy equation~(\ref{5}), on the center-of-mass binding energy $E$ and
the total gravitational-wave energy flux ${\cal L}$.

In summary, the theoretical problem posed by inspiralling compact
binaries is two-fold: On the one hand $E$, and on the other hand ${\cal
L}$, are to be deduced from general relativity with the 3PN precision or
better. To obtain $E$ we must control the 3PN equations of motion of the
binary in the case of general, not necessarily circular, orbits. As for
${\cal L}$ it necessitates the application of a 3PN wave generation
formalism (actually, things are more complicated because the equations
of motion are also needed during the computation of the flux). It is
quite interesting that such a high order approximation as the 3PN one
should be needed in preparation for LIGO and VIRGO data analysis. As we
shall see, the signal from compact binaries contains at the 3PN order
the signature of several non-linear effects which are specific to
general relativity. Therefore, we have here the possibility of probing,
experimentally, some aspects of the non-linear structure of Einstein's
theory~\cite{BSat94, BSat95}.


\subsection{Post-Newtonian equations of motion and radiation}
\label{subsec:1.3}

By equations of motion we mean the explicit expression of the
accelerations of the bodies in terms of the positions and velocities. In
Newtonian gravity, writing the equations of motion for a system of $N$
particles is trivial; in general relativity, even writing the equations
in the case $N=2$ is difficult. The first relativistic term, at the 1PN
order, was derived by Lorentz and Droste~\cite{LD17}. Subsequently,
Einstein, Infeld and Hoffmann~\cite{EIH} obtained the 1PN corrections by
means of their famous ``surface-integral'' method, in which the
equations of motion are deduced from the \emph{vacuum} field equations,
and which are therefore applicable to any compact objects (be they
neutron stars, black holes, or, perhaps, naked singularities). The
1PN-accurate equations were also obtained, for the motion of the centers
of mass of extended bodies, by Petrova~\cite{Petrova} and
Fock~\cite{Fock39} (see also Ref.~\cite{Papa51}).

The 2PN approximation was tackled by Ohta et al.~\cite{OO73,
OO74a, OO74b}, who considered the post-Newtonian iteration of the
Hamiltonian of $N$ point-particles. We refer here to the Hamiltonian as
the Fokker-type Hamiltonian, which is obtained from the
matter-plus-field Arnowitt--Deser--Misner (ADM) Hamiltonian by
eliminating the field degrees of freedom. The result for the 2PN and
even 2.5PN equations of binary motion in harmonic coordinates was
obtained by Damour and Deruelle~\cite{DD81a, DD81b, Dthese, D82,
D83houches}, building on a non-linear iteration of the metric of two
particles initiated in Ref.~\cite{BeDD81}. The corresponding result for
the ADM-Hamiltonian of two particles at the 2PN order was given in
Ref.~\cite{DS85} (see also Refs.~\cite{S85, S86}). Kopeikin~\cite{Kop85}
derived the 2.5PN equations of motion for two extended compact objects.
The 2.5PN-accurate harmonic-coordinate equations as well as the complete
gravitational field (namely the metric $g_{\alpha\beta}$) generated by
two point masses were computed in Ref.~\cite{BFP98}, following a method
based on previous work on wave generation~\cite{B95}.

Up to the 2PN level the equations of motion are conservative. Only at
the 2.5PN order appears the first non-conservative effect, associated
with the gravitational radiation reaction. The (harmonic-coordinate)
equations of motion up to that level, as derived by Damour and
Deruelle~\cite{DD81a, DD81b, Dthese, D82, D83houches}, have been used
for the study of the radiation damping of the binary pulsar -- its
orbital ${\dot P}$~\cite{D83houches, D83, DT91}. It is important to
realize that the 2.5PN equations of motion have been proved to hold in
the case of binary systems of strongly self-gravitating
bodies~\cite{D83houches}. This is \emph{via} an ``effacing'' principle
(in the terminology of Damour~\cite{D83houches}) for the internal
structure of the bodies. As a result, the equations depend only on the
``Schwarzschild'' masses, $m_1$ and $m_2$, of the compact objects.
Notably their compactness parameters $K_1$ and $K_2$, defined by
Equation~(\ref{7}), do not enter the equations of motion, as has been
explicitly verified up to the 2.5PN order by Kopeikin et
al.~\cite{Kop85, GKop86}, who made a ``physical'' computation, \emph{\`a
la} Fock, taking into account the internal structure of two
self-gravitating extended bodies. The 2.5PN equations of motion have
also been established by Itoh, Futamase and Asada~\cite{IFA00, IFA01},
who use a variant of the surface-integral approach of Einstein, Infeld
and Hoffmann~\cite{EIH}, that is valid for compact bodies, independently
of the strength of the internal gravity.

The present state of the art is the 3PN approximation\epubtkFootnote{Let
us mention that the 3.5PN terms in the equations of motion are also
known, both for point-particle binaries~\cite{IW93, IW95, JaraS97, PW02,
KFS03, NB05} and extended fluid bodies~\cite{B93, B97}; they correspond
to 1PN ``relative'' corrections in the radiation reaction force. Known
also is the contribution of wave tails in the equations of motion, which
arises at the 4PN order and represents a 1.5PN modification of the
gravitational radiation damping~\cite{BD88}.}. To this order the
equations have been worked out independently by two groups, by means of
different methods, and with equivalent results. On the one hand,
Jaranowski and Sch\"afer~\cite{JaraS98, JaraS99, JaraS00}, and Damour,
Jaranowski, and Sch\"afer~\cite{DJSpoinc, DJSequiv, DJSdim}, following
the line of research of Refs.~\cite{OO73, OO74a, OO74b, DS85}, employ
the ADM-Hamiltonian formalism of general relativity; on the other hand,
Blanchet and Faye~\cite{BF00, BFeom, BFreg, BFregM}, and de Andrade,
Blanchet, and Faye~\cite{ABF01}, founding their approach on the
post-Newtonian iteration initiated in Ref.~\cite{BFP98}, compute
directly the equations of motion (instead of a Hamiltonian) in harmonic
coordinates. The end results have been shown~\cite{DJSequiv, ABF01} to
be physically equivalent in the sense that there exists a unique
``contact'' transformation of the dynamical variables that changes the
harmonic-coordinates Lagrangian obtained in Ref.~\cite{ABF01} into a new
Lagrangian, whose Legendre transform coincides exactly with the
Hamiltonian given in Ref.~\cite{DJSpoinc}. The 3PN equations of motion,
however, depend on one unspecified numerical coefficient,
$\omega_\mathrm{static}$ in the ADM-Hamiltonian formalism and $\lambda$ in the
harmonic-coordinates approach, which is due to some incompleteness of
the Hadamard self-field regularization method. This coefficient has been
fixed by means of a \emph{dimensional regularization}, both within the
ADM-Hamiltonian formalism~\cite{DJSdim}, and the harmonic-coordinates
equations of motion~\cite{BDE04}. The works~\cite{DJSdim, BDE04} have
demonstrated the power of dimensional regularization and its perfect
adequateness for the problem of the interaction between point masses in
general relativity. Furthermore, an important work by Itoh and
Futamase~\cite{itoh1, itoh2} (using the same surface-integral method as
in Refs.~\cite{IFA00, IFA01}) succeeded in obtaining the complete 3PN
equations of motion in harmonic coordinates directly, i.e.\ without
ambiguity and containing the correct value for the parameter $\lambda$.

So far the status of the post-Newtonian equations of motion is quite
satisfying. There is mutual agreement between all the results obtained
by means of different approaches and techniques, whenever it is possible
to compare them: point particles described by Dirac delta-functions,
extended post-Newtonian fluids, surface-integrals methods, mixed
post-Minkowskian and post-Newtonian expansions, direct post-Newtonian
iteration and matching, harmonic coordinates versus ADM-type
coordinates, and different processes or variants of the regularization
of the self field of point particles. In Part~\ref{part:b} of this
article, we shall present the complete results for the 3PN equations of
motion, and for the associated Lagrangian and Hamiltonian formulations
(from which we deduce the center-of-mass energy $E$).

The second sub-problem, that of the computation of the energy flux
${\cal L}$, has been carried out by application of the wave-generation
formalism described previously. Following earliest computations at the
1PN level~\cite{WagW76, BS89}, at a time when the post-Newtonian
corrections in ${\cal L}$ had a purely academic interest, the energy
flux of inspiralling compact binaries was completed to the 2PN order by
Blanchet, Damour and Iyer~\cite{BDI95, GopuI97}, and, independently, by
Will and Wiseman~\cite{WW96}, using their own formalism (see
Refs.~\cite{BDIWW95, BIWW96} for joint reports of these calculations).
The preceding approximation, 1.5PN, which represents in fact the
dominant contribution of tails in the wave zone, had been obtained in
Refs.~\cite{Wiseman93, BS93} by application of the formula for tail
integrals given in Ref.~\cite{BD92}. Higher-order tail effects at the
2.5PN and 3.5PN orders, as well as a crucial contribution of tails
generated by the tails themselves (the so-called ``tails of tails'') at
the 3PN order, were obtained by Blanchet~\cite{B96, B98tail}. However,
unlike the 1.5PN, 2.5PN, and 3.5PN orders that are entirely composed of
tail terms, the 3PN approximation also involves, besides the tails of
tails, many non-tail contributions coming from the relativistic
corrections in the (source) multipole moments of the binary. These have
been ``almost'' completed in Refs.~\cite{BIJ02, BFIJ02, BI04mult}, in
the sense that the result still involves one unknown numerical
coefficient, due to the use of the Hadamard regularization, which is a
combination of the parameter $\lambda$ in the equations of motion, and a
new parameter $\theta$ coming from the computation of the 3PN quadrupole
moment. The latter parameter is itself a linear combination of three
unknown parameters, $\theta=\xi+2\kappa+\zeta$. We shall review the
computation of the three parameters $\xi$, $\kappa$, and $\zeta$ by means
of dimensional regularization~\cite{BDEI04, BDEI05dr}. In
Part~\ref{part:b} of this article, we shall present the most up-to-date
results for the 3.5PN energy flux and orbital phase, deduced from the
energy balance equation~(\ref{5}), supposed to be valid at this order.

The post-Newtonian flux ${\cal L}$, which comes from a ``standard''
post-Newtonian calculation, is in complete agreement (up to the 3.5PN
order) with the result given by the very different technique of linear
black-hole perturbations, valid in the ``test-mass'' limit where the
mass of one of the bodies tends to zero (limit $\nu\to 0$, where
$\nu=\mu /m$). Linear black-hole perturbations, triggered by the
geodesic motion of a small mass around the black hole, have been applied
to this problem by Poisson~\cite{P93} at the 1.5PN order (following the
pioneering work of Galt'sov et al.~\cite{Galtsov}), and by Tagoshi
and Nakamura~\cite{TNaka94}, using a numerical code, up to the 4PN
order. This technique has culminated with the beautiful analytical
methods of Sasaki, Tagoshi and Tanaka~\cite{Sasa94, TSasa94, TTS96} (see
also Ref.~\cite{MSSTT}), who solved the problem up to the extremely high
5.5PN order.

\newpage


\part{Post-Newtonian Sources}
\label{part:a}


\section{Einstein's Field Equations}
\label{sec:2}

The field equations of general relativity form a system of ten
second-order partial differential equations obeyed by the space-time
metric $g_{\alpha\beta}$,
\begin{equation}
  G^{\alpha\beta}[g,\partial g,\partial^2g] =
  \frac{8\pi G}{c^4} T^{\alpha\beta}[g],
  \label{8_1}
\end{equation}
where the Einstein curvature tensor $G^{\alpha\beta}\equiv
R^{\alpha\beta}-\frac{1}{2}R \, g^{\alpha\beta}$ is generated, through
the gravitational coupling $\kappa=8\pi G/c^4$, by the matter
stress-energy tensor $T^{\alpha\beta}$. Among these ten equations, four
govern, \emph{via} the contracted Bianchi identity, the evolution of the
matter system,
\begin{equation}
  \nabla_\mu G^{\alpha\mu}\equiv 0
  \quad \Longrightarrow \quad
  \nabla_\mu T^{\alpha\mu}=0.
  \label{9}
\end{equation}
The space-time geometry is constrained by the six remaining equations,
which place six independent constraints on the ten components of the
metric $g_{\alpha\beta}$, leaving four of them to be fixed by a choice
of a coordinate system.

In most of this paper we adopt the conditions of \emph{harmonic}, or de
Donder, coordinates. We define, as a basic variable, the
gravitational-field amplitude
\begin{equation}
  h^{\alpha\beta} = \sqrt{-g}\, g^{\alpha\beta} - \eta^{\alpha\beta},
  \label{10}
\end{equation}
where $g^{\alpha\beta}$ denotes the contravariant metric (satisfying
$g^{\alpha\mu}g_{\mu\beta}=\delta^\alpha_\beta$), where $g$ is the
determinant of the covariant metric, $g = \mathrm{det}(
g_{\alpha\beta})$, and where $\eta^{\alpha\beta}$ represents an
auxiliary Minkowskian metric. The harmonic-coordinate condition, which
accounts exactly for the four equations~(\ref{9}) corresponding to the
conservation of the matter tensor, reads
\begin{equation}
  \partial_\mu h^{\alpha\mu} = 0.
  \label{11}
\end{equation}
Equations~(\ref{10}, \ref{11}) introduce into the definition of our
coordinate system a preferred Minkowskian structure, with Minkowski
metric $\eta_{\alpha\beta}$. Of course, this is not contrary to the
spirit of general relativity, where there is only one physical metric
$g_{\alpha\beta}$ without any flat prior geometry, because the
coordinates are not governed by geometry (so to speak), but rather are
chosen by researchers when studying physical phenomena and doing
experiments. Actually, the coordinate condition~(\ref{11}) is especially
useful when we view the gravitational waves as perturbations of
space-time propagating on the fixed Minkowskian manifold with the
background metric $\eta_{\alpha\beta}$. This view is perfectly
legitimate and represents a fruitful and rigorous way to think of the
problem when using approximation methods. Indeed, the metric
$\eta_{\alpha\beta}$, originally introduced in the coordinate
condition~(\ref{11}), does exist at any \emph{finite} order of
approximation (neglecting higher-order terms), and plays in a sense the
role of some ``prior'' flat geometry.

The Einstein field equations in harmonic coordinates can be written in
the form of inhomogeneous flat d'Alembertian equations,
\begin{equation}
  \Box h^{\alpha\beta} = \frac{16\pi G}{c^4} \tau^{\alpha\beta},
  \label{12}
\end{equation}
where $\Box\equiv\Box_\eta =\eta^{\mu\nu}\partial_\mu\partial_\nu$. The
source term $\tau^{\alpha\beta}$ can rightly be interpreted as the
stress-energy pseudo-tensor (actually, $\tau^{\alpha\beta}$ is a Lorentz
tensor) of the matter fields, described by $T^{\alpha\beta}$, \emph{and}
the gravitational field, given by the gravitational source term
$\Lambda^{\alpha\beta}$, i.e.
\begin{equation}
  \tau^{\alpha\beta} = |g| T^{\alpha\beta}+
  \frac{c^4}{16\pi G}\Lambda^{\alpha\beta}.
  \label{13}
\end{equation}
The exact expression of $\Lambda^{\alpha\beta}$, including all
non-linearities, reads\epubtkFootnote{See also Equation~(\ref{Lambdad}) for
the expression in $d+1$ space-time dimensions.}
\begin{eqnarray}
  \Lambda^{\alpha\beta} & = & - h^{\mu\nu} \partial^2_{\mu\nu}
  h^{\alpha\beta}+\partial_\mu h^{\alpha\nu} \partial_\nu h^{\beta\mu}
  +\frac{1}{2}g^{\alpha\beta}g_{\mu\nu}\partial_\lambda h^{\mu\tau}
  \partial_\tau h^{\nu\lambda}
  \nonumber \\
  & & - g^{\alpha\mu}g_{\nu\tau}\partial_\lambda h^{\beta\tau} \partial_\mu
  h^{\nu\lambda} -g^{\beta\mu}g_{\nu\tau}\partial_\lambda h^{\alpha\tau}
  \partial_\mu h^{\nu\lambda} +g_{\mu\nu}g^{\lambda\tau}\partial_\lambda
  h^{\alpha\mu} \partial_\tau h^{\beta\nu}
  \nonumber \\
  & & + \frac{1}{8}(2g^{\alpha\mu}g^{\beta\nu}-g^{\alpha\beta}g^{\mu\nu})
  (2g_{\lambda\tau}g_{\epsilon\pi}-g_{\tau\epsilon}g_{\lambda\pi})
  \partial_\mu h^{\lambda\pi} \partial_\nu h^{\tau\epsilon}.
  \label{14}
\end{eqnarray}%
As is clear from this expression, $\Lambda^{\alpha\beta}$ is made of
terms at least quadratic in the gravitational-field strength $h$ and its
first and second space-time derivatives. In the following, for the
highest post-Newtonian order that we consider (3PN), we need the
quadratic, cubic and quartic pieces of $\Lambda^{\alpha\beta}$. With
obvious notation, we can write them as
\begin{equation}
  \Lambda^{\alpha\beta} = N^{\alpha\beta} [h, h] + M^{\alpha\beta} [h,
  h, h] + L^{\alpha\beta}[h, h, h, h] + {\cal O}(h^5).
  \label{14_1}
\end{equation}
These various terms can be straightforwardly computed from
Equation~(\ref{14}); see Equations (3.8) in Ref.~\cite{BFeom} for explicit
expressions.

As said above, the condition~(\ref{11}) is equivalent to the matter
equations of motion, in the sense of the conservation of the total
pseudo-tensor $\tau^{\alpha\beta}$,
\begin{equation}
  \partial_\mu \tau^{\alpha\mu}=0
  \quad \Longleftrightarrow \quad
  \nabla_\mu T^{\alpha\mu}=0.
  \label{15}
\end{equation}
In this article, we look for the solutions of the field
equations~(\ref{12}, \ref{13}, \ref{14}, \ref{15}) under the following
four hypotheses:
\begin{enumerate}
\item The matter stress-energy tensor $T^{\alpha\beta}$ is of spatially
  compact support, i.e.\ can be enclosed into some time-like world tube,
  say $r\leq a$, where $r=|{\bf x}|$ is the harmonic-coordinate radial
  distance. Outside the domain of the source, when $r> a$, the
  gravitational source term, according to Equation~(\ref{15}), is
  divergence-free,
  \begin{equation}
    \partial_\mu \Lambda^{\alpha\mu} = 0 \qquad (\mathrm{when\ }r>a).
    \label{16}
  \end{equation} 
\item The matter distribution inside the source is
  smooth\epubtkFootnote{$\mathbb{N}$, $\mathbb{Z}$, $\mathbb{R}$, and
  $\mathbb{C}$ are the usual sets of non-negative integers, integers,
  real numbers, and complex numbers; $C^p (\Omega)$ is the set of
  $p$-times continuously differentiable functions on the open domain
  $\Omega$ ($p\leq +\infty$).}: $T^{\alpha\beta}\in C^\infty
  ({\mathbb{R}}^3)$. We have in mind a smooth hydrodynamical ``fluid''
  system, without any singularities nor shocks (\emph{a priori}), that is
  described by some Eulerian equations including high relativistic
  corrections. In particular, we exclude from the start any black holes
  (however we shall return to this question when we find a model for
  describing compact objects).
\item The source is post-Newtonian in the sense of the existence of the
  small parameter defined by Equation~(\ref{1}). For such a source we assume
  the legitimacy of the method of matched asymptotic expansions for
  identifying the inner post-Newtonian field and the outer multipolar
  decomposition in the source's exterior near zone.
\item The gravitational field has been independent of time (stationary)
  in some remote past, i.e.\ before some finite instant $-{\cal T}$ in
  the past, in the sense that
  \begin{equation}
    \frac{\partial}{\partial t}
    \left[h^{\alpha\beta}({\bf x}, t)\right] = 0
    \qquad \mathrm{when\ } t\leq -{\cal T}.
    \label{17}
  \end{equation}
\end{enumerate}
The latter condition is a means to impose, by brute force, the famous
\emph{no-incoming} radiation condition, ensuring that the matter source
is isolated from the rest of the Universe and does not receive any
radiation from infinity. Ideally, the no-incoming radiation condition
should be imposed at past null infinity. We shall later argue (see
Section~\ref{sec:6}) that our condition of stationarity in the
past, Equation~(\ref{17}), although much weaker than the real no-incoming
radiation condition, does not entail any physical restriction on the
general validity of the formulas we derive.

Subject to the condition~(\ref{17}), the Einstein differential field
equations~(\ref{12}) can be written equivalently into the form of the
integro-differential equations
\begin{equation}
  h^{\alpha\beta} = \frac{16\pi G}{c^4} \Box^{-1}_\mathrm{ret}
  \tau^{\alpha\beta},
  \label{18}
\end{equation}
containing the usual retarded inverse d'Alembertian operator, given by
\begin{equation}
  (\Box^{-1}_\mathrm{ret} f)({\bf x}, t) \equiv -\frac{1}{4\pi} \int
  \!\!\! \int \!\!\! \int \frac{d^3{\bf x}'}{|{\bf x}-{\bf x}'|} f ({\bf
  x}', t-|{\bf x}-{\bf x}'|/c),
  \label{19}
\end{equation}
extending over the whole three-dimensional space ${\mathbb{R}}^3$.

\newpage


\section{Linearized Vacuum Equations}
\label{sec:3}

In what follows we solve the field equations~(\ref{11}, \ref{12}), in
the \emph{vacuum} region outside the compact-support source, in the form
of a formal non-linearity or \emph{post-Minkowskian} expansion,
considering the field variable $h^{\alpha\beta}$ as a non-linear metric
perturbation of Minkowski space-time. At the linearized level (or
first-post-Minkowskian approximation), we write:
\begin{equation}
  h^{\alpha\beta}_\mathrm{ext} = G h^{\alpha\beta}_1 + {\cal O}(G^2),
  \label{20}
\end{equation}
where the subscript ``ext'' reminds us that the solution is valid only in
the exterior of the source, and where we have introduced Newton's
constant $G$ as a book-keeping parameter, enabling one to label very
conveniently the successive post-Minkowskian approximations. Since
$h^{\alpha\beta}$ is a dimensionless variable, with our convention the
linear coefficient $h^{\alpha\beta}_1$ in Equation~(\ref{20}) has the
dimension of the inverse of $G$ -- a mass squared in a system of units
where $\hbar=c=1$. In vacuum, the harmonic-coordinate metric coefficient
$h^{\alpha\beta}_1$ satisfies
\begin{eqnarray}
  \Box h_1^{\alpha\beta}&=&0,
  \label{21a}
  \\
  \partial_\mu h_1^{\alpha\mu}&=&0.
  \label{21b}
\end{eqnarray}%
We want to solve those equations by means of an infinite multipolar
series valid outside a time-like world tube containing the source.
Indeed the multipole expansion is the correct method for describing the
physics of the source as seen from its exterior ($r>a$). On the other
hand, the post-Minkowskian series is physically valid in the weak-field
region, which surely includes the exterior of any source, starting at a
sufficiently large distance. For post-Newtonian sources the exterior
weak-field region, where both multipole and post-Minkowskian expansions
are valid, simply coincides with the exterior $r>a$. It is therefore
quite natural, and even, one would say inescapable when considering
general sources, to combine the post-Minkowskian approximation with the
multipole decomposition. This is the original idea of the
``double-expansion'' series of Bonnor~\cite{Bo59}, which combines the
$G$-expansion (or $m$-expansion in his notation) with the $a$-expansion
(equivalent to the multipole expansion, since the $l$th order multipole
moment scales like $a^l$ with the source radius).

The multipolar-post-Minkowskian method will be implemented
systematically, using STF-harmonics to describe the multipole
expansion~\cite{Th80}, and looking for a definite \emph{algorithm} for
the approximation scheme~\cite{BD86}. The solution of the system of
equations~(\ref{21a}, \ref{21b}) takes the form of a series of retarded
multipolar waves\epubtkFootnote{Our notation is the following:
$L=i_1i_2\dots i_l$ denotes a multi-index, made of $l$ (spatial)
indices. Similarly we write for instance $P=j_1\dots j_p$ (in practice,
we generally do not need to consider the carrier letter $i$ or $j$), or
$aL-1=ai_1\dots i_{l-1}$. Always understood in expressions such as
Equation~(\ref{22}) are $l$ summations over the $l$ indices $i_1, \dots, i_l$
ranging from 1 to 3. The derivative operator $\partial_L$ is a
short-hand for $\partial_{i_1}\dots\partial_{i_l}$. The function $K_L$
is \emph{symmetric and trace-free} (STF) with respect to the $l$ indices
composing $L$. This means that for any pair of indices $i_p, i_q\in L$,
we have $K_{\dots i_p\dots i_q\dots}=K_{\dots i_q\dots i_p\dots}$ and
that $\delta_{i_pi_q}K_{\dots i_p\dots i_q\dots}=0$ (see
Ref.~\cite{Th80} and Appendices A and B in Ref.~\cite{BD86} for reviews
about the STF formalism). The STF projection is denoted with a hat, so
$K_L\equiv {\hat K}_L$, or sometimes with carets around the indices,
$K_L\equiv K_{\langle L \rangle}$. In particular, ${\hat n}_L=n_{\langle
L \rangle}$ is the STF projection of the product of unit vectors
$n_L=n_{i_1}\dots n_{i_l}$; an expansion into STF tensors ${\hat
n}_L={\hat n}_L(\theta,\phi)$ is equivalent to the usual expansion in
spherical harmonics $\mathrm{Y}_{lm}=\mathrm{Y}_{lm}(\theta,\phi)$.
Similarly, we denote $x_L=x_{i_1}\dots x_{i_l}=r^l n_{L}$ and ${\hat
x}_L=x_{\langle L \rangle}$. Superscripts like $(p)$ indicate $p$
successive time-derivations.}
\begin{equation}
  h_1^{\alpha\beta}=\sum_{l=0}^{+\infty}
  \partial_L\left(\frac{K_L^{\alpha\beta}(t-r/c)}{r}\right),
  \label{22}
\end{equation}
where $r=|{\bf x}|$, and where the functions $K_L^{\alpha\beta}\equiv
K_{i_1\dots i_l}^{\alpha\beta}$ are smooth functions of the retarded
time $u\equiv t-r/c$ [$K_L(u)\in C^\infty (\mathbb{R})$], which become
constant in the past, when $t\leq -{\cal T}$. It is evident, since a
monopolar wave satisfies $\Box (K_L (u) / r) = 0$ and the d'Alembertian
commutes with the multi-derivative $\partial_L$, that Equation~(\ref{22})
represents the most general solution of the wave equation~(\ref{21a})
(see Section~\ref{sec:2} in Ref.~\cite{BD86} for a proof based on the
Euler--Poisson--Darboux equation). The gauge condition~(\ref{21b}),
however, is not fulfilled in general, and to satisfy it we must
algebraically decompose the set of functions $K^{00}_L$, $K^{0i}_L$,
$K^{ij}_L$ into ten tensors which are STF with respect to all their
indices, including the spatial indices $i$, $ij$. Imposing the
condition~(\ref{21b}) reduces the number of independent tensors to six,
and we find that the solution takes an especially simple ``canonical''
form, parametrized by only two moments, plus some arbitrary linearized
gauge transformation~\cite{Th80, BD86}.

\begin{theorem}
  The most general solution of the linearized field
  equations~(\ref{21a}, \ref{21b}), outside some time-like world tube
  enclosing the source ($r>a$), and stationary in the past (see
  Equation~(\ref{17})), reads
  \begin{equation}
    h^{\alpha\beta}_1 = k^{\alpha\beta}_1 +
    \partial^\alpha\varphi^\beta_1 +
    \partial^\beta\varphi^\alpha_1 -
    \eta^{\alpha\beta}\partial_\mu\varphi^\mu_1.
    \label{23}
  \end{equation}
  The first term depends on two STF-tensorial multipole moments,
  $\mathrm{I}_L(u)$ and $\mathrm{J}_L(u)$, which are arbitrary functions
  of time except for the laws of conservation of the monopole:
  $\mathrm{I} = \mathrm{const}$, and dipoles: $\mathrm{I}_i =
  \mathrm{const}$, $\mathrm{J}_i = \mathrm{const}$. It is given by
  \begin{equation}
    \begin{array}{rcl}
      k^{00}_1 &=& \displaystyle -\frac{4}{c^2}\sum_{l\geq 0}
      \frac{(-)^l}{l !} \partial_L \left( \frac{1}{r} \mathrm{I}_L
      (u)\right),
      \\ [1.5 em]
      k^{0i}_1 &=& \displaystyle
      \frac{4}{c^3}\sum_{l\geq 1} \frac{(-)^l}{l !} \left\{
      \partial_{L-1} \left( \frac{1}{r} \mathrm{I}_{iL-1}^{(1)}
      (u)\right) + \frac{l}{l+1} \varepsilon_{iab} \partial_{aL-1}
      \left( \frac{1}{r} \mathrm{J}_{bL-1} (u)\right)\right\},
      \\ [1.5 em]
      k^{ij}_1 &=& \displaystyle -\frac{4}{c^4}\sum_{l\geq 2}
      \frac{(-)^l}{l !} \left\{ \partial_{L-2} \left( \frac{1}{r}
      \mathrm{I}_{ijL-2}^{(2)} (u)\right) + \frac{2l}{l+1}
      \partial_{aL-2} \left( \frac{1}{r} \varepsilon_{ab(i}
      \mathrm{J}_{j)bL-2}^{(1)} (u)\right)\right\}.
    \end{array}
    \label{24}
  \end{equation}
  The other terms represent a linearized gauge transformation, with
  gauge vector $\varphi^\alpha_1$ of the type~(\ref{22}), and
  parametrized for four other multipole moments, say $\mathrm{W}_L(u)$,
  $\mathrm{X}_L(u)$, $\mathrm{Y}_L(u)$ and $\mathrm{Z}_L(u)$.
  \label{th1}
\end{theorem}

\noindent
The conservation of the lowest-order moments gives the constancy of the
total mass of the source, $\mathrm{M} \equiv \mathrm{I} = \mathrm{const}$,
center-of-mass position\epubtkFootnote{The constancy of the center of
mass $\mathrm{X}_i$ -- rather than a linear variation with time -- results
from our assumption of stationarity before the date $-{\cal T}$. Hence,
$\mathrm{P}_i = 0$.}, $\mathrm{X}_i \equiv \mathrm{I}_i / \mathrm{I} =
\mathrm{const}$, total linear momentum $\mathrm{P}_i \equiv \mathrm{I}_i^{(1)} = 0$,
and total angular momentum, $\mathrm{S}_i \equiv \mathrm{J}_i = \mathrm{const}$.
It is always possible to achieve $\mathrm{X}_i=0$ by translating the origin
of our coordinates to the center of mass. The total mass $\mathrm{M}$ is
the ADM mass of the Hamiltonian formulation of
general relativity. Note that the quantities $\mathrm{M}$, $\mathrm{X}_i$,
$\mathrm{P}_i$ and $\mathrm{S}_i$ include the contributions due to the waves
emitted by the source. They describe the ``initial'' state of the
source, before the emission of gravitational radiation.

The multipole functions $\mathrm{I}_L(u)$ and $\mathrm{J}_L(u)$, which
thoroughly encode the physical properties of the source at the
linearized level (because the other moments $\mathrm{W}_L, \dots,
\mathrm{Z}_L$ parametrize a gauge transformation), will be referred to as the
\emph{mass-type} and \emph{current-type} source multipole moments. Beware,
however, that at this stage the moments are not specified in terms of
the stress-energy tensor $T^{\alpha\beta}$ of the source: the above
theorem follows merely from the algebraic and differential properties of
the vacuum equations outside the source.

For completeness, let us give the components of the gauge-vector
$\varphi^\alpha_1$ entering Equation~(\ref{23}):
\begin{equation}
  \begin{array}{rcl}
    \varphi^0_1 &=& \displaystyle {4\over c^3}\sum_{l\geq 0} {(-)^l\over
    l !} \partial_L \left( {1\over r} \mathrm{W}_L (u)\right),
    \\ [1.5 em]
    \varphi^i_1 &=& \displaystyle -{4\over c^4}\sum_{l\geq 0}
    {(-)^l\over l !} \partial_{iL} \left( {1\over r} \mathrm{X}_L
    (u)\right)
    \\ [1.5 em]
    && \displaystyle -{4\over c^4}\sum_{l\geq 1}
    {(-)^l\over l !} \left\{ \partial_{L-1} \left( {1\over r}
    \mathrm{Y}_{iL-1} (u)\right) + {l\over l+1} \varepsilon_{iab}
    \partial_{aL-1} \left( {1\over r} \mathrm{Z}_{bL-1}
    (u)\right)\right\}.
  \end{array}
  \label{25}
\end{equation}
Because the theory is covariant with respect to non-linear
diffeomorphisms and not merely with respect to linear gauge
transformations, the moments $\mathrm{W}_L, \dots, \mathrm{Z}_L$ do
play a physical role starting at the non-linear level, in the following
sense. If one takes these moments equal to zero and continues the
calculations one ends up with a metric depending on $\mathrm{I}_L$ and
$\mathrm{J}_L$ only, but that metric will not describe the same physical
source as the one constructed from the six moments $\mathrm{I}_L,
\dots, \mathrm{Z}_L$. In other words, the two non-linear metrics
associated with the sets of multipole moments $\{\mathrm{I}_L,
\mathrm{J}_L, 0, \dots, 0\}$ and $\{\mathrm{I}_L, \mathrm{J}_L,
\mathrm{W}_L, \dots, \mathrm{Z}_L\}$ are not isometric. We point
out in Section~\ref{subsec:4.2} below that the full set of moments
$\{\mathrm{I}_L, \mathrm{J}_L, \mathrm{W}_L, \dots,
\mathrm{Z}_L\}$ is in fact physically equivalent to some reduced set
$\{\mathrm{M}_L, \mathrm{S}_L, 0, \dots, 0\}$, but with some moments
$\mathrm{M}_L$, $\mathrm{S}_L$ that differ from $\mathrm{I}_L$,
$\mathrm{J}_L$ by non-linear corrections (see Equation~(\ref{77})). All the
multipole moments $\mathrm{I}_L$, $\mathrm{J}_L$, $\mathrm{W}_L$,
$\mathrm{X}_L$, $\mathrm{Y}_L$, $\mathrm{Z}_L$ will be computed in
Section~\ref{sec:5}.

\newpage


\section{Non-linear Iteration of the Field Equations}
\label{sec:4}

By Theorem~\ref{th1} we know the most general solution of the linearized
equations in the exterior of the source. We then tackle the problem of
the post-Minkowskian iteration of that solution. We consider the full
post-Minkowskian series
\begin{equation}
  h^{\alpha\beta}_\mathrm{ext}=\sum_{n=1}^{+\infty}G^nh^{\alpha\beta}_n,
  \label{26}
\end{equation}
where the first term is composed of the result given by Equations~(\ref{23},
\ref{24}, \ref{25}). In this article, we shall always understand the
infinite sums such as the one in Equation~(\ref{26}) in the sense of
\emph{formal} power series, i.e.\ as an ordered collection of
coefficients, e.g., $\left(h^{\alpha\beta}_n\right)_{n\in
\mathbb{N}}$. We do not attempt to control the mathematical nature of
the series and refer to the mathematical-physics literature for
discussion (in the present context, see Refs.~\cite{CS79, DS90, Rend90,
Rend92, Rend94}).


\subsection{The post-Minkowskian solution}
\label{subsec:4.1}

We insert the ansatz~(\ref{26}) into the vacuum Einstein field
equations~(\ref{11}, \ref{12}), i.e.\ with $\tau^{\alpha\beta} = c^4 /
(16\pi G) \Lambda^{\alpha\beta}$, and we equate term by term the factors
of the successive powers of our book-keeping parameter $G$. We get an
infinite set of equations for each of the $h^{\alpha\beta}_n$'s:
$\forall n \geq 2$,
\begin{eqnarray}
  \Box h_n^{\alpha\beta}&=&\Lambda_n^{\alpha\beta}[h_1, h_2,\dots, h_{n-1}],
  \label{27a}
  \\
  \partial_\mu h_n^{\alpha\mu}&=&0.
  \label{27b}
\end{eqnarray}%
The right-hand side of the wave equation~(\ref{27a}) is obtained from
inserting the previous iterations, up to the order $n-1$, into the
gravitational source term. In more details, the series of
equations~(\ref{27a}) reads
\begin{eqnarray}
  \Box h_2^{\alpha\beta}&=&N^{\alpha\beta}[h_1, h_1],
  \label{28a}
  \\
  \Box h_3^{\alpha\beta}&=&M^{\alpha\beta}[h_1, h_1, h_1]+
  N^{\alpha\beta}[h_1, h_2]+N^{\alpha\beta}[h_2, h_1],
  \label{28b}
  \\
  \Box h_4^{\alpha\beta}&=& L^{\alpha\beta}[h_1, h_1, h_1, h_1]
  \nonumber \\
  && +\,M^{\alpha\beta}[h_1, h_1, h_2]+
  M^{\alpha\beta}[h_1, h_2, h_1]+M^{\alpha\beta}[h_2, h_1, h_1]
  \nonumber \\ && +\,N^{\alpha\beta}[h_2, h_2]+N^{\alpha\beta}[h_1,
  h_3]+ N^{\alpha\beta}[h_3, h_1]
  \nonumber \\
  && \vdots
  \label{28c}
\end{eqnarray}%
The quadratic, cubic and quartic pieces of $\Lambda^{\alpha\beta}$ are
defined by Equation~(\ref{14_1}).

Let us now proceed by induction. Some $n$ being given, we assume that we
succeeded in constructing, from the linearized coefficient $h_1$, the
sequence of post-Minkowskian coefficients $h_2, h_3, \dots,
h_{n-1}$, and from this we want to infer the next coefficient $h_n$.
The right-hand side of Equation~(\ref{27a}), $\Lambda_n^{\alpha\beta}$, is
known by induction hypothesis. Thus the problem is that of solving a
wave equation whose source is given. The point is that this wave
equation, instead of being valid everywhere in ${\mathbb{R}}^3$, is
correct only outside the matter ($r>a$), and it makes no sense to solve
it by means of the usual retarded integral. Technically speaking, the
right-hand side of Equation~(\ref{27a}) is composed of the product of many
multipole expansions, which are singular at the origin of the spatial
coordinates $r=0$, and which make the retarded integral divergent at
that point. This does not mean that there are no solutions to the wave
equation, but simply that the retarded integral does not constitute the
appropriate solution in that context.

What we need is a solution which takes the same structure as the source
term $\Lambda_n^{\alpha\beta}$, i.e.\ is expanded into multipole
contributions, with a singularity at $r=0$, and satisfies the
d'Alembertian equation as soon as $r>0$. Such a particular solution can
be obtained, following the suggestion in Ref.~\cite{BD86}, by means of a
mathematical trick in which one first ``regularizes'' the source term
$\Lambda_n^{\alpha\beta}$ by multiplying it by the factor $r^B$, where
$B\in \mathbb{C}$. Let us assume, for definiteness, that
$\Lambda^{\alpha\beta}_n$ is composed of multipolar pieces with maximal
multipolarity $l_\mathrm{max}$. This means that we start the iteration
from the linearized metric~(\ref{23}, \ref{24}, \ref{25}) in which the
multipolar sums are actually finite\epubtkFootnote{This assumption is
justified because we are ultimately interested in the radiation field at
some given \emph{finite} post-Newtonian precision like 3PN, and because
only a finite number of multipole moments can contribute at any finite
order of approximation. With a finite number of multipoles in the
linearized metric~(\ref{23}, \ref{24}, \ref{25}), there is a maximal
multipolarity $l_\mathrm{max}(n)$ at any post-Minkowskian order $n$,
which grows linearly with $n$.}. The divergences when $r\to 0$ of the
source term are typically power-like, say $1/r^k$ (there are also powers
of the logarithm of $r$), and with the previous assumption there will
exist a maximal order of divergency, say $k_\mathrm{max}$. Thus, when
the real part of $B$ is large enough, i.e.\ $\Re \, (B) >
k_\mathrm{max}-3$, the ``regularized'' source term $r^B
\Lambda^{\alpha\beta}_n$ is regular enough when $r\to 0$ so that one can
perfectly apply the retarded integral operator. This defines the
$B$-dependent retarded integral
\begin{equation}
  I^{\alpha\beta}(B) \equiv \Box^{-1}_\mathrm{ret} \left[ {\widetilde
  r}^B \Lambda^{\alpha\beta}_n \right],
  \label{29}
\end{equation}
where the symbol $\Box^{-1}_\mathrm{ret}$ stands for the retarded
integral~(\ref{19}). It is convenient to introduce inside the
regularizing factor some arbitrary constant length scale $r_0$ in order
to make it dimensionless. Everywhere in this article we pose
\begin{equation}
  {\widetilde r}\equiv \frac{r}{r_0}.
  \label{29_1}
\end{equation}
The fate of the constant $r_0$ in a detailed calculation will be
interesting to follow, as we shall see, because it provides some check
that the calculation is going well. Now the point for our purpose is
that the function $I^{\alpha\beta}(B)$ on the complex plane, which was
originally defined only when $\Re \, (B) > k_\mathrm{max}-3$, admits a
unique \emph{analytic continuation} to all values of $B\in \mathbb{C}$
except at some integer values. Furthermore, the analytic continuation of
$I^{\alpha\beta}(B)$ can be expanded, when $B\to 0$ (namely the limit of
interest to us) into a Laurent expansion involving in general some
multiple poles. The key idea, as we shall prove, is that the \emph{finite
part}, or the coefficient of the zeroth power of $B$ in that expansion,
represents the particular solution we are looking for. We write the
Laurent expansion of $I^{\alpha\beta}(B)$, when $B\to 0$, in the form
\begin{equation}
  I^{\alpha\beta}(B) = \sum_{p=p_0}^{+\infty} \iota^{\alpha\beta}_p B^p,
  \label{30}
\end{equation}
where $p\in \mathbb{Z}$, and the various coefficients
$\iota^{\alpha\beta}_p$ are functions of the field point $({\bf x}, t)$.
When $p_0\leq -1$ there are poles; $-p_0$, which depends on $n$, refers
to the maximal order of the poles. By applying the box operator onto
both sides of Equation~(\ref{30}), and equating the different powers of $B$, we
arrive at
\begin{equation}
  \begin{array}{rcl}
    p_0\leq p\leq -1 \quad \Longrightarrow \quad \Box
    \iota^{\alpha\beta}_p &=& 0,
    \\ [0.5 em]
    p\geq 0 \quad \Longrightarrow \quad \Box \iota^{\alpha\beta}_p &=&
    \displaystyle \frac{(\ln r)^p}{p!}\Lambda^{\alpha\beta}_n.
  \end{array}
\end{equation}
As we see, the case $p=0$ shows that the finite-part coefficient in
Equation~(\ref{30}), namely $\iota^{\alpha\beta}_0$, is a particular solution
of the requested equation: $\Box
\iota^{\alpha\beta}_0=\Lambda^{\alpha\beta}_n$. Furthermore, we can
prove that this term, by its very construction, owns the same structure
made of a multipolar expansion singular at $r=0$.

Let us forget about the intermediate name $\iota^{\alpha\beta}_0$, and
denote, from now on, the latter solution by $u^{\alpha\beta}_n\equiv
\iota^{\alpha\beta}_0$, or, in more explicit terms,
\begin{equation}
  u^{\alpha\beta}_n = {\cal FP}_{B=0}\, \Box^{-1}_\mathrm{ret}
  \left[ {\widetilde r}^B \Lambda^{\alpha\beta}_n \right],
  \label{32}
\end{equation}
where the finite-part symbol ${\cal FP}_{B=0}$ means the previously
detailed operations of considering the analytic continuation, taking the
Laurent expansion, and picking up the finite-part coefficient when $B\to
0$. The story is not complete, however, because $u^{\alpha\beta}_n$ does
not fulfill the constraint of harmonic coordinates~(\ref{27b}); its
divergence, say $w_n^\alpha=\partial_\mu u_n^{\alpha\mu}$, is different
from zero in general. From the fact that the source term is
divergence-free in vacuum, $\partial_\mu \Lambda_n^{\alpha\mu}=0$ (see
Equation~(\ref{16})), we find instead
\begin{equation}
  w^\alpha_n = {\cal FP}_{B=0}\, \Box^{-1}_\mathrm{ret} \left[ B \,
  {\widetilde r}^B \frac{n_i}{r} \Lambda^{\alpha i}_n \right].
  \label{33}
\end{equation}
The factor $B$ comes from the differentiation of the regularization
factor ${\widetilde r}^B$. So, $w^\alpha_n$ is zero only in the special
case where the Laurent expansion of the retarded integral in
Equation~(\ref{33}) does not develop any simple pole when $B\to 0$.
Fortunately, when it does, the structure of the pole is quite easy to
control. We find that it necessarily consists of a solution of the
\emph{source-free} d'Alembertian equation, and, what is more (from its
stationarity in the past), the solution is a retarded one. Hence, taking
into account the index structure of $w^\alpha_n$, there must exist four
STF-tensorial functions of the retarded time $u=t-r/c$, say $N_L(u)$,
$P_L(u)$, $Q_L(u)$ and $R_L(u)$, such that
\begin{equation}
  \begin{array}{rcl}
    w^0_n &=& \displaystyle \sum_{l = 0}^{+\infty}\partial_L
    \left[r^{-1} N_L(u)\right],
    \\ [1.5 em]
    w^i_n &=& \displaystyle \sum_{l = 0}^{+\infty}\partial_{iL}
    \left[ r^{-1} P_L (u) \right] + \sum_{l = 1}^{+\infty} \left\{
    \partial_{L-1} \left[ r^{-1} Q_{iL-1} (u) \right] +
    \varepsilon_{iab} \partial_{aL-1} \left[r^{-1}
    R_{bL-1} (u) \right] \right\}.
  \end{array}
  \label{34}
\end{equation}
From that expression we are able to find a new object, say
$v_n^{\alpha\beta}$, which takes the same structure as $w^\alpha_n$ (a
retarded solution of the source-free wave equation) and, furthermore,
whose divergence is exactly the opposite of the divergence of
$u_n^{\alpha\beta}$, i.e.\ $\partial_\mu v_n^{\alpha\mu}=-w_n^\alpha$.
Such a $v_n^{\alpha\beta}$ is not unique, but we shall see that it is
simply necessary to make a choice for $v_n^{\alpha\beta}$ (the simplest
one) in order to obtain the general solution. The formulas that we adopt
are
\begin{equation}
  \begin{array}{rcl}
    v^{00}_n &=& \displaystyle - r^{-1} N^{(-1)} + \partial_a \left[
    r^{-1} \left(- N^{(-1)}_a+ C^{(-2)}_a -3P_a\right) \right],
    \\ [1.5 em]
    v^{0i}_n &=& \displaystyle r^{-1} \left( - Q^{(-1)}_i +3
    P^{(1)}_i\right) - \varepsilon_{iab} \partial_a \left[ r^{-1}
    R^{(-1)}_b \right] - \sum_{l = 2}^{+\infty}\partial_{L-1} \left[
    r^{-1} N_{iL-1} \right],
    \\ [1.5 em]
    v^{ij}_n &=& \displaystyle -
    \delta_{ij} r^{-1} P + \sum_{l = 2}^{+\infty} \biggl\{ 2
    \delta_{ij}\partial_{L-1} \left[ r^{-1} P_{L-1}\right] - 6
    \partial_{L-2(i} \left[ r^{-1} P_{j)L-2}\right]
    \\ [1.5 em]
    && \qquad \qquad \qquad \quad ~ \displaystyle + \partial_{L-2}
    \left[ r^{-1} (N^{(1)}_{ijL-2} + 3 P^{(2)}_{ijL-2} - Q_{ijL-2})
    \right] - 2 \partial_{aL-2}\left[ r^{-1} \varepsilon_{ab(i}
    R_{j)bL-2} \right] \biggr\}. \hspace{-2.5 em}
  \end{array}
  \label{35}
\end{equation}
Notice the presence of anti-derivatives, denoted, e.g., by
$N^{(-1)}(u)=\int_{-\infty}^u dv N(v)$; there is no problem with the
limit $v\to -\infty$ since all the corresponding functions are zero when
$t\leq -{\cal T}$. The choice made in Equations~(\ref{35}) is dictated by the
fact that the $00$ component involves only some monopolar and dipolar
terms, and that the spatial trace $ii$ is monopolar: $v^{ii}_n = -3
r^{-1} P$. Finally, if we pose
\begin{equation}
  h_n^{\alpha\beta}=u_n^{\alpha\beta}+v_n^{\alpha\beta},
  \label{36}
\end{equation}
we see that we solve at once the d'Alembertian equation~(\ref{27a})
\emph{and} the coordinate condition~(\ref{27b}). That is, we have succeeded in
finding a solution of the field equations at the $n$th post-Minkowskian
order. By induction the same method applies to \emph{any} order $n$, and,
therefore, we have constructed a complete post-Minkowskian
series~(\ref{26}) based on the linearized approximation
$h^{\alpha\beta}_1$ given by Equations~(\ref{23}, \ref{24}, \ref{25}). The
previous procedure constitutes an \emph{algorithm}, which could be
implemented by an algebraic computer programme.


\subsection{Generality of the solution}
\label{subsec:4.2}

We have a solution, but is that a general solution? The answer, yes, is
provided by the following result~\cite{BD86}:

\begin{theorem}
  The most general solution of the harmonic-coordinates Einstein field
  equations in the vacuum region outside an isolated source, admitting
  some post-Minkowskian and multipolar expansions, is given by the
  previous construction as $h^{\alpha\beta}=\sum_{n=1}^{+\infty}
  G^nh_n^{\alpha\beta}[\mathrm{I}_L,\mathrm{J}_L,\dots,\mathrm{Z}_L]$.
  It depends on two sets of arbitrary STF-tensorial functions of time
  $\mathrm{I}_L(u)$ and $\mathrm{J}_L(u)$ (satisfying the conservation
  laws) defined by Equations~(\ref{24}), and on four supplementary functions
  $\mathrm{W}_L(u), \dots, \mathrm{Z}_L(u)$ parametrizing the gauge
  vector~(\ref{25}).
  \label{th2}
\end{theorem}

\noindent 
The proof is quite easy. With Equation~(\ref{36}) we obtained a
\emph{particular} solution of the system of equations~(\ref{27a}, \ref{27b}).
To it we should add the most general solution of the corresponding
\emph{homogeneous} system of equations, which is obtained by setting
$\Lambda^{\alpha\beta}_n=0$ into Equations~(\ref{27a}, \ref{27b}). But this
homogeneous system of equations is nothing but the \emph{linearized}
vacuum field equations~(\ref{21a}, \ref{21b}), for which we know the
most general solution $h_1^{\alpha\beta}$ given by Equations~(\ref{23}, \ref{24},
\ref{25}). Thus, we must add to our ``particular'' solution
$h^{\alpha\beta}_n$ a general homogeneous solution that is necessarily
of the type $h_1^{\alpha\beta}[\delta \mathrm{I}_L,\dots,\delta
\mathrm{Z}_L]$, where $\delta \mathrm{I}_L, \dots, \delta
\mathrm{Z}_L$ denote some ``corrections'' to the multipole moments at
the $n$th post-Minkowskian order. It is then clear, since precisely the
linearized metric is a linear functional of all these moments, that the
previous corrections to the moments can be absorbed into a re-definition
of the original ones $\mathrm{I}_L,\dots,\mathrm{Z}_L$ by posing
\begin{eqnarray}
  \mathrm{I}_L^\mathrm{new}&=&\mathrm{I}_L+G^{n-1}\delta \mathrm{I}_L,
  \\
  &\vdots&
  \nonumber \\
  \mathrm{Z}_L^\mathrm{new}&=&\mathrm{Z}_L+G^{n-1}\delta \mathrm{Z}_L.
\end{eqnarray}%
After re-arranging the metric in terms of these new moments, taking into
account the fact that the precision of the metric is limited to the
$n$th post-Minkowskian order, and dropping the superscript ``new'', we
find exactly the same solution as the one we had before (indeed, the
moments are arbitrary functions of time) -- hence the proof.

The six sets of multipole moments $\mathrm{I}_L(u), \dots,
\mathrm{Z}_L(u)$ contain the physical information about \emph{any}
isolated source as seen in its exterior. However, as we now discuss, it
is always possible to find \emph{two}, and only two, sets of multipole
moments, $\mathrm{M}_L(u)$ and $\mathrm{S}_L(u)$, for parametrizing the
most general isolated source as well. The route for constructing such a
general solution is to get rid of the moments
$\mathrm{W}_L,\mathrm{X}_L, \mathrm{Y}_L,\mathrm{Z}_L$ at the linearized
level by performing the linearized gauge transformation $\delta x^\alpha
= \varphi^\alpha_1$, where $\varphi^\alpha_1$ is the gauge vector given
by Equations~(\ref{25}). So, at the linearized level, we have only the two
types of moments $\mathrm{M}_L$ and $\mathrm{S}_L$, parametrizing
$k^{\alpha\beta}_1$ by the same formulas as in Equations~(\ref{24}). We must
be careful to denote these moments with some names different from
$\mathrm{I}_L$ and $\mathrm{J}_L$ because they will ultimately
correspond to a different physical source. Then we apply exactly the
same post-Minkowskian algorithm, following the formulas~(\ref{32},
\ref{33}, \ref{34}, \ref{35}, \ref{36}) as we did above, but starting
from the gauge-transformed linear metric $k^{\alpha\beta}_1$ instead of
$h^{\alpha\beta}_1$. The result of the iteration is therefore some
$k^{\alpha\beta}=\sum_{n=1}^{+\infty}G^nk^{\alpha\beta}_n[\mathrm{M}_L,\mathrm{S}_L]$.
Obviously this post-Minkowskian algorithm yields some simpler
calculations as we have only two multipole moments to iterate. The point
is that one can show that the resulting metric
$k^{\alpha\beta}[\mathrm{M}_L,\mathrm{S}_L]$ is \emph{isometric} to the
original one
$h^{\alpha\beta}[\mathrm{I}_L,\mathrm{J}_L,\dots,\mathrm{Z}_L]$ if and
only if $\mathrm{M}_L$ and $\mathrm{S}_L$ are related to the moments
$\mathrm{I}_L,\mathrm{J}_L,\dots,\mathrm{Z}_L$ by some (quite involved)
non-linear equations. Therefore, the most general solution of the field
equations, modulo a coordinate transformation, can be obtained by
starting from the linearized metric
$k^{\alpha\beta}_1[\mathrm{M}_L,\mathrm{S}_L]$ instead of the more
complicated $k^{\alpha\beta}_1[\mathrm{I}_L,\mathrm{J}_L]+
\partial^\alpha\varphi^\beta_1 + \partial^\beta\varphi^\alpha_1 -
\eta^{\alpha\beta}\partial_\mu\varphi^\mu_1$, and continuing the
post-Minkowskian calculation.

So why not consider from the start that the best description of the
isolated source is provided by only the two types of multipole moments,
$\mathrm{M}_L$ and $\mathrm{S}_L$, instead of the six, $\mathrm{I}_L,
\mathrm{J}_L, \dots, \mathrm{Z}_L$? The reason is that we shall
determine (in Theorem~\ref{th6} below) the explicit closed-form
expressions of the six moments $\mathrm{I}_L, \mathrm{J}_L, \dots,
\mathrm{Z}_L$, but that, by contrast, it seems to be impossible to
obtain some similar closed-form expressions for $\mathrm{M}_L$ and
$\mathrm{S}_L$. The only thing we can do is to write down the explicit
non-linear algorithm that computes $\mathrm{M}_L$, $\mathrm{S}_L$
starting from $\mathrm{I}_L, \mathrm{J}_L, \dots, \mathrm{Z}_L$.
In consequence, it is better to view the moments $\mathrm{I}_L,
\mathrm{J}_L, \dots, \mathrm{Z}_L$ as more ``fundamental'' than
$\mathrm{M}_L$ and $\mathrm{S}_L$, in the sense that they appear to be
more tightly related to the description of the source, since they admit
closed-form expressions as some explicit integrals over the source.
Hence, we choose to refer collectively to the six moments
$\mathrm{I}_L, \mathrm{J}_L, \dots, \mathrm{Z}_L$ as \emph{the}
multipole moments of the source. This being said, the moments
$\mathrm{M}_L$ and $\mathrm{S}_L$ are often useful in practical
computations because they yield a simpler post-Minkowskian iteration.
Then, one can generally come back to the more fundamental source-rooted
moments by using the fact that $\mathrm{M}_L$ and $\mathrm{S}_L$ differ
from the corresponding $\mathrm{I}_L$ and $\mathrm{J}_L$ only by
high-order post-Newtonian terms like 2.5PN; see Ref.~\cite{B96} and
Equation~(\ref{77}) below. Indeed, this is to be expected because the
physical difference between both types of moments stems only from
non-linearities.


\subsection{Near-zone and far-zone structures}
\label{subsec:4.3}

In our presentation of the post-Minkowskian algorithm~(\ref{32},
\ref{33}, \ref{34}, \ref{35}, \ref{36}) we have omitted a crucial
recursive hypothesis, which is required in order to prove that at each
post-Minkowskian order $n$, the inverse d'Alembertian operator can be
applied in the way we did (and notably that the $B$-dependent retarded
integral can be analytically continued down to a neighbourhood of
$B=0$). This hypothesis is that the ``near-zone'' expansion, i.e.\
when $r\to 0$, of each one of the post-Minkowskian coefficients
$h^{\alpha\beta}_n$ has a certain structure. This hypothesis is
established as a theorem once the mathematical induction succeeds.

\begin{theorem}
  The general structure of the expansion of the post-Minkowskian
  exterior metric in the near-zone (when $r\to 0$) is of the type:
  $\forall N \in \mathbb{N}$,
  \begin{equation}
    h_n({\bf x}, t) =
    \sum {\hat n}_L r^m (\ln r)^p F_{L, m, p, n}(t)+o(r^{N}),
    \label{38}
  \end{equation}
  where $m\in \mathbb{Z}$, with $m_0 \leq m \leq N$ (and $m_0$ becoming
  more and more negative as $n$ grows), $p \in \mathbb{N}$ with $p \leq
  n-1$. The functions $F_{L, m, p, n}$ are multilinear functionals of
  the source multipole moments $\mathrm{I}_L, \dots ,\mathrm{Z}_L$.
  \label{th3}
\end{theorem}

\noindent 
For the proof see Ref.~\cite{BD86}\epubtkFootnote{The $o$ and ${\cal O}$
Landau symbols for remainders have their standard meaning.}. As we see,
the near-zone expansion involves, besides the simple powers of $r$, some
powers of the logarithm of $r$, with a maximal power of $n-1$. As a
corollary of that theorem, we find (by restoring all the powers of $c$
in Equation~(\ref{38}) and using the fact that each $r$ goes into the
combination $r/c$), that the general structure of the post-Newtonian
expansion ($c\to +\infty$) is necessarily of the type
\begin{equation}
  h_n(c) \simeq \!\! \sum_{p, q \in \mathbb{N}}\frac{(\ln c)^p}{c^q},
  \label{39}
\end{equation}
where $p\leq n-1$ (and $q\geq 2$). The post-Newtonian expansion proceeds
not only with the normal powers of $1/c$ but also with powers of the
logarithm of $c$~\cite{BD86}.
 
Paralleling the structure of the near-zone expansion, we have a similar
result concerning the structure of the \emph{far-zone} expansion at
Minkowskian future null infinity, i.e.\ when $r\to +\infty$ with
$u=t-r/c= \mathrm{const}$: $\forall N \in \mathbb{N}$,
\begin{equation}
  h_n({\bf x}, t) = \sum \frac{{\hat n}_L (\ln r)^p}{r^k}
  G_{L, k, p, n}(u)+o\left(\frac{1}{r^N}\right),
  \label{40}
\end{equation}
where $k, p\in \mathbb{N}$, with $1\leq k\leq N$, and where, likewise in
the near-zone expansion~(\ref{38}), some powers of logarithms, such that
$p \leq n-1$, appear. The appearance of logarithms in the far-zone
expansion of the harmonic-coordinates metric has been known since the
work of Fock~\cite{Fock}. One knows also that this is a coordinate
effect, because the study of the ``asymptotic'' structure of space-time
at future null infinity by Bondi et al.~\cite{BBM62},
Sachs~\cite{Sachs62}, and Penrose~\cite{P63, P65}, has revealed the
existence of other coordinate systems that avoid the appearance of any
logarithms: the so-called \emph{radiative} coordinates, in which the
far-zone expansion of the metric proceeds with simple powers of the
inverse radial distance. Hence, the logarithms are simply an artifact of
the use of harmonic coordinates~\cite{IW68, Madore}. The following
theorem, proved in Ref.~\cite{B87}, shows that our general construction
of the metric in the exterior of the source, when developed at future
null infinity, is consistent with the Bondi--Sachs--Penrose~\cite{BBM62,
Sachs62, P63, P65} approach to gravitational radiation.

\begin{theorem}
  The most general multipolar-post-Minkowskian solution, stationary in
  the past (see Equation~(\ref{17})), admits some radiative coordinates
  $(T,{\bf X})$, for which the expansion at future null infinity,
  $R\to+\infty$ with $U\equiv T-R/c= \mathrm{const}$, takes the form
  \begin{equation}
    H_n({\bf X},T) = \sum \frac{{\hat N}_L}{R^k} K_{L, k, n}(U)+{\cal
    O}\left(\frac{1}{R^N}\right).
    \label{41}
  \end{equation} 
  The functions $K_{L, k, n}$ are computable functionals of the source
  multipole moments. In radiative coordinates the retarded time
  $U=T-R/c$ is a null coordinate in the asymptotic limit. The metric
  $H_\mathrm{ext}^{\alpha\beta}=\sum_{n\geq 1}G^nH^{\alpha\beta}_n$ is
  asymptotically simple in the sense of Penrose~\cite{P63, P65},
  perturbatively to any post-Minkowskian order.
  \label{th4}
\end{theorem}

\noindent 
\emph{Proof}: We introduce a linearized ``radiative'' metric by
performing a gauge transformation of the harmonic-coordinates metric
defined by Equations~(\ref{23}, \ref{24}, \ref{25}), namely
\begin{equation}
  H^{\alpha\beta}_1 = h^{\alpha\beta}_1 + \partial^\alpha\xi^\beta_1 +
  \partial^\beta\xi^\alpha_1 - \eta^{\alpha\beta} \partial_\mu \xi^\mu_1,
  \label{42}
\end{equation}
where the gauge vector $\xi^\alpha_1$ is
\begin{equation}
  \xi^\alpha_1 = 2 \mathrm{M} \, \eta^{0\alpha}\ln \left(\frac{r}{r_0}\right).
  \label{43}
\end{equation}
This gauge transformation is non-harmonic:
\begin{equation}
  \partial_\mu H^{\alpha\mu}_1 = \Box \xi^\alpha_1 =
  \frac{2\mathrm{M}}{r^2}\eta^{0\alpha}.
\end{equation}
Its effect is to ``correct'' for the well-known logarithmic deviation of
the harmonic coordinates' retarded time with respect to the true
space-time characteristic or light cones. After the change of gauge, the
coordinate $u=t-r/c$ coincides with a null coordinate at the linearized
level\epubtkFootnote{In this proof the coordinates are considered as
dummy variables denoted $(t, r)$. At the end, when we obtain the
radiative metric, we shall denote the associated radiative coordinates
by $(T, R)$.}. This is the requirement to be satisfied by a linearized
metric so that it can constitute the linearized approximation to a full
(post-Minkowskian) radiative field~\cite{Madore}. One can easily show
that, at the dominant order when $r\to +\infty$,
\begin{equation}
  k_\mu k_\nu H^{\mu\nu}_1 = {\cal O}\left(\frac{1}{r^2}\right),
  \label{44}
\end{equation}
where $k^\alpha = \left( 1, {\bf n} \right)$ is the outgoing Minkowskian
null vector. Given any $n\geq 2$, let us recursively assume that we have
obtained all the previous radiative post-Minkowskian coefficients
$H^{\alpha\beta}_m$, i.e.\ $\forall m\leq n-1$, and that all of them
satisfy
\begin{equation}
  k_\mu k_\nu H^{\mu\nu}_m = {\cal O}\left(\frac{1}{r^2}\right).
  \label{45}
\end{equation}
From this induction hypothesis one can prove that the $n$th
post-Minkowskian source term
$\Lambda^{\alpha\beta}_n=\Lambda^{\alpha\beta}_n(H_1,\dots, H_{n-1})$ is
such that
\begin{equation}
  \Lambda^{\alpha\beta}_n =
  \frac{k^\alpha k^\beta}{r^2}\sigma_n\left(u,{\bf n}\right) +
  {\cal O}\left(\frac{1}{r^3}\right).
  \label{46}
\end{equation}
To the leading order this term takes the classic form of the
stress-energy tensor for a swarm of massless particles, with $\sigma_n$
being related to the power in the waves. One can show that all the
problems with the appearance of logarithms come from the retarded
integral of the terms in Equation~(\ref{46}) that behave like $1/r^2$: See
indeed the integration formula~(\ref{87}), which behaves like $\ln r/r$
at infinity. But now, thanks to the particular index structure of the
term~(\ref{46}), we can correct for the effect by adjusting the gauge at
the $n$th post-Minkowskian order. We pose, as a gauge vector,
\begin{equation}
  \xi^\alpha_n = {\cal FP}\, \Box^{-1}_\mathrm{ret}
  \left[\frac{k^\alpha}{2r^2} \int_{-\infty}^u \!\!\!\!\! dv \,
  \sigma_n (v,{\bf n})\right],
  \label{47}
\end{equation}
where ${\cal FP}$ refers to the same finite part operation as in
Equation~(\ref{32}). This vector is such that the logarithms that will appear
in the corresponding gauge terms \emph{cancel out} the logarithms coming
from the retarded integral of the source term~(\ref{46}); see
Ref.~\cite{B87} for the details. Hence, to the $n$th post-Minkowskian
order, we define the radiative metric as
\begin{equation}
  H_n^{\alpha\beta}=U_n^{\alpha\beta}+V_n^{\alpha\beta}+
  \partial^\alpha\xi_n^\beta+\partial^\beta\xi_n^\alpha-
  \eta^{\alpha\beta}\partial_\mu\xi_n^\mu.
  \label{48}
\end{equation}
Here $U_n^{\alpha\beta}$ and $V_n^{\alpha\beta}$ denote the quantities
that are the analogues of $u_n^{\alpha\beta}$ and $v_n^{\alpha\beta}$,
which were introduced into the harmonic-coordinates algorithm: See
Equations~(\ref{32}, \ref{33}, \ref{34}, \ref{35}). In particular, these
quantities are constructed in such a way that the sum
$U_n^{\alpha\beta}+V_n^{\alpha\beta}$ is divergence-free, so we see that
the radiative metric does not obey the harmonic-gauge condition:
\begin{equation}
  \partial_\mu H_n^{\alpha\mu}=\Box \xi^\alpha_n = 
  \frac{k^\alpha}{2 r^2} \int_{-\infty}^u \!\!\!\!\! dv \, \sigma_n (v,{\bf n}).
  \label{49}
\end{equation}
The far-zone expansion of the latter metric is of the type~(\ref{41}),
i.e.\ is free of any logarithms, and the retarded time in these
coordinates tends asymptotically toward a null coordinate at infinity.
The property of asymptotic simplicity, in the mathematical form given by
Geroch and Horowitz~\cite{GH78}, is proved by introducing the conformal
factor $\Omega=1/r$ in radiative coordinates (see Ref.~\cite{B87}).
Finally, it can be checked that the metric so constructed, which is a
functional of the source multipole moments $\mathrm{I}_L, \dots,
\mathrm{Z}_L$ (from the definition of the algorithm), is as general as
the general harmonic-coordinate metric of Theorem~\ref{th2}, since it
merely differs from it by a coordinate transformation $(t,{\bf
x})\longrightarrow (T,{\bf X})$, where $(t,{\bf x})$ are the harmonic
coordinates and $(T,{\bf X})$ the radiative ones, together with a
re-definition of the multipole moments.


\subsection{The radiative multipole moments}
\label{subsec:4.4}

The leading-order term $1/R$ of the metric in radiative coordinates,
neglecting ${\cal O}(1/R^2)$, yields the operational definition of two
sets of STF \emph{radiative} multipole moments, mass-type
$\mathrm{U}_L(U)$ and current-type $\mathrm{V}_L(U)$. \emph{By
definition}, we have
\begin{eqnarray}
  H^\mathrm{TT}_{ij} (U,{\bf N}) &=& \frac{4G}{c^2R} {\cal P}_{ijab}
  ({\bf N}) \sum^{+\infty}_{l=2}\frac{1}{c^l l !} \left\{ N_{L-2}
  \mathrm{U}_{abL-2}(U) - \frac{2l}{c(l+1)} N_{cL-2} \varepsilon_{cd(a}
  \mathrm{V}_{b)dL-2}(U) \right\}
  \nonumber \\
  & & + \, {\cal O}\left(\frac{1}{R^2}\right).
  \label{50}
\end{eqnarray}%
This multipole decomposition represents the generalization, up to any
post-Newtonian order (witness the factors of $1/c$ in front of each of
the multipolar pieces) of the quadrupole-moment formalism reviewed in
Equation~(\ref{2}). The corresponding total gravitational flux reads
\begin{eqnarray}
  {\cal L} (U) &=& \sum^{+\infty}_{l=2} \frac{G}{c^{2l+1}}
  \biggl\{\! \frac{(l+1)(l+2)}{(l-1)l l!(2l+1)!!} \mathrm{U}^{(1)}_L(U)
  \mathrm{U}^{(1)}_L(U) + \frac{4l (l+2)}{c^2(l-1)(l+1)!(2l+1)!!}
  \mathrm{V}^{(1)}_L(U) \mathrm{V}^{(1)}_L(U) \!\biggr\}.
  \nonumber \\
  \label{51}
\end{eqnarray}%
Notice that the meaning of such formulas is rather empty, because we do
not know yet how the radiative moments are given in terms of the actual
source parameters. Only at the Newtonian level do we know this relation,
which from the comparison with the quadrupole formalism of
Equations~(\ref{2}, \ref{3}, \ref{4}) reduces to
\begin{equation}
  \mathrm{U}_{ij}(U) = \mathrm{Q}_{ij}^{(2)}(U) + {\cal
  O}\left(\frac{1}{c^2}\right),
  \label{52}
\end{equation}
where $\mathrm{Q}_{ij}$ is the Newtonian quadrupole given by
Equation~(\ref{3}). Fortunately, we are not in such bad shape because we have
learned from Theorem~\ref{th4} the general method that permits us to
compute the radiative multipole moments $\mathrm{U}_L$, $\mathrm{V}_L$
in terms of the source moments $\mathrm{I}_L, \mathrm{J}_L, \dots,
\mathrm{Z}_L$. Therefore, what is missing is the explicit dependence of
the \emph{source} multipole moments as functions of the actual parameters
of some isolated source. We come to grips with this question in the next
Section~\ref{sec:5}.

\newpage


\section{Exterior Field of a Post-Newtonian Source}
\label{sec:5}

By Theorem~\ref{th2} we control the most general class of solutions of
the vacuum equations outside the source, in the form of non-linear
functionals of the source multipole moments. For instance, these
solutions include the Schwarzschild and Kerr solutions, as well as all
their perturbations. By Theorem~\ref{th4} we learned how to construct
the radiative moments at infinity. We now want to understand how a
specific choice of stress-energy tensor $T^{\alpha\beta}$ (i.e.\ a
choice of some physical model describing the source) selects a
particular physical exterior solution among our general class.


\subsection{The matching equation}
\label{subsec:5.1}

We shall provide the answer in the case of a post-Newtonian source for
which the post-Newtonian parameter $1/c$ defined by Equation~(\ref{1}) is
small. The fundamental fact that permits the connection of the exterior
field to the inner field of the source is the existence of a
``matching'' region, in which both the multipole and the post-Newtonian
expansions are valid. This region is nothing but the exterior near zone,
such that $r>a$ (exterior) \emph{and} $r\ll\lambda$ (near zone). It
always exists around post-Newtonian sources.

Let us denote by ${\cal M}(h)$ the multipole expansion of $h$ (for
simplicity, we suppress the space-time indices). By ${\cal M}(h)$ we
really mean the multipolar-post-Minkowskian exterior metric that we have
constructed in Sections~\ref{sec:3} and~\ref{sec:4}:
\begin{equation}
  {\cal M}(h)  \equiv h_\mathrm{ext} = \sum_{n=1}^{+\infty} G^n
  h_n[\mathrm{I}_L,\dots,\mathrm{Z}_L].
  \label{53}
\end{equation}
Of course, $h$ agrees with its own multipole expansion in the exterior
of the source,
\begin{equation}
  r > a \quad\Longrightarrow\quad {\cal M}(h) = h.
  \label{54}
\end{equation}
By contrast, inside the source, $h$ and ${\cal M}(h)$ disagree with each
other because $h$ is a fully-fledged solution of the field equations
with matter source, while ${\cal M}(h)$ is a vacuum solution becoming
singular at $r=0$. Now let us denote by $\overline h$ the post-Newtonian
expansion of $h$. We have already anticipated the general structure of
this expansion as given in Equation~(\ref{39}). In the matching region, where
both the multipolar and post-Newtonian expansions are valid, we write
the numerical equality
\begin{equation}
  a < r \ll\lambda \quad\Longrightarrow\quad {\cal M}(h) =
  \overline h.
  \label{55}
\end{equation}
This ``numerical'' equality is viewed here in a sense of formal
expansions, as we do not control the convergence of the series. In fact,
we should be aware that such an equality, though quite natural and even
physically obvious, is probably not really justified within the
approximation scheme (mathematically speaking), and we take it as part
of our fundamental assumptions.

We now transform Equation~(\ref{55}) into a \emph{matching equation}, by
replacing in the left-hand side ${\cal M}(h)$ by its near-zone
re-expansion $\overline{{\cal M}(h)}$, and in the right-hand side
$\overline h$ by its multipole expansion ${\cal M}(\overline h)$. The
structure of the near-zone expansion ($r\to 0$) of the exterior
multipolar field has been found in Equation~(\ref{38}). We denote the
corresponding infinite series $\overline{{\cal M}(h)}$ with the same
overbar as for the post-Newtonian expansion because it is really an
expansion when $r/c\to 0$, equivalent to an expansion when $c\to\infty$.
Concerning the multipole expansion of the post-Newtonian metric, ${\cal
M}(\overline h)$, we simply postulate its existence. Therefore, the
matching equation is the statement that
\begin{equation}
  \overline{{\cal M}(h)} = {\cal M}(\overline h),
  \label{56}
\end{equation}
by which we really mean an infinite set of \emph{functional} identities,
valid $\forall ({\bf x}, t) \in {\mathbb{R}}^3_* \times \mathbb{R}$,
between the coefficients of the series in both sides of the equation.
Note that such a meaning is somewhat different from that of a
\emph{numerical} equality like Equation~(\ref{55}), which is valid only when ${\bf
x}$ belongs to some limited spatial domain. The matching
equation~(\ref{56}) tells us that the formal \emph{near-zone} expansion
of the multipole decomposition is \emph{identical}, term by term, to the
multipole expansion of the post-Newtonian solution. However, the former
expansion is nothing but the formal \emph{far-zone} expansion, when
$r\to\infty$, of each of the post-Newtonian coefficients. Most
importantly, it is possible to write down, within the present formalism,
the general structure of these identical expansions as a consequence of
Theorem~\ref{th3}, Equation~(\ref{38}):
\begin{equation}
  \overline{{\cal M}(h)} =
  \!\sum {\hat n}_L r^m (\ln r)^p F_{L, m, p}(t) =
  {\cal M}(\overline h),
  \label{57}
\end{equation}
where the functions $F_{L, m, p} = \sum_{n\geq 1} G^n F_{L, m, p, n}$.
The latter expansion can be interpreted either as the singular
re-expansion of the multipole decomposition when $r\to 0$ (first
equality in Equation~(\ref{57})), or the singular re-expansion of the
post-Newtonian series when $r\to +\infty$ (second equality). We
recognize the beauty of singular perturbation theory, where two
asymptotic expansions, taken formally outside their respective domains
of validity, are matched together. Of course, the method works because
there exists, physically, an overlapping region in which the two
approximation series are expected to be numerically close to the exact
solution.


\subsection{General expression of the multipole expansion}
\label{subsec:5.2}

\begin{theorem}
  Under the hypothesis of matching, Equation~(\ref{56}), the multipole
  expansion of the solution of the Einstein field equation outside a
  post-Newtonian source reads
  \begin{equation}    
    {\cal M}(h^{\alpha\beta}) = {\cal FP}_{B=0}\, \Box^{-1}_\mathrm{ret}
    [\widetilde r^B {\cal M}(\Lambda^{\alpha\beta})] - \frac{4G}{c^4}
    \sum^{+\infty}_{l=0} \frac{(-)^l}{l!} \partial_L
    \left\{ \frac{1}{r} {\cal H}^{\alpha\beta}_L (t-r/c) \right\},
    \label{58}
  \end{equation}
  where the ``multipole moments'' are given by
  \begin{equation}
    {\cal H}^{\alpha\beta}_L (u) = {\cal FP}_{B=0} \int d^3 {\bf x} \,
    |\widetilde {\bf x}|^B x_L \, {\overline \tau}^{\alpha\beta}({\bf
    x}, u).
    \label{59}
  \end{equation}
  Here, ${\overline \tau}^{\alpha\beta}$ denotes the post-Newtonian
  expansion of the stress-energy pseudo-tensor defined by
  Equation~(\ref{13}).
  \label{th5}
\end{theorem}

\noindent 
\emph{Proof}~\cite{B95, B98mult}: First notice where the physical
restriction of considering a post-Newtonian source enters this theorem:
the multipole moments~(\ref{59}) depend on the \emph{post-Newtonian}
expansion ${\overline \tau}^{\alpha\beta}$, rather than on
$\tau^{\alpha\beta}$ itself. Consider $\Delta^{\alpha\beta}$, namely the
difference between $h^{\alpha\beta}$, which is a solution of the field
equations everywhere inside and outside the source, and the first term
in Equation~(\ref{58}), namely the finite part of the retarded integral of
the multipole expansion ${\cal M}(\Lambda^{\alpha\beta})$:
\begin{equation}
  \Delta^{\alpha\beta} \equiv h^{\alpha\beta} - {\cal FP}\,
  \Box^{-1}_\mathrm{ret} [ {\cal M}(\Lambda^{\alpha\beta})].
  \label{60}
\end{equation}
From now on we shall generally abbreviate the symbols concerning the
finite-part operation at $B=0$ by a mere ${\cal FP}$. According to
Equation~(\ref{18}), $h^{\alpha\beta}$ is given by the retarded integral of
the pseudo-tensor $\tau^{\alpha\beta}$. So,
\begin{equation}
  \Delta^{\alpha\beta} = \frac{16\pi G}{c^4} \Box^{-1}_\mathrm{ret}
  \tau^{\alpha\beta} - {\cal FP}\, \Box^{-1}_\mathrm{ret}
  \left[ {\cal M}(\Lambda^{\alpha\beta}) \right].
  \label{61}
\end{equation}
In the second term the finite part plays a crucial role because the
multipole expansion ${\cal M}(\Lambda^{\alpha\beta})$ is singular at
$r=0$. By contrast, the first term in Equation~(\ref{61}), as it stands, is
well-defined because we are considering only some smooth field
distributions: $\tau^{\alpha\beta}\in C^\infty({\mathbb{R}}^4)$. There
is no need to include a finite part ${\cal FP}$ in the first term, but
\emph{a contrario} there is no harm to add one in front of it, because
for convergent integrals the finite part simply gives back the value of
the integral. The advantage of adding ``artificially'' the ${\cal FP}$
in the first term is that we can re-write Equation~(\ref{61}) into the much
more interesting form
\begin{equation}
  \Delta^{\alpha\beta} = \frac{16\pi G}{c^4} {\cal FP}\,
  \Box^{-1}_\mathrm{ret} \left[ \tau^{\alpha\beta} - {\cal
  M}(\tau^{\alpha\beta})\right],
  \label{62}
\end{equation}
in which we have also used the fact that ${\cal M}
(\Lambda^{\alpha\beta})=16\pi G / c^4 \cdot {\cal
M}(\tau^{\alpha\beta})$ because $T^{\alpha\beta}$ has a compact support.
The interesting point about Equation~(\ref{62}) is that
$\Delta^{\alpha\beta}$ appears now to be the (finite part of a) retarded
integral of a source with spatially \emph{compact} support. This follows
from the fact that the pseudo-tensor agrees numerically with its own
multipole expansion when $r>a$ (same equation as~(\ref{54})). Therefore,
${\cal M}(\Delta^{\alpha\beta})$ can be obtained from the known formula
for the multipole expansion of the retarded solution of a wave equation
with compact-support source. This formula, given in Appendix~B of
Ref.~\cite{BD89}, yields the second term in Equation~(\ref{58}),
\begin{equation}
  {\cal M}(\Delta^{\alpha\beta}) = - \frac{4G}{c^4} \sum^{+\infty}_{l=0}
  \frac{(-)^l}{l!} \partial_L \left\{ \frac{1}{r} {\cal
  H}^{\alpha\beta}_L (u) \right\},
  \label{63}
\end{equation}
but in which the moments do not yet match the result~(\ref{59});
instead,
\begin{equation}
  {\cal H}^{\alpha\beta}_L = {\cal FP} \int d^3 {\bf x} \, x_L
  \left[\tau^{\alpha\beta} - {\cal M}(\tau^{\alpha\beta})\right].
  \label{64}
\end{equation}
The reason is that we have not yet applied the assumption of a
post-Newtonian source. Such sources are entirely covered by their own
near zone (i.e.\ $a\ll\lambda$), and, in addition, the
integral~(\ref{64}) has a compact support limited to the domain of the
source. In consequence, we can replace the integrand in Equation~(\ref{64})
by its post-Newtonian expansion, valid over all the near zone, i.e.
\begin{equation}
  {\cal H}^{\alpha\beta}_L = {\cal FP} \int d^3 {\bf x} \, x_L
  \left[{\overline\tau}^{\alpha\beta} -
  \overline{{\cal M}(\tau^{\alpha\beta})}\right].
  \label{65}
\end{equation} 
Strangely enough, we do not get the expected result because of the
presence of the second term in Equation~(\ref{65}). Actually, this term is a
bit curious, because the object $\overline{{\cal
M}(\tau^{\alpha\beta})}$ it contains is only known in the form of the
formal series whose structure is given by the first equality in
Equation~(\ref{57}) (indeed $\tau$ and $h$ have the same type of structure).
Happily (because we would not know what to do with this term in
applications), we are now going to prove that the second term in
Equation~(\ref{65}) is in fact \emph{identically zero}. The proof is based on
the properties of the analytic continuation as applied to the formal
structure~(\ref{57}) of $\overline{{\cal M}(\tau^{\alpha\beta})}$. Each
term of this series yields a contribution to Equation~(\ref{65}) that takes
the form, after performing the angular integration, of the integral
${\cal FP}_{B=0}\int_0^{+\infty} dr \, r^{B+b} (\ln r)^p$, and
multiplied by some function of time. We want to prove that the radial
integral $\int_0^{+\infty} dr \, r^{B+b} (\ln r)^p$ is zero by analytic
continuation ($\forall B\in\mathbb{C}$). First we can get rid of the
logarithms by considering some repeated differentiations with respect to
$B$; thus we need only to consider the simpler integral
$\int_0^{+\infty} dr \, r^{B+b}$. We split the integral into a
``near-zone'' integral $\int_0^{\cal R} dr \, r^{B+b}$ and a
``far-zone'' one $\int_{\cal R}^{+\infty} dr \, r^{B+b}$, where ${\cal
R}$ is some constant radius. When $\Re \, (B)$ is a large enough
\emph{positive} number, the value of the near-zone integral is ${\cal
R}^{B+b+1}/(B+b+1)$, while when $\Re \, (B)$ is a large \emph{negative}
number, the far-zone integral reads the opposite, $-{\cal
R}^{B+b+1}/(B+b+1)$. Both obtained values represent the unique analytic
continuations of the near-zone and far-zone integrals for any
$B\in\mathbb{C}$ except $-b-1$. The complete integral $\int_0^{+\infty}
dr \, r^{B+b}$ is equal to the sum of these analytic continuations, and
is therefore identically zero ($\forall B\in\mathbb{C}$, including the
value $-b-1$). At last we have completed the proof of Theorem~\ref{th5}:
\begin{equation}
  {\cal H}^{\alpha\beta}_L =
  {\cal FP} \int d^3 {\bf x}  \, x_L {\overline\tau}^{\alpha\beta}.
  \label{66}
\end{equation} 

The latter proof makes it clear how crucial the analytic-continuation
finite part ${\cal FP}$ is, which we recall is the same as in our
iteration of the exterior post-Minkowskian field (see Equation~(\ref{32})).
Without a finite part, the multipole moment~(\ref{66}) would be strongly
divergent, because the pseudo-tensor ${\overline \tau}^{\alpha\beta}$
has a non-compact support owing to the contribution of the gravitational
field, and the multipolar factor $x_L$ behaves like $r^l$ when $r\to
+\infty$. In applications (Part~\ref{part:b} of this article) we must
carefully follow the rules for handling the ${\cal FP}$ operator.

The two terms in the right-hand side of Equation~(\ref{58}) depend separately
on the length scale $r_0$ that we have introduced into the definition of
the finite part, through the analytic-continuation factor ${\widetilde
r}^B=(r/r_0)^B$ (see Equation~(\ref{29_1})). However, the sum of these two
terms, i.e.\ the exterior multipolar field ${\cal M}(h)$ itself, is
independent of $r_0$. To see this, the simplest way is to differentiate
formally ${\cal M}(h)$ with respect to $r_0$. The independence of the
field upon $r_0$ is quite useful in applications, since in general many
intermediate calculations do depend on $r_0$, and only in the final
stage does the cancellation of the $r_0$'s occur. For instance, we shall
see that the source quadrupole moment depends on $r_0$ starting from the
3PN level~\cite{BIJ02}, but that this $r_0$ is compensated by another
$r_0$ coming from the non-linear ``tails of tails'' at the 3PN order.


\subsection{Equivalence with the Will--Wiseman formalism}
\label{subsec:5.3}

Recently, Will and Wiseman~\cite{WW96} (see also Refs.~\cite{W99,
PW00}), extending previous work of Epstein and Wagoner~\cite{EW75} and
Thorne~\cite{Th80}, have obtained a different-looking multipole
decomposition, with different definitions for the multipole moments of a
post-Newtonian source. They find, instead of our multipole decomposition
given by Equation~(\ref{58}),
\begin{equation}
  {\cal M}(h^{\alpha\beta}) = \Box^{-1}_\mathrm{ret} [{\cal
  M}(\Lambda^{\alpha\beta})]_{\bigl|_{\cal R}}\!\!\! - \frac{4G}{c^4}
  \sum^{+\infty}_{l=0} \frac{(-)^l}{l!} \partial_L \left\{ \frac{1}{r}
  {\cal W}^{\alpha\beta}_L (t-r/c) \right\}.
  \label{67}
\end{equation}
There is no ${\cal FP}$ operation in the first term, but instead the
retarded integral is \emph{truncated}, as indicated by the subscript
${\cal R}$, to extend only in the ``far zone'': i.e.\ $|{\bf x}'|>{\cal
R}$ in the notation of Equation~(\ref{19}), where ${\cal R}$ is a constant
radius enclosing the source (${\cal R}> a$). The near-zone part of the
retarded integral is thereby removed, and there is no problem with the
singularity of the multipole expansion ${\cal M}(\Lambda^{\alpha\beta})$
at the origin. The multipole moments ${\cal W}_L$ are then given, in
contrast with our result~(\ref{59}), by an integral extending over the
``near zone'' only:
\begin{equation}
  {\cal W}^{\alpha\beta}_L (u) = \int_{|{\bf x}|<{\cal R}}d^3 {\bf x}~ x_L
  \, {\overline \tau}^{\alpha\beta}({\bf x}, u).
  \label{68}
\end{equation} 
Since the integrand is compact-supported there is no problem with the
bound at infinity and the integral is well-defined (no need of a ${\cal
FP}$).

Let us show that the two different formalisms are equivalent. We compute
the difference between our moment ${\cal H}_L$, defined by
Equation~(\ref{59}), and the Will--Wiseman moment ${\cal W}_L$, given by
Equation~(\ref{68}). For the comparison we split ${\cal H}_L$ into far-zone
and near-zone integrals corresponding to the radius ${\cal R}$. Since
the finite part ${\cal FP}$ present in ${\cal H}_L$ deals only with the
bound at infinity, it can be removed from the near-zone integral, which
is then seen to be exactly equal to ${\cal W}_L$. So the difference
between the two moments is simply given by the far-zone integral:
\begin{equation}
  {\cal H}^{\alpha\beta}_L(u) - {\cal W}^{\alpha\beta}_L(u) ={\cal FP}  
  \int_{|{\bf x}|>{\cal R}}d^3 {\bf x} \, x_L 
  {\overline \tau}^{\alpha\beta}({\bf x}, u).
  \label{69}
\end{equation}
Next, we transform this expression. Successively we write ${\overline
\tau}^{\alpha\beta}={\cal M}({\overline \tau}^{\alpha\beta})$ because we
are outside the source, and ${\cal M}({\overline
\tau}^{\alpha\beta})=\overline{{\cal M}(\tau^{\alpha\beta})}$ from the
matching equation~(\ref{56}). At this stage, we recall from our
reasoning right after Equation~(\ref{65}) that the finite part of an integral
over the whole space ${\mathbb{R}}^3$ of a quantity having the same
structure as $\overline{{\cal M}(\tau^{\alpha\beta})}$ is identically
zero by analytic continuation. The main trick of the proof is made
possible by this fact, as it allows us to transform the far-zone
integration $|{\bf x}|>{\cal R}$ in Equation~(\ref{69}) into a
\emph{near-zone} one $|{\bf x}|<{\cal R}$, at the price of changing the
overall sign in front of the integral. So,
\begin{equation}
  {\cal H}^{\alpha\beta}_L(u) - {\cal W}^{\alpha\beta}_L(u) =
  - {\cal FP} \int_{|{\bf x}|<{\cal R}}d^3 {\bf x} \, x_L
  \overline{{\cal M}(\tau^{\alpha\beta})}({\bf x}, u).
  \label{70}
\end{equation}
Finally, it is straightforward to check that the right-hand side of this
equation, when summed up over all multipolarities $l$, accounts exactly
for the near-zone part that was removed from the retarded integral of
${\cal M}(\Lambda^{\alpha\beta})$ (first term in Equation~(\ref{67})), so
that the ``complete'' retarded integral as given by the first term in
our own definition~(\ref{58}) is exactly reconstituted. In conclusion,
the formalism of Ref.~\cite{WW96} is equivalent to the one of
Refs.~\cite{B95, B98mult}.


\subsection{The source multipole moments}
\label{subsec:5.4}

In principle the bridge between the exterior gravitational field
generated by the post-Newtonian source and its inner field is provided
by Theorem~\ref{th5}; however, we still have to make the connection with
the explicit construction of the general multipolar and post-Minkowskian
metric in Sections~\ref{sec:3} and~\ref{sec:4}. Namely, we must find the
expressions of the six STF source multipole moments $\mathrm{I}_L$,
$\mathrm{J}_L, \dots, \mathrm{Z}_L$ parametrizing the linearized
metric~(\ref{23}, \ref{24}, \ref{25}) at the basis of that
construction\epubtkFootnote{Recall that in actual applications we need
mostly the mass-type moment $\mathrm{I}_L$ and current-type one
$\mathrm{J}_L$, because the other moments parametrize a linearized gauge
transformation.}.

To do this we first find the equivalent of the multipole expansion given
in Theorem~\ref{th5}, which was parametrized by non-trace-free multipole
functions ${\cal H}^{\alpha\beta}_L(u)$, in terms of new multipole
functions ${\cal F}^{\alpha\beta}_L(u)$ that are STF in all their
indices $L$. The result (which follows from Equation~(B.14a) in~\cite{BD89})
is
\begin{equation}    
  {\cal M}(h^{\alpha\beta}) = {\cal FP}\, \Box^{-1}_\mathrm{ret}
  [ {\cal M}(\Lambda^{\alpha\beta})] - \frac{4G}{c^4}
  \sum^{+\infty}_{l=0} \frac{(-)^l}{l!} \partial_L
  \left\{ \frac{1}{r} {\cal F}^{\alpha\beta}_L (t-r/c) \right\},
  \label{multSTF}
\end{equation}
where the STF multipole functions (witness the multipolar factor
$\hat{x}_L\equiv \mathrm{STF}[x_L]$) read
\begin{equation}
  {\cal F}^{\alpha\beta}_L (u) = {\cal FP} \int d^3 {\bf x} \,
  \hat{x}_L \,\int^1_{-1} dz \,\delta_l(z) \,{\overline
    \tau}^{\alpha\beta}({\bf x}, u+z |{\bf x}|/c).
  \label{FL}
\end{equation}
Notice the presence of an extra integration variable $z$, ranging from
$-1$ to 1. The $z$-integration involves the weighting
function\epubtkFootnote{This function approaches the Dirac
delta-function (hence its name) in the limit of large multipoles:
$\lim_{\, l \to +\infty}\delta_l (z)=\delta (z)$. Indeed the source
looks more and more like a point mass as we increase the multipolar
order $l$.}
\begin{equation}
  \delta_l (z) = \frac{(2l+1)!!}{2^{l+1} l!} (1-z^2)^l,
  \label{73}
\end{equation}
which is normalized in such a way that
\begin{equation}
  \int^1_{-1} dz \, \delta_l (z) = 1.
  \label{74}
\end{equation}
The next step is to impose the harmonic-gauge conditions~(\ref{11}) onto
the multipole decomposition~(\ref{multSTF}), and to decompose the
multipole functions ${\cal F}^{\alpha\beta}_L(u)$ into STF irreducible
pieces with respect to both $L$ and their space-time indices
$\alpha\beta$. This technical part of the calculation is identical to
the one of the STF irreducible multipole moments of linearized
gravity~\cite{DI91b}. The formulas needed in this decomposition read
\begin{equation}
  \begin{array}{rcl}
  {\cal F}^{00}_L &=& R_L,
  \\ [1.5 em]
  {\cal F}^{0i}_L &=& ^{(+)}T_{iL} + \varepsilon_{ai<i_l}
  {}^{(0)}T_{L-1>a} +\delta_{i<i_l} {}^{(-)}T_{L-1>},
  \\ [1.5 em]
  {\cal F}^{ij}_L &=& ^{(+2)}U_{ijL} + \displaystyle{\mathop{STF}_L}\
  \displaystyle{\mathop{STF}_{ij}}\,
  [\varepsilon_{aii_l} {}^{(+1)}U_{ajL-1} +\delta_{ii_l}
  {}^{(0)}U_{jL-1}
  \\ [1.5 em]
  && +\delta_{ii_l} \varepsilon_{aji_{l-1}} {}^{(-1)}U_{aL-2}
  +\delta_{ii_l} \delta_{ji_{l-1}} {}^{(-2)}U_{L-2} ] + \delta_{ij}
  V_L, \hspace{-2.5 em}
  \end{array}
\label{FLirred}
\end{equation}
where the ten tensors $R_L, {}^{(+)}T_{L+1}, \dots, {}^{(-2)}U_{L-2},
V_L$ are STF, and are uniquely given in terms of the ${\cal
F}^{\alpha\beta}_L$'s by some inverse formulas. Finally, the latter
decompositions lead to the following theorem.

\begin{theorem}
  The STF multipole moments $\mathrm{I}_L$ and $\mathrm{J}_L$ of a
  post-Newtonian source are given, formally up to any post-Newtonian
  order, by ($l\geq 2$)
  \begin{equation}
    \begin{array}{rcl}
      \mathrm{I}_L(u)&=& \displaystyle {\cal FP} \int d^3{\bf x}
      \int^1_{-1} dz\biggl\{ \delta_l\hat x_L\Sigma -
      \frac{4(2l+1)}{c^2(l+1)(2l+3)} \delta_{l+1} \hat x_{iL}
      \Sigma^{(1)}_i
      \\ [1.5 em]
      && \qquad \qquad \qquad \qquad
      \displaystyle + \frac{2(2l+1)}{c^4(l+1)(l+2)(2l+5)} \delta_{l+2}
      \hat x_{ijL} \Sigma^{(2)}_{ij} \biggr\} ({\bf x}, u+z |{\bf
      x}|/c),
      \\ [1.5 em]
      \mathrm{J}_L(u)&=& \displaystyle {\cal FP}
      \int d^3{\bf x}\int^1_{-1} dz \, \varepsilon_{ab \langle i_l}
      \biggl\{ \delta_l {\hat x}_{L-1 \rangle a} \Sigma_b -
      \frac{2l+1}{c^2(l+2)(2l+3)} \delta_{l+1} \hat x_{L-1 \rangle ac}
      \Sigma^{(1)}_{bc} \biggr\} ({\bf x}, u+z |{\bf x}|/c).
      \hspace{-2.5 em}
    \end{array}
    \label{71}
  \end{equation}
  These moments are the ones that are to be inserted into the
  linearized metric $h^{\alpha\beta}_1$ that represents the lowest
  approximation to the post-Minkowskian field
  $h_\mathrm{ext}^{\alpha\beta}=\sum_{n\geq 1}G^nh^{\alpha\beta}_n$
  defined in Section~\ref{sec:4}.
  \label{th6}
\end{theorem}

\noindent 
In these formulas the notation is as follows: Some convenient source
densities are defined from the post-Newtonian expansion of the
pseudo-tensor $\tau^{\alpha\beta}$ by
\begin{equation}
  \begin{array}{rcl}
    \Sigma &=& \displaystyle
    \frac{\overline\tau^{00}+\overline\tau^{ii}}{c^2},
    \\ [1.0 em]
    \Sigma_i &=& \displaystyle \frac{\overline\tau^{0i}}{c},
    \\ [1.0 em]
    \Sigma_{ij} &=& \displaystyle \overline{\tau}^{ij}
  \end{array}
  \label{72}
\end{equation}
(where $\overline{\tau}^{ii} \equiv\delta_{ij}\overline\tau^{ij}$). As
indicated in Equations~(\ref{71}) these quantities are to be evaluated at the
spatial point ${\bf x}$ and at time $u+z|{\bf x}|/c$.

For completeness, we give also the formulas for the four auxiliary
source moments $\mathrm{W}_L, \dots, \mathrm{Z}_L$, which parametrize
the gauge vector $\varphi^\alpha_1$ as defined in Equations~(\ref{25}):
\begin{eqnarray}
  \mathrm{W}_L(u)&=& {\cal FP} \int d^3{\bf x} \int^1_{-1}
  dz\biggl\{ {2l+1\over (l+1)(2l+3)} \delta_{l+1} \hat
  x_{iL} \Sigma_i - {2l+1\over2c^2(l+1)(l+2)(2l+5)}
  \delta_{l+2} \hat x_{ijL} {\Sigma}_{ij}^{(1)} \biggr\},
  \nonumber \\
  \label{76a}
  \\ [1.0 em]
  \mathrm{X}_L(u) &=& {\cal FP} \int d^3{\bf x} \int^1_{-1}
  dz\biggl\{  {2l+1\over 2(l+1)(l+2)(2l+5)} \delta_{l+2}
  \hat x_{ijL} \Sigma_{ij} \biggr\},
  \label{76b}
  \\ [1.0 em]
  \mathrm{Y}_L(u)&=&{\cal FP} \int d^3{\bf x} \int^1_{-1}
  dz\biggl\{ -\delta_l \hat x_L \Sigma_{ii} + {3(2l+1)\over
    (l+1)(2l+3)} \delta_{l+1} \hat x_{iL} {\Sigma}_i^{(1)}
  \nonumber \\ [1.0 em]
  && \qquad \qquad \qquad \qquad \;
  - {2(2l+1)\over c^2(l+1)(l+2)(2l+5)}
  \delta_{l+2} \hat x_{ijL} {\Sigma}_{ij}^{(2)} \biggr\},
  \label{76c}
  \\ [1.0 em]
  \mathrm{Z}_L(u) &=& {\cal FP} \int d^3{\bf x} \int^1_{-1}
  dz \, \varepsilon_{ab \langle i_l}\biggl\{- {2l+1\over (l+2)(2l+3)} 
  \delta_{l+1} \hat x_{L-1 \rangle bc} \Sigma_{ac} \biggr\}.
  \label{76d}
\end{eqnarray}%
As discussed in Section~\ref{sec:4}, one can always find two
intermediate ``packages'' of multipole moments, $\mathrm{M}_L$ and
$\mathrm{S}_L$, which are some non-linear functionals of the source
moments~(\ref{71}) and Equations~(\ref{76a}, \ref{76b}, \ref{76c}, \ref{76d}), and
such that the exterior field depends only on them, modulo a change of
coordinates (see, e.g., Equation~(\ref{77}) below).

In fact, all these source moments make sense only in the form of a
post-Newtonian expansion, so in practice we need to know how to expand
all the $z$-integrals as series when $c\to +\infty$. Here is the
appropriate formula:
\begin{equation}
  \int^1_{-1} dz~ \delta_l(z) \tau({\bf x}, u+z|{\bf x}|/c) =
  \sum_{k=0}^{+\infty}\frac{(2l+1)!!}{2^kk!(2l+2k+1)!!}
  \biggl(\frac{|{\bf x}|}{c}\frac{\partial}{\partial u}\biggr)^{\!2k}
  \tau({\bf x}, u).
  \label{75}
\end{equation}
Since the right-hand side involves only even powers of $1/c$, the same
result holds equally well for the ``advanced'' variable $u+z|{\bf x}|/c$
or the ``retarded'' one $u-z|{\bf x}|/c$. Of course, in the Newtonian
limit, the moments $\mathrm{I}_L$ and $\mathrm{J}_L$ (and also
$\mathrm{M}_L$, $\mathrm{S}_L$) reduce to the standard expressions. For
instance, we have
\begin{equation}
  \mathrm{I}_L(u) = \mathrm{Q}_L(u) + {\cal O}\left(\frac{1}{c^2}\right),
  \label{76_1}
\end{equation}
where $\mathrm{Q}_L$ is the Newtonian mass-type multipole moment (see
Equation~(\ref{3})). (The moments $\mathrm{W}_L, \dots, \mathrm{Z}_L$
have also a Newtonian limit, but it is not particularly illuminating.)

Needless to say, the formalism becomes prohibitively difficult to apply
at very high post-Newtonian approximations. Some post-Newtonian order
being given, we must first compute the relevant relativistic corrections
to the pseudo stress-energy-tensor $\tau^{\alpha\beta}$ (this
necessitates solving the field equations inside the matter, see
Section~\ref{subsec:5.5}) before inserting them into the source
moments~(\ref{71}, \ref{72}, \ref{73}, \ref{74}, \ref{75}, \ref{76a},
\ref{76b}, \ref{76c}, \ref{76d}). The formula~(\ref{75}) is used to
express all the terms up to that post-Newtonian order by means of more
tractable integrals extending over ${\mathbb{R}}^3$. Given a specific
model for the matter source we then have to find a way to compute all
these spatial integrals (we do it in Section~\ref{sec:10} in the case of
point-mass binaries). Next, we must substitute the source multipole
moments into the linearized metric~(\ref{23}, \ref{24}, \ref{25}), and
iterate them until all the necessary multipole interactions taking place
in the radiative moments $\mathrm{U}_L$ and $\mathrm{V}_L$ are under control.
In fact, we shall work out these multipole interactions for general
sources in the next section up to the 3PN order. Only at this point does
one have the physical radiation field at infinity, from which we can
build the templates for the detection and analysis of gravitational
waves. We advocate here that the complexity of the formalism reflects
simply the complexity of the Einstein field equations. It is probably
impossible to devise a different formalism, valid for general sources
devoid of symmetries, that would be substantially simpler.


\subsection{Post-Newtonian field in the near zone}
\label{subsec:5.5}

Theorem~\ref{th6} solves in principle the question of the generation of
gravitational waves by extended post-Newtonian sources. However, note
that this result has to be completed by the definition of an explicit
algorithm for the post-Newtonian iteration, analogous to the
post-Minkowskian algorithm we defined in Section~\ref{sec:4}, so that
the source multipole moments, which contain the full post-Newtonian
expansion of the pseudo-tensor $\tau^{\alpha\beta}$, can be completely
specified. Such a systematic post-Newtonian iteration scheme, valid
(formally) to any post-Newtonian order, has been implemented~\cite{PB02,
BFN05} using matched asymptotic expansions. The solution of this problem
yields, in particular, some general expression, valid up to any order,
of the terms associated with the gravitational radiation reaction force
inside the post-Newtonian source\epubtkFootnote{An alternative approach
to the problem of radiation reaction, besides the matching procedure, is
to work only within a post-Minkowskian iteration scheme (which does not
expand the retardations): see, e.g., Ref.~\cite{CKMR01}.}.

Before proceeding, let us recall that the ``standard'' post-Newtonian
approximation, as it was used until, say, the early 1980's (see for
instance Refs.~\cite{ADec75, K80a, K80b, PapaL81}), is plagued with some
apparently inherent difficulties, which crop up at some high
post-Newtonian order. The first problem is that in higher approximations
some \emph{divergent} Poisson-type integrals appear. Indeed the
post-Newtonian expansion replaces the resolution of a hyperbolic-like
d'Alembertian equation by a perturbatively equivalent hierarchy of
elliptic-like Poisson equations. Rapidly it is found during the
post-Newtonian iteration that the right-hand side of the Poisson
equations acquires a non-compact support (it is distributed over all
space), and that as a result the standard Poisson integral diverges at
the bound of the integral at spatial infinity, i.e.\ $r\equiv
|{\mathbf x}|\to +\infty$, with $t=\mathrm{const}$.

The second problem is related with the \emph{a priori} limitation of the
approximation to the near zone, which is the region surrounding the
source of small extent with respect to the wavelength of the emitted
radiation: $r\ll \lambda$. The post-Newtonian expansion assumes from the
start that all retardations $r/c$ are small, so it can rightly be viewed
as a formal \emph{near-zone} expansion, when $r\to 0$. In particular, the
fact which makes the Poisson integrals to become typically divergent,
namely that the coefficients of the post-Newtonian series blow up at
``spatial infinity'', when $r\to +\infty$, has nothing to do with the
actual behaviour of the field at infinity. However, the serious
consequence is that it is not possible, \emph{a priori}, to implement
within the post-Newtonian iteration the physical information that the
matter system is isolated from the rest of the universe. Most
importantly, the no-incoming radiation condition, imposed at past null
infinity, cannot be taken into account, \emph{a priori}, into the scheme.
In a sense the post-Newtonian approximation is not ``self-supporting'',
because it necessitates some information taken from outside its own
domain of validity.

Here we present, following Refs.~\cite{PB02, BFN05}, a solution of both
problems, in the form of a general expression for the near-zone
gravitational field, developed to any post-Newtonian order, which has
been determined from implementing the matching equation~(\ref{56}). This
solution is free of the divergences of Poisson-type integrals we
mentionned above, and it incorporates the effects of gravitational
radiation reaction appropriate to an isolated system.

\begin{theorem}
  The expression of the post-Newtonian field in the near zone of a
  post-Newtonian source, satisfying correct boundary conditions at
  infinity (no incoming radiation), reads
  \begin{equation}
    \overline{h}^{\alpha\beta} = \frac{16\pi G}{c^4}
    \left[ {\cal FP}\,\Box^{-1}_\mathrm{ret} \left[
    \overline{\tau}^{\alpha\beta}\right] + \sum^{+\infty}_{l=0}
    \frac{(-)^l}{l!} \partial_L \left\{ \frac{{\cal R}^{\alpha\beta}_L
    (t-r/c)-{\cal R}^{\alpha\beta}_L(t+r/c)}{2\,r} \right\} \right].
    \label{hnz}
  \end{equation}
  The first term represents a particular solution of the hierarchy of
  post-Newtonian equations, while the second one is a homogeneous
  multipolar solution of the wave equation, of the ``anti-symmetric''
  type that is regular at the origin $r=0$ located in the source.
  \label{th7}
\end{theorem}

\noindent
More precisely, the flat retarded d'Alembertian operator in
Equation~(\ref{hnz}) is given by the standard expression~(\ref{19}) but with
all retardations expanded ($r/c\to 0$), and with the finite part ${\cal
FP}$ procedure involved for dealing with the bound at infinity of the
Poisson-type integrals (so that all the integrals are well-defined at
any order of approximation),
\begin{equation}
  {\cal FP}\,\Box^{-1}_\mathrm{ret}
  \left[ \overline{\tau}^{\alpha\beta} \right] \equiv
  -\frac{1}{4\pi} \sum_{n=0}^{+\infty}\frac{(-)^n}{n!}
  \left(\frac{\partial}{c\,\partial t}\right)^{\!n}\,{\cal FP}\int
  d^3\mathbf{x}'\,\vert\mathbf{x} - \mathbf{x}'\vert^{n-1}\,
  \overline{\tau}^{\alpha\beta}(\mathbf{x}',t).
  \label{FPret}
\end{equation}
The existence of the solution~(\ref{FPret}) shows that the problem of
divergences of the post-Newtonian expansion is simply due to the fact
that the standard Poisson integral does not constitute the correct
solution of the Poisson equation in the context of post-Newtonian
expansions. So the problem is purely of a technical nature, and is
solved once we succeed in finding the appropriate solution to the
Poisson equation.

Theorem~\ref{th7} is furthermore to be completed by the information
concerning the multipolar functions ${\cal R}^{\alpha\beta}_L(u)$
parametrizing the anti-symmetric homogeneus solution, the second term of
Equation~(\ref{hnz}). Note that this homogeneous solution represents the
unique one for which the matching equation~(\ref{56}) is satisfied. The
result is
\begin{equation}
  {\cal R}^{\alpha\beta}_L(u) =
  -\frac{1}{4\pi}{\cal FP}\int d^3{\bf x}'\,{\hat x}'_L
  \int_1^{+\infty} \!\!\!\!\! dz\,\gamma_l(z)\,{\cal M} 
  (\tau^{\alpha\beta})({\bf x}',u-z |{\bf x}'|/c),
  \label{Rab}
\end{equation}
where ${\cal M}(\tau^{\alpha\beta})$ denotes the multipole expansion of
the pseudo-tensor (in the sense of Equation~(\ref{53})), and where we denote
$\gamma_l(z)=-2\delta_l(z)$, with $\delta_l(z)$ being given
by Equation~(\ref{73})\epubtkFootnote{Notice that the normalization
$\int_1^{+\infty}dz \,\gamma_l(z)=1$ holds as a consequence of the
corresponding normalization~(\ref{74}) for $\delta_l(z)$, together with
the fact that $\int_{-\infty}^{+\infty}dz \,\gamma_l(z)=0$ by analytic
continuation in the variable $l\in\mathbb{C}$.}. 

Importantly, we find that the post-Newtonian expansion
$\overline{h}^{\alpha\beta}$ given by Theorem~\ref{th7} is a functional
not only of the related expansion of the pseudo-tensor,
$\overline{\tau}^{\alpha\beta}$, but also, by Equation~(\ref{Rab}), of its
multipole expansion ${\cal M}(\tau^{\alpha\beta})$, which is valid in
the exterior of the source, and in particular in the asymptotic regions
far from the source. This can be understood by the fact that the
post-Newtonian solution~(\ref{hnz}) depends on the boundary conditions
imposed at infinity, that describe a matter system isolated from the
rest of the universe.

Equation~(\ref{hnz}) is interesting for providing a practical recipe for
performing the post-Newtonian iteration \emph{ad infinitum}. Moreover, it
gives some insights on the structure of radiation reaction terms. Recall
that the anti-symmetric waves, regular in the source, are associated
with radiation reaction effects. More precisely, it has been
shown~\cite{PB02} that the specific anti-symmetric wave given by the
second term of Equation~(\ref{hnz}) is linked with some \emph{non-linear}
contribution due to gravitational wave tails in the radiation reaction
force. Such a contribution constitutes a generalization of the
tail-transported radiation reaction term at the 4PN order, i.e.\
1.5PN order relative to the dominant radiation reaction order, as
determined in Ref.~\cite{BD88}. This term is in fact required by energy
conservation and the presence of tails in the wave zone (see, e.g.,
Equation~(\ref{78}) below). Hence, the second term of Equation~(\ref{hnz}) is
dominantly of order 4PN and can be neglected in computations of the
radiation reaction up to 3.5PN order (as in Ref.~\cite{NB05}). The usual
radiation reaction terms, up to 3.5PN order, which are \emph{linear} in
the source multipole moments (for instance the usual radiation reaction
term at 2.5PN order), are contained in the first term of
Equation~(\ref{hnz}), and are given by the terms with odd powers of $1/c$ in
the post-Newtonian expansion~(\ref{FPret}). It can be shown~\cite{BFN05}
that such terms take also the form of some anti-symmetric multipolar
wave, which turn out to be parametrized by the same moments as in the
exterior field, namely the moments which are the STF analogues of
Equations~(\ref{59}).

\newpage


\section{Non-linear Multipole Interactions}
\label{sec:6}

We shall now show that the radiative mass-type quadrupole moment
$\mathrm{U}_{ij}$ includes a quadratic tail at the relative 1.5PN order (or
$1/c^3$), corresponding to the interaction of the mass $\mathrm{M}$ of the
source and its quadrupole moment $\mathrm{I}_{ij}$. This is due to the
back-scattering of quadrupolar waves off the Schwarzschild curvature
generated by $\mathrm{M}$. Next, $\mathrm{U}_{ij}$ includes a so-called
non-linear memory integral at the 2.5PN order, due to the quadrupolar
radiation of the stress-energy distribution of linear quadrupole waves
themselves, i.e.\ of multipole interactions $\mathrm{I}_{ij}\times
\mathrm{I}_{kl}$. Finally, we have also a cubic tail, or ``tail of tail'',
arising at the 3PN order, and associated with the multipole interaction
$\mathrm{M}^2\times \mathrm{I}_{ij}$. The result for $\mathrm{U}_{ij}$ is better
expressed in terms of the intermediate quadrupole moment $\mathrm{M}_{ij}$
already discussed in Section~\ref{subsec:4.2}. This moment
reads~\cite{B96}
\begin{equation}
  \mathrm{M}_{ij} = \mathrm{I}_{ij}-{4G\over c^5} \left[
  \mathrm{W}^{(2)} \mathrm{I}_{ij}-\mathrm{W}^{(1)}
  \mathrm{I}_{ij}^{(1)}\right] + {\cal O}\left(\frac{1}{c^7}\right),
  \label{77}
\end{equation}
where $\mathrm{W}$ means $\mathrm{W}_L$ as given by Equation~(\ref{76a}) in
the case $l=0$ (of course, in Equation~(\ref{77}) we need only the Newtonian
value of $\mathrm{W}$). The difference between the two moments
$\mathrm{M}_{ij}$ and $\mathrm{I}_{ij}$ is a small 2.5PN quantity.
Henceforth, we shall express many of the results in terms of the mass
moments $\mathrm{M}_L$ and the corresponding current ones
$\mathrm{S}_L$. The complete formula for the radiative quadrupole, valid
through the 3PN order, reads~\cite{B98quad, B98tail}
\begin{eqnarray}
  \mathrm{U}_{ij}(U) &=& \mathrm{M}^{(2)}_{ij}(U) +
  \frac{2G\mathrm{M}}{c^3} \int^{+\infty}_0 d\tau \,
  \mathrm{M}^{(4)}_{ij} (U-\tau) \left[ \ln \left( \frac{c\tau}{2r_0}
  \right) + \frac{11}{12} \right]
  \nonumber \\
  && +\frac{G}{c^5}
  \Biggl\{ - \frac{2}{7} \int^{+\infty}_0 d\tau \, \mathrm{M}^{(3)}_{a
  \langle i}(U-\tau) \mathrm{M}^{(3)}_{j \rangle a}(U-\tau)
  \nonumber \\
  && \qquad ~\, - \frac{2}{7} \mathrm{M}^{(3)}_{a \langle
  i}\mathrm{M}^{(2)}_{j \rangle a} - \frac{5}{7} \mathrm{M}^{(4)}_{a
  \langle i} \mathrm{M}^{(1)}_{j \rangle a} + \frac{1}{7}
  \mathrm{M}^{(5)}_{a \langle i} \mathrm{M}_{j \rangle a} + \frac{1}{3}
  \varepsilon_{ab \langle i} \mathrm{M}^{(4)}_{j \rangle a} \mathrm{S}_b
  \Biggr\}
  \nonumber \\
  && + \frac{2G^2 \mathrm{M}^2}{c^6}
  \int^{+\infty}_0 d\tau \, \mathrm{M}^{(5)}_{ij}(U-\tau) \left[ \ln^2
  \left( \frac{c\tau}{2r_0} \right) + \frac{57}{70} \ln \left(
  \frac{c\tau}{2r_0} \right) + \frac{124627}{44100} \right]
  \nonumber \\
  && +\, {\cal O}\left( \frac{1}{c^7} \right).
  \label{78}
\end{eqnarray}%
The retarded time in radiative coordinates is denoted $U=T-R/c$. The
constant $r_0$ is the one that enters our definition of the finite-part
operation ${\cal FP}$ (see Equation~(\ref{29_1})). The ``Newtonian'' term in
Equation~(\ref{78}) contains the Newtonian quadrupole moment
$\mathrm{Q}_{ij}$ (see Equation~(\ref{76_1})). The dominant radiation tail at
the 1.5PN order was computed within the present formalism in
Ref.~\cite{BD92}. The 2.5PN non-linear memory integral -- the first term
inside the coefficient of $G/c^5$ -- has been obtained using both
post-Newtonian methods~\cite{B90, WW91, Th92, BD92, B98quad} and
rigorous studies of the field at future null infinity~\cite{Chr91}. The
other multipole interactions at the 2.5PN order can be found in
Ref.~\cite{B98quad}. Finally the ``tail of tail'' integral appearing at
the 3PN order has been derived in this formalism in Ref.~\cite{B98tail}.
Be careful to note that the latter post-Newtonian orders correspond to
``relative'' orders when counted in the local radiation-reaction force,
present in the equations of motion: For instance, the 1.5PN tail
integral in Equation~(\ref{78}) is due to a 4PN radiative effect in the
equations of motion~\cite{BD88}; similarly, the 3PN tail-of-tail
integral is (presumably) associated with some radiation-reaction terms
occuring at the 5.5PN order.

Notice that all the radiative multipole moments, for any $l$, get some
tail-induced contributions. They are computed at the 1.5PN level in
Appendix~C of Ref.~\cite{B95}. We find
\begin{equation}
  \begin{array}{rcl}
    \mathrm{U}_L(U) &=& \displaystyle \mathrm{M}^{(l)}_L(U) +
    \frac{2G\mathrm{M}}{c^3} \int^{+\infty}_0 \!\!\!\!\! d\tau \,
    \mathrm{M}^{(l+2)}_L (U-\tau) \left[ \ln \left( \frac{c\tau}{2r_0}
    \right) + \kappa_l \right] + {\cal O}\left( \frac{1}{c^5} \right),
    \\ [1.5 em]
    \mathrm{V}_L(U) &=& \displaystyle \mathrm{S}^{(l)}_L(U) +
    \frac{2G\mathrm{M}}{c^3} \int^{+\infty}_0 \!\!\!\!\! d\tau \,
    \mathrm{S}^{(l+2)}_L (U-\tau) \left[ \ln \left( \frac{c\tau}{2r_0}
    \right) + \pi_l \right] + {\cal O}\left( \frac{1}{c^5} \right),
  \end{array}
  \label{79}
\end{equation}
where the constants $\kappa_l$ and $\pi_l$ are given by
\begin{equation}
  \begin{array}{rcl}
    \kappa_l &=& \displaystyle
    \frac{2l^2+5l+4}{l(l+1)(l+2)}+\sum_{k=1}^{l-2}\frac{1}{k},
    \\ [1.5 em]
    \pi_l &=& \displaystyle
    \frac{l-1}{l(l+1)}+\sum_{k=1}^{l-1}\frac{1}{k}.
  \end{array}
  \label{80}
\end{equation}
Recall that the retarded time $U$ in radiative coordinates is given by
\begin{equation}
  U=t-\frac{r}{c}-\frac{2GM}{c^3}\ln\left(\frac{r}{r_0}\right)+
  {\cal O}\left(G^2\right),
  \label{80_1}
\end{equation}
where $(t, r)$ are harmonic coordinates; recall the gauge vector
$\xi^\alpha_1$ in Equation~(\ref{43}). Inserting $U$ as given by
Equation~(\ref{80_1}) into Equations~(\ref{79}) we obtain the radiative moments
expressed in terms of source-rooted coordinates $(t, r)$, e.g.,
\begin{equation}
  \mathrm{U}_L = \mathrm{M}^{(l)}_L(t-r/c) +
  \frac{2G\mathrm{M}}{c^3} \int^{+\infty}_0 \!\!\!\!\! d\tau \,
  \mathrm{M}^{(l+2)}_L (t-\tau-r/c) \left[ \ln \left( \frac{c\tau}{2r}
  \right) + \kappa_l \right] + {\cal O}\left( \frac{1}{c^5} \right).
\end{equation}
This expression no longer depends on the constant $r_0$ (i.e.\ the $r_0$
gets replaced by $r$)\epubtkFootnote{At the 3PN order (taking into
account the tails of tails), we find that $r_0$ does not completely
cancel out after the replacement of $U$ by the right-hand side of
Equation~(\ref{80_1}). The reason is that the moment $\mathrm{M}_L$ also
depends on $r_0$ at the 3PN order. Considering also the latter
dependence we can check that the 3PN radiative moment $\mathrm{U}_L$ is
actually free of the unphysical constant $r_0$.}. If we now change the
harmonic coordinates $(t, r)$ to some new ones, such as, for instance,
some ``Schwarzschild-like'' coordinates $(t', r')$ such that $t'=t$ and
$r'=r+G\mathrm{M}/c^2$, we get
\begin{equation}
  \mathrm{U}_L = \mathrm{M}^{(l)}_L(t'-r'/c) +
  \frac{2G\mathrm{M}}{c^3} \int^{+\infty}_0 \!\!\!\!\! d\tau \,
  \mathrm{M}^{(l+2)}_L (t'-\tau-r'/c) \left[ \ln \left( \frac{c\tau}{2r'}
  \right) + \kappa'_l \right] + {\cal O}\left( \frac{1}{c^5}
  \right),
\end{equation}
where $\kappa'_l=\kappa_l+1/2$. Therefore the constant $\kappa_l$ (and
$\pi_l$ as well) depends on the choice of source-rooted coordinates $(t,
r)$: For instance, we have $\kappa_2=11/12$ in harmonic coordinates (see
Equation~(\ref{78})), but $\kappa'_2=17/12$ in Schwarzschild
coordinates~\cite{BS93}.

The tail integrals in Equations~(\ref{78}, \ref{79}) involve all the instants
from $-\infty$ in the past up to the current time $U$. However, strictly
speaking, the integrals must not extend up to minus infinity in the
past, because we have assumed from the start that the metric is
stationary before the date $-{\cal T}$; see Equation~(\ref{17}). The range of
integration of the tails is therefore limited \emph{a priori} to the time
interval [$-{\cal T}$, $U$]. But now, once we have derived the tail
integrals, thanks in part to the technical assumption of stationarity in
the past, we can argue that the results are in fact valid in more
general situations for which the field has \emph{never} been stationary.
We have in mind the case of two bodies moving initially on some unbound
(hyperbolic-like) orbit, and which capture each other, because of the
loss of energy by gravitational radiation, to form a bound system at our
current epoch. In this situation we can check, using a simple Newtonian
model for the behaviour of the quadrupole moment
$\mathrm{M}_{ij}(U-\tau)$ when $\tau\to +\infty$, that the tail
integrals, when assumed to extend over the whole time interval
[$-\infty$, $U$], remain perfectly well-defined (i.e.\ convergent) at
the integration bound $\tau=+\infty$. We regard this fact as a solid
\emph{a posteriori} justification (though not a proof) of our \emph{a
priori} too restrictive assumption of stationarity in the past. This
assumption does not seem to yield any physical restriction on the
applicability of the final formulas.

To obtain the result~(\ref{78}), we must implement in details the
post-Minkows-kian algorithm presented in Section~\ref{subsec:4.1}. Let
us outline here this computation, limiting ourselves to the interaction
between one or two masses $\mathrm{M}\equiv
\mathrm{M}_\mathrm{ADM}\equiv \mathrm{I}$ and the time-varying
quadrupole moment $\mathrm{M}_{ab}(u)$ (that is related to the source
quadrupole $\mathrm{I}_{ab}(u)$ by Equation~(\ref{77})). For these moments
the linearized metric~(\ref{23}, \ref{24}, \ref{25}) reads
\begin{equation}
  h^{\alpha\beta}_1 = h^{\alpha\beta}_{(\mathrm{M})} +
  h^{\alpha\beta}_{(\mathrm{M}_{ab})},
  \label{81}
\end{equation}
where the monopole part is nothing but the linearized piece of the
Schwarzschild metric in harmonic coordinates,
\begin{equation}
  \begin{array}{rcl}
    h^{00}_{(\mathrm{M})} &=& - 4 r^{-1} \mathrm{M},
    \\ [0.8 em]
    h^{0i}_{(\mathrm{M})} &=& 0,
    \\ [0.8 em]
    h^{ij}_{(\mathrm{M})} &=& 0,
  \end{array}
\end{equation}
and the quadrupole part is
\begin{equation}
  \begin{array}{rcl}
    h^{00}_{(\mathrm{M}_{ab})} &=& - 2 \partial_{ab} 
    \left[ r^{-1} \mathrm{M}_{ab}(u) \right],
    \\ [0.8 em]
    h^{0i}_{(\mathrm{M}_{ab})} &=& 2 \partial_a 
    \left[ r^{-1} \mathrm{M}^{(1)}_{ai}(u) \right],
    \\ [0.8 em]
    h^{ij}_{(\mathrm{M}_{ab})} &=& - 2 r^{-1} \mathrm{M}^{(2)}_{ij}(u).
  \end{array}
\end{equation}
(We pose $c=1$ until the end of this section.) Consider next the
quadratically non-linear metric $h^{\alpha\beta}_2$ generated by these
moments. Evidently it involves a term proportional to $\mathrm{M}^2$,
the mixed term corresponding to the interaction $\mathrm{M} \times
\mathrm{M}_{ab}$, and the self-interaction term of $\mathrm{M}_{ab}$.
Say,
\begin{equation}
  h^{\alpha\beta}_2 = h^{\alpha\beta}_{(\mathrm{M}^2)} +
  h^{\alpha\beta}_{(\mathrm{M}\mathrm{M}_{ab})} +
  h^{\alpha\beta}_{(\mathrm{M}_{ab} \mathrm{M}_{cd})}.
  \label{84}
\end{equation}
The first term represents the quadratic piece of the Schwarzschild
metric,
\begin{equation}
  \begin{array}{rcl}
  h^{00}_{(\mathrm{M}^2)} &=& - 7 r^{-2} \mathrm{M}^2,
  \\ [0.8 em]
  h^{0i}_{(\mathrm{M}^2)} &=& 0,
  \\ [0.8 em]
  h^{ij}_{(\mathrm{M}^2)} &=& - n_{ij} r^{-2} \mathrm{M}^2.
  \end{array}
\end{equation}
The second term in Equation~(\ref{84}) represents the dominant non-static
multipole interaction, that is between the mass and the quadrupole
moment, and that we now compute\epubtkFootnote{The computation of the
third term in Equation~(\ref{84}), which corresponds to the interaction
between two quadrupoles, $\mathrm{M}_{ab}\times \mathrm{M}_{cd}$, can be
found in Ref.~\cite{B98quad}.}. We apply Equations~(\ref{32},
\ref{33}, \ref{34}, \ref{35}, \ref{36}) in Section~\ref{sec:4}. First we
obtain the source for this term, viz.
\begin{equation}
  \Lambda^{\alpha\beta}_{(\mathrm{M}\mathrm{M}_{ab})} =
  N^{\alpha\beta}[h_{(\mathrm{M})}, h_{(\mathrm{M}_{ab})}]+
  N^{\alpha\beta}[h_{(\mathrm{M}_{ab})}, h_{(\mathrm{M})}],
  \label{86}
\end{equation}
where $N^{\alpha\beta}(h, h)$ denotes the quadratic-order part of the
gravitational source, as defined by Equation~(\ref{14_1}). To integrate this
term we need some explicit formulas for the retarded integral of an
extended (non-compact-support) source having some definite multipolarity
$l$. A thorough account of the technical formulas necessary for handling
the quadratic and cubic interactions is given in the appendices of
Refs.~\cite{B98quad} and~\cite{B98tail}. For the present computation the
crucial formula corresponds to a source term behaving like $1/r^2$:
\begin{equation}
  \Box^{-1}_\mathrm{ret} \left[\frac{\hat n_L}{r^2} F (t- r) \right] =
  - \hat n_L \int^{+\infty}_1 \!\!\!\! dx  \, Q_{l}(x) F (t-r x),
  \label{87}
\end{equation}
where $Q_l$ is the Legendre function of the second
  kind\epubtkFootnote{The function $Q_l$ is given in terms of the
  Legendre polynomial $P_l$ by
  \begin{displaymath}
    Q_{l} (x) = \frac{1}{2} \int^1_{-1} \frac{dz \, P_l (z)}{x- z}
    = \frac{1}{2} P_l (x) \,  \mathrm{ln} \left(\frac{x+1}{x-1} \right)-
    \sum^{l}_{ j=1} \frac{1}{j} P_{l -j}(x) P_{j-1}(x).
  \end{displaymath}
In the complex plane there is a branch cut from $-\infty$ to 1. The
first equality is known as the Neumann formula for the Legendre
function.}. With the help of this and other formulas we obtain the
object $u^{\alpha\beta}_2$ given by Equation~(\ref{32}). Next we compute the
divergence $w^{\alpha}_2=\partial_\mu u^{\alpha\mu}_2$, and obtain the
supplementary term $v^{\alpha\beta}_2$ by applying Equations~(\ref{35}).
Actually, we find for this particular interaction $w^{\alpha}_2=0$ and
thus also $v^{\alpha\beta}_2=0$. Following Equation~(\ref{36}), the result is
the sum of $u^{\alpha\beta}_2$ and $v^{\alpha\beta}_2$, and we get
\begin{equation}
  \begin{array}{rcl}
    \mathrm{M}^{-1} h^{00}_{(\mathrm{M}\mathrm{M}_{ab})} &=& \displaystyle
    n_{ab} r^{-4}
    \left[ -21 \mathrm{M}_{ab} -21 r \mathrm{M}^{(1)}_{ab} + 7 r^2
    \mathrm{M}^{(2)}_{ab} + 10 r^3 \mathrm{M}^{(3)}_{ab} \right]
    \\ [1.5 em]
    && \displaystyle
    +\, 8 n_{ab}\int^{+\infty}_1 \!\!\!\!\! dx \, Q_2 (x)
    \mathrm{M}^{(4)}_{ab}(t - rx),
    \\ [1.5 em]
    \mathrm{M}^{-1} h^{0i}_{(\mathrm{M}\mathrm{M}_{ab})} &=& \displaystyle
    n_{iab} r^{-3}
    \left[ -\mathrm{M}^{(1)}_{ab} - r \mathrm{M}^{(2)}_{ab} - \frac{1}{3}
    r^2 \mathrm{M}^{(3)}_{ab} \right]
    \\ [1.5 em]
    && \displaystyle
    +\, n_a r^{-3} \left[ -5 \mathrm{M}^{(1)}_{ai} - 5 r
    \mathrm{M}^{(2)}_{ai} + \frac{19}{3} r^2 \mathrm{M}^{(3)}_{ai} \right]
    \\ [1.5 em]
    && \displaystyle
    +\, 8 n_a \int^{+\infty}_1 \!\!\!\!\! dx \, Q_1 (x)
    \mathrm{M}^{(4)}_{ai} (t - rx),
    \\ [1.5 em]
    \mathrm{M}^{-1} h^{ij}_{(\mathrm{M}\mathrm{M}_{ab})} &=& \displaystyle
    n_{ijab} r^{-4} \left[ -\frac{15}{2} \mathrm{M}_{ab} - \frac{15}{2}
    r \mathrm{M}^{(1)}_{ab} - 3 r^2 \mathrm{M}^{(2)}_{ab} - \frac{1}{2}
    r^3 \mathrm{M}^{(3)}_{ab} \right]
    \\ [1.5 em]
    && \displaystyle
    +\, \delta_{ij} n_{ab} r^{-4} \left[ -\frac{1}{2} \mathrm{M}_{ab} -
    \frac{1}{2} r \mathrm{M}^{(1)}_{ab} - 2 r^2 \mathrm{M}^{(2)}_{ab} -
    \frac{11}{6} r^3 \mathrm{M}^{(3)}_{ab} \right]
    \\ [1.5 em]
    && \displaystyle
    +\, n_{a(i}r^{-4} \left[ 6 \mathrm{M}_{j)a} + 6 r
    \mathrm{M}^{(1)}_{j)a} + 6 r^2 \mathrm{M}^{(2)}_{j)a} + 4 r^3
    \mathrm{M}^{(3)}_{j)a} \right]
    \\ [1.5 em]
    && \displaystyle
    +\, r^{-4} \left[ - \mathrm{M}_{ij} - r\mathrm{M}^{(1)}_{ij} - 4 r^2
    \mathrm{M}^{(2)}_{ij} - \frac{11}{3} r^3 \mathrm{M}^{(3)}_{ij} \right]
    \\ [1.5 em]
    && \displaystyle
    +\, 8 \int^{+\infty}_1 \!\!\!\!\! dx \, Q_0 (x)
    \mathrm{M}^{(4)}_{ij} (t - rx).
  \end{array}
  \label{88}
\end{equation}
The metric is composed of two types of terms: ``instantaneous'' ones
depending on the values of the quadrupole moment at the retarded time
$u=t-r$, and ``non-local'' or tail integrals, depending on all previous
instants $t-rx\leq u$.

Let us investigate now the cubic interaction between \emph{two} mass
monopoles $\mathrm{M}$ with the quadrupole $\mathrm{M}_{ab}$. Obviously,
the source term corresponding to this interaction reads
\begin{eqnarray}
  \Lambda^{\alpha\beta}_{(\mathrm{M}^2\mathrm{M}_{ab})} &=&
  N^{\alpha\beta}[h_{(\mathrm{M})}, h_{(\mathrm{M}\mathrm{M}_{ab})}]+
  N^{\alpha\beta}[h_{(\mathrm{M}\mathrm{M}_{ab})}, h_{(\mathrm{M})}] +
  N^{\alpha\beta}[h_{(\mathrm{M}^2)}, h_{(\mathrm{M}_{ab})}]+
  N^{\alpha\beta}[h_{(\mathrm{M}_{ab})}, h_{(\mathrm{M}^2)}]
  \nonumber \\
  && +\, M^{\alpha\beta}[h_{(\mathrm{M})}, h_{(\mathrm{M})},
  h_{(\mathrm{M}_{ab})}]+ M^{\alpha\beta}[h_{(\mathrm{M})},
  h_{(\mathrm{M}_{ab})}, h_{(\mathrm{M})}] +
  M^{\alpha\beta}[h_{(\mathrm{M}_{ab})}, h_{(\mathrm{M})},
  h_{(\mathrm{M})}]
  \label{89}
\end{eqnarray}%
(see Equation~(\ref{28b})). Notably, the $N$-terms in Equation~(\ref{89}) involve
the interaction between a linearized metric, $h_{(\mathrm{M})}$ or
$h_{(\mathrm{M}_{ab})}$, and a quadratic one, $h_{(\mathrm{M}^2)}$ or
$h_{(\mathrm{M}\mathrm{M}_{ab})}$. So, included into these terms are the
tails present in the quadratic metric $h_{(\mathrm{M}\mathrm{M}_{ab})}$
computed previously with the result~(\ref{88}). These tails will produce
in turn some ``tails of tails'' in the cubic metric
$h_{(\mathrm{M}^2\mathrm{M}_{ab})}$. The rather involved computation
will not be detailed here (see Ref.~\cite{B98tail}). Let us just mention
the most difficult of the needed integration
formulas\epubtkFootnote{Equation~(\ref{90}) has been obtained using a not so
well known mathematical relation between the Legendre functions and
polynomials:
  \begin{displaymath}
    \frac{1}{2} \int^1_{-1} \frac{dz \, P_l (z)}{\sqrt{(xy - z)^2 - (x^2 -
        1) (y^2 - 1)}} = Q_l (x) P_l (y)
  \end{displaymath}
(where $1 \leq y < x$ is assumed). See Appendix~A in Ref.~\cite{B98tail}
for the proof. This relation constitutes a generalization of the Neumann
formula (see footnote after Equation~(\ref{87})).}:
\begin{eqnarray}
  && {\cal FP} \; \Box^{-1}_\mathrm{ret} \!
  \left[\frac{\hat{n}_L}{r} \int^{+\infty}_1 \!\!\!\!\! dx \,
  Q_m(x) F (t - rx) \right] =
  \hat{n}_L \int^{+\infty}_1 \!\!\!\!\! dy \, F^{(-1)} (t - ry)
  \nonumber \\
  && \qquad \qquad \;
  \times \left\{ Q_l (y) \! \int^y_1 dx \,
  Q_m (x) \frac{dP_l}{dx} (x) +
  P_l (y) \! \int^{+\infty}_y \!\!\!\!\! dx \,
  Q_m (x) \frac{dQ_l}{dx} (x) \right\}\!, \qquad
  \label{90}
\end{eqnarray}%
where $F^{(-1)}$ is the time anti-derivative of $F$. With this formula
and others given in Ref.~\cite{B98tail} we are able to obtain the closed
algebraic form of the metric $h^{\alpha\beta}_{(\mathrm{M}^2
\mathrm{M}_{ab})}$, at the leading order in the distance to the source.
The net result is
\begin{equation}
  \begin{array}{rcl}
    \mathrm{M}^{-2} h^{00}_{(\mathrm{M}^2 \mathrm{M}_{ab})} &=& \displaystyle
    \frac{n_{ab}}{r} \int^{+\infty}_0 \!\!\!\!\! d\tau \,
    \mathrm{M}^{(5)}_{ab} \left[ -4 \ln^2 \left( \frac{\tau}{2r}
    \right) -4 \ln \left( \frac{\tau}{2r} \right) +
    \frac{116}{21} \ln \left( \frac{\tau}{2r_0} \right) -
    \frac{7136}{2205} \right]
    \\ [1.2 em]
    && \displaystyle
    +\, {\cal O}\left(\frac{1}{r^2} \right),
    \\ [1.2 em]
    \mathrm{M}^{-2} h^{0i}_{(\mathrm{M}^2 \mathrm{M}_{ab})} &=& \displaystyle
    \frac{\hat{n}_{iab}}{r} \int^{+\infty}_0 \!\!\!\!\! d\tau \,
    \mathrm{M}^{(5)}_{ab} \left[ -\frac{2}{3} \ln \left(
    \frac{\tau}{2r} \right) - \frac{4}{105} \ln \left(
    \frac{\tau}{2r_0} \right) - \frac{716}{1225} \right]
    \\ [1.2 em]
    && \displaystyle
    + \frac{n_a}{r} \int^{+\infty}_0 \!\!\!\!\! d\tau \,
    \mathrm{M}^{(5)}_{ai} \left[ -4 \ln^2 \left( \frac{\tau}{2r}
    \right) - \frac{18}{5} \ln \left( \frac{\tau}{2r} \right) +
    \frac{416}{75} \ln \left( \frac{\tau}{2r_0}  \right) -
    \frac{22724}{7875} \right]
    \\ [1.2 em]
    && \displaystyle
    +\, {\cal O}\left(\frac{1}{r^2} \right),
    \\ [1.2 em]
    \mathrm{M}^{-2} h^{ij}_{(\mathrm{M}^2 \mathrm{M}_{ab})} &=& \displaystyle
    \frac{\hat{n}_{ijab}}{r} \int^{+\infty}_0 \!\!\!\!\! d\tau \,
    \mathrm{M}^{(5)}_{ab} \left[ - \ln \left( \frac{\tau}{2r} \right) -
    \frac{191}{210} \right]
    \\ [1.2 em]
    && \displaystyle
    + \frac{\delta_{ij} n_{ab}}{r} \int^{+\infty}_0 \!\!\!\!\!
    d\tau \, \mathrm{M}^{(5)}_{ab} \left[ -\frac{80}{21} \ln \left(
    \frac{\tau}{2r} \right) - \frac{32}{21} \ln \left(
    \frac{\tau}{2r_0} \right) - \frac{296}{35} \right]
    \\ [1.2 em]
    && \displaystyle
    + \frac{\hat{n}_{a(i}}{r} \int^{+\infty}_0 \!\!\!\!\! d\tau \,
    \mathrm{M}^{(5)}_{j)a} \left[ \frac{52}{7} \ln \left(
    \frac{\tau}{2r} \right) + \frac{104}{35} \ln \left(
    \frac{\tau}{2r_0} \right) + \frac{8812}{525} \right]
    \\ [1.2 em]
    && \displaystyle
    + \frac{1}{r} \int^{+\infty}_0 \!\!\!\!\! d\tau \,
    \mathrm{M}^{(5)}_{ij} \left[ -4 \ln^2 \left( \frac{\tau}{2r}
    \right) - \frac{24}{5} \ln \left( \frac{\tau}{2r} \right) +
    \frac{76}{15} \ln \left( \frac{\tau}{2r_0}
    \right) - \frac{198}{35} \right]
    \\ [1.2 em]
    && \displaystyle
    +\, {\cal O}\left(\frac{1}{r^2} \right),
    \end{array}
  \label{91}
\end{equation}
where all the moments $\mathrm{M}_{ab}$ are evaluated at the instant
$t-r-\tau$ (recall that $c=1$). Notice that some of the logarithms in
Equations~(\ref{91}) contain the ratio $\tau /r$ while others involve $\tau
/r_0$. The indicated remainders ${\cal O}(1/r^2)$ contain some
logarithms of $r$; in fact they should be more accurately written as
$o(r^{\epsilon-2})$ for some $\epsilon\ll 1$.

The presence of logarithms of $r$ in Equations~(\ref{91}) is an artifact of
the harmonic coordinates $x^\alpha$, and we need to gauge them away by
introducing the radiative coordinates $X^\alpha$ at future null infinity
(see Theorem~\ref{th4}). As it turns out, it is sufficient for the
present calculation to take into account the ``linearized'' logarithmic
deviation of the light cones in harmonic coordinates so that $X^\alpha =
x^\alpha + G\xi^\alpha_1 +{\cal O}(G^2)$, where $\xi^\alpha_1$ is the
gauge vector defined by Equation~(\ref{43}) (see also Equation~(\ref{80_1})). With
this coordinate change one removes all the logarithms of $r$ in
Equations~(\ref{91}). Hence, we obtain the radiative metric
\begin{equation}
  \begin{array}{rcl}
    \mathrm{M}^{-2} H^{00}_{(\mathrm{M}^2 \mathrm{M}_{ab})} &=&
    \displaystyle
    {N_{ab}\over R} \int^{+\infty}_0 \!\!\!\!\! d\tau \,
    \mathrm{M}^{(5)}_{ab} \left[ - 4 \ln^2 \left( {\tau\over 2r_0}
    \right) + {32\over 21} \ln \left( {\tau\over 2r_0} \right) -
    {7136\over 2205} \right]
    \\ [1.5 em]
    && \displaystyle
    +\, {\cal O} \left( \frac{1}{R^2} \right),
    \\ [1.5 em]
    \mathrm{M}^{-2} H^{0i}_{(\mathrm{M}^2 \mathrm{M}_{ab})} &=&
    \displaystyle
    {\hat{N}_{iab}\over R} \int^{+\infty}_0 \!\!\!\!\! d\tau \,
    \mathrm{M}^{(5)}_{ab} \left[ - {74\over 105} \ln \left(
    {\tau\over 2r_0} \right) - {716\over 1225} \right]
    \\ [1.5 em]
    && \displaystyle
    +\, {N_a\over R} \int^{+\infty}_0 \!\!\!\!\! d\tau \,
    \mathrm{M}^{(5)}_{ai} \left[ - 4 \ln^2 \left( {\tau\over 2r_0}
    \right) + {146\over 75} \ln \left( {\tau\over 2r_0} \right) - 
    {22724\over 7875} \right]
    \\ [1.5 em]
    && \displaystyle
    +\, {\cal O} \left( \frac{1}{R^2} \right),
    \\ [1.5 em]
    \mathrm{M}^{-2} H^{ij}_{(\mathrm{M}^2 \mathrm{M}_{ab})} &=&
    \displaystyle
    {\hat{N}_{ijab}\over R} \int^{+\infty}_0 \!\!\!\!\! d\tau \,
    \mathrm{M}^{(5)}_{ab} \left[ - \ln \left( {\tau\over 2r_0}
    \right) - {191\over 210} \right]
    \\ [1.5 em]
    && \displaystyle
    +\, {\delta_{ij} N_{ab}\over R} \int^{+\infty}_0 \!\!\!\!\! d\tau
    \, \mathrm{M}^{(5)}_{ab} \left[ - {16\over 3} \ln \left(
    {\tau\over 2r_0} \right) - {296\over 35} \right]
    \\ [1.5 em]
    && \displaystyle
    +\, {\hat{N}_{a(i}\over R} \int^{+\infty}_0 \!\!\!\!\! d\tau \,
    \mathrm{M}^{(5)}_{j)a} \left[ {52\over 5} \ln \left(
    {\tau\over 2r_0} \right) + {8812\over 525} \right]
    \\ [1.5 em]
    && \displaystyle
    +\, {1\over R}
    \int^{+\infty}_0 \!\!\!\!\! d\tau \, \mathrm{M}^{(5)}_{ij}
    \left[ - 4 \ln^2 \left( {\tau\over 2r_0} \right) +
    {4\over 15} \ln \left( {\tau\over 2r_0} \right) - {198\over 35}
    \right]
    \\ [1.5 em]
    && \displaystyle
    +\, {\cal O} \left( \frac{1}{R^2} \right),
  \end{array}
\end{equation}
where the moments are evaluated at time $U-\tau\equiv T-R-\tau$. It is
trivial to compute the contribution of the radiative moments
$\mathrm{U}_L(U)$ and $\mathrm{V}_L(U)$ corresponding to that metric. We
find the ``tail of tail'' term reported in Equation~(\ref{78}).

\newpage


\section{The Third Post-Newtonian Metric}
\label{sec:7}

The detailed calculations that are called for in applications
necessitate having at one's disposal some explicit expressions of the
metric coefficients $g_{\alpha\beta}$, in harmonic coordinates, at the
highest possible post-Newtonian order. The 3PN metric that we present
below\epubtkFootnote{Actually, such a metric is valid up to 3.5PN order.}
is expressed by means of some particular retarded-type potentials, $V$,
$V_i$, ${\hat W}_{ij}$, etc., whose main advantages are to
somewhat minimize the number of terms, so that even at the 3PN order the
metric is still tractable, and to delineate the different problems
associated with the computation of different categories of terms. Of
course, these potentials have no physical significance by themselves.
The basic idea in our post-Newtonian iteration is to use whenever
possible a ``direct'' integration, with the help of some formulas like
$\Box^{-1}_\mathrm{ret}(\partial_\mu V\partial^\mu V+V\Box V) = V^2 / 2$.
The 3PN harmonic-coordinates metric (issued from Ref.~\cite{BFeom})
reads
\begin{equation}
  \begin{array}{rcl}
  g_{00} &=& \displaystyle - 1 + \frac{2}{c^2} V - \frac{2}{c^4} V^2 +
  \frac{8}{c^6} \left( \hat{X} + V_i V_i + \frac{V^3}{6} \right)
  + \frac{32}{c^8} \left( \hat{T} - \frac{1}{2} V
  \hat{X} + \hat{R}_i V_i - \frac{1}{2} V V_i V_i - \frac{1}{48} V^4
  \right)
  \\ [1.0 em]
  & & \displaystyle + {\cal O} \left( \frac{1}{c^{10}} \right),
  \\ [1.5 em]
  g_{0i} &=& \displaystyle - \frac{4}{c^3} V_i - \frac{8}{c^5} \hat{R}_i -
  \frac{16}{c^7} \left( \hat{Y}_i + \frac{1}{2} \hat{W}_{ij} V_j +
  \frac{1}{2} V^2 V_i \right) + {\cal O}\left(\frac{1}{c^9}\right),
  \\ [1.5 em]
  g_{ij} &=& \displaystyle \delta_{ij} \left[ 1 + \frac{2}{c^2}
  V + \frac{2}{c^4} V^2 + \frac{8}{c^6} \left(\hat{X} + V_k V_k +
  \frac{V^3}{6} \right) \right] +
  \frac{4}{c^4} \hat{W}_{ij} + \frac{16}{c^6} \left( \hat{Z}_{ij} +
  \frac{1}{2} V \hat{W}_{ij} - V_i V_j \right)
  \\ [1.0 em]
  && \displaystyle
  + {\cal O}\left( \frac{1}{c^8} \right).
  \end{array}
  \label{93}
\end{equation}
All the potentials are generated by the matter stress-energy tensor
$T^{\alpha\beta}$ through the definitions (analogous to
Equations~(\ref{72}))
\begin{equation}
  \begin{array}{rcl}
    \sigma &=& \displaystyle
    \frac{T^{00}+T^{ii}}{c^2},
    \\ [1.0 em]
    \sigma_{i} &=& \displaystyle
    \frac{T^{0i}}{c},
    \\ [1.0 em]
    \sigma_{ij} &=& \displaystyle
    T^{ij}.
  \end{array}
  \label{94}
\end{equation}
$V$ and $V_i$ represent some retarded versions of the Newtonian and
gravitomagnetic potentials,
\begin{equation}
  \begin{array}{rcl}
    V &=& \displaystyle
    \Box^{-1}_\mathrm{ret} \left[ -4 \pi G \sigma \right],
    \\ [1.0 em]
    V_i &=& \displaystyle
    \Box^{-1}_\mathrm{ret} \left[ -4 \pi G \sigma_i \right].
  \end{array}
  \label{95}
\end{equation}
From the 2PN order we have the potentials
\begin{equation}
  \begin{array}{rcl}
    \hat{X} &=& \displaystyle
    \Box^{-1}_\mathrm{ret}
    \left[ - 4 \pi G V \sigma_{ii}+\hat{W}_{ij} \partial_{ij}
    V+2 V_i \partial_t \partial_i V+ V \partial_t^2 V +
    \frac{3}{2} (\partial_t V)^2 - 2 \partial_i V_j \partial_j V_i \right],
    \\ [1.5 em]
    \hat{R}_i &=& \displaystyle
    \Box^{-1}_\mathrm{ret}
    \left[ -4 \pi G (V \sigma_i-V_i \sigma)-2 \partial_k V
    \partial_i V_k-\frac{3}{2} \partial_t V \partial_i V \right],
    \\ [1.5 em]
    \hat{W}_{ij} &=& \displaystyle
    \Box^{-1}_\mathrm{ret}
    \left[ -4 \pi G (\sigma_{ij}-\delta_{ij} \sigma_{kk})-
    \partial_i V \partial_j V \right].
  \end{array}
  \label{96}
\end{equation}
Some parts of these potentials are directly generated by compact-support
matter terms, while other parts are made of non-compact-support products
of $V$-type potentials. There exists also a very important cubically
non-linear term generated by the coupling between $\hat{W}_{ij}$ and
$V$, the second term in the ${\hat X}$-potential. At the 3PN level we
have the most complicated of these potentials, namely
\begin{equation}
  \begin{array}{rcl}
    \hat{T} &=& \displaystyle
    \Box^{-1}_\mathrm{ret} \biggl[ -4 \pi G \left(
    \frac{1}{4} \sigma_{ij} \hat{W}_{ij}+\frac{1}{2} V^2 \sigma_{ii} +
    \sigma V_i V_i \right) + \hat{Z}_{ij} \partial_{ij} V + \hat{R}_i \partial_t
    \partial_i V - 2 \partial_i V_j \partial_j \hat{R}_i -
    \partial_i V_j \partial_t \hat{W}_{ij}
    \\ [1.5 em]
    && \qquad \:\: \displaystyle
    +\, V V_i \partial_t \partial_i V + 2 V_i \partial_j V_i \partial_j
    V + \frac{3}{2} V_i \partial_t V \partial_i V+\frac{1}{2} V^2
    \partial^2_t V + \frac{3}{2} V (\partial_t V)^2 -
    \frac{1}{2} (\partial_t V_i)^2 \biggr]\!,
    \\ [1.5 em]
    \hat{Y}_i &=& \displaystyle
    \Box^{-1}_\mathrm{ret} \biggl[ -4 \pi G \left(
    - \sigma \hat{R}_i - \sigma V V_i + \frac{1}{2} \sigma_k
    \hat{W}_{ik} + \frac{1}{2} \sigma_{ik} V_k + \frac{1}{2}
    \sigma_{kk} V_i \right)
    \\ [1.5 em]
    && \qquad \:\: \displaystyle
    +\, \hat{W}_{kl} \partial_{kl} V_i - \partial_t \hat{W}_{ik}
    \partial_k V + \partial_i \hat{W}_{kl} \partial_k V_l -
    \partial_k \hat{W}_{il} \partial_l V_k -
    2 \partial_k V \partial_i \hat{R}_k - \frac{3}{2} V_k \partial_i V
    \partial_k V 
    \\ [1.5 em]
    && \qquad \:\: \displaystyle
    -\, \frac{3}{2} V \partial_t V \partial_i V -
    2 V \partial_k V \partial_k V_i + V \partial^2_t V_i +
    2 V_k \partial_k \partial_t V_i \biggr],
    \\ [1.5 em]
    \hat{Z}_{ij} &=& \displaystyle
    \Box^{-1}_\mathrm{ret} \biggl[ -4 \pi G V
    \left( \sigma_{ij} - \delta_{ij} \sigma_{kk} \right) -
    2 \partial_{(i} V \partial_t V_{j)} + \partial_i V_k
    \partial_j V_k + \partial_k V_i \partial_k V_j -
    2 \partial_{(i} V_k \partial_k V_{j)}
    \\ [1.5 em]
    && \qquad \:\: \displaystyle
    -\, \frac{3}{4} \delta_{ij} (\partial_t V)^2 -
    \delta_{ij} \partial_k V_m (\partial_k V_m-\partial_m V_k) \biggr],
  \end{array}
  \label{97}
\end{equation}
which involve many types of compact-support contributions, as well as
quadratic-order and cubic-order parts; but, surprisingly, there are
\emph{no} quartically non-linear terms\epubtkFootnote{It has been possible to
``integrate directly'' all the quartic contributions in the 3PN metric.
See the terms composed of $V^4$ and $V\hat{X}$ in the first of
Equations~(\ref{93}).}.

The above potentials are not independent. They are linked together by
some differential identities issued from the harmonic gauge conditions,
which are equivalent, \emph{via} the Bianchi identities, to the equations
of motion of the matter fields (see Equation~(\ref{15})). These identities
read
\begin{equation}
  \begin{array}{rcl}
    0 &=& \displaystyle
    \partial_t \left\{ V +\frac{1}{c^2} \left[ \frac{1}{2}
    \hat{W}_{kk}+ 2 V^2 \right] + \frac{4}{c^4} \left[ \hat{X} +
    \frac{1}{2} \hat{Z}_{kk}+\frac{1}{2} V \hat{W}_{kk} +
    \frac{2}{3} V^3 \right] \right\}
    \\ [1.5 em]
    && \displaystyle
    +\, \partial_i \left\{ V_i + \frac{2}{c^2} \left[\hat{R}_i +
    V V_i \right] + \frac{4}{c^4} \left[\hat{Y}_i - \frac{1}{2}
    \hat{W}_{ij} V_j + \frac{1}{2} \hat{W}_{kk} V_i + V \hat{R}_i +
    V^2 V_i\right] \right\}
    \\ [1.5 em]
    && \displaystyle
    +\, {\cal O}\left(\frac{1}{c^6}\right),
    \\ [1.5 em]
    0 &=& \displaystyle
    \partial_t \left\{ V_i + \frac{2}{c^2} \left[ \hat{R}_i +
    V V_i \right] \right\} + \partial_j \left\{ \hat{W}_{ij} -
    \frac{1}{2} \hat{W}_{kk}
    \delta_{ij} + \frac{4}{c^2} \left[\hat{Z}_{ij} - \frac{1}{2}
    \hat{Z}_{kk} \delta_{ij} \right] \right\}
    \\ [1.5 em]
    && \displaystyle +\, {\cal O}\left(\frac{1}{c^4}\right).
  \end{array}
\end{equation}

It is important to remark that the above 3PN metric represents the inner
post-Newtonian field of an \emph{isolated} system, because it contains,
to this order, the correct radiation-reaction terms corresponding to
outgoing radiation. These terms come from the expansions of the
retardations in the retarded-type potentials~(\ref{95}, \ref{96},
\ref{97}).

\newpage


\part{Compact Binary Systems}
\label{part:b}

The problem of the motion and gravitational radiation of compact objects
in post-Newtonian approximations of general relativity is of crucial
importance, for at least three reasons. First, the motion of $N$ objects
at the 1PN level ($1/c^2$), according to the Einstein--Infeld--Hoffmann
equations~\cite{EIH}, is routinely taken into account to describe the
Solar System dynamics (see Ref.~\cite{NSW}). Second, the gravitational
radiation-reaction force, which appears in the equations of motion at
the 2.5PN order, has been experimentally verified, by the observation of
the secular acceleration of the orbital motion of the binary pulsar
PSR~1913+16~\cite{TFMc79, TW82, T93}.

Last but not least, the forthcoming detection and analysis of
gravitational waves emitted by inspiralling compact binaries -- two
neutron stars or black holes driven into coalescence by emission of
gravitational radiation -- will necessitate the prior knowledge of the
equations of motion and radiation field up to high post-Newtonian order.
As discussed in the introduction in Section~\ref{sec:1} (see around
Equations~(\ref{6}, \ref{7},
\ref{8})), the appropriate theoretical description of inspiralling
compact binaries is by two structureless point-particles, characterized
solely by their masses $m_1$ and $m_2$ (and possibly their spins), and
moving on a quasi-circular orbit. Strategies to detect and analyze the
very weak signals from compact binary inspiral involve matched filtering
of a set of accurate theoretical template waveforms against the output
of the detectors. Several analyses~\cite{3mn, CFPS93, FCh93, CF94,
TNaka94, P95, PW95, KKS95, DIS98, DIS00, BCV03a, BCV03b, DIJS03, AIRS05,
AISS05} have shown that, in order to get sufficiently accurate
theoretical templates, one must include post-Newtonian effects up to the
3PN level at least.

To date, the templates have been completed through 3.5PN order for the
phase evolution~\cite{BDIWW95, BFIJ02, BDEI04}, and 2.5PN order for the
amplitude corrections~\cite{BIWW96, ABIQ04}. Spin effects are known for
the dominant relativistic spin-orbit coupling term at 1.5PN order and
the spin-spin coupling term at 2PN order~\cite{KWW93, ACST94, K95,
cho98}, and also for the next-to-leading spin-orbit coupling at 2.5PN
order~\cite{OTO98, TOO01, FBB06spin, BBF06spin}.

\newpage


\section{Regularization of the Field of Point Particles}
\label{sec:8}

Our aim is to compute the metric (and its gradient needed in the
equations of motion) at the 3PN order for a system of two point-like
particles. \emph{A priori} one is not allowed to use directly the metric
expressions~(\ref{93}), as they have been derived under the assumption
of a continuous (smooth) matter distribution. Applying them to a system
of point particles, we find that most of the integrals become divergent
at the location of the particles, i.e.\ when ${\bf x}\to{\bf
y}_1(t)$ or ${\bf y}_2(t)$, where ${\bf y}_1(t)$ and ${\bf y}_2(t)$
denote the two trajectories. Consequently, we must supplement the
calculation by a prescription for how to remove the ``infinite part'' of
these integrals. At this stage different choices for a ``self-field''
regularization (which will take care of the infinite self-field of point
particles) are possible. In this section we review:
\begin{enumerate}
\item Hadamard's self-field regularization, which has proved to be very
  convenient for doing practical computations (in particular, by
  computer), but suffers from the important drawback of yielding some
  ambiguity parameters, which cannot be determined within this
  regularization, at the 3PN order;
\item Dimensional self-field regularization, an extremely powerful
  regularization which is free of any ambiguities (at least up to the
  3PN level), and permits therefore to uniquely fix the values of the
  ambiguity parameters coming from Hadamard's regularization. However,
  dimensional regularization has not yet been implemented to the present
  problem in the general case (i.e.\ for an arbitrary space
  dimension $d\in\mathbb{C}$).
\end{enumerate}
The why and how the final results are unique and independent of the
employed self-field regularization (in agreement with the physical
expectation) stems from the effacing principle of general
relativity~\cite{D83houches} -- namely that the internal structure of
the compact bodies makes a contribution only at the formal 5PN
approximation. However, we shall review several alternative
computations, independent of the self-field regularization, which
confirm the end results.


\subsection{Hadamard self-field regularization}
\label{subsec:8.1}

In most practical computations we employ the Hadamard
regularization~\cite{Hadamard, Schwartz} (see Ref.~\cite{Sellier} for an
entry to the mathematical literature). Let us present here an account of
this regularization, as well as a theory of generalized functions (or
pseudo-functions) associated with it, following the investigations
detailed in Refs.~\cite{BFreg, BFregM}.

Consider the class ${\cal F}$ of functions $F({\bf x})$ which are smooth
($C^\infty$) on ${\mathbb{R}}^3$ \emph{except} for the two points ${\bf
y}_1$ and ${\bf y}_2$, around which they admit a power-like singular
expansion of the type\epubtkFootnote{The function $F(\mathbf{x})$
depends also on time $t$, through for instance its dependence on the
velocities $\mathbf{v}_1(t)$ and $\mathbf{v}_2(t)$, but the (coordinate)
$t$ time is purely ``spectator'' in the regularization process, and thus
will not be indicated.}
\begin{equation}
  \forall n\in {\mathbb{N}},
  \qquad
  F({\bf x}) = \!\!\!\!\! \sum_{a_0\leq a\leq n} \!\!\!
  r_1^a \, \mathop{f}_{1}{}_{\!\!a}({\bf n}_1)+o(r_1^n),
  \label{99}
\end{equation}
and similarly for the other point 2. Here $r_1=|{\bf x}-{\bf y}_1|\to
0$, and the coefficients ${}_1f_a$ of the various powers of $r_1$ depend
on the unit direction ${\bf n}_1=({\bf x}-{\bf y}_1)/r_1$ of approach to
the singular point. The powers $a$ of $r_1$ are real, range in discrete
steps (i.e.\ $a\in (a_i)_{i\in \mathbb{N}}$), and are bounded from
below ($a_0\leq a$). The coefficients ${}_1f_a$ (and ${}_2f_a$) for
which $a<0$ can be referred to as the \emph{singular} coefficients of
$F$. If $F$ and $G$ belong to ${\cal F}$ so does the ordinary product
$FG$, as well as the ordinary gradient $\partial_iF$. We define the
Hadamard \emph{partie finie} of $F$ at the location of the point 1 where
it is singular as
\begin{equation}
  (F)_1= \int \frac{d\Omega_1}{ 4\pi}\,\mathop{f}_{1}{}_{\!\!0}({\bf n}_1),
  \label{100}
\end{equation}
where $d\Omega_1= d\Omega ({\bf n}_1)$ denotes the solid angle element
centered on ${\bf y}_1$ and of direction ${\bf n}_1$. Notice that
because of the angular integration in Equation~(\ref{100}), the
Hadamard \emph{partie finie} is ``non-distributive'' in the sense that 
\begin{equation}
  (FG)_1\not= (F)_1(G)_1
  \qquad
  \mathrm{in\ general}.
  \label{100'}
\end{equation}
The non-distributivity of Hadamard's partie finie is the main source of
the appearance of ambiguity parameters at the 3PN order, as discussed in
Section~\ref{subsec:8.2}.

The second notion of Hadamard \emph{partie finie} ($\pf$) concerns that
of the integral $\int d^3{\bf x} \, F$, which is generically divergent
at the location of the two singular points ${\bf y}_1$ and ${\bf y}_2$
(we assume that the integral converges at infinity). It is defined by
\begin{equation}
  \pf_{s_1 s_2} \int d^3{\bf x} \, F = \lim_{s \to 0} \,
  \left\{\int_{{\cal S}(s)} \!\!\! d^3{\bf x} \, F + 4\pi\sum_{a+3<
  0}{\frac{s^{a+3}}{a+3}} \left( \frac{F}{r_1^a} \right)_1 + 4 \pi \ln
  \left(\frac{s}{s_1}\right) \left(r_1^3 F\right)_1 + 1\leftrightarrow
  2\right\}.
  \label{101}
\end{equation}
The first term integrates over a domain ${\cal S}(s)$ defined as
${\mathbb{R}}^3$ from which the two spherical balls $r_1\leq s$ and
$r_2\leq s$ of radius $s$ and centered on the two singularities, denoted
${\cal B}({\bf y}_1, s)$ and ${\cal B}({\bf y}_2, s)$, are excised:
${\cal S}(s)\equiv {\mathbb{R}}^3\setminus {\cal B}({\bf y}_1, s)\cup
{\cal B}({\bf y}_2, s)$. The other terms, where the value of a function
at point 1 takes the meaning~(\ref{100}) are such that they cancel out
the divergent part of the first term in the limit where $s\to 0$ (the
symbol $1\leftrightarrow 2$ means the same terms but corresponding to
the other point 2). The Hadamard partie-finie integral depends on two
strictly positive constants $s_1$ and $s_2$, associated with the
logarithms present in Equation~(\ref{101}). These constants will ultimately
yield some gauge-type constants, denoted by $r'_1$ and $r'_2$, in the
3PN equations of motion and radiation field. See Ref.~\cite{BFreg} for
alternative expressions of the partie-finie integral.

We now come to a specific variant of Hadamard's regularization called
the extended Hadamard regularization and defined in Refs.~\cite{BFreg,
BFregM}. The basic idea is to associate to any $F\in {\cal F}$ a
\emph{pseudo-function}, called the \emph{partie finie} pseudo-function $\pf F$,
namely a linear form acting on functions $G$ of ${\cal F}$, and which is
defined by the duality bracket
\begin{equation}
  \forall G\in {\cal F},
  \qquad
  \langle \pf F, G \rangle = \pf \int d^3{\bf x} \, FG.
  \label{102}
\end{equation}
When restricted to the set ${\cal D}$ of smooth functions (i.e.\
$C^\infty(\mathbb{R}^4)$) with compact support (obviously we have ${\cal
D}\subset {\cal F}$), the pseudo-function $\pf F$ is a distribution in
the sense of Schwartz~\cite{Schwartz}. The product of pseudo-functions
coincides, by definition, with the ordinary pointwise product, namely
$\pf F.\pf G=\pf (FG)$. In practical computations, we use an interesting
pseudo-function, constructed on the basis of the Riesz delta
function~\cite{Riesz}, which plays a role analogous to the Dirac measure
in distribution theory, $\delta_1({\bf x})\equiv \delta({\bf x}-{\bf
y}_1)$. This is the so-called delta-pseudo-function $\pf \delta_1$
defined by
\begin{equation}
  \forall F\in {\cal F},
  \qquad
  \langle \pf \delta_1, F \rangle =\pf \int d^3{\bf x} \, \delta_1 F=(F)_1,
  \label{103}
\end{equation}
where $(F)_1$ is the \emph{partie finie} of $F$ as given by
Equation~(\ref{100}). From the product of $\pf \delta_1$ with any $\pf F$ we
obtain the new pseudo-function $\pf (F\delta_1)$, that is such that
\begin{equation}
  \forall G\in {\cal F},
  \qquad
  \langle \pf (F\delta_1),G \rangle =(FG)_1.
  \label{104}
\end{equation}
As a general rule, we are not allowed, in consequence of the
``non-distributivity'' of the Hadamard partie finie, Equation~(\ref{100'}),
to replace $F$ within the pseudo-function $\pf (F\delta_1)$ by its
regularized value: $\pf (F\delta_1)\not= (F)_1\,\pf \delta_1$ in
general. It should be noticed that the object $\pf (F\delta_1)$ has no
equivalent in distribution theory.

Next, we treat the spatial derivative of a pseudo-function of the type
$\pf F$, namely $\partial_i(\pf F)$. Essentially, we require (in
Ref.~\cite{BFreg}) that the so-called rule of integration by parts
holds. By this we mean that we are allowed to freely operate by parts
any duality bracket, with the all-integrated (``surface'') terms always
zero, as in the case of non-singular functions. This requirement is
motivated by our will that a computation involving singular functions be
as much as possible the same as if we were dealing with regular
functions. Thus, by definition,
\begin{equation}
  \forall F, G\in {\cal F},\quad \langle \partial_i(\pf F),G \rangle = -
  \langle \partial_i(\pf G),F \rangle.
  \label{105}
\end{equation}
Furthermore, we assume that when all the singular coefficients of $F$
vanish, the derivative of $\pf F$ reduces to the ordinary derivative,
i.e.\ $\partial_i(\pf F)=\pf (\partial_iF)$. Then it is trivial to
check that the rule~(\ref{105}) contains as a particular case the
standard definition of the distributional derivative~\cite{Schwartz}.
Notably, we see that the integral of a gradient is always zero: $\langle
\partial_i(\pf F),1 \rangle =0$. This should certainly be the case if we
want to compute a quantity (e.g., a Hamiltonian density) which is
defined only modulo a total divergence. We pose
\begin{equation}
  \partial_i (\pf  F) = \pf (\partial_i F) + \mathrm{D}_i[F],
  \label{106}
\end{equation}
where $\pf (\partial_i F)$ represents the ``ordinary'' derivative and
$\mathrm{D}_i[F]$ the distributional term. The following solution of the
basic relation~(\ref{105}) was obtained in Ref.~\cite{BFreg}:
\begin{equation}
  \mathrm{D}_i[F] = 4\pi \, \pf  \Biggl( n_1^i \biggl[
  \frac{1}{2} \, r_1 \, \mathop{f}_{1}{}_{\!\!-1}+\sum_{k\geq 0}
  \frac{1}{r_1^k} \, \mathop{f}_{1}{}_{\!\!-2-k}\biggl] \delta_1 \Biggr) +
  1 \leftrightarrow 2,
  \label{107}
\end{equation}
where for simplicity we assume that the powers $a$ in the
expansion~(\ref{99}) of $F$ are relative integers. The distributional
term~(\ref{107}) is of the form $\pf (G\delta_1)$ (plus
$1\leftrightarrow 2$). It is generated solely by the singular
coefficients of $F$ (the sum over $k$ in Equation~(\ref{107}) is always
finite since there is a maximal order $a_0$ of divergency in
Equation~(\ref{99})). The formula for the distributional term associated with
the $l$th distributional derivative, i.e.\
$\mathrm{D}_L[F]=\partial_L\pf F-\pf \partial_L F$, where $L=i_1i_2\dots
i_l$, reads
\begin{equation}
  \mathrm{D}_{L}[F]=\sum_{k=1}^l\partial_{i_1\dots i_{k-1}}
  \mathrm{D}_{i_{k}}[\partial_{i_{k+1}\dots i_{l}}F].
  \label{108}
\end{equation}
We refer to Theorem~4 in Ref.~\cite{BFreg} for the definition of another
derivative operator, representing the most general derivative satisfying
the same properties as the one defined by Equation~(\ref{107}), and, in
addition, the commutation of successive derivatives (or Schwarz
lemma)\epubtkFootnote{It was shown in Ref.~\cite{BFeom} that using one
or the other of these derivatives results in some equations of motion
that differ by a mere coordinate transformation. This result indicates
that the distributional derivatives introduced in Ref.~\cite{BFreg}
constitute merely some technical tools which are devoid of physical
meaning.}.

The distributional derivative~(\ref{106}, \ref{107}, \ref{108}) does not
satisfy the Leibniz rule for the derivation of a product, in accordance
with a general result of Schwartz~\cite{Schwartz54}. Rather, the
investigation~\cite{BFreg} suggests that, in order to construct a
consistent theory (using the ``ordinary'' product for pseudo-functions),
the Leibniz rule should be weakened, and replaced by the rule of
integration by part, Equation~(\ref{105}), which is in fact nothing but an
``integrated'' version of the Leibniz rule. However, the loss of the
Leibniz rule \emph{stricto sensu} constitutes one of the reasons for the
appearance of the ambiguity parameters at 3PN order.

The Hadamard regularization $(F)_1$ is defined by Equation~(\ref{100}) in a
preferred spatial hypersurface $ t = \mathrm{const}$ of a coordinate
system, and consequently is not \emph{a priori} compatible with the
Lorentz invariance. Thus we expect that the equations of motion in
harmonic coordinates (which manifestly preserve the global Lorentz
invariance) should exhibit at some stage a violation of the Lorentz
invariance due to the latter regularization. In fact this occurs exactly
at the 3PN order. Up to the 2.5PN level, the use of the regularization
$(F)_1$ is sufficient to get some unambiguous equations of motion which
are Lorentz invariant~\cite{BFP98}. To deal with the problem at 3PN
order, a Lorentz-invariant variant of the regularization, denoted
$[F]_1$, was introduced in Ref.~\cite{BFregM}. It consists of performing
the Hadamard regularization within the spatial hypersurface that is
geometrically orthogonal (in a Minkowskian sense) to the four-velocity
of the particle. The regularization $[F]_1$ differs from the simpler
regularization $(F)_1$ by relativistic corrections of order $1/c^2$ at
least. See Ref.~\cite{BFregM} for the formulas defining this
regularization in the form of some infinite power series in $1/c^2$. The
regularization $[F]_1$ plays a crucial role in obtaining the equations
of motion at the 3PN order in Refs.~\cite{BF00, BFeom}. In particular,
the use of the Lorentz-invariant regularization $[F]_1$ permits to
obtain the value of the ambiguity parameter $\omega_\mathrm{kinetic}$ in
Equation~(\ref{109a}) below.


\subsection{Hadamard regularization ambiguities}
\label{subsec:8.2}

The ``standard'' Hadamard regularization yields some ambiguous results
for the computation of certain integrals at the 3PN order, as Jaranowski
and Sch\"afer~\cite{JaraS98, JaraS99, JaraS00} first noticed in their
computation of the equations of motion within the ADM-Hamiltonian
formulation of general relativity. By standard Hadamard regularization
we mean the regularization based solely on the definitions of the partie
finie of a singular function, Equation~(\ref{100}), and the partie finie of a
divergent integral, Equation~(\ref{101}) (i.e.\ without using a theory
of pseudo-functions and generalized distributional derivatives as
proposed in Refs.~\cite{BFreg, BFregM}). It was shown in
Refs.~\cite{JaraS98, JaraS99, JaraS00} that there are \emph{two and only
two} types of ambiguous terms in the 3PN Hamiltonian, which were then
parametrized by two unknown numerical coefficients $\omega_\mathrm{static}$
and $\omega_\mathrm{kinetic}$.

Motivated by the previous result, Blanchet and Faye~\cite{BFreg, BFregM}
introduced their ``extended'' Hadamard regularization, the one we
outlined in Section~\ref{subsec:8.1}. This new regularization is
mathematically well-defined and free of ambiguities; in particular it
yields unique results for the computation of any of the integrals
occuring in the 3PN equations of motion. Unfortunately, the extended
Hadamard regularization turned out to be in a sense incomplete, because
it was found~\cite{BF00, BFeom} that the 3PN equations of motion involve
\emph{one and only one} unknown numerical constant, called $\lambda$,
which cannot be determined within the method. The comparison of this
result with the work of Jaranowski and Sch\"afer~\cite{JaraS98,
JaraS99}, on the basis of the computation of the invariant energy of
compact binaries moving on circular orbits, showed~\cite{BF00} that
\begin{eqnarray}
  \omega_\mathrm{kinetic} &=& \frac{41}{24},
  \label{109a}
  \\
  \omega_\mathrm{static} &=& - \frac{11}{3} \lambda - \frac{1987}{840}.
  \label{109b}
\end{eqnarray}%
Therefore, the ambiguity $\omega_\mathrm{kinetic}$ is fixed, while
$\lambda$ is equivalent to the other ambiguity $\omega_\mathrm{static}$.
Notice that the value~(\ref{109a}) for the kinetic ambiguity parameter
$\omega_\mathrm{kinetic}$, which is in factor of some velocity dependent
terms, is the only one for which the 3PN equations of motion are Lorentz
invariant. Fixing up this value was possible because the extended
Hadamard regularization~\cite{BFreg, BFregM} was defined in such a way
that it keeps the Lorentz invariance.

Damour, Jaranowski, and Sch\"afer~\cite{DJSpoinc} recovered the value of
$\omega_\mathrm{kinetic}$ given in Equation~(\ref{109a}) by directly proving
that this value is the unique one for which the global Poincar\'e
invariance of the ADM-Hamiltonian formalism is verified. Since the
coordinate conditions associated with the ADM formalism do not
manifestly respect the Poincar\'e symmetry, they had to prove that the
3PN Hamiltonian is compatible with the existence of generators for the
Poincar\'e algebra. By contrast, the harmonic-coordinate conditions
preserve the Poincar\'e invariance, and therefore the associated
equations of motion at 3PN order should be manifestly Lorentz-invariant,
as was indeed found to be the case in Refs.~\cite{BF00, BFeom}.

The appearance of one and only one physical unknown coefficient
$\lambda$ in the equations of motion constitutes a quite striking fact,
that is related specifically with the use of a Hadamard-type
regularization\epubtkFootnote{Note also that the harmonic-coordinates
3PN equations of motion as they have been obtained in Refs.~\cite{BF00,
BFeom} depend, in addition to $\lambda$, on two arbitrary constants
$r'_1$ and $r'_2$ parametrizing some logarithmic terms. These constants
are closely related to the constants $s_1$ and $s_2$ in the
\emph{partie-finie} integral~(\ref{101}); see Ref.~\cite{BFeom} for the
precise definition. However, $r'_1$ and $r'_2$ are not ``physical'' in
the sense that they can be removed by a coordinate transformation.}.
Technically speaking, the presence of the ambiguity parameter $\lambda$
is associated with the non-distributivity of Hadamard's regularization,
in the sense of Equation~(\ref{100'}). Mathematically speaking, $\lambda$ is
probably related to the fact that it is impossible to construct a
distributional derivative operator, such as Equations~(\ref{106}, \ref{107},
\ref{108}), satisfying the Leibniz rule for the derivation of the
product~\cite{Schwartz54}. The Einstein field equations can be written
in many different forms, by shifting the derivatives and operating
some terms by parts with the help of the Leibniz rule. All these forms
are equivalent in the case of regular sources, but since the derivative
operator~(\ref{106}, \ref{107}, \ref{108}) violates the Leibniz rule
they become inequivalent for point particles. Finally, physically
speaking, let us argue that $\lambda$ has its root in the fact that in
a complete computation of the equations of motion valid for two regular
\emph{extended} weakly self-gravitating bodies, many non-linear
integrals, when taken \emph{individually}, start depending, from the 3PN
order, on the internal structure of the bodies, even in the
``compact-body'' limit where the radii tend to zero. However, when
considering the full equations of motion, we expect that all the terms
depending on the internal structure can be removed, in the compact-body
limit, by a coordinate transformation (or by some appropriate shifts of
the central world lines of the bodies), and that finally $\lambda$ is
given by a pure number, for instance a rational fraction, independent of
the details of the internal structure of the compact bodies. From this
argument (which could be justified by the effacing principle in general
relativity) the value of $\lambda$ is necessarily the one we compute
below, Equation~(\ref{lambda}), and will be valid for any compact objects,
for instance black holes.
 
The ambiguity parameter $\omega_\mathrm{static}$, which is in factor of
some static, velocity-independent term, and hence cannot be derived by
invoking Lorentz invariance, was computed by Damour, Jaranowski, and
Sch\"afer~\cite{DJSdim} by means of \emph{dimensional regularization},
instead of some Hadamard-type one, within the ADM-Hamiltonian formalism.
Their result is
\begin{equation}
  \omega_\mathrm{static}=0.
  \label{omegas}
\end{equation}
As Damour et al.~\cite{DJSdim} argue, clearing up the static
ambiguity is made possible by the fact that dimensional regularization,
contrary to Hadamard's regularization, respects all the basic properties
of the algebraic and differential calculus of ordinary functions:
associativity, commutativity and distributivity of point-wise addition
and multiplication, Leibniz's rule, and the Schwarz lemma. In this
respect, dimensional regularization is certainly better than Hadamard's
one, which does not respect the distributivity of the product (recall
Equation~(\ref{100'})) and unavoidably violates at some stage the Leibniz
rule for the differentiation of a product.

The ambiguity parameter $\lambda$ is fixed from the result
(\ref{omegas}) and the necessary link~(\ref{109b}) provided by the
equivalence between the harmonic-coordinates and ADM-Hamiltonian
formalisms~\cite{BF00, DJSequiv}. However, $\lambda$ was also been
computed directly by Blanchet, Damour, and Esposito-Far\`ese~\cite{BDE04}
applying dimensional regularization to the 3PN equations of motion in
harmonic coordinates (in the line of Refs.~\cite{BF00, BFeom}). The end
result,
\begin{equation}
  \lambda=-\frac{1987}{3080},
  \label{lambda}
\end{equation}
is in full agreement with Equation~(\ref{omegas})\epubtkFootnote{One may
wonder why the value of $\lambda$ is a complicated rational fraction
while $\omega_\mathrm{static}$ is so simple. This is because
$\omega_\mathrm{static}$ was introduced precisely to measure the
amount of ambiguities
of certain integrals, while, by contrast, $\lambda$ was introduced as an
unknown constant entering the relation between the arbitrary scales
$r'_1 , r'_2$ on the one hand, and $s_1 , s_2$ on the other hand, which
has \emph{a priori} nothing to do with ambiguities of integrals.}.
Besides the independent confirmation of the value of
$\omega_\mathrm{static}$ or $\lambda$, the work~\cite{BDE04} provides also a
confirmation of the \emph{consistency} of dimensional regularization,
because the explicit calculations are entirely different from the ones
of Ref.~\cite{DJSdim}: Harmonic coordinates are used instead of ADM-type
ones, the work is at the level of the equations of motion instead of the
Hamiltonian, and a different form of Einstein's field equations is solved by
a different iteration scheme.

Let us comment here that the use of a self-field regularization, be it
dimensional or based on Hadamard's partie finie, signals a somewhat
unsatisfactory situation on the physical point of view, because,
ideally, we would like to perform a complete calculation valid for
extended bodies, taking into account the details of the internal
structure of the bodies (energy density, pressure, internal velocity
field, etc.). By considering the limit where the radii of the
objects tend to zero, one should recover the same result as obtained by
means of the point-mass regularization. This would demonstrate the
suitability of the regularization. This program was undertaken at the
2PN order by Kopeikin et al.~\cite{Kop85, GKop86} who derived the
equations of motion of two extended fluid balls, and obtained equations
of motion depending only on the two masses $m_1$ and $m_2$ of the
compact bodies\epubtkFootnote{See some comments on this work in
Ref.~\cite{D300}, pp.~168\,--\,169.}. At the 3PN order we expect that the
extended-body program should give the value of the regularization
parameter $\lambda$ (maybe after some gauge transformation to remove the
terms depending on the internal structure). Ideally, its value should be
confirmed by independent and more physical methods (like those of
Refs.~\cite{ThH85, Kop88, DSX91}).

An important work, in several respects more physical than the formal use
of regularizations, is the one of Itoh and Futamase~\cite{itoh1, itoh2},
following previous investigations in Refs.~\cite{IFA00, IFA01}. These
authors derived the 3PN equations of motion in harmonic coordinates by
means of a particular variant of the famous ``surface-integral'' method
introduced long ago by Einstein, Infeld, and Hoffmann~\cite{EIH}. The aim
is to describe extended relativistic compact binary systems in the
strong-field point particle limit defined in Ref.~\cite{F87}. This
approach is very interesting because it is based on the physical notion
of extended compact bodies in general relativity, and is free of the
problems of ambiguities due to the Hadamard self-field regularization.
The end result of Refs.~\cite{itoh1, itoh2} is in agreement with the 3PN
harmonic coordinates equations of motion~\cite{BF00, BFeom} and,
moreover, is unambiguous, as it does determine the ambiguity parameter
$\lambda$ to exactly the value~(\ref{lambda}).

We next consider the problem of the binary's radiation field, where the
same phenomenon occurs, with the appearance of some Hadamard
regularization ambiguity parameters at 3PN order. More precisely,
Blanchet, Iyer, and Joguet~\cite{BIJ02}, in their computation of the 3PN
compact binary's \emph{mass quadrupole moment} $\mathrm{I}_{ij}$, found it
necessary to introduce \emph{three} Hadamard regularization constants
$\xi$, $\kappa$, and $\zeta$, which are additional to and independent of
the equation-of-motion related constant $\lambda$. The total
gravitational-wave flux at 3PN order, in the case of circular orbits,
was found to depend on a single combination of the latter constants,
$\theta = \xi+2\kappa+\zeta$, and the binary's orbital phase, for
circular orbits, involves only the linear combination of $\theta$ and
$\lambda$ given by $\hat{\theta} = \theta-7\lambda/3$, as shown
in~\cite{BFIJ02}.

Dimensional regularization (instead of Hadamard's) has next been applied
by Blanchet, Damour, Esposito-Far\`ese, and Iyer~\cite{BDEI04,BDEI05dr}
to the computation of the 3PN radiation field of compact binaries,
leading to the following unique values for the ambiguity
parameters\epubtkFootnote{The result for $\xi$ happens to be amazingly
related to the one for $\lambda$ by a cyclic permutation of digits;
compare $ 3\xi=-9871/3080 $ with $ \lambda= -1987/3080 $.}:
\begin{equation}
  \begin{array}{rcl}
    \xi &=& \displaystyle - \frac{9871}{9240},
    \\ [1.0 em]
    \kappa &=& 0,
    \\ [1.0 em]
    \zeta &=& \displaystyle - \frac{7}{33}.
  \end{array}
  \label{xikappazeta}
\end{equation}
These values represent the end result of dimensional regularization.
However, several alternative calculations provide a check, independent
of dimensional regularization, for all the
parameters~(\ref{xikappazeta}). Blanchet and Iyer~\cite{BI04mult}
compute the 3PN binary's \emph{mass dipole moment} $\mathrm{I}_{i}$ using
Hadamard's regularization, and identify $\mathrm{I}_{i}$ with the 3PN
\emph{center of mass vector position} $\mathrm{G}_{i}$, already known as a
conserved integral associated with the Poincar\'e invariance of the 3PN
equations of motion in harmonic coordinates~\cite{ABF01}. This yields
$\xi + \kappa = - 9871/9240$ in agreement with Equation~(\ref{xikappazeta}).
Next, we consider~\cite{BDI04zeta} the limiting physical situation where
the mass of one of the particles is exactly zero (say, $m_2=0$), and the
other particle moves with uniform velocity. Technically, the 3PN
quadrupole moment of a \emph{boosted} Schwarzschild black hole is
computed and compared with the result for $\mathrm{I}_{ij}$ in the limit
$m_2=0$. The result is $\zeta = - 7/33$, and represents a direct
verification of the global Poincar\'e invariance of the wave generation
formalism (the parameter $\zeta$ represents the analogue for the
radiation field of the equation-of-motion related parameter
$\omega_\mathrm{kinetic}$)\epubtkFootnote{The work~\cite{BDI04zeta}
  provided also some
new expressions for the multipole moments of an isolated post-Newtonian
source, alternative to those given by Theorem~\ref{th6}, in the form of
\emph{surface integrals} extending on the outer part of the source's near
zone.}. Finally, $\kappa = 0$ is proven~\cite{BDEI05dr} by showing that
there are no dangerously divergent ``diagrams'' corresponding to
non-zero $\kappa$-values, where a diagram is meant here in the sense of
Ref.~\cite{Dgef96}.

The determination of the parameters~(\ref{xikappazeta}) completes the
problem of the general relativistic prediction for the templates of
inspiralling compact binaries up to 3PN order (and actually up to 3.5PN
order as the corresponding tail terms have already been
determined~\cite{B98tail}). The relevant combination of the
parameters~(\ref{xikappazeta}) entering the 3PN energy flux in the case
of circular orbits is now fixed to be
\begin{equation}
  \theta\equiv\xi+2\kappa+\zeta=-\frac{11831}{9240}.
  \label{theta}
\end{equation}
Numerically, $\theta\simeq -1.28041$. The orbital phase of compact
binaries, in the adiabatic inspiral regime (i.e.\ evolving by
radiation reaction), involves at 3PN order a combination of parameters
which is determined as
\begin{equation}
  \hat{\theta}\equiv \theta-\frac{7}{3}\lambda=\frac{1039}{4620}.
  \label{thetahat}
\end{equation}
The fact that the numerical value of this parameter is quite small,
$\hat{\theta}\simeq 0.22489$, indicates, following measurement-accuracy
analyses~\cite{BCV03a, BCV03b, DIJS03}, that the 3PN (or, even better, 3.5PN)
order should provide an excellent approximation for both the on-line
search and the subsequent off-line analysis of gravitational wave
signals from inspiralling compact binaries in the LIGO and VIRGO
detectors.


\subsection{Dimensional regularization of the equations of motion}
\label{subsec:8.3}

As reviewed in Section~\ref{subsec:8.2}, work at 3PN order using
Hadamard's self-field regularization showed the appearance of ambiguity
parameters, due to an incompleteness of the Hadamard regularization
employed for curing the infinite self field of point particles. We give
here more details on the determination using \emph{dimensional
regularization} of the ambiguity parameter $\lambda$ which appeared in
the 3PN equations of motion (recall that $\lambda$ is equivalent to the
static ambiguity parameter $\omega_\mathrm{static}$, see Equation~(\ref{109b})).

Dimensional regularization was invented as a means to preserve the gauge
symmetry of perturbative quantum field theories~\cite{tHooft, Bollini,
Breitenlohner, Collins}. Our basic problem here is to respect the gauge
symmetry associated with the diffeomorphism invariance of the classical
general relativistic description of interacting point masses. Hence, we
use dimensional regularization not merely as a trick to compute some
particular integrals which would otherwise be divergent, but as a
powerful tool for solving in a consistent way the Einstein field
equations with singular point-mass sources, while preserving its crucial
symmetries. In particular, we shall prove that dimensional
regularization determines the kinetic ambiguity parameter
$\omega_\mathrm{kinetic}$ (and its radiation-field analogue $\zeta$),
and is therefore able to correctly keep track of the global
Lorentz--Poincar\'e invariance of the gravitational field of isolated
systems.

The Einstein field equations in $d+1$ space-time dimensions, relaxed by
the condition of harmonic coordinates $\partial_\mu h^{\alpha\mu}=0$,
take exactly the same form as given in Equations~(\ref{8_1}, \ref{13}). In
particular $\Box$ denotes the flat space-time d'Alembertian operator in
$d+1$ dimensions. The gravitational constant $G$ is related to the usual
three-dimensional Newton's constant $G_N$ by
\begin{equation}
  G=G_N\,\ell_0^{d-3},
  \label{G}
\end{equation}
where $\ell_0$ denotes an arbitrary length scale. The explicit
expression of the gravitational source term $\Lambda^{\alpha\beta}$
involves some $d$-dependent coefficients, and is given by
\begin{eqnarray}
  \Lambda^{\alpha\beta} = &-& h^{\mu\nu} \partial^2_{\mu\nu}
  h^{\alpha\beta}+\partial_\mu h^{\alpha\nu} \partial_\nu h^{\beta\mu}
  +\frac{1}{2}g^{\alpha\beta}g_{\mu\nu}\partial_\lambda h^{\mu\tau}
  \partial_\tau h^{\nu\lambda}
  \nonumber \\
  &-&g^{\alpha\mu}g_{\nu\tau}\partial_\lambda h^{\beta\tau} \partial_\mu
  h^{\nu\lambda} -g^{\beta\mu}g_{\nu\tau}\partial_\lambda h^{\alpha\tau}
  \partial_\mu h^{\nu\lambda} +g_{\mu\nu}g^{\lambda\tau}\partial_\lambda
  h^{\alpha\mu} \partial_\tau h^{\beta\nu}
  \nonumber \\
  &+&\frac{1}{4}(2g^{\alpha\mu}g^{\beta\nu}-g^{\alpha\beta}g^{\mu\nu})
  \left(g_{\lambda\tau}g_{\epsilon\pi}-
  \frac{1}{d-1}g_{\tau\epsilon}g_{\lambda\pi}\right)
  \partial_\mu h^{\lambda\pi} \partial_\nu h^{\tau\epsilon}.
  \label{Lambdad}
\end{eqnarray}%
When $d=3$ we recover Equation~(\ref{14}). In the following we assume, as
usual in dimensional regularization, that the dimension of space is a
complex number, $d\in\mathbb{C}$, and prove many results by invoking
complex analytic continuation in $d$. We shall pose $\varepsilon\equiv
d-3$. 

We parametrize the 3PN metric in $d$ dimensions by means of
straightforward $d$-dimensional generalizations of the retarded
potentials $V$, $V_i$, $\hat{W}_{ij}$, $\hat{R}_i$, and $\hat{X}$ of
Section~\ref{sec:7}. Those are obtained by post-Newtonian iteration of
the $d$-dimensional field equations, starting from the following
definitions of matter source densities
\begin{equation}
  \begin{array}{rcl}
    \sigma &=& \displaystyle
    \frac{2}{d-1}\frac{(d-2)T^{00}+T^{ii}}{c^2},
    \\ [1.0 em]
    \sigma_i &=& \displaystyle \frac{T^{0i}}{c},
    \\ [1.0 em]
    \sigma_{ij} &=& \displaystyle T^{ij},
  \end{array}
  \label{72d}
\end{equation}
which generalize Equations~(\ref{94}). As a result all the expressions of
Section~\ref{sec:7} acquire some explicit $d$-dependent coefficients.
For instance we find~\cite{BDE04}
\begin{equation}
  \begin{array}{rcl}
    V &=& \displaystyle \Box^{-1}_\mathrm{ret}
    \left[-4 \pi G \sigma\right],
    \\ [1.5 em]
    \hat{W}_{ij} &=& \displaystyle \Box^{-1}_\mathrm{ret}
    \left[ -4 \pi G \left(\sigma_{ij}-\delta_{ij}
    \frac{\sigma_{kk}}{d-2}\right)- \frac{d-1}{2(d-2)} \partial_i V
    \partial_j V \right].
  \end{array}
  \label{96d}
\end{equation}
Here $\Box^{-1}_\mathrm{ret}$ means the retarded integral in $d+1$
space-time dimensions, which admits, though, no simple expression in
physical $(t,\mathbf{x})$ space.

As reviewed in Section~\ref{subsec:8.1}, the generic functions we have
to deal with in 3 dimensions, say $F(\mathbf{x})$, are smooth on
$\mathbb{R}^3$ except at $\mathbf{y}_1$ and $\mathbf{y}_2$, around which
they admit singular Laurent-type expansions in powers and inverse powers
of $r_1\equiv\vert\mathbf{x}-\mathbf{y}_1\vert$ and
$r_2\equiv\vert\mathbf{x}-\mathbf{y}_2\vert$, given by Equation~(\ref{99}).
In $d$ spatial dimensions, there is an analogue of the function $F$,
which results from the post-Newtonian iteration process performed in $d$
dimensions as we just outlined. Let us call this function
$F^{(d)}(\mathbf{x})$, where $\mathbf{x}\in\mathbb{R}^d$. When
$r_1\rightarrow 0$ the function $F^{(d)}$ admits a singular expansion
which is a little bit more complicated than in 3 dimensions, as it
reads
\begin{equation}
  F^{(d)}(\mathbf{x}) = \!\!\!\!
  \sum_{\substack{p_0 \leq p \leq N \\ q_0 \leq q\leq q_1}} \!\!\!\!
  r_1^{p+q\varepsilon}
  \mathop{f}_1{}_{p,q}^{(\varepsilon)}(\mathbf{n}_1) + o(r_1^N).
  \label{Fdx}
\end{equation}
The coefficients $\mathop{f}_1{}_{p,q}^{(\varepsilon)}(\mathbf{n}_1)$
depend on $\varepsilon = d-3$, and the powers of $r_1$ involve the
relative integers $p$ and $q$ whose values are limited by some $p_0$,
$q_0$, and $q_1$ as indicated. Here we will be interested in functions
$F^{(d)}(\mathbf{x})$ which have no poles as $\varepsilon \rightarrow 0$
(this will always be the case at 3PN order). Therefore, we can deduce
from the fact that $F^{(d)}(\mathbf{x})$ is continuous at $d=3$ the
constraint
\begin{equation}
  \sum_{q=q_0}^{q_1}\mathop{f}_1{}_{p,q}^{(\varepsilon=0)}
  (\mathbf{n}_1) = \mathop{f}_1{}_{\!p}(\mathbf{n}_1).
  \label{constr}
\end{equation}

For the problem at hand, we essentially have to deal with the
regularization of Poisson integrals, or iterated Poisson integrals (and
their gradients needed in the equations of motion), of the generic
function $F^{(d)}$. The Poisson integral of $F^{(d)}$, in $d$
dimensions, is given by the Green's function for the Laplace operator,
\begin{equation}
  P^{(d)}({\mathbf{x}}')=
  \Delta^{-1} \left[ F^{(d)}({\mathbf{x}}) \right] \equiv
  -\frac{\tilde{k}}{4\pi} \int
  \frac{d^d{\mathbf{x}}}{\vert{\mathbf{x}}-{\mathbf{x}}'\vert^{d-2}}
  F^{(d)}({\mathbf{x}}),
  \label{Pdx}
\end{equation}
where $\tilde{k}$ is a constant related to the usual Eulerian
$\Gamma$-function by\epubtkFootnote{We have $\lim_{d\rightarrow
3}\tilde{k}=1$. Notice that $\tilde{k}$ is closely linked to the volume
$\Omega_{d-1}$ of the sphere with $d-1$ dimensions (i.e.\
embedded into Euclidean $d$-dimensional space):
$$\tilde{k}\,\Omega_{d-1}=\frac{4\pi}{d-2}.$$}
\begin{equation}
  \tilde{k}=\frac{\Gamma\left(\frac{d-2}{2}\right)}{\pi^{\frac{d-2}{2}}}.
  \label{ktilde}
\end{equation}
We need to evaluate the Poisson integral at the point ${\mathbf{x}}' =
{\mathbf{y}}_1$ where it is singular; this is quite easy in dimensional
regularization, because the nice properties of analytic continuation
allow simply to get $[P^{(d)} ({\mathbf{x}}')]_{\mathbf{x}' =
{\mathbf{y}}_1}$ by replacing ${\mathbf{x}}'$ by ${\mathbf{y}}_1$ in
the explicit integral form~(\ref{Pdx}). So we simply have
\begin{equation}
  P^{(d)}({\mathbf{y}}_1)=-\frac{\tilde{k}}{4\pi}
  \int\frac{d^d{\mathbf{x}}}{r_1^{d-2}}F^{(d)}({\mathbf{x}}).
  \label{Pd}
\end{equation}

It is not possible at present to compute the equations of motion in the
general $d$-dimensional case, but only in the limit where
$\varepsilon\rightarrow 0$~\cite{DJSdim, BDE04}. The main technical step
of our strategy consists of computing, in the limit
$\varepsilon\rightarrow 0$, the \emph{difference} between the
$d$-dimensional Poisson potential~(\ref{Pd}), and its Hadamard
3-dimensional counterpart given by $(P)_1$, where the Hadamard partie
finie is defined by Equation~(\ref{100}). Actually, we must be very precise
when defining the Hadamard partie finie of a Poisson integral. Indeed,
the definition~(\ref{100}) \emph{stricto sensu} is applicable when the
expansion of the function $F$, when $r_1\rightarrow 0$, does not involve
logarithms of $r_1$; see Equation~(\ref{99}). However, the Poisson integral
$P(\mathbf{x}')$ of $F(\mathbf{x})$ will typically involve such
logarithms at the 3PN order, namely some $\ln r'_1$ where $r'_1\equiv
\vert\mathbf{x}'-\mathbf{y}_1\vert$ formally tends to zero (hence $\ln
r'_1$ is formally infinite). The proper way to define the Hadamard
partie finie in this case is to include the $\ln r'_1$ into its
definition, so we arrive at~\cite{BFreg}
\begin{equation}
  (P)_1 = -\frac{1}{4\pi}\pf_{r_1',s_2}
  \int\frac{d^3{\mathbf{x}}}{r_1} F({\mathbf{x}}) - (r_1^2\,F)_1.
  \label{P1}
\end{equation}
The first term follows from Hadamard's partie finie
integral~(\ref{101}); the second one is given by Equation~(\ref{100}). Notice
that in this result the constant $s_1$ entering the partie finie
integral~(\ref{101}) has been ``replaced'' by $r'_1$, which plays the
role of a new regularization constant (together with $r'_2$ for the
other particle), and which ultimately parametrizes the final Hadamard
regularized 3PN equations of motion. It was shown that $r'_1$ and $r'_2$
are unphysical, in the sense that they can be removed by a coordinate
transformation~\cite{BF00, BFeom}. On the other hand, the constant $s_2$
remaining in the result~(\ref{P1}) is the source for the appearance of
the physical ambiguity parameter $\lambda$, as it will be related to it
by Equation~(\ref{lnr2s2}). Denoting the difference between the dimensional
and Hadamard regularizations by means of the script letter
${\cal D}$, we pose (for the result concerning the point 1)
\begin{equation}
  {\cal D}P_1\equiv P^{(d)}({\mathbf{y}}_1)-(P)_1.
  \label{DP1}
\end{equation}
That is, ${\cal D}P_1$ is what we shall have to \emph{add} to the
Hadamard-regularization result in order to get the $d$-dimensional
result. However, we shall only compute the first two terms of the
Laurent expansion of ${\cal D}P_1$ when $\varepsilon \rightarrow
0$, say $a_{-1} \, \varepsilon^{-1} + a_0 + {\cal O}
(\varepsilon)$. This is the information we need to clear up the
ambiguity parameter. We insist that the difference ${\cal D}P_1$
comes exclusively from the contribution of terms developing some poles
$\propto 1/\varepsilon$ in the $d$-dimensional calculation.

Next we outline the way we obtain, starting from the computation of the
``difference'', the 3PN equations of motion in dimensional
regularization, and show how the ambiguity parameter $\lambda$ is
determined. By contrast to $r'_1$ and $r'_2$ which are pure gauge,
$\lambda$ is a genuine physical ambiguity, introduced in
Refs.~\cite{BFreg, BFeom} as the \emph{single} unknown numerical constant
parametrizing the ratio between $s_2$ and $r'_2$ (where $s_2$ is the
constant left in Equation~(\ref{P1})) as
\begin{equation}
  \ln \left( \frac{r_2'}{s_2} \right) = 
  \frac{159}{308}+\lambda \frac{m_1+m_2}{m_2}
  \qquad
  (\mbox{and }1\leftrightarrow 2),
  \label{lnr2s2}
\end{equation}
where $m_1$ and $m_2$ are the two masses. The terms corresponding to the
$\lambda$-ambiguity in the acceleration $\mathbf{a}_1=d\mathbf{v}_1/dt$
of particle 1 read simply 
\begin{equation}
  \Delta\mathbf{a}_1 [\lambda] =
  -\frac{44\lambda}{3}\,\frac{G_N^4\,m_1\,m_2^2\,(m_1+m_2)}{r_{12}^5\,c^6}\,
  \mathbf{n}_{12},
  \label{Dealtaa1}
\end{equation}
where the relative distance between particles is denoted
$\mathbf{y}_1-\mathbf{y}_2\equiv r_{12}\,\mathbf{n}_{12}$ (with
$\mathbf{n}_{12}$ being the unit vector pointing from particle 2 to particle
1). We start from the end result of Ref.~\cite{BFeom} for the 3PN
harmonic coordinates acceleration $\mathbf{a}_1$ in Hadamard's
regularization, abbreviated as HR. Since the result was obtained by
means of the specific extended variant of Hadamard's regularization (in
short EHR, see Section~\ref{subsec:8.1}) we write it as
\begin{equation}
  \mathbf{a}_1^\mathrm{(HR)} =
  \mathbf{a}_1^\mathrm{(EHR)} + \Delta\mathbf{a}_1 [\lambda],
  \label{a1HR}
\end{equation}
where $\mathbf{a}_1^\mathrm{(EHR)}$ is a fully determined functional of the
masses $m_1$ and $m_2$, the relative distance $r_{12}\,\mathbf{n}_{12}$,
the coordinate velocities $\mathbf{v}_1$ and $\mathbf{v}_2$, and also
the gauge constants $r_1'$ and $r_2'$. The only ambiguous term is the
second one and is given by Equation~(\ref{Dealtaa1}).

Our strategy is to express both the dimensional and Hadamard
regularizations in terms of their common ``core'' part, obtained by
applying the so-called ``pure-Hadamard--Schwartz'' (pHS) regularization.
Following the definition of Ref.~\cite{BDE04}, the pHS regularization is
a specific, minimal Hadamard-type regularization of integrals, based on
the partie finie integral~(\ref{101}), together with a minimal treatment
of ``contact'' terms, in which the definition~(\ref{101}) is applied
separately to each of the elementary potentials $V, V_i, \dots $ (and
gradients) that enter the post-Newtonian metric in the form given
in Section~\ref{sec:7}. Furthermore, the regularization of a product of
these potentials is assumed to be distributive, i.e.\ $(FG)_1=
(F)_1(G)_1$ in the case where $F$ and $G$ are given by such elementary
potentials (this is in contrast with Equation~(\ref{100'})). The pHS
regularization also assumes the use of standard Schwartz distributional
derivatives~\cite{Schwartz}. The interest of the pHS regularization is
that the dimensional regularization is equal to it plus the
``difference''; see Equation~(\ref{a1DimReg}).

To obtain the pHS-regularized acceleration we need to substract from the
EHR result a series of contributions, which are specific consequences of
the use of EHR~\cite{BFreg,BFregM}. For instance, one of these
contributions corresponds to the fact that in the EHR the distributional
derivative is given by Equations~(\ref{106}, \ref{107}) which differs from
the Schwartz distributional derivative in the pHS regularization. Hence
we define
\begin{equation}
  {\mathbf{a}}_1^\mathrm{(pHS)} = {\mathbf{a}}_1^\mathrm{(EHR)} -
  \sum_{A}\delta_A{\mathbf{a}}_1,
  \label{accpH}
\end{equation}
where the $\delta_A{\mathbf{a}}_1$'s denote the extra terms following
from the EHR prescriptions. The pHS-regularized
acceleration~(\ref{accpH}) constitutes essentially the result of the
first stage of the calculation of ${\mathbf{a}}_1$, as reported in
Ref.~\cite{FayeThesis}.

The next step consists of evaluating the Laurent expansion, in powers of
$\varepsilon = d-3$, of the difference between the dimensional
regularization and the pHS (3-dimensional) computation. As we reviewed
above, this difference makes a contribution only when a term generates a
\emph{pole} $\sim 1/\varepsilon$, in which case the dimensional
regularization adds an extra contribution, made of the pole and the
finite part associated with the pole (we consistently neglect all terms
${\cal O}(\varepsilon)$). One must then be especially wary of
combinations of terms whose pole parts finally cancel (``cancelled
poles'') but whose dimensionally regularized finite parts generally do
not, and must be evaluated with care. We denote the above defined
difference by
\begin{equation}
  {\cal D}{\mathbf{a}}_1 = \sum {\cal D}P_1.
  \label{deltaacc}
\end{equation}
It is made of the sum of all the individual differences of Poisson or
Poisson-like integrals as computed in Equation~(\ref{DP1}). The total
difference~(\ref{deltaacc}) depends on the Hadamard regularization
scales $r_1'$ and $s_2$ (or equivalently on $\lambda$ and $r_1'$,
$r_2'$), and on the parameters associated with dimensional
regularization, namely $\varepsilon$ and the characteristic length scale
$\ell_0$ introduced in Equation~(\ref{G}). Finally, our main result is the
explicit computation of the $\varepsilon$-expansion of the dimensional
regularization (DR) acceleration as
\begin{equation}
  {\mathbf{a}}_1^\mathrm{(DR)} = {\mathbf{a}}_1^\mathrm{(pHS)} +
  {\cal D}{\mathbf{a}}_1.
  \label{a1DimReg}
\end{equation}
With this result we can prove two theorems~\cite{BDE04}:

\begin{theorem}
  The pole part $\propto 1/\varepsilon$ of the DR
  acceleration~(\ref{a1DimReg}) can be re-absorbed (i.e.\
  renormalized) into some shifts of the two ``bare'' world-lines:
  $\mathbf{y}_1 \rightarrow \mathbf{y}_1+\bm{\xi}_1$ and $\mathbf{y}_2
  \rightarrow \mathbf{y}_2+\bm{\xi}_2$, with, say, $\bm{\xi}_{1,2} \propto
  1/\varepsilon$, so that the result, expressed in terms of the
  ``dressed'' quantities, is finite when $\varepsilon\rightarrow 0$.
  \label{th8}
\end{theorem}

\noindent
The situation in harmonic coordinates is to be contrasted with the
calculation in ADM-type coordinates within the Hamiltonian formalism,
where it was shown that all pole parts directly cancel out in the total
3PN Hamiltonian: No renormalization of the world-lines is
needed~\cite{DJSdim}. A central result is then as follows:

\begin{theorem}
  The renormalized (finite) DR acceleration is physically equivalent
  to the Hadamard-regularized (HR) acceleration (end result of
  Ref.~\cite{BFeom}), in the sense that
  \begin{equation}
    {\mathbf{a}}_1^{\mathrm{(HR)}} = \lim_{\varepsilon\rightarrow 0} \,
    \left[ {\mathbf{a}}_1^{\mathrm{(DR)}} + \delta_{\bm{\xi}} \,
    {\mathbf{a}}_1 \right],
    \label{eta}
  \end{equation}
  where $\delta_{\bm{\xi}} \, {\mathbf{a}}_1$ denotes the effect of the
  shifts on the acceleration, if and only if the HR ambiguity parameter
  $\lambda$ entering the harmonic-coordinates equations of motion takes
  the unique value~(\ref{lambda}).
  \label{th9}
\end{theorem}

\noindent
The precise shifts $\bm{\xi}_1$ and $\bm{\xi}_2$ needed in
Theorem~\ref{th9} involve not only a pole contribution $\propto
1/\varepsilon$ (which would define a renormalization by minimal
subtraction (MS)), but
also a finite contribution when $\varepsilon\rightarrow 0$. Their
explicit expressions read\epubtkFootnote{When working at the level of
the equations of motion (not considering the metric outside the
world-lines), the effect of shifts can be seen as being induced by a
coordinate transformation of the bulk metric as in Ref.~\cite{BFeom}.}:
\begin{equation}
  \bm{\xi}_1=\frac{11}{3}\frac{G_N^2\,m_1^2}{c^6}
  \left[ \frac{1}{\varepsilon}-2\ln\left(
  \frac{r'_1\overline{q}^{1/2}}{\ell_0}\right) -\frac{327}{1540}\right]
  {\mathbf{a}}_{N1}
  \qquad
  (\mbox{together with }1\leftrightarrow 2),
\end{equation}
where $G_N$ is Newton's constant, $\ell_0$ is the characteristic length
scale of dimensional regularization (cf.\ Equation~(\ref{G})),
${\mathbf{a}}_{N1}$ is the Newtonian acceleration of the particle 1 in
$d$ dimensions, and $\overline{q}\equiv 4\pi e^C$ depends on Euler's
constant $C=0.577\cdots$.


\subsection{Dimensional regularization of the radiation field}
\label{subsec:8.4}

We now address the similar problem concerning the binary's radiation
field (3PN beyond the Einstein quadrupole formalism), for which three
ambiguity parameters, $\xi$, $\kappa$, $\zeta$, have been shown to
appear~\cite{BIJ02, BI04mult} (see Section~\ref{subsec:8.2}).

To apply dimensional regularization, we must use as in
Section~\ref{subsec:8.3} the $d$-dimensional post-Newtonian iteration
[leading to equations such as~(\ref{96d})]; and, crucially, we have to
generalize to $d$ dimensions some key results of the wave generation
formalism of Part~\ref{part:a}. Essentially we need the $d$-dimensional
analogues of the multipole moments of an isolated source $\mathrm{I}_L$ and
$\mathrm{J}_L$, Equations~(\ref{71}). The result we find in the case of the
mass-type moments is
\begin{eqnarray}
  \mathrm{I}^{(d)}_L(t)&=& \frac{d-1}{2(d-2)}\,{\cal FP}
  \int d^d\mathbf{x}\, \biggl\{
  \hat{x}_L\,\mathop{\Sigma}_{[l]}(\mathbf{x},t) -
  \frac{4(d+2l-2)}{c^2(d+l-2)(d+2l)} \,\hat{x}_{aL}\,
  \mathop{\Sigma}_{[l+1]}{}_{\!\!a}^{\!\!\!(1)}(\mathbf{x},t)
  \nonumber \\
  && \qquad \qquad \qquad \qquad \quad \:
  + \frac{2(d+2l-2)}{c^4(d+l-1)(d+l-2)(d+2l+2)}\hat{x}_{abL}
  \,\mathop{\Sigma}_{[l+2]}{}_{\!\!ab}^{\!\!\!(2)}(\mathbf{x},t)\biggr\}, \quad
  \label{ILd}
\end{eqnarray}%
where we denote (generalizing Equations~(\ref{72}))
\begin{equation}
  \begin{array}{rcl}
    \Sigma &=& \displaystyle
    \frac{2}{d-1}\frac{(d-2)\overline\tau^{00}+\overline\tau^{ii}}{c^2},
    \\ [1.0 em]
    \Sigma_i &=& \displaystyle \frac{\overline\tau^{0i}}{c},
    \\ [1.0 em]
    \Sigma_{ij} &=& \displaystyle \overline{\tau}^{ij},
  \end{array}
  \label{Sigd}
\end{equation}
and where for any source densities the underscript $[l]$ means the
infinite series
\begin{eqnarray}
  \mathop{\Sigma}_{[l]}(\mathbf{x},t) =
  \sum_{k=0}^{+\infty}\frac{1}{2^{2k}k!}
  \frac{\Gamma\left(\frac{d}{2}+l\right)}{\Gamma\left(\frac{d}{2}+l+k\right)}
  \left(\frac{\vert \mathbf{x}\vert}{c}\frac{\partial}{\partial t}\right)^{2k}
  \!\!\Sigma(\mathbf{x},t).
  \label{series}
\end{eqnarray}%
The latter definition represents the $d$-dimensional version of the
post-Newtonian expansion series~(\ref{75}). At Newtonian order,
Equation~(\ref{ILd}) reduces to the standard result $\mathrm{I}_L^{(d)}=\int
d^d\mathbf{x}\,\rho\,\hat{x}_L+{\cal O}(c^{-2})$ with $\rho=T^{00}/c^2$.

The ambiguity parameters $\xi$, $\kappa$, and $\zeta$ come from the
Hadamard regularization of the mass quadrupole moment $\mathrm{I}_{ij}$ at
the 3PN order. The terms corresponding to these ambiguities were found
to be
\begin{equation}
  \Delta \mathrm{I}_{ij}[\xi, \kappa, \zeta] =
  \frac{44}{3}\frac{G_N^2\,m_1^3}{c^6} \left[ \left( \xi + \kappa
  \frac{m_1+m_2}{m_1} \right) y_1^{\langle i}a_1^{j\rangle} +\,\zeta
  \,v_1^{\langle i}v_1^{j\rangle} \right] + 1\leftrightarrow 2,
  \label{amb}
\end{equation}
where $\mathbf{y}_1$, $\mathbf{v}_1$, and $\mathbf{a}_1$ denote the first
particle's position, velocity, and acceleration. We recall that the
brackets $\langle\,\rangle$ surrounding indices refer to the
symmetric-trace-free (STF) projection. Like in Section~\ref{subsec:8.3},
we express both the Hadamard and dimensional results in terms of the
more basic pHS regularization. The first step
of the calculation~\cite{BI04mult} is therefore to relate the
Hadamard-regularized quadrupole moment $I_{ij}^{(\mathrm{HR})}$, for
general orbits, to its pHS part:
\begin{equation}
  \mathrm{I}_{ij}^{(\mathrm{HR})} =
  \mathrm{I}_{ij}^{(\mathrm{pHS})} + \Delta \mathrm{I}_{ij}
  \left[ \xi+\frac{1}{22},\kappa,\zeta+\frac{9}{110} \right].
  \label{IijH}
\end{equation}
In the right-hand side we find both the pHS part, and the effect of
adding the ambiguities, with some numerical shifts of the ambiguity
parameters coming from the difference between the specific Hadamard-type
regularization scheme used in Ref.~\cite{BIJ02} and the pHS one. The pHS
part is free of ambiguities but depends on the gauge constants $r_1'$
and $r_2'$ introduced in the harmonic-coordinates equations of
motion~\cite{BF00, BFeom}.

We next use the $d$-dimensional moment~(\ref{ILd}) to compute the
difference between the dimensional regularization (DR) result and the
pHS one~\cite{BDEI04, BDEI05dr}. As in the work on equations of motion,
we find that the ambiguities arise solely from the terms in the
integration regions near the particles (i.e.\
$r_1=\vert\mathbf{x}-\mathbf{y}_1\vert \rightarrow 0$ or
$r_2=\vert\mathbf{x}-\mathbf{y}_2\vert \rightarrow 0$) that give rise to
poles $\propto 1/\varepsilon$, corresponding to logarithmic ultra-violet
(UV) divergences in 3 dimensions. The infra-red (IR) region at infinity
(i.e.\ $\vert\mathrm{x}\vert\rightarrow +\infty$) does not contribute
to the difference $\mathrm{DR} - \mathrm{pHS}$. The compact-support terms in the integrand
of Equation~(\ref{ILd}), proportional to the matter source densities $\sigma$,
$\sigma_a$, and $\sigma_{ab}$, are also found not to contribute to the
difference. We are therefore left with evaluating the difference linked
with the computation of the \emph{non-compact} terms in the expansion of
the integrand in~(\ref{ILd}) near the singularities that produce poles
in $d$ dimensions.

Let $F^{(d)}(\mathbf{x})$ be the non-compact part of the integrand of
the quadrupole moment~(\ref{ILd}) (with indices $L\equiv ij$), where
$F^{(d)}$ includes the appropriate multipolar factors such as
$\hat{x}_{ij}$, so that
\begin{equation}
  \mathrm{I}^{(d)}_{ij} = \int d^d\mathbf{x}\,F^{(d)}(\mathbf{x}).
  \label{ILFd}
\end{equation}
We do not indicate that we are considering here only the non-compact
part of the moments. Near the singularities the function
$F^{(d)}(\mathbf{x})$ admits a singular expansion of the
type~(\ref{Fdx}). In practice, the various coefficients
${}_1f_{p,q}^{(\varepsilon)}$ are computed by specializing the general expressions of
the non-linear retarded potentials $V, V_a, \hat{W}_{ab}, \dots $
(valid for general extended sources) to the point particles case in
$d$ dimensions. On the other hand, the analogue of Equation~(\ref{ILFd}) in 3
dimensions is
\begin{equation}
  \mathrm{I}_{ij} = \pf \int d^3\mathbf{x}\,F(\mathbf{x}),
  \label{ILF}
\end{equation}
where $\pf$ refers to the Hadamard partie finie defined in
Equation~(\ref{101}). The difference ${\cal D}\mathrm{I}$ between the DR
evaluation of the $d$-dimensional integral~(\ref{ILFd}), and its
corresponding three-dimensional evaluation, i.e.\ the partie
finie~(\ref{ILF}), reads then
\begin{equation}
  {\cal D}\mathrm{I}_{ij} = \mathrm{I}^{(d)}_{ij} - \mathrm{I}_{ij}.
  \label{DIL}
\end{equation}
Such difference depends only on the UV behaviour of the integrands, and
can therefore be computed ``locally'', i.e.\ in the vicinity of the
particles, when $r_1 \rightarrow 0$ and $r_2 \rightarrow 0$. We find
that Equation~(\ref{DIL}) depends on two constant scales $s_1$ and $s_2$
coming from Hadamard's partie finie~(\ref{101}), and on the constants
belonging to dimensional regularization, which are $\varepsilon=d-3$ and
the length scale $\ell_0$ defined by Equation~(\ref{G}). The dimensional
regularization of the 3PN quadrupole moment is then obtained as the sum
of the pHS part, and of the difference computed according to
Equation~(\ref{DIL}), namely
\begin{equation}
  \mathrm{I}_{ij}^{(\mathrm{DR})} = \mathrm{I}_{ij}^{(\mathrm{pHS})} +
  {\cal D}\mathrm{I}_{ij}.
  \label{IijDR}
\end{equation}
An important fact, hidden in our too-compact notation~(\ref{IijDR}), is
that the sum of the two terms in the right-hand side of Equation~(\ref{IijDR})
does not depend on the Hadamard regularization scales $s_1$ and $s_2$.
Therefore it is possible without changing the sum to re-express these
two terms (separately) by means of the constants $r'_1$ and $r'_2$
instead of $s_1$ and $s_2$, where $r'_1$, $r'_2$ are the two fiducial
scales entering the Hadamard-regularization result~(\ref{IijH}). This
replacement being made the pHS term in Equation~(\ref{IijDR}) is exactly the
same as the one in Equation~(\ref{IijH}). At this stage all elements are in
place to prove the following theorem~\cite{BDEI04, BDEI05dr}:

\begin{theorem}
  The DR quadrupole moment~(\ref{IijDR}) is physically equivalent to the
  Hadamard-regularized one (end result of Refs.~\cite{BIJ02, BI04mult}),
  in the sense that
  \begin{equation}
    \mathrm{I}_{ij}^{(\mathrm{HR})} =
    \lim_{\varepsilon\rightarrow 0}
    \left[ \mathrm{I}_{ij}^{(\mathrm{DR})} +
    \delta_{\bm{\xi}}\mathrm{I}_{ij}\right],
    \label{shift}
  \end{equation}
  where $\delta_{\bm{\xi}}\mathrm{I}_{ij}$ denotes the effect of the same
  shifts as determined in Theorems~\ref{th8} and~\ref{th9}, if and only if
  the HR ambiguity parameters $\xi$, $\kappa$, and $\zeta$ take the unique
  values~(\ref{xikappazeta}). Moreover, the poles $1/\varepsilon$
  separately present in the two terms in the brackets of Equation~(\ref{shift})
  cancel out, so that the physical (``dressed'') DR quadrupole moment is
  finite and given by the limit when $\varepsilon\rightarrow 0$ as shown
  in Equation~(\ref{shift}).
  \label{th10}
\end{theorem}

\noindent
This theorem finally provides an unambiguous determination of the 3PN
radiation field by dimensional regularization. Furthermore, as reviewed
in Section~\ref{subsec:8.2}, several checks of this calculation could be
done, which provide, together with comparisons with alternative
methods~\cite{DJSdim, BDE04, itoh1, itoh2}, independent confirmations
for the four ambiguity parameters $\lambda$, $\xi$, $\kappa$, and
$\zeta$, and confirm the consistency of dimensional regularization and
its validity for describing the general-relativistic dynamics of compact
bodies.

\newpage


\section{Newtonian-like Equations of Motion}
\label{sec:9}


\subsection{The 3PN acceleration and energy}
\label{subsec:9.1}

We present the acceleration of one of the particles, say the particle 1,
at the 3PN order, as well as the 3PN energy of the binary, which is
conserved in the absence of radiation reaction. To get this result we
used essentially a ``direct'' post-Newtonian method (issued from
Ref.~\cite{BFP98}), which consists of reducing the 3PN metric of an
extended regular source, worked out in Equations~(\ref{93}), to the case
where the matter tensor is made of delta functions, and then curing the
self-field divergences by means of the Hadamard regularization
technique. The equations of motion are simply the geodesic equations
associated with the regularized metric (see Ref.~\cite{BFregM} for a
proof). The Hadamard ambiguity parameter $\lambda$ is computed from
dimensional regularization in Section~\ref{subsec:8.3}. We also add the
3.5PN terms which are known from Refs.~\cite{IW93, IW95, JaraS97, PW02,
KFS03, NB05}.

Though the successive post-Newtonian approximations are really a
consequence of general relativity, the final equations of motion must be
interpreted in a Newtonian-like fashion. That is, once a convenient
general-relativistic (Cartesian) coordinate system is chosen, we should
express the results in terms of the \emph{coordinate} positions,
velocities, and accelerations of the bodies, and view the trajectories
of the particles as taking place in the absolute Euclidean space of
Newton. But because the equations of motion are actually relativistic,
they must
\begin{enumerate}
  \renewcommand{\labelenumi}{(\roman{enumi})}
\item stay manifestly invariant -- at least in harmonic coordinates --
  when we perform a global post-Newtonian-expanded Lorentz
  transformation,
\item possess the correct ``perturbative'' limit, given by the
  geodesics of the (post-Newtonian-expanded) Schwarzschild metric,
  when one of the masses tends to zero, and
\item be conservative, i.e.\ to admit a Lagrangian or
  Hamiltonian formulation, when the gravitational radiation reaction
  is turned off.
\end{enumerate}

We denote by $r_{12}=|{\bf y}_1(t)-{\bf y}_2(t)|$ the
harmonic-coordinate distance between the two particles, with ${\bf
y}_1=(y_1^i)$ and ${\bf y}_2=(y_2^i)$, by
$n_{12}^i=(y_1^i-y_2^i)/r_{12}$ the corresponding unit direction, and by
$v_1^i=dy_1^i/dt$ and $a_1^i=dv_1^i/dt$ the coordinate velocity and
acceleration of the particle 1 (and \emph{idem} for 2). Sometimes we pose
$v_{12}^i=v_1^i-v_2^i$ for the relative velocity. The usual Euclidean
scalar product of vectors is denoted with parentheses, e.g.,
$(n_{12}v_1)={\bf n}_{12}\cdot{\bf v}_1$ and $(v_1v_2)={\bf
v}_1\cdot{\bf v}_2$. The equations of the body 2 are obtained by
exchanging all the particle labels $1\leftrightarrow 2$ (remembering
that $n_{12}^i$ and $v_{12}^i$ change sign in this operation):
\begin{eqnarray}
  a_1^i &=& -\frac{G m_2 n_{12}^i}{r_{12}^2}
  \nonumber \\
  && + \frac{1}{c^2} \Bigg\{ \left[
  \frac{5 G^2 m_1 m_2}{r_{12}^3} + \frac{4 G^2 m_2^2}{r_{12}^3} +
  \frac{G m_2}{r_{12}^2} \left( \frac{3}{2}
  (n_{12}v_2)^2 - v_1^2 + 4 (v_1v_2) - 2 v_2^2 \right) \right] n_{12}^i
  \nonumber \\
  && \qquad \quad \:
  + \frac{G m_2}{r_{12}^2} \left( 4 (n_{12}v_1) - 3 (n_{12}v_2)
  \right) v_{12}^i \Bigg\}
  \nonumber \\
  && + \frac{1}{c^4} \Bigg\{ \bigg[ \! -
  \frac{57 G^3 m_1^2 m_2}{4 r_{12}^4} -
  \frac{69 G^3 m_1 m_2^2}{2 r_{12}^4} -
  \frac{9 G^3 m_2^3}{r_{12}^4}
  \nonumber \\
  && \qquad \quad \:
  + \frac{G m_2}{r_{12}^2} \bigg( \!\! - \frac{15}{8}
  (n_{12}v_2)^4 + \frac{3}{2} (n_{12}v_2)^2 v_1^2 -
  6 (n_{12}v_2)^2 (v_1v_2) - 2 (v_1v_2)^2 + \frac{9}{2} (n_{12}v_2)^2 v_2^2
  \nonumber \\
  && \qquad \qquad \qquad \quad
  + 4 (v_1v_2) v_2^2 - 2 v_2^4 \bigg)
  \nonumber \\
  && \qquad \quad \:
  + \frac{G^2 m_1 m_2}{r_{12}^3} \left( \frac{39}{2}
  (n_{12}v_1)^2 - 39 (n_{12}v_1) (n_{12}v_2) + \frac{17}{2}
  (n_{12}v_2)^2 - \frac{15}{4} v_1^2 - \frac{5}{2} (v_1v_2) +
  \frac{5}{4} v_2^2 \right)
  \nonumber \\
  && \qquad \quad \:
  + \frac{G^2 m_2^2}{r_{12}^3} \left( 2 (n_{12}v_1)^2 -
  4 (n_{12}v_1) (n_{12}v_2) - 6 (n_{12}v_2)^2 -
  8 (v_1v_2) + 4 v_2^2 \right) \bigg] n_{12}^i
  \nonumber \\
  && \qquad \quad \:
  + \bigg[ \frac{G^2 m_2^2}{r_{12}^3} \left( - 2 (n_{12}v_1) -
  2 (n_{12}v_2) \right) + \frac{G^2 m_1 m_2}{r_{12}^3} \left( \!\! - \frac{63}{4}
  (n_{12}v_1) + \frac{55}{4} (n_{12}v_2) \right)
  \nonumber \\
  && \qquad \qquad ~
  + \frac{G m_2}{r_{12}^2} \bigg( \!\! - 6 (n_{12}v_1) (n_{12}v_2)^2 +
  \frac{9}{2} (n_{12}v_2)^3 + (n_{12}v_2) v_1^2 - 4 (n_{12}v_1) (v_1v_2)
  \nonumber \\
  && \qquad \qquad \qquad \qquad \,\:\:
  +\, 4 (n_{12}v_2) (v_1v_2) + 4 (n_{12}v_1) v_2^2 -
  5 (n_{12}v_2) v_2^2 \bigg) \bigg] v_{12}^i \Bigg\}
  \nonumber \\
  && + \frac{1}{c^5} \Bigg\{ \left[
  \frac{208 G^3 m_1 m_2^2}{15 r_{12}^4} (n_{12}v_{12}) -
  \frac{24 G^3 m_1^2 m_2}{5 r_{12}^4} (n_{12}v_{12}) +
  \frac{12 G^2 m_1 m_2}{5 r_{12}^3} (n_{12}v_{12}) v_{12}^2
  \right] n_{12}^i
  \nonumber \\
  && \qquad \quad \;\,
  + \left[ \frac{8 G^3 m_1^2 m_2}{5 r_{12}^4} -
  \frac{32 G^3 m_1 m_2^2}{5 r_{12}^4} -
  \frac{4 G^2 m_1 m_2}{5 r_{12}^3} v_{12}^2 \right] v_{12}^i \Bigg\}
  \nonumber \\
  && + \frac{1}{c^6} \Bigg\{ \bigg[ \frac{G m_2}{r_{12}^2}
  \bigg( \frac{35}{16} (n_{12}v_2)^6 - \frac{15}{8} (n_{12}v_2)^4
  v_1^2 + \frac{15}{2} (n_{12}v_2)^4 (v_1v_2) + 3 (n_{12}v_2)^2 (v_1v_2)^2 
  \nonumber \\
  && \qquad \qquad \qquad \;
  - \frac{15}{2} (n_{12}v_2)^4 v_2^2 + \frac{3}{2} (n_{12}v_2)^2 v_1^2
  v_2^2 - 12 (n_{12}v_2)^2 (v_1v_2) v_2^2 - 2 (v_1v_2)^2 v_2^2 
  \nonumber \\
  && \qquad \qquad \qquad \;
  + \frac{15}{2} (n_{12}v_2)^2 v_2^4 + 4 (v_1v_2) v_2^4 - 2 v_2^6 \bigg)
  \nonumber \\
  && \qquad \quad \:
  + \frac{G^2 m_1 m_2}{r_{12}^3} \bigg( \!\! - \frac{171}{8}
  (n_{12}v_1)^4 + \frac{171}{2} (n_{12}v_1)^3 (n_{12}v_2) -
  \frac{723}{4} (n_{12}v_1)^2 (n_{12}v_2)^2 
  \nonumber \\
  && \qquad \qquad \qquad \qquad ~\:
  + \frac{383}{2} (n_{12}v_1) (n_{12}v_2)^3 -
  \frac{455}{8} (n_{12}v_2)^4 + \frac{229}{4} (n_{12}v_1)^2 v_1^2
  \nonumber \\
  && \qquad \qquad \qquad \qquad ~\:
  - \frac{205}{2} (n_{12}v_1) (n_{12}v_2) v_1^2 + \frac{191}{4}
  (n_{12}v_2)^2 v_1^2 - \frac{91}{8} v_1^4 - \frac{229}{2} (n_{12}v_1)^2 (v_1v_2) 
  \nonumber \\
  && \qquad \qquad \qquad \qquad ~\:
  + 244 (n_{12}v_1) (n_{12}v_2) (v_1v_2) - \frac{225}{2}
  (n_{12}v_2)^2 (v_1v_2) + \frac{91}{2} v_1^2 (v_1v_2) 
  \nonumber \\
  && \qquad \qquad \qquad \qquad ~\:
  - \frac{177}{4} (v_1v_2)^2 + \frac{229}{4} (n_{12}v_1)^2 v_2^2 -
  \frac{283}{2} (n_{12}v_1) (n_{12}v_2) v_2^2
  \nonumber \\
  && \qquad \qquad \qquad \qquad ~\:
  + \frac{259}{4} (n_{12}v_2)^2 v_2^2 - \frac{91}{4} v_1^2 v_2^2 +
  43 (v_1v_2) v_2^2 - \frac{81}{8} v_2^4 \bigg)
  \nonumber \\
  && \qquad \quad \:
  + \frac{G^2 m_2^2}{r_{12}^3} \bigg( \!\! - 6 (n_{12}v_1)^2
  (n_{12}v_2)^2 + 12 (n_{12}v_1) (n_{12}v_2)^3 + 6 (n_{12}v_2)^4
  \nonumber \\
  && \qquad \qquad \qquad \quad \,\:
  +\, 4 (n_{12}v_1) (n_{12}v_2) (v_1v_2) +
  12 (n_{12}v_2)^2 (v_1v_2) + 4 (v_1v_2)^2 
  \nonumber \\
  && \qquad \qquad \qquad \quad \,\:
  -\, 4 (n_{12}v_1) (n_{12}v_2) v_2^2 -
  12 (n_{12}v_2)^2 v_2^2 - 8 (v_1v_2) v_2^2 + 4 v_2^4 \bigg)
  \nonumber \\
  && \qquad \quad \:
  + \frac{G^3 m_2^3}{r_{12}^4} \left( \!\! - (n_{12}v_1)^2 +
  2 (n_{12}v_1) (n_{12}v_2) + \frac{43}{2} (n_{12}v_2)^2 +
  18 (v_1v_2) - 9 v_2^2 \right)
  \nonumber \\
  && \qquad \quad \:
  + \frac{G^3 m_1 m_2^2}{r_{12}^4} \bigg( \frac{415}{8}
  (n_{12}v_1)^2 - \frac{375}{4} (n_{12}v_1) (n_{12}v_2) +
  \frac{1113}{8} (n_{12}v_2)^2 - \frac{615}{64} (n_{12}v_{12})^2 \pi^2 
  \nonumber \\
  && \qquad \qquad \qquad \qquad ~\,\,
  + 18 v_1^2 +
  \frac{123}{64} \pi^2 v_{12}^2 + 33 (v_1v_2)
  - \frac{33}{2} v_2^2 \bigg)
  \nonumber \\
  && \qquad \quad \:
  + \frac{G^3 m_1^2 m_2}{r_{12}^4} \bigg( \!\! - \frac{45887}{168}
  (n_{12}v_1)^2 + \frac{24025}{42} (n_{12}v_1) (n_{12}v_2) -
  \frac{10469}{42} (n_{12}v_2)^2 + \frac{48197}{840} v_1^2 
  \nonumber \\
  && \qquad \qquad \qquad \qquad ~\:
  - \frac{36227}{420} (v_1v_2) + \frac{36227}{840} v_2^2 + 110 (n_{12}v_{12})^2
  \ln \left( \frac{r_{12}}{r'_1} \right) -
  22 v_{12}^2 \ln \left( \frac{r_{12}}{r'_1} \right) \bigg)
  \nonumber \\
  && \qquad \quad \:
  + \frac{16 G^4 m_2^4}{r_{12}^5} + \frac{G^4 m_1^2 m_2^2}{r_{12}^5}
  \left( 175 - \frac{41}{16} \pi^2 \right)
  + \frac{G^4 m_1^3 m_2}{r_{12}^5} \left( \!\! - \frac{3187}{1260} +
  \frac{44}{3} \ln \left( \frac{r_{12}}{r'_1} \right) \right)
  \nonumber \\
  &&\qquad \quad \:
  + \frac{G^4 m_1 m_2^3}{r_{12}^5} \left( \frac{110741}{630}
  - \frac{41}{16} \pi^2 - \frac{44}{3}
  \ln \left( \frac{r_{12}}{r'_2} \right) \right) \bigg] n_{12}^i
  \nonumber \\
  && \qquad \quad \:
  + \bigg[ \frac{G m_2}{r_{12}^2} \bigg( \frac{15}{2} (n_{12}v_1)
  (n_{12}v_2)^4 - \frac{45}{8} (n_{12}v_2)^5 - \frac{3}{2}
  (n_{12}v_2)^3 v_1^2 + 6 (n_{12}v_1) (n_{12}v_2)^2 (v_1v_2) 
  \nonumber \\
  && \qquad \qquad \qquad \quad \:\:\:
  -\, 6 (n_{12}v_2)^3 (v_1v_2) - 2 (n_{12}v_2) (v_1v_2)^2 -
  12 (n_{12}v_1)(n_{12}v_2)^2 v_2^2 + 12 (n_{12}v_2)^3 v_2^2
  \nonumber \\
  && \qquad \qquad \qquad \quad \:\:\:
  + (n_{12}v_2) v_1^2 v_2^2 - 4 (n_{12}v_1) (v_1v_2) v_2^2 +
  8 (n_{12}v_2) (v_1v_2) v_2^2 + 4 (n_{12}v_1) v_2^4
  \nonumber \\
  && \qquad \qquad \qquad \quad \:\:\:
  -\, 7 (n_{12}v_2) v_2^4 \bigg)
  \nonumber \\
  && \qquad \qquad ~
  + \frac{G^2 m_2^2}{r_{12}^3} \bigg( \!\! - 2 (n_{12}v_1)^2
  (n_{12}v_2) + 8 (n_{12}v_1) (n_{12}v_2)^2 + 2 (n_{12}v_2)^3 +
  2 (n_{12}v_1) (v_1v_2) 
  \nonumber \\
  && \qquad \qquad \qquad \qquad \quad
  + 4 (n_{12}v_2) (v_1v_2) - 2 (n_{12}v_1) v_2^2 - 4 (n_{12}v_2) v_2^2 \bigg)
  \nonumber \\
  && \qquad \qquad ~
  + \frac{G^2 m_1 m_2}{r_{12}^3} \bigg( \!\! - \frac{243}{4}
  (n_{12}v_1)^3 + \frac{565}{4} (n_{12}v_1)^2 (n_{12}v_2) -
  \frac{269}{4} (n_{12}v_1) (n_{12}v_2)^2 
  \nonumber \\
  && \qquad \qquad \qquad \qquad \qquad \;
  - \frac{95}{12} (n_{12}v_2)^3 + \frac{207}{8} (n_{12}v_1) v_1^2 -
  \frac{137}{8} (n_{12}v_2) v_1^2 - 36 (n_{12}v_1) (v_1v_2)
  \nonumber \\
  && \qquad \qquad \qquad \qquad \qquad \;
  + \frac{27}{4} (n_{12}v_2) (v_1v_2) + \frac{81}{8} (n_{12}v_1)
  v_2^2 + \frac{83}{8} (n_{12}v_2) v_2^2 \bigg)
  \nonumber \\
  && \qquad \qquad ~
  + \frac{G^3 m_2^3}{r_{12}^4} \left( 4 (n_{12}v_1) +
  5 (n_{12}v_2) \right)
  \nonumber \\
  && \qquad \qquad ~
  + \frac{G^3 m_1 m_2^2}{r_{12}^4} \left( \!\! - \frac{307}{8}
  (n_{12}v_1) + \frac{479}{8} (n_{12}v_2) + \frac{123}{32}
  (n_{12}v_{12}) \pi^2 \right)
  \nonumber \\
  && \qquad \qquad ~
  + \frac{G^3 m_1^2 m_2}{r_{12}^4} \left( \frac{31397}{420}
  (n_{12}v_1) - \frac{36227}{420} (n_{12}v_2) -
  44 (n_{12}v_{12}) \ln \left( \frac{r_{12}}{r'_1} \right)
  \right) \bigg] v_{12}^i \Bigg\}
  \nonumber \\
  && + \frac{1}{c^7} \Bigg\{ \bigg[ \frac{G^4 m_1^3m_2}{r_{12}^5} \left(
  \frac{3992}{105} (n_{12} v_1) - \frac{4328}{105} (n_{12} v_2) \right)
  \nonumber \\
  && \qquad \quad \;
  + \frac{G^4 m_1^2 m_2^2}{r_{12}^6}
  \left( \!\! - \frac{13576}{105} (n_{12} v_1) + \frac{2872}{21} (n_{12} v_2)
  \right) - \frac{3172}{21} \frac{G^4 m_1 m_2^3}{r_{12}^6} (n_{12} v_{12})
  \nonumber \\
  && \qquad \quad \;
  + \frac{G^3 m_1^2m_2}{r_{12}^4} \bigg(
  48 (n_{12} v_1)^3 - \frac{696}{5} (n_{12} v_1)^2 (n_{12} v_2) +
  \frac{744}{5} (n_{12} v_1) (n_{12} v_2)^2 - \frac{288}{5} (n_{12} v_2)^3
  \nonumber \\
  && \qquad \qquad \qquad \qquad \quad
  - \frac{4888}{105} (n_{12} v_1) v_1^2 + \frac{5056}{105}
  (n_{12} v_2) v_1^2 + \frac{2056}{21} (n_{12} v_1) (v_1 v_2)
  \nonumber \\
  && \qquad \qquad \qquad \qquad \quad
  - \frac{2224}{21} (n_{12} v_2) (v_1 v_2) - \frac{1028}{21}
  (n_{12} v_1) v_2^2 + \frac{5812}{105} (n_{12} v_2) v_2^2 \bigg)
  \nonumber \\
  && \qquad \quad \;
  + \frac{G^3 m_1 m_2^2}{r_{12}^4} \bigg( \!\! - \frac{582}{5}
  (n_{12} v_1)^3 + \frac{1746}{5} (n_{12} v_1)^2 (n_{12} v_2) -
  \frac{1954}{5} (n_{12} v_1) (n_{12} v_2)^2
  \nonumber \\
  && \qquad \qquad \qquad \qquad ~
  + 158 (n_{12} v_2)^3 +\frac{3568}{105} (n_{12} v_{12}) v_1^2 -
  \frac{2864}{35} (n_{12} v_1) (v_1 v_2)
  \nonumber \\
  && \qquad \qquad \qquad \qquad ~
  + \frac{10048}{105} (n_{12} v_2) (v_1 v_2) +\frac{1432}{35} (n_{12} v_1)
  v_2^2 - \frac{5752}{105} (n_{12} v_2) v_2^2 \bigg)
  \nonumber \\
  && \qquad \quad \;
  + \frac{G^2 m_1 m_2}{r_{12}^3} \bigg( \!\! - 56 (n_{12} v_{12})^5 +
  60 (n_{12} v_{1})^3 v_{12}^2 - 180 (n_{12} v_{1})^2 (n_{12} v_{2}) v_{12}^2
  \nonumber \\
  && \qquad \qquad \qquad \qquad ~\;
  + 174 (n_{12} v_{1}) (n_{12} v_{2})^2 v_{12}^2 - 54 (n_{12} v_{2})^3
  v_{12}^2 - \frac{246}{35} (n_{12} v_{12}) v_1^4 
  \nonumber \\
  && \qquad \qquad \qquad \qquad ~\;
  + \frac{1068}{35} (n_{12} v_1) v_1^2 (v_1 v_2) - \frac{984}{35}
  (n_{12} v_2) v_1^2 (v_1 v_2) - \frac{1068}{35} (n_{12} v_1) (v_1 v_2)^2 
  \nonumber \\
  && \qquad \qquad \qquad \qquad ~\;
  + \frac{180}{7} (n_{12} v_2) (v_1 v_2)^2 - \frac{534}{35}
  (n_{12} v_1) v_1^2 v_2^2 + \frac{90}{7} (n_{12} v_2) v_1^2 v_2^2 
  \nonumber \\
  && \qquad \qquad \qquad \qquad ~\;
  + \frac{984}{35} (n_{12} v_1) (v_1 v_2) v_2^2 - \frac{732}{35}
  (n_{12} v_2) (v_1 v_2) v_2^2 - \frac{204}{35} (n_{12} v_1) v_2^4 
  \nonumber \\
  && \qquad \qquad \qquad \qquad ~\;
  + \frac{24}{7} (n_{12} v_2) v_2^4 \bigg) \bigg] n_{12}^i
  \nonumber \\
  && \qquad ~\,
  + \bigg[ \! - \frac{184}{21} \frac{G^4 m_1^3 m_2}{r_{12}^5} +
  \frac{6224}{105} \frac{G^4 m_1^2 m_2^2}{r_{12}^6} + \frac{6388}{105}
  \frac{G^4 m_1 m_2^3}{r_{12}^6}
  \nonumber \\
  && \qquad \qquad \;
  + \frac{G^3 m_1^2 m_2}{r_{12}^4} \bigg( \frac{52}{15} (n_{12} v_1)^2 -
  \frac{56}{15} (n_{12} v_1) (n_{12} v_2) - \frac{44}{15}
  (n_{12} v_2)^2 - \frac{132}{35} v_1^2 + \frac{152}{35} (v_1 v_2)
  \nonumber \\
  && \qquad \qquad \qquad \qquad \qquad
  - \frac{48}{35} v_2^2 \bigg) \nonumber \\
  && \qquad \qquad \;
  + \frac{G^3 m_1 m_2^2}{r_{12}^4} \bigg( \frac{454}{15} (n_{12} v_1)^2 -
  \frac{372}{5} (n_{12} v_1) (n_{12} v_2) + \frac{854}{15}
  (n_{12} v_2)^2 - \frac{152}{21} v_1^2
  \nonumber \\
  && \qquad \qquad \qquad \qquad \qquad
  + \frac{2864}{105} (v_1 v_2) - \frac{1768}{105} v_2^2 \bigg)
  \nonumber \\
  && \qquad \qquad \;
  + \frac{G^2 m_1 m_2}{r_{12}^3} \bigg( 60 (n_{12} v_{12})^4 -
  \frac{348}{5} (n_{12} v_1)^2 v_{12}^2 + \frac{684}{5} (n_{12} v_1)
  (n_{12} v_2) v_{12}^2
  \nonumber \\
  && \qquad \qquad \qquad \qquad \qquad
  - 66 (n_{12} v_2)^2 v_{12}^2 + \frac{334}{35} v_1^4 -
  \frac{1336}{35} v_1^2 (v_1 v_2) + \frac{1308}{35} (v_1 v_2)^2 +
  \frac{654}{35} v_1^2 v_2^2
  \nonumber \\
  && \qquad \qquad \qquad \qquad \qquad
  - \frac{1252}{35} (v_1 v_2) v_2^2 + \frac{292}{35} v_2^4 \bigg)
  \bigg] v_{12}^i \Bigg\}
  \nonumber \\
  && +\, {\cal O} \left( \frac{1}{c^8} \right).
  \label{112}
\end{eqnarray}%
The 2.5PN and 3.5PN terms are associated with gravitational radiation
reaction. The 3PN harmonic-coordinates equations of motion depend on two
arbitrary length scales $r'_1$ and $r'_2$ associated with the logarithms
present at the 3PN order\epubtkFootnote{Notice also the dependence upon
$\pi^2$. Technically, the $\pi^2$ terms arise from non-linear
interactions involving some integrals such as
\begin{displaymath}
  \frac{1}{\pi}\int\frac{d^3{\bf x}}{r_1^2r_2^2} =
  \frac{\pi^2}{r_{12}}.
\end{displaymath}
}. It has been proved in Ref.~\cite{BFeom} that $r'_1$ and $r'_2$ are
merely linked with the choice of coordinates -- we can refer to $r'_1$
and $r'_2$ as ``gauge constants''. In our approach~\cite{BF00, BFeom},
the harmonic coordinate system is not uniquely fixed by the coordinate
condition $\partial_\mu h^{\alpha\mu}=0$. In fact there are infinitely
many harmonic coordinate systems that are local. For general smooth
sources, as in the general formalism of Part~\ref{part:a}, we expect the
existence and uniqueness of a global harmonic coordinate system. But
here we have some point-particles, with delta-function singularities,
and in this case we do not have the notion of a global coordinate system.
We can always change the harmonic coordinates by means of the gauge
vector $\eta^\alpha=\delta x^\alpha$, satisfying $\Delta\eta^\alpha=0$
except at the location of the two particles (we assume that the
transformation is at the 3PN level, so we can consider simply a
flat-space Laplace equation). More precisely, we can show that the
logarithms appearing in Equation~(\ref{112}), together with the constants
$r'_1$ and $r'_2$ therein, can be removed by the coordinate
transformation associated with the 3PN gauge vector (with $r_1 = |{\bf
x}-{\bf y}_1(t)|$ and $r_2 = |{\bf x}-{\bf y}_2(t)|$):
\begin{equation}
  \eta^\alpha=-\frac{22}{3}\frac{G^2 m_1 m_2}{c^6}\partial^\alpha
  \left[\frac{G m_1}{r_2}\ln\left(\frac{r_{12}}{r'_1}\right)+
  \frac{G m_2}{r_1}\ln\left(\frac{r_{12}}{r'_2}\right)\right].
  \label{113}
\end{equation}
Therefore, the ``ambiguity'' in the choice of the constants $r'_1$ and
$r'_2$ is completely innocuous on the physical point of view, because
the physical results must be gauge invariant. Indeed we shall verify
that $r'_1$ and $r'_2$ cancel out in our final results.

When retaining the ``even'' relativistic corrections at the 1PN, 2PN and
3PN orders, and neglecting the ``odd'' radiation reaction terms at the
2.5PN and 3.5PN orders, we find that the equations of motion admit a
conserved energy (and a Lagrangian, as we shall see), and that energy
can be straightforwardly obtained by guess-work starting from
Equation~(\ref{112}), with the result
\begin{eqnarray}
  E &=& \frac{ m_1 v_1^2}{2} - \frac{G m_1 m_2}{2 r_{12}}
  \nonumber \\
  && + \frac{1}{c^2} \left\{ \frac{G^2 m_1^2 m_2}{2 r_{12}^2} +
  \frac{3 m_1 v_1^4}{8} + \frac{G m_1 m_2}{r_{12}}
  \left(\!\! - \frac{1}{4} (n_{12}v_1)
  (n_{12}v_2) + \frac{3}{2} v_1^2 - \frac{7}{4} (v_1v_2) \right)
  \right\}
  \nonumber \\
  && + \frac{1}{c^4} \Bigg\{ \! - \frac{G^3 m_1^3 m_2}{2 r_{12}^3} -
  \frac{19 G^3 m_1^2 m_2^2}{8 r_{12}^3} + \frac{5 m_1 v_1^6}{16}
  \nonumber \\
  && \qquad ~\,
  + \frac{G m_1 m_2}{r_{12}} \bigg( \frac{3}{8} (n_{12}v_1)^3
  (n_{12}v_2) + \frac{3}{16} (n_{12}v_1)^2 (n_{12}v_2)^2 -
  \frac{9}{8} (n_{12}v_1) (n_{12}v_2) v_1^2 
  \nonumber \\
  && \qquad \qquad \qquad \qquad \:
  - \frac{13}{8} (n_{12}v_2)^2 v_1^2 + \frac{21}{8} v_1^4 +
  \frac{13}{8} (n_{12}v_1)^2 (v_1v_2) + \frac{3}{4} (n_{12}v_1)
  (n_{12}v_2) (v_1v_2)
  \nonumber \\
  && \qquad \qquad \qquad \qquad \:
  - \frac{55}{8} v_1^2 (v_1v_2) + \frac{17}{8} (v_1v_2)^2 +
  \frac{31}{16} v_1^2 v_2^2 \bigg)
  \nonumber \\
  && \qquad ~\,
  + \frac{G^2 m_1^2 m_2}{r_{12}^2} \left( \frac{29}{4}
  (n_{12}v_1)^2 - \frac{13}{4} (n_{12}v_1) (n_{12}v_2) +
  \frac{1}{2}(n_{12}v_2)^2 - \frac{3}{2} v_1^2 +
  \frac{7}{4} v_2^2 \right) \Bigg\}
  \nonumber \\
  && + \frac{1}{c^6} \Bigg\{ \frac{35 m_1 v_1^8}{128}
  \nonumber \\
  && \qquad ~\,
  + \frac{G m_1 m_2}{r_{12}} \bigg( \!\! - \frac{5}{16}
  (n_{12}v_1)^5 (n_{12}v_2) - \frac{5}{16} (n_{12}v_1)^4
  (n_{12}v_2)^2
  - \frac{5}{32} (n_{12}v_1)^3 (n_{12}v_2)^3
  \nonumber \\
  && \qquad \qquad \qquad \qquad
  + \frac{19}{16} (n_{12}v_1)^3 (n_{12}v_2) v_1^2 +
  \frac{15}{16} (n_{12}v_1)^2 (n_{12}v_2)^2 v_1^2 +
  \frac{3}{4} (n_{12}v_1) (n_{12}v_2)^3 v_1^2
  \nonumber \\
  && \qquad \qquad \qquad \qquad
  + \frac{19}{16} (n_{12}v_2)^4 v_1^2 -
  \frac{21}{16} (n_{12}v_1) (n_{12}v_2) v_1^4 - 2 (n_{12}v_2)^2 v_1^4 
  \nonumber \\
  && \qquad \qquad \qquad \qquad
  + \frac{55}{16} v_1^6 - \frac{19}{16} (n_{12}v_1)^4 (v_1v_2) -
  (n_{12}v_1)^3 (n_{12}v_2) (v_1v_2)
  \nonumber \\
  && \qquad \qquad \qquad \qquad
  - \frac{15}{32} (n_{12}v_1)^2 (n_{12}v_2)^2 (v_1v_2) +
  \frac{45}{16} (n_{12}v_1)^2 v_1^2 (v_1v_2)
  \nonumber \\
  && \qquad \qquad \qquad \qquad
  + \frac{5}{4} (n_{12}v_1) (n_{12}v_2) v_1^2 (v_1v_2) +
  \frac{11}{4}(n_{12}v_2)^2 v_1^2 (v_1v_2) - \frac{139}{16} v_1^4 (v_1v_2)
  \nonumber \\
  && \qquad \qquad \qquad \qquad
  - \frac{3}{4} (n_{12}v_1)^2 (v_1v_2)^2 + \frac{5}{16} (n_{12}v_1)
  (n_{12}v_2) (v_1v_2)^2 + \frac{41}{8} v_1^2 (v_1v_2)^2 + \frac{1}{16} (v_1v_2)^3
  \nonumber \\
  && \qquad \qquad \qquad \qquad
   - \frac{45}{16} (n_{12}v_1)^2 v_1^2 v_2^2 -
   \frac{23}{32} (n_{12}v_1) (n_{12}v_2) v_1^2 v_2^2 +
  \frac{79}{16} v_1^4 v_2^2 - \frac{161}{32} v_1^2 (v_1v_2) v_2^2 \bigg)
  \nonumber \\
  && \qquad ~\,
  + \frac{G^2 m_1^2 m_2}{r_{12}^2} \bigg( \!\! - \frac{49}{8}
  (n_{12}v_1)^4 + \frac{75}{8} (n_{12}v_1)^3 (n_{12}v_2) -
  \frac{187}{8} (n_{12}v_1)^2 (n_{12}v_2)^2
  \nonumber \\
  && \qquad \qquad \qquad \qquad ~
  + \frac{247}{24} (n_{12}v_1) (n_{12}v_2)^3 + \frac{49}{8}
  (n_{12}v_1)^2 v_1^2 + \frac{81}{8} (n_{12}v_1) (n_{12}v_2) v_1^2
  \nonumber \\
  && \qquad \qquad \qquad \qquad ~
  - \frac{21}{4} (n_{12}v_2)^2 v_1^2 + \frac{11}{2} v_1^4 -
  \frac{15}{2} (n_{12}v_1)^2 (v_1v_2) -
  \frac{3}{2} (n_{12}v_1) (n_{12}v_2) (v_1v_2)
  \nonumber \\
  && \qquad \qquad \qquad \qquad ~
  + \frac{21}{4} (n_{12}v_2)^2 (v_1v_2) - 27 v_1^2 (v_1v_2) +
  \frac{55}{2} (v_1v_2)^2 + \frac{49}{4} (n_{12}v_1)^2 v_2^2 
  \nonumber \\
  && \qquad \qquad \qquad \qquad ~
  - \frac{27}{2} (n_{12}v_1) (n_{12}v_2) v_2^2 +
  \frac{3}{4} (n_{12}v_2)^2 v_2^2 + \frac{55}{4} v_1^2 v_2^2 -
  28 (v_1v_2) v_2^2 + \frac{135}{16} v_2^4 \bigg)
  \nonumber \\
  && \qquad ~\,
  + \frac{3 G^4 m_1^4 m_2}{8 r_{12}^4} +
  \frac{G^4 m_1^3 m_2^2}{r_{12}^4} \left( \frac{9707}{420} 
  - \frac{22}{3} \ln \left( \frac{r_{12}}{r'_1}
  \right) \! \right)
  \nonumber \\
  && \qquad ~\,
  + \frac{G^3 m_1^2 m_2^2}{r_{12}^3} \bigg( \frac{547}{12}
  (n_{12}v_1)^2 - \frac{3115}{48} (n_{12}v_1) (n_{12}v_2) -
  \frac{123}{64} (n_{12}v_1)(n_{12}v_{12}) \pi^2 - \frac{575}{18} v_1^2
  \nonumber \\
  && \qquad \qquad \qquad \qquad ~
  + \frac{41}{64} \pi^2 (v_1v_{12}) + \frac{4429}{144} (v_1v_2) \bigg)
  \nonumber \\
  && \qquad ~\,
  + \frac{G^3 m_1^3 m_2}{r_{12}^3} \bigg( \!\! - \frac{44627}{840}
  (n_{12}v_1)^2 + \frac{32027}{840} (n_{12}v_1) (n_{12}v_2) +
  \frac{3}{2} (n_{12}v_2)^2 + \frac{24187}{2520} v_1^2
  \nonumber \\
  && \qquad \qquad \qquad \qquad ~
  - \frac{27967}{2520} (v_1v_2) + \frac{5}{4} v_2^2 + 22 (n_{12}v_1)(n_{12}v_{12})
  \ln \!\left(\!\frac{r_{12}}{r'_1}\!\right)\! - \frac{22}{3} (v_1v_{12})
  \ln \!\left(\!\frac{r_{12}}{r'_1}\!\right) \! \bigg) \!\! \Bigg\}
  \nonumber \\
  && + 1 \leftrightarrow 2 + {\cal O}\left(\frac{1}{c^7} \right).
  \label{114}
\end{eqnarray}%
To the terms given above, we must add the terms 
corresponding to the relabelling $1\leftrightarrow 2$.
Actually, this energy is not conserved because of 
the radiation reaction. Thus its time derivative, as computed by means 
of the 3PN equations of motion themselves (i.e.\ order-reducing all
the accelerations), is purely equal to the 2.5PN effect,
\begin{eqnarray}
  \frac{d E}{dt} &=& \frac{4}{5} \frac{G^2m_1^2m_2}{c^5 r_{12}^3}
  \!\left[(v_1v_{12}) \left(\!\! - v_{12}^2 + 2\frac{Gm_1}{r_{12}} \!-
  8 \frac{Gm_2}{r_{12}} \!\right) + (n_{12}v_1)(n_{12}v_{12}) \left(\!
  3v_{12}^2 - 6\frac{Gm_1}{r_{12}} \!+
  \frac{52}{3}\frac{Gm_2}{r_{12}}\!\right)\!\right]
  \nonumber \\
  && + 1 \leftrightarrow 2 + {\cal O}\left(\frac{1}{c^7}\right).
  \label{115}
\end{eqnarray}%
The resulting ``balance equation'' can be better expressed by
transfering to the left-hand side certain 2.5PN terms so that the
right-hand side takes the familiar form of a total energy flux. Posing
\begin{equation}
  {\widetilde E} = E+\frac{4 G^2 m_1^2 m_2}{5 c^5 r_{12}^2}
  (n_{12}v_1) \left[ v_{12}^2 -\frac{2 G (m_1-m_2)}{r_{12}} \right] +
  1\leftrightarrow 2,
  \label{116}
\end{equation}
we find agreement with the standard Einstein quadrupole
formula~(\ref{4}, \ref{5}):
\begin{equation}
  \frac{d\widetilde{E}}{dt} =
  -\frac{G}{5 c^5} \frac{d^3 \mathrm{Q}_{ij}}{dt^3}
  \frac{d^3 \mathrm{Q}_{ij}}{dt^3} + {\cal O}\left( \frac{1}{c^7} \right),
  \label{117}
\end{equation}
where the Newtonian trace-free quadrupole moment is $\mathrm{Q}_{ij}=m_1
(y_1^i y_1^j-\frac{1}{3}\delta^{ij} {\bf y}_1^2)+1 \leftrightarrow 2$.
We refer to Iyer and Will~\cite{IW93, IW95} for the discussion of the
energy balance equation at the next 3.5PN order. As we can see, the
3.5PN equations of motion~(\ref{112}) are highly relativistic when
describing the \emph{motion}, but concerning the \emph{radiation} they are
in fact 1PN, because they contain merely the radiation reaction force at
the 2.5PN\,+\,3.5PN orders.


\subsection{Lagrangian and Hamiltonian formulations}
\label{subsec:9.2}

The conservative part of the equations of motion in harmonic
coordinates~(\ref{112}) is derivable from a \emph{generalized}
Lagrangian, depending not only on the positions and velocities of the
bodies, but also on their accelerations: $a_1^i=dv_1^i/dt$ and
$a_2^i=dv_2^i/dt$. As shown by Damour and Deruelle~\cite{DD81b}, the
accelerations in the harmonic-coordinates Lagrangian occur already
from the 2PN order. This fact is in accordance with a general result
of Martin and Sanz~\cite{MS} that $N$-body equations of motion cannot
be derived from an ordinary Lagrangian beyond the 1PN level, provided
that the gauge conditions preserve the Lorentz invariance. Note that
we can always arrange for the dependence of the Lagrangian upon the
accelerations to be \emph{linear}, at the price of adding some
so-called ``multi-zero'' terms to the Lagrangian, which do not modify
the equations of motion (see, e.g., Ref.~\cite{DS85}). At the 3PN 
level, we find that the Lagrangian also depends on accelerations. It is
notable that these accelerations are sufficient -- there is no need to
include derivatives of accelerations. Note also that the Lagrangian
is not unique because we can always add to it a total time derivative
$dF/dt$, where $F$ depends on the positions and velocities, without
changing the dynamics. We find~\cite{ABF01}
\begin{eqnarray}
  L^\mathrm{harm} &=& \frac{G m_1 m_2}{2 r_{12}} + \frac{m_1 v_1^2}{2}
  \nonumber \\
  && + \frac{1}{c^2} \left\{ \! - \frac{G^2 m_1^2 m_2}{2 r_{12}^2} +
  \frac{m_1 v_1^4}{8} + \frac{G m_1 m_2}{r_{12}} \left( \!\! - \frac{1}{4} (n_{12}v_1)
  (n_{12}v_2) + \frac{3}{2} v_1^2 - \frac{7}{4} (v_1v_2) \right) \! \right\}
  \nonumber \\
  && + \frac{1}{c^4} \Bigg\{ \frac{G^3 m_1^3 m_2}{2 r_{12}^3} +
  \frac{19 G^3 m_1^2 m_2^2}{8 r_{12}^3}
  \nonumber \\
  && \qquad ~\,
  + \frac{G^2 m_1^2 m_2}{r_{12}^2} \left( \frac{7}{2}
  (n_{12}v_1)^2 - \frac{7}{2} (n_{12}v_1) (n_{12}v_2) +
  \frac{1}{2}(n_{12}v_2)^2 + \frac{1}{4} v_1^2 -
  \frac{7}{4} (v_1v_2) + \frac{7}{4} v_2^2 \right)
  \nonumber \\
  && \qquad ~\,
  + \frac{G m_1 m_2}{r_{12}} \bigg( \frac{3}{16} (n_{12}v_1)^2
  (n_{12}v_2)^2 - \frac{7}{8} (n_{12}v_2)^2 v_1^2 + \frac{7}{8} v_1^4 +
  \frac{3}{4} (n_{12}v_1) (n_{12}v_2) (v_1v_2) 
  \nonumber \\
  && \qquad \qquad \qquad \qquad \:
  - 2 v_1^2 (v_1v_2) + \frac{1}{8} (v_1v_2)^2 + \frac{15}{16} v_1^2 v_2^2 \bigg) +
  \frac{m_1 v_1^6}{16}
  \nonumber \\
  && \qquad ~\,
  + G m_1 m_2 \left( \!\! - \frac{7}{4} (a_1 v_2) (n_{12}v_2) -
  \frac{1}{8} (n_{12} a_1) (n_{12}v_2)^2 + \frac{7}{8}
  (n_{12} a_1) v_2^2 \right) \! \Bigg\}
  \nonumber \\
  && + \frac{1}{c^6} \Bigg\{ \frac{G^2 m_1^2 m_2}{r_{12}^2}
  \bigg( \frac{13}{18} (n_{12}v_1)^4 + \frac{83}{18} (n_{12}v_1)^3
  (n_{12}v_2) - \frac{35}{6} (n_{12}v_1)^2 (n_{12}v_2)^2 - \frac{245}{24}
  (n_{12}v_1)^2 v_1^2
  \nonumber \\
  && \qquad \qquad \qquad \quad ~
  + \frac{179}{12} (n_{12}v_1) (n_{12}v_2) v_1^2 - \frac{235}{24}
  (n_{12}v_2)^2 v_1^2 + \frac{373}{48} v_1^4 + \frac{529}{24} (n_{12}v_1)^2 (v_1v_2)
  \nonumber \\
  && \qquad \qquad \qquad \quad ~
  - \frac{97}{6} (n_{12}v_1) (n_{12}v_2) (v_1v_2) - \frac{719}{24} v_1^2 (v_1v_2) +
  \frac{463}{24} (v_1v_2)^2 - \frac{7}{24} (n_{12}v_1)^2 v_2^2
  \nonumber \\
  && \qquad \qquad \qquad \quad ~
  - \frac{1}{2} (n_{12}v_1) (n_{12}v_2) v_2^2 + \frac{1}{4}
  (n_{12}v_2)^2 v_2^2 + \frac{463}{48} v_1^2 v_2^2 - \frac{19}{2} (v_1v_2) v_2^2 +
  \frac{45}{16} v_2^4 \bigg)
  \nonumber \\
  && \qquad ~\,
  + \frac{5m_1 v_1^8}{128}
  \nonumber \\
  && \qquad ~\,
  + G m_1 m_2 \bigg(\frac{3}{8} (a_1 v_2) (n_{12}v_1)
  (n_{12}v_2)^2 + \frac{5}{12} (a_1 v_2) (n_{12}v_2)^3 +
  \frac{1}{8} (n_{12} a_1) (n_{12}v_1) (n_{12}v_2)^3 
  \nonumber \\
  && \qquad \qquad \qquad \qquad
  + \frac{1}{16} (n_{12} a_1) (n_{12}v_2)^4 + \frac{11}{4} (a_1 v_1)
  (n_{12}v_2) v_1^2 - (a_1 v_2) (n_{12}v_2) v_1^2
  \nonumber \\
  && \qquad \qquad \qquad \qquad
  - 2 (a_1 v_1) (n_{12}v_2) (v_1v_2) + \frac{1}{4} (a_1 v_2)
  (n_{12}v_2) (v_1v_2)
  \nonumber \\
  && \qquad \qquad \qquad \qquad
  + \frac{3}{8} (n_{12} a_1) (n_{12}v_2)^2 (v_1v_2) -
  \frac{5}{8} (n_{12} a_1) (n_{12}v_1)^2 v_2^2 +
  \frac{15}{8} (a_1 v_1) (n_{12}v_2) v_2^2 
  \nonumber \\
  && \qquad \qquad \qquad \qquad
  - \frac{15}{8} (a_1 v_2) (n_{12}v_2) v_2^2 -
  \frac{1}{2} (n_{12} a_1) (n_{12}v_1) (n_{12}v_2) v_2^2
  \nonumber \\
  && \qquad \qquad \qquad \qquad
  - \frac{5}{16} (n_{12} a_1) (n_{12}v_2)^2 v_2^2 \bigg)
  \nonumber \\
  && \qquad ~\,
  + \frac{G^2 m_1^2 m_2}{r_{12}} \bigg( \!\!
  - \frac{235}{24} (a_2 v_1) (n_{12}v_1) -
  \frac{29}{24} (n_{12} a_2) (n_{12}v_1)^2 -
  \frac{235}{24} (a_1 v_2) (n_{12}v_2) 
  \nonumber \\
  && \qquad \qquad \qquad \qquad ~
  - \frac{17}{6} (n_{12} a_1) (n_{12}v_2)^2 +
  \frac{185}{16} (n_{12} a_1) v_1^2 - \frac{235}{48} (n_{12} a_2) v_1^2 
  \nonumber \\
  && \qquad \qquad \qquad \qquad ~
  - \frac{185}{8} (n_{12} a_1)
  (v_1v_2) + \frac{20}{3} (n_{12} a_1) v_2^2 \bigg)
  \nonumber \\
  && \qquad ~\,
  + \frac{G m_1 m_2}{r_{12}} \bigg( \!\! -
  \frac{5}{32} (n_{12}v_1)^3 (n_{12}v_2)^3 +
  \frac{1}{8} (n_{12}v_1) (n_{12}v_2)^3 v_1^2 +
  \frac{5}{8} (n_{12}v_2)^4 v_1^2 
  \nonumber \\
  && \qquad \qquad \qquad \qquad
  - \frac{11}{16} (n_{12}v_1)
  (n_{12}v_2) v_1^4 + \frac{1}{4} (n_{12}v_2)^2 v_1^4 +
  \frac{11}{16} v_1^6
  \nonumber \\
  && \qquad \qquad \qquad \qquad
  - \frac{15}{32} (n_{12}v_1)^2 (n_{12}v_2)^2 (v_1v_2) +
  (n_{12}v_1) (n_{12}v_2) v_1^2 (v_1v_2)
  \nonumber \\
  && \qquad \qquad \qquad \qquad
  + \frac{3}{8} (n_{12}v_2)^2 v_1^2 (v_1v_2)
  - \frac{13}{16} v_1^4 (v_1v_2) +
  \frac{5}{16} (n_{12}v_1) (n_{12}v_2) (v_1v_2)^2
  \nonumber \\
  && \qquad \qquad \qquad \qquad
  + \frac{1}{16} (v_1v_2)^3 - \frac{5}{8} (n_{12}v_1)^2
  v_1^2 v_2^2 - \frac{23}{32} (n_{12}v_1) (n_{12}v_2) v_1^2 v_2^2 +
  \frac{1}{16} v_1^4 v_2^2 
  \nonumber \\
  && \qquad \qquad \qquad \qquad
  - \frac{1}{32} v_1^2 (v_1v_2) v_2^2 \bigg)
  \nonumber \\
  && \qquad ~\,
  - \frac{3 G^4 m_1^4 m_2}{8 r_{12}^4} + \frac{G^4 m_1^3 m_2^2}{r_{12}^4}
  \left( \!\! - \frac{9707}{420} +
  \frac{22}{3} \ln \left(\frac{r_{12}}{r'_1} \right) \!\! \right)
  \nonumber \\
  && \qquad ~\,
  + \frac{G^3 m_1^2 m_2^2}{r_{12}^3} \bigg( \frac{383}{24}
  (n_{12}v_1)^2 - \frac{889}{48} (n_{12}v_1) (n_{12}v_2) -
  \frac{123}{64} (n_{12}v_1)(n_{12}v_{12}) \pi^2 - \frac{305}{72} v_1^2
  \nonumber \\
  && \qquad \qquad \qquad \qquad ~
  + \frac{41}{64} \pi^2 (v_1v_{12}) + \frac{439}{144} (v_1v_2) \bigg)
  \nonumber \\
  && \qquad ~\,
  + \frac{G^3 m_1^3 m_2}{r_{12}^3} \bigg( \!\!
  - \frac{8243}{210} (n_{12}v_1)^2 +
  \frac{15541}{420} (n_{12}v_1) (n_{12}v_2) + \frac{3}{2} (n_{12}v_2)^2 +
  \frac{15611}{1260} v_1^2 
  \nonumber \\
  && \qquad \qquad \qquad \qquad ~
  - \frac{17501}{1260} (v_1v_2) + \frac{5}{4} v_2^2 +
  22 (n_{12}v_1)(n_{12}v_{12}) \ln \left( \frac{r_{12}}{r'_1} \right) 
  \nonumber \\
  && \qquad \qquad \qquad \qquad ~
  - \frac{22}{3} (v_1v_{12})
  \ln \left( \frac{r_{12}}{r'_1} \right) \!\! \bigg) \! \Bigg\}
  \nonumber \\
  && + 1 \leftrightarrow 2 + {\cal O}\left(\frac{1}{c^7}\right).
  \label{118}
\end{eqnarray}%
Witness the accelerations occuring at the 2PN and 3PN orders; see also
the gauge-dependent logarithms of $r_{12}/r'_1$ and $r_{12}/r'_2$. We
refer to~\cite{ABF01} for the explicit expressions of the ten conserved
quantities corresponding to the integrals of energy (also given in
Equation~(\ref{114})), linear and angular momenta, and center-of-mass
position. Notice that while it is strictly forbidden to replace the
accelerations by the equations of motion in the Lagrangian, this can and
\emph{should} be done in the final expressions of the conserved integrals
derived from that Lagrangian.

Now we want to exhibit a transformation of the particles 
dynamical variables -- or
\emph{contact} transformation, as it is called in the jargon -- 
which transforms the 3PN harmonic-coordinates Lagrangian~(\ref{118})
into a new Lagrangian, valid in some ADM or ADM-like coordinate system, 
and such that the associated Hamiltonian
coincides with the 3PN Hamiltonian that has been obtained by 
Damour, Jaranowski, and Sch\"afer~\cite{DJSpoinc}.
In ADM coordinates the Lagrangian will be ``ordinary'', 
depending only on the positions and velocities of the bodies.
Let this contact transformation be $Y_1^i(t)=y_1^i(t)+\delta 
y_1^i(t)$ and $1\leftrightarrow 2$, 
where $Y_1^i$ and $y_1^i$ denote the trajectories in ADM and harmonic
coordinates, respectively. For this transformation to be able to
remove all the accelerations in the initial Lagrangian $L^\mathrm{harm}$
up to the 3PN order, we determine~\cite{ABF01} it to be necessarily of
the form
\begin{equation}
  \delta y_1^i=\frac{1}{m_1}
  \left[\frac{\partial L^\mathrm{harm}}{\partial a_1^i}+
  \frac{\partial F}{\partial v_1^i}+\frac{1}{c^6}X_1^i\right]+
  {\cal O}\left(\frac{1}{c^8}\right)
  \label{119}
\end{equation}
(and \emph{idem} $1\leftrightarrow 2$), 
where $F$ is a freely adjustable function of the 
positions and velocities, made of 2PN and 3PN terms, 
and where $X_1^i$ represents a special correction 
term, that is purely of order 3PN. The point is that once 
the function $F$ is specified there is a \emph{unique} 
determination of the correction term $X_1^i$ for the 
contact transformation to work (see Ref.~\cite{ABF01} 
for the details). Thus, the freedom we have is entirely coded into
the function $F$, and the work then consists in showing 
that there exists a unique choice of $F$ for which 
our Lagrangian $L^\mathrm{harm}$ is physically equivalent, \emph{via} the
contact transformation~(\ref{119}), to the ADM Hamiltonian of
Ref.~\cite{DJSpoinc}. An interesting point is that not only
the transformation must remove all the accelerations in $L^\mathrm{harm}$, 
but it should also cancel out all the logarithms
$\ln(r_{12}/r'_1)$ and $\ln(r_{12}/r'_2)$, because there are no
logarithms in ADM coordinates. The result we find, which can be
checked to be in full agreement with the expression of the gauge
vector in Equation~(\ref{113}), is that $F$ involves the logarithmic terms
\begin{equation}
  F = \frac{22}{3}\frac{G^3 m_1 m_2}{c^6r_{12}^2}
  \left[m_1^2(n_{12}v_1)\ln\left(\frac{r_{12}}{r'_1}\right)-
  m_2^2(n_{12}v_2)\ln\left(\frac{r_{12}}{r'_2}\right)\right]+\dots,
  \label{120}
\end{equation}
together with many other non-logarithmic terms (indicated by dots) 
that are entirely specified by the isometry
of the harmonic and ADM descriptions of the motion. For this particular choice
of $F$ the ADM Lagrangian reads
\begin{equation}
  L^\mathrm{ADM}=L^\mathrm{harm}+\frac{\delta L^\mathrm{harm}}{\delta y_1^i}\delta
  y_1^i+\frac{\delta L^\mathrm{harm}}{\delta y_2^i}\delta
  y_2^i+\frac{dF}{dt}+{\cal O}\left(\frac{1}{c^8}\right).
  \label{121}
\end{equation}
Inserting into this equation all our explicit expressions we find 
\begin{eqnarray}
  L^\mathrm{ADM} &=& \frac{G m_1 m_2}{2 R_{12}} + \frac{1}{2} m_1 V_1^2
  \nonumber \\
  && + \frac{1}{c^2} \left\{ \! - \frac{G^2 m_1^2 m_2}{2 R_{12}^2} +
  \frac{1}{8} m_1 V_1^4 + \frac{G m_1 m_2}{R_{12}} \left( \!\! -
  \frac{1}{4} (N_{12}V_1)
  (N_{12}V_2) + \frac{3}{2} V_1^2 - \frac{7}{4} (V_1V_2) \right)
  \! \right\}
  \nonumber \\
  && + \frac{1}{c^4} \Bigg\{ \! \frac{G^3 m_1^3 m_2}{4 R_{12}^3} +
  \frac{5 G^3 m_1^2 m_2^2}{8R_{12}^3} + \frac{m_1 V_1^6}{16}
  \nonumber \\
  && \qquad ~
  + \frac{G^2 m_1^2 m_2}{R_{12}^2} \left( \frac{15}{8}
  (N_{12}V_1)^2 + \frac{11}{8} V_1^2 - \frac{15}{4} (V_1V_2) +
  2 V_2^2 \right)
  \nonumber \\
  && \qquad ~
  + \frac{G m_1 m_2}{R_{12}} \bigg( \frac{3}{16} (N_{12}V_1)^2
  (N_{12}V_2)^2 - \frac{1}{4} (N_{12}V_1) (N_{12}V_2) V_1^2 -
  \frac{5}{8} (N_{12}V_2)^2 V_1^2 + \frac{7}{8} V_1^4 
  \nonumber \\
  && \qquad \qquad \qquad \qquad
  + \frac{3}{4} (N_{12}V_1) (N_{12}V_2) (V_1V_2) -
  \frac{7}{4} V_1^2 (V_1V_2) + \frac{1}{8} (V_1V_2)^2 +
  \frac{11}{16} V_1^2 V_2^2 \bigg) \! \Bigg\}
  \nonumber \\
  && + \frac{1}{c^6} \Bigg\{ \! \frac{5 m_1 V_1^8}{128} -
  \frac{G^4 m_1^4 m_2}{8 R_{12}^4} + \frac{G^4 m_1^3 m_2^2}{R_{12}^4}
  \left( \!\! - \frac{227}{24} 
  + \frac{21}{32} \pi^2 \right)
  \nonumber \\
  && \qquad ~
  + \frac{G m_1 m_2}{R_{12}} \bigg( \!\! - \frac{5}{32}
  (N_{12}V_1)^3 (N_{12}V_2)^3 + \frac{3}{16} (N_{12}V_1)^2 
  (N_{12}V_2)^2 V_1^2
  \nonumber \\
  && \qquad \qquad \qquad \qquad
  + \frac{9}{16} (N_{12}V_1) (N_{12}V_2)^3 V_1^2 -
  \frac{3}{16} (N_{12}V_1) (N_{12}V_2) V_1^4 - \frac{5}{16} (N_{12}V_2)^2 V_1^4 
  \nonumber \\
  && \qquad \qquad \qquad \qquad
  + \frac{11}{16} V_1^6 - \frac{15}{32} (N_{12}V_1)^2 (N_{12}V_2)^2 (V_1V_2) +
  \frac{3}{4} (N_{12}V_1) (N_{12}V_2) V_1^2 (V_1V_2) 
  \nonumber \\
  && \qquad \qquad \qquad \qquad
  - \frac{1}{16} (N_{12}V_2)^2 V_1^2 (V_1V_2) -
  \frac{21}{16} V_1^4 (V_1V_2) + \frac{5}{16} (N_{12}V_1) (N_{12}V_2) (V_1V_2)^2
  \nonumber \\
  && \qquad \qquad \qquad \qquad
  + \frac{1}{8} V_1^2 (V_1V_2)^2 + \frac{1}{16} (V_1V_2)^3 -
  \frac{5}{16} (N_{12}V_1)^2 V_1^2 V_2^2
  \nonumber \\
  && \qquad \qquad \qquad \qquad
  - \frac{9}{32} (N_{12}V_1) (N_{12}V_2) V_1^2 V_2^2 +
  \frac{7}{8} V_1^4 V_2^2 - \frac{15}{32} V_1^2 (V_1V_2) V_2^2 \bigg)
  \nonumber \\
  && \qquad ~
  + \frac{G^2 m_1^2 m_2}{R_{12}^2} \bigg( \!\! - \frac{5}{12}
  (N_{12}V_1)^4 - \frac{13}{8} (N_{12}V_1)^3 (N_{12}V_2) -
  \frac{23}{24} (N_{12}V_1)^2 (N_{12}V_2)^2 
  \nonumber \\
  && \qquad \qquad \qquad \qquad \;
  + \frac{13}{16} (N_{12}V_1)^2 V_1^2 + \frac{1}{4} (N_{12}V_1)
  (N_{12}V_2) V_1^2 + \frac{5}{6}
  (N_{12}V_2)^2 V_1^2 + \frac{21}{16} V_1^4
  \nonumber \\
  && \qquad \qquad \qquad \qquad \;
  - \frac{1}{2} (N_{12}V_1)^2 (V_1V_2) + \frac{1}{3} 
  (N_{12}V_1) (N_{12}V_2) (V_1V_2) - \frac{97}{16} V_1^2 (V_1V_2) 
  \nonumber \\
  && \qquad \qquad \qquad \qquad \;
  + \frac{341}{48} (V_1V_2)^2 + \frac{29}{24} (N_{12}V_1)^2 V_2^2 -
  (N_{12}V_1) (N_{12}V_2) V_2^2 + \frac{43}{12} V_1^2 V_2^2 
  \nonumber \\
  && \qquad \qquad \qquad \qquad \;
  - \frac{71}{8} (V_1V_2) V_2^2 + \frac{47}{16} V_2^4 \bigg)
  \nonumber \\
  && \qquad ~
  + \frac{G^3 m_1^2 m_2^2}{R_{12}^3} \bigg( \frac{73}{16}
  (N_{12}V_1)^2 - 11 (N_{12}V_1) (N_{12}V_2) +
  \frac{3}{64} \pi^2 (N_{12}V_1)(N_{12}V_{12}) 
  \nonumber \\
  && \qquad \qquad \qquad \qquad \:\:
  - \frac{265}{48} V_1^2 - \frac{1}{64} \pi^2 (V_1V_{12}) +
  \frac{59}{8} (V_1V_2) \bigg)
  \nonumber \\
  && \qquad ~
  + \frac{G^3 m_1^3 m_2}{R_{12}^3} \left( \!\! - 5 (N_{12}V_1)^2 -
  \frac{1}{8} (N_{12}V_1) (N_{12}V_2) + \frac{173}{48} V_1^2 -
  \frac{27}{8} (V_1V_2) + \frac{13}{8} V_2^2 \right) \! \Bigg\}
  \nonumber \\
  && + 1 \leftrightarrow 2 + {\cal O} \left( \frac{1}{c^7} \right).
  \label{122}
\end{eqnarray}%
The notation is the same as in Equation~(\ref{118}), except that we use
upper-case letters to denote the ADM-coordinates positions and
velocities; thus, for instance ${\bf N}_{12}=({\bf Y}_1-{\bf
Y}_2)/R_{12}$ and $(N_{12}V_1)={\bf N}_{12}\cdot{\bf V}_1$. The Hamiltonian
is simply deduced from the latter Lagrangian by applying the usual
Legendre transformation. Posing $P_1^i=\partial L^\mathrm{ADM}/\partial
V_1^i$ and $1\leftrightarrow 2$, we get~\cite{JaraS98, JaraS99, JaraS00,
  DJSpoinc, ABF01}\epubtkFootnote{Note that in the result published in
  Ref.~\cite{DJSpoinc} the following terms are missing:
  \begin{displaymath}
    \frac{G^2}{c^6r_{12}^2}
    \bigg(-\frac{55}{12}m_1-\frac{193}{48}m_2\bigg)
    \frac{(N_{12}P_2)^2 P_1^2}{m_1m_2}+1\leftrightarrow 2.
  \end{displaymath}
  This misprint has been corrected in an Erratum~\cite{DJSpoinc}.}
\begin{eqnarray}
  H^\mathrm{ADM} &=& - \frac{G m_1 m_2}{2 R_{12}} + \frac{P_1^2}{2 m_1}
  \nonumber \\
  && + \frac{1}{c^2} \left\{ \! - \frac{P_1^4}{8 m_1^3} +
  \frac{G^2 m_1^2 m_2}{2 R_{12}^2}
  + \frac{G m_1 m_2}{R_{12}} \left( \frac{1}{4}
  \frac{(N_{12}P_1) (N_{12}P_2)}{ m_1 m_2} - \frac{3}{2}
  \frac{P_1^2}{m_1^2} + \frac{7}{4} \frac{(P_1P_2)}{m_1 m_2} \right) \right\}
  \nonumber \\
  && + \frac{1}{c^4} \Bigg\{ \! \frac{P_1^6}{16 m_1^5} -
  \frac{G^3 m_1^3 m_2}{4 R_{12}^3} - \frac{5 G^3 m_1^2 m_2^2}{8R_{12}^3}
  \nonumber \\
  && \qquad ~
  + \frac{G^2 m_1^2 m_2}{R_{12}^2} \left( \! - \frac{3}{2}
  \frac{(N_{12}P_1) (N_{12}P_2)}{m_1 m_2} + \frac{19}{4}
  \frac{P_1^2}{m_1^2} - \frac{27}{4} \frac{(P_1P_2)}{m_1 m_2} +
  \frac{5 P_2^2}{2 m_2^2} \right)
  \nonumber \\
  && \qquad ~
  + \frac{G m_1 m_2}{R_{12}} \bigg( \!\! - \frac{3}{16}
  \frac{(N_{12}P_1)^2 (N_{12}P_2)^2}{m_1^2 m_2^2} + \frac{5}{8}
  \frac{(N_{12}P_2)^2 P_1^2}{m_1^2 m_2^2}
  \nonumber \\
  && \qquad \qquad \qquad \qquad
  + \frac{5}{8} \frac{P_1^4}{m_1^4} - \frac{3}{4}
  \frac{(N_{12}P_1) (N_{12}P_2) (P_1P_2)}{m_1^2 m_2^2}
  - \frac{1}{8} \frac{(P_1P_2)^2}{m_1^2 m_2^2} - \frac{11}{16}
  \frac{P_1^2 P_2^2}{m_1^2 m_2^2} \bigg) \Bigg\}
  \nonumber \\
  && + \frac{1}{c^6} \Bigg\{ \! - \frac{5 P_1^8}{128 m_1^7} +
  \frac{G^4 m_1^4 m_2}{8 R_{12}^4} + \frac{G^4 m_1^3 m_2^2}{R_{12}^4}
  \left( \frac{227}{24} - \frac{21}{32} \pi^2 \right)
  \nonumber \\
  && \qquad ~
  + \frac{G^3 m_1^2 m_2^2}{R_{12}^3} \bigg( \!\! - \frac{43}{16}
  \frac{(N_{12}P_1)^2}{m_1^2} + \frac{119}{16}
  \frac{(N_{12}P_1) (N_{12}P_2)}{m_1 m_2} -
  \frac{3}{64} \pi^2 \frac{(N_{12}P_1)^2}{m_1^2} 
  \nonumber \\
  && \qquad \qquad \qquad \qquad \;
  + \frac{3}{64}
  \pi^2 \frac{(N_{12}P_1) (N_{12}P_2)}{m_1 m_2} - \frac{473}{48}
  \frac{P_1^2}{m_1^2} + \frac{1}{64} \pi^2 \frac{P_1^2}{m_1^2} +
  \frac{143}{16} \frac{(P_1P_2)}{m_1 m_2} 
  \nonumber \\
  && \qquad \qquad \qquad \qquad \;
  - \frac{1}{64} \pi^2 \frac{(P_1P_2)}{m_1 m_2} \bigg)
  \nonumber \\
  && \qquad ~
  + \frac{G^3 m_1^3 m_2}{R_{12}^3} \left( \frac{5}{4}
  \frac{(N_{12}P_1)^2}{m_1^2} + \frac{21}{8}
  \frac{(N_{12}P_1) (N_{12}P_2)}{m_1 m_2}
  - \frac{425}{48} \frac{P_1^2}{m_1^2} + \frac{77}{8}
  \frac{(P_1 P_2)}{m_1 m_2} - \frac{25 P_2^2}{8 m_2^2} \right)
  \nonumber \\
  && \qquad ~
  + \frac{G^2 m_1^2 m_2}{R_{12}^2} \bigg( \frac{5}{12}
  \frac{(N_{12}P_1)^4}{m_1^4} - \frac{3}{2}
  \frac{(N_{12}P_1)^3 (N_{12}P_2)}{m_1^3 m_2} +
  \frac{10}{3} \frac{(N_{12}P_1)^2 (N_{12}P_2)^2}{m_1^2 m_2^2} 
  \nonumber \\
  && \qquad \qquad \qquad \qquad ~
  + \frac{17}{16} \frac{(N_{12}P_1)^2 P_1^2}{m_1^4}
  - \frac{15}{8} \frac{(N_{12}P_1) (N_{12}P_2) P_1^2}{m_1^3 m_2} -
  \frac{55}{12} \frac{(N_{12}P_2)^2 P_1^2}{m_1^2 m_2^2}
  \nonumber \\
  && \qquad \qquad \qquad \qquad ~
  + \frac{P_1^4}{16 m_1^4} - \frac{11}{8}
  \frac{(N_{12}P_1)^2 (P_1P_2)}{m_1^3 m_2} +
  \frac{125}{12} \frac{(N_{12}P_1) (N_{12}P_2) (P_1P_2)}{m_1^2 m_2^2}
  \nonumber \\
  && \qquad \qquad \qquad \qquad ~
  - \frac{115}{16} \frac{P_1^2 (P_1P_2)}{m_1^3 m_2} + \frac{25}{48}
  \frac{(P_1P_2)^2}{m_1^2 m_2^2} - \frac{193}{48}
  \frac{(N_{12}P_1)^2 P_2^2}{m_1^2 m_2^2} +
  \frac{371}{48} \frac{P_1^2 P_2^2}{m_1^2 m_2^2} 
  \nonumber \\
  && \qquad \qquad \qquad \qquad ~
  - \frac{27}{16} \frac{P_2^4}{m_2^4} \bigg)
  \nonumber \\
  && \qquad ~
  + \frac{G m_1 m_2}{R_{12}} \bigg( \frac{5}{32}
  \frac{(N_{12}P_1)^3 (N_{12}P_2)^3}{m_1^3 m_2^3} + \frac{3}{16}
  \frac{(N_{12}P_1)^2 (N_{12}P_2)^2 P_1^2}{m_1^4 m_2^2}
  \nonumber \\
  && \qquad \qquad \qquad \qquad \,
  - \frac{9}{16}
  \frac{(N_{12}P_1) (N_{12}P_2)^3 P_1^2}{m_1^3 m_2^3} -
  \frac{5}{16} \frac{(N_{12}P_2)^2 P_1^4}{m_1^4 m_2^2} -
  \frac{7}{16} \frac{P_1^6}{m_1^6}
  \nonumber \\
  && \qquad \qquad \qquad \qquad \,
  + \frac{15}{32}
  \frac{(N_{12}P_1)^2 (N_{12}P_2)^2 (P_1P_2)}{m_1^3 m_2^3}
  + \frac{3}{4} \frac{(N_{12}P_1) (N_{12}P_2) P_1^2 (P_1P_2)}{m_1^4 m_2^2}
  \nonumber \\
  && \qquad \qquad \qquad \qquad \,
  + \frac{1}{16} \frac{(N_{12}P_2)^2 P_1^2 (P_1P_2)}{m_1^3 m_2^3}
  - \frac{5}{16} \frac{(N_{12}P_1) (N_{12}P_2) (P_1P_2)^2}{m_1^3 m_2^3}
  \nonumber \\
  && \qquad \qquad \qquad \qquad \,
  + \frac{1}{8} \frac{P_1^2 (P_1P_2)^2}{m_1^4 m_2^2} - \frac{1}{16}
  \frac{(P_1P_2)^3}{m_1^3 m_2^3} -\frac{5}{16}
  \frac{(N_{12}P_1)^2 P_1^2 P_2^2}{m_1^4 m_2^2}
  \nonumber \\
  && \qquad \qquad \qquad \qquad \,
  + \frac{7}{32}
  \frac{(N_{12}P_1) (N_{12}P_2) P_1^2 P_2^2}{m_1^3 m_2^3} +
  \frac{1}{2} \frac{P_1^4 P_2^2}{ m_1^4 m_2^2} + \frac{1}{32}
  \frac{P_1^2 (P_1P_2) P_2^2}{m_1^3 m_2^3} \bigg) \Bigg\}
  \nonumber \\
  && + 1 \leftrightarrow 2 + {\cal O}\left( \frac{1}{c^7} \right).
  \label{123}
\end{eqnarray}%
Arguably, the results given by the ADM-Hamiltonian formalism (for the
problem at hand) look simpler than their harmonic-coordinate
counterparts. Indeed, the ADM Lagrangian is ordinary -- no accelerations
-- and there are no logarithms nor associated gauge constants $r'_1$ and
$r'_2$. But of course, one is free to describe the binary motion in
whatever coordinates one likes, and the two formalisms,
harmonic~(\ref{118}) and ADM~(\ref{122}, \ref{123}), describe rigorously
the same physics. On the other hand, the higher complexity of the
harmonic-coordinates Lagrangian~(\ref{118}) enables one to perform more
tests of the computations, notably by inquiring about the future of the
constants $r'_1$ and $r'_2$, that we know \emph{must} disappear from
physical quantities such as the center-of-mass energy and the total
gravitational-wave flux.


\subsection{Equations of motion in the center-of-mass frame}
\label{subsec:9.2'}

In this section we translate the origin of coordinates to the binary's
center-of-mass by imposing that the binary's dipole $\mathrm{I}_i = 0$
(notation of Part~\ref{part:a}). Actually the dipole moment is computed
as the center-of-mass conserved integral associated with the boost
symmetry of the 3PN equations of motion and Lagrangian~\cite{ABF01,
BI03CM}. This condition results in the (3PN-accurate, say) relationship
between the individual positions in the center-of-mass frame $y_1^i$ and
$y_2^i$, and the relative position $x^i\equiv y_{1}^i-y_{2}^i$ and
velocity $v^i\equiv v_{1}^i-v_{2}^i = d x^i/dt$ (formerly denoted
$y_{12}^i$ and $v_{12}^i$). We shall also use the orbital separation
$r\equiv\vert{\bf x}\vert$, together with ${\bf n}={\bf x}/r$ and
$\dot{r}\equiv{\bf n}\cdot{\bf v}$. Mass parameters are the total mass
$m=m_1+m_2$ ($m\equiv \mathrm{M}$ in the notation of Part~\ref{part:a}),
the mass difference $\delta m=m_1-m_2$, the reduced mass $\mu=m_1m_2/m$,
and the very useful symmetric mass ratio
\begin{equation}
  \nu\equiv \frac{\mu}{m}\equiv \frac{m_1m_2}{(m_1+m_2)^2}.
  \label{nu}
\end{equation}
The usefulness of this ratio lies in its interesting range 
of variation: $0<\nu\leq 1/4$, with
$\nu=1/4$ in the case of equal masses, and $\nu\to 0$ in the
``test-mass'' limit for one of the bodies. 

The 3PN and even 3.5PN center-of-mass equations of motion are obtained
by replacing in the general-frame 3.5PN equations of motion~(\ref{112})
the positions and velocities by their center-of-mass expressions,
applying as usual the order-reduction of all accelerations where
necessary. We write the relative acceleration in the center-of-mass
frame in the form
\begin{equation}
  \frac{d v^i}{dt}=-\frac{G m}{r^2}
  \left[ (1+{\cal A})\,n^i + {\cal B}\,v^i \right] +
  {\cal O}\left( \frac{1}{c^8} \right),
  \label{eomstruct}
\end{equation}
and find~\cite{BI03CM} that the coefficients ${\cal A}$ and ${\cal B}$
are
\begin{eqnarray}
  {\cal A} &=& \frac{1}{c^2} \left\{ -\frac{3 \dot{r}^2 \nu}{2} +
  v^2 + 3 \nu v^2 - \frac{G m}{r} \left( 4 + 2 \nu \right) \right\}
  \nonumber \\
  && + \frac{1}{c^4} \bigg\{ \frac{15 \dot{r}^4 \nu}{8} -
  \frac{45 \dot{r}^4 \nu^2}{8} - \frac{9 \dot{r}^2 \nu v^2}{2} +
  6 \dot{r}^2 \nu^2 v^2 + 3 \nu v^4 - 4 \nu^2 v^4
  \nonumber \\
  && \qquad ~\,
  + \frac{G m}{r} \left( -2 \dot{r}^2 - 25 \dot{r}^2 \nu -
  2 \dot{r}^2 \nu^2 - \frac{13 \nu v^2}{2} + 2 \nu^2 v^2 \right) +
  \frac{G^2 m^2}{r^2} \left( 9 + \frac{87 \nu}{4} \right) \bigg\}
  \nonumber \\
  && + \frac{1}{c^5} \left\{ - \frac{24 \dot{r} \nu v^2}{5}
  \frac{G m}{r} - \frac{136 \dot{r} \nu}{15} \frac{G^2 m^2}{r^2} \right\}
  \nonumber \\
  && + \frac{1}{c^6} \bigg\{ \! -\frac{35 \dot{r}^6 \nu}{16} +
  \frac{175 \dot{r}^6 \nu^2}{16} - \frac{175 \dot{r}^6 \nu^3}{16} +
  \frac{15 \dot{r}^4 \nu v^2}{2} - \frac{135 \dot{r}^4 \nu^2 v^2}{4} +
  \frac{255 \dot{r}^4 \nu^3 v^2}{8} 
  \nonumber \\
  && \qquad ~\,
  - \frac{15 \dot{r}^2 \nu v^4}{2} + \frac{237 \dot{r}^2 \nu^2 v^4}{8} -
  \frac{45 \dot{r}^2 \nu^3 v^4}{2} + \frac{11 \nu v^6}{4} -
  \frac{49 \nu^2 v^6}{4} + 13 \nu^3 v^6
  \nonumber \\
  && \qquad ~\,
  + \frac{G m}{r} \bigg( 79 \dot{r}^4 \nu - \frac{69 \dot{r}^4 \nu^2}{2} -
  30 \dot{r}^4 \nu^3 - 121 \dot{r}^2 \nu v^2 + 16 \dot{r}^2 \nu^2 v^2 +
  20 \dot{r}^2 \nu^3 v^2 + \frac{75 \nu v^4}{4} 
  \nonumber\\
  && \qquad \qquad \qquad \;
  + 8 \nu^2 v^4 - 10 \nu^3 v^4 \bigg)
  \nonumber \\
  && \qquad ~\,
  + \frac{G^2 m^2}{r^2} \bigg( \dot{r}^2 + \frac{32573 \dot{r}^2 \nu}{168} +
  \frac{11 \dot{r}^2 \nu^2}{8} - 7 \dot{r}^2 \nu^3 +
  \frac{615 \dot{r}^2 \nu \pi^2}{64} - \frac{26987 \nu v^2}{840} + \nu^3 v^2 
  \nonumber \\
  && \qquad \qquad \qquad \quad \;
  - \frac{123 \nu \pi^2 v^2}{64} - 110 \dot{r}^2 \nu
  \ln \left( \frac{r}{r'_0} \right) + 22 \nu v^2
  \ln \left( \frac{r}{r'_0} \right) \! \bigg)
  \nonumber \\
  && \qquad ~\,
  + \frac{G^3 m^3}{r^3} \left( -16 - \frac{437 \nu}{4} -
  \frac{71 \nu^2}{2} + \frac{41 \nu {\pi }^2}{16} \right) \bigg\}
  \nonumber \\
  && + \frac{1}{c^7} \bigg\{ \frac{G m}{r} \left( \frac{366}{35}
  \nu v^4 + 12 \nu^2 v^4 - 114 v^2 \nu \dot{r}^2 - 12 \nu^2 v^2
  \dot{r}^2 + 112 \nu \dot{r}^4\right)
  \nonumber \\
  && \qquad ~\,
  + \frac{G^2 m^2}{r^2} \left( \frac{692}{35} \nu v^2 -
  \frac{724}{15} v^2 \nu^2 + \frac{294}{5} \nu \dot{r}^2 +
  \frac{376}{5} \nu^2 \dot{r}^2 \right)
  \nonumber \\
  && \qquad ~\,
  + \frac{G^3 m^3}{r^3} \left( \frac{3956}{35} \nu +
  \frac{184}{5} \nu^2 \right) \bigg\},
  \label{ABcoeff_1}
  \\ [1 em]
  {\cal B} &=& \frac{1}{c^2} \left\{ -4 \dot{r} + 2 \dot{r} \nu \right\}
  \nonumber \\
  && + \frac{1}{c^4} \left\{ \frac{9 \dot{r}^3 \nu}{2} +
  3 \dot{r}^3 \nu^2 - \frac{15 \dot{r} \nu v^2}{2} -
  2 \dot{r} \nu^2 v^2 + \frac{G m}{r} \left( 2 \dot{r} +
  \frac{41 \dot{r} \nu}{2} + 4 \dot{r} \nu^2 \right) \right\}
  \nonumber \\
  && + \frac{1}{c^5} \left\{ \frac{8 \nu v^2}{5} \frac{G m}{r} +
  \frac{24 \nu}{5} \frac{G^2 m^2}{r^2} \right\}
  \nonumber \\
  && + \frac{1}{c^6} \bigg\{ \! - \frac{45 \dot{r}^5 \nu}{8} +
  15 \dot{r}^5 \nu^2 + \frac{15 \dot{r}^5 \nu^3}{4} +
  12 \dot{r}^3 \nu v^2 - \frac{111 \dot{r}^3 \nu^2 v^2}{4} -
  12 \dot{r}^3 \nu^3 v^2 - \frac{65 \dot{r} \nu v^4}{8} 
  \nonumber \\
  && \qquad ~\,
  + 19 \dot{r} \nu^2 v^4 + 6 \dot{r} \nu^3 v^4
  \nonumber \\
  && \qquad ~\,
  + \frac{G m}{r} \left( \frac{329 \dot{r}^3 \nu}{6} +
  \frac{59 \dot{r}^3 \nu^2}{2} + 18 \dot{r}^3 \nu^3 - 15 \dot{r}
  \nu v^2 - 27 \dot{r} \nu^2 v^2 - 10 \dot{r} \nu^3 v^2 \right)
  \nonumber \\
  && \qquad ~\,
  + \frac{G^2 m^2}{r^2} \left( - 4 \dot{r} -
  \frac{18169 \dot{r} \nu}{840} + 25 \dot{r} \nu^2 + 8 \dot{r} \nu^3 -
  \frac{123 \dot{r} \nu \pi^2}{32} + 44 \dot{r} \nu
  \ln \left( \frac{r}{r'_0} \right) \right) \! \bigg\}
  \nonumber \\
  && + \frac{1}{c^7} \bigg\{ \frac{G m}{r} \left( - \frac{626}{35}
  \nu v^4 - \frac{12}{5} \nu^2 v^4 + \frac{678}{5} \nu v^2 \dot{r}^2 +
  \frac{12}{5} \nu^2 v^2 \dot{r}^2 - 120 \nu \dot{r}^4 \right)
  \nonumber \\
  && \qquad ~\,
  + \frac{G^2 m^2}{r^2} \left( \frac{164}{21} \nu v^2 +
  \frac{148}{5} \nu^2 v^2 - \frac{82}{3} \nu \dot{r}^2 -
  \frac{848}{15} \nu^2 \dot{r}^2 \right)
  \nonumber \\
  && \qquad ~\,
  + \frac{G^3 m^3}{r^3} \left( - \frac{1060}{21}\nu -
  \frac{104}{5}\nu^2 \right) \bigg\}.
  \label{ABcoeff_2}
\end{eqnarray}%
Up to the 2.5PN order the result agrees with the calculation
of~\cite{LW90}. The 3.5PN term is issued from Refs.~\cite{IW93, IW95,
JaraS97, PW02, KFS03, NB05}. At the 3PN order we have some
gauge-dependent logarithms containing a constant $r'_0$ which is the
``logarithmic barycenter'' of the two constants $r'_1$ and $r'_2$:
\begin{equation}
  \ln r'_0 = X_1 \ln r'_1 + X_2 \ln r'_2.
  \label{lnr0}
\end{equation}
The logarithms in Equations~(\ref{ABcoeff_1}, \ref{ABcoeff_2}), together with the constant
$r'_0$ therein, can be removed by applying the gauge
transformation~(\ref{113}), while still staying within the class of
harmonic coordinates. The resulting modification of the equations of
motion will affect only the coefficients of the 3PN order in
Equations~(\ref{ABcoeff_1}, \ref{ABcoeff_2}), let us denote them by
${\cal A}_\mathrm{3PN}$ and
${\cal B}_\mathrm{3PN}$. The new values of these coefficients, say ${\cal
A}'_\mathrm{3PN}$ and ${\cal B}'_\mathrm{3PN}$, obtained after removal of the
logarithms by the latter harmonic gauge transformation, are
then~\cite{MW03}
\begin{eqnarray}
  {\cal A}'_\mathrm{3PN}&=&
  \frac{1}{c^6} \bigg\{ \! -\frac{35 \dot{r}^6 \nu}{16} +
  \frac{175 \dot{r}^6 \nu^2}{16} - \frac{175 \dot{r}^6 \nu^3}{16} +
  \frac{15 \dot{r}^4 \nu v^2}{2} - \frac{135 \dot{r}^4 \nu^2 v^2}{4} +
  \frac{255 \dot{r}^4 \nu^3 v^2}{8} - \frac{15 \dot{r}^2 \nu v^4}{2} 
  \nonumber \\
  && \qquad
  + \frac{237 \dot{r}^2 \nu^2 v^4}{8} -
  \frac{45 \dot{r}^2 \nu^3 v^4}{2} + \frac{11 \nu v^6}{4} -
  \frac{49 \nu^2 v^6}{4} + 13 \nu^3 v^6
  \nonumber \\
  && \qquad
  + \frac{G m}{r} \bigg( 79 \dot{r}^4 \nu -
  \frac{69 \dot{r}^4 \nu^2}{2} - 30 \dot{r}^4 \nu^3 -
  121 \dot{r}^2 \nu v^2 + 16 \dot{r}^2 \nu^2 v^2 +
  20 \dot{r}^2 \nu^3 v^2 + \frac{75 \nu v^4}{4} 
  \nonumber \\
  && \qquad \qquad \quad \;\,
  + 8 \nu^2 v^4 - 10 \nu^3 v^4 \bigg)
  \nonumber \\
  && \qquad
  + \frac{G^2 m^2}{r^2} \bigg( \dot{r}^2 +
  \frac{22717 \dot{r}^2 \nu}{168} + \frac{11 \dot{r}^2 \nu^2}{8} -
  7 \dot{r}^2 \nu^3 + \frac{615 \dot{r}^2 \nu \pi^2}{64} -
  \frac{20827 \nu v^2}{840} + \nu^3 v^2 
  \nonumber \\
  && \qquad \qquad \qquad \;
  - \frac{123 \nu \pi^2 v^2}{64} \bigg)
  \nonumber \\
  && \qquad
  + \frac{G^3 m^3}{r^3} \left( -16 - \frac{1399 \nu}{12} -
  \frac{71 \nu^2}{2} + \frac{41 \nu {\pi }^2}{16} \right) \! \bigg\},
  \label{ABcoeffprime_1}
  \\ [1 em]
  {\cal B}'_\mathrm{3PN}&=&
  \frac{1}{c^6} \bigg\{ \! - \frac{45 \dot{r}^5 \nu}{8} +
  15 \dot{r}^5 \nu^2 + \frac{15 \dot{r}^5 \nu^3}{4} +
  12 \dot{r}^3 \nu v^2 - \frac{111 \dot{r}^3 \nu^2 v^2}{4} -
  12 \dot{r}^3 \nu^3 v^2 - \frac{65 \dot{r} \nu v^4}{8} 
  \nonumber \\
  && \qquad
  + 19 \dot{r} \nu^2 v^4 + 6 \dot{r} \nu^3 v^4
  \nonumber \\
  && \qquad
  + \frac{G m}{r}\left( \frac{329 \dot{r}^3 \nu}{6} +
  \frac{59 \dot{r}^3 \nu^2}{2} + 18 \dot{r}^3 \nu^3 -
  15 \dot{r} \nu v^2 - 27 \dot{r} \nu^2 v^2 -
  10 \dot{r} \nu^3 v^2 \right)
  \nonumber \\
  && \qquad
  + \frac{G^2 m^2}{r^2} \left( -4 \dot{r} -
  \frac{5849 \dot{r} \nu}{840} + 25 \dot{r} \nu^2 +
  8 \dot{r} \nu^3 - \frac{123 \dot{r} \nu \pi^2}{32} \right) \bigg\}.
  \label{ABcoeffprime_2}
\end{eqnarray}%
These gauge-transformed coefficients are useful because they do not
yield the usual complications associated with logarithms. However, they
must be handled with care in applications such as~\cite{MW03}, since one
must ensure that all other quantities in the problem (energy, angular
momentum, gravitational-wave fluxes, etc.) are defined in the same
specific harmonic gauge avoiding logarithms. In the following we shall
no longer use the coordinate system leading to
Equations~(\ref{ABcoeffprime_1}, \ref{ABcoeffprime_2}).
Therefore all expressions we shall derive below, notably all those
concerning the radiation field, are valid in the ``standard'' harmonic
coordinate system in which the equations of motion are given
by Equation~(\ref{112}) or~(\ref{ABcoeff_1}, \ref{ABcoeff_2}).


\subsection{Equations of motion and energy for circular orbits}
\label{subsec:9.3}

Most inspiralling compact binaries will have been circularized by the
time they become visible by the detectors LIGO and VIRGO. In the case of
orbits that are circular -- apart from the gradual 2.5PN
radiation-reaction inspiral -- the complicated equations of motion
simplify drastically, since we have $\dot{r}=(nv)={\cal O}(1/c^5)$, and
the remainder can always be neglected at the 3PN level. In the case of
circular orbits, up to the 2.5PN order, the relation between
center-of-mass variables and the relative ones
reads~\cite{B96}\epubtkFootnote{Actually, in the present computation we
do not need the radiation-reaction 2.5PN term in these relations; we
give it only for completeness.}
\begin{equation}
  \begin{array}{rcl}
    m \, y_1^i &=& \displaystyle x^i \left[ m_2 +3\gamma^2 \nu \delta m
    \right] - \frac{4}{5}\frac{G^2 m^2 \nu\delta m}{r c^5} v^i +
    {\cal O}\left( \frac{1}{c^6} \right),
    \\ [1.5 em]
    m \, y_2^i &=&
    \displaystyle x^i \left[ - m_1+3\gamma^2 \nu \delta m \right] -
    \frac{4}{5}\frac{G^2 m^2 \nu\delta m}{r c^5} v^i +
    {\cal O} \left( \frac{1}{c^6} \right).
  \end{array}
  \label{124}
\end{equation}
To display conveniently the successive post-Newtonian corrections, we
employ the post-Newtonian parameter
\begin{equation}
  \gamma \equiv \frac{G m}{r c^2} = {\cal O}\left(\frac{1}{c^2}\right).
  \label{126}
\end{equation}
Notice that there are no corrections of order 1PN in
Equations~(\ref{124}) for circular orbits; the dominant term
is of order 2PN, i.e.\ proportional to $\gamma^2={\cal O}(1/c^4)$.

The relative acceleration $a^i\equiv a_1^i-a_2^i$ of two bodies moving
on a circular orbit at the 3PN order is then given by
\begin{equation}
  a^i = -\omega^2 x^i- \frac{32}{5}\frac{G^3m^3\nu}{c^5r^4}v^i +
  {\cal O}\left(\frac{1}{c^7}\right),
  \label{127}
\end{equation}
where $x^i\equiv y_1^i-y_2^i$ is the relative separation (in harmonic
coordinates) and $\omega$ denotes the angular frequency of the circular
motion. The second term in Equation~(\ref{127}), opposite to the velocity
$v^i\equiv v_1^i-v_2^i$, is the 2.5PN radiation reaction force (we
neglect here its 3.5PN extension), which comes from the reduction of the
coefficient of $1/c^5$ in Equations~(\ref{ABcoeff_1},
\ref{ABcoeff_2}). The main content of the 3PN
equations~(\ref{127}) is the relation between the frequency $\omega$ and
the orbital separation $r$, that we find to be given by the generalized
version of Kepler's third law~\cite{BF00, BFeom}:
\begin{eqnarray}
  \omega^2 &=& \frac{Gm}{r^3} \bigg\{ 1+(-3+\nu) \gamma + \left( 6 +
  \frac{41}{4}\nu + \nu^2 \right) \gamma^2
  \nonumber \\
  && \qquad ~ + \left( -10 + \left[- \frac{75707}{840} + \frac{41}{64}
  \pi^2 + 22 \ln \left( \frac{r}{r'_0}\right) \right]\nu +
  \frac{19}{2}\nu^2 + \nu^3 \right) \gamma^3 \biggr\}
  \nonumber \\
  && +\, {\cal O}\left(\frac{1}{c^8}\right).
  \label{128}
\end{eqnarray}%
The length scale $r'_0$ is given in terms of the two gauge-constants
$r'_1$ and $r'_2$ by Equation~(\ref{lnr0}). As for the energy, it is
immediately obtained from the circular-orbit reduction of the general
result~(\ref{114}). We have
\begin{eqnarray}
  E &=& -\frac{\mu c^2 \gamma}{2} \biggl\{ 1 + \left( - \frac{7}{4} +
  \frac{1}{4} \nu \right) \gamma + \left( - \frac{7}{8} + \frac{49}{8}
  \nu + \frac{1}{8} \nu^2 \right) \gamma^2 \nonumber
  \\
  && \qquad \qquad + \left(-\frac{235}{64} + \left[\frac{46031}{2240}
  - \frac{123}{64} \pi^2 + \frac{22}{3} \ln \left( \frac{r}{r_0'}
  \right) \right] \nu + \frac{27}{32} \nu^2 + \frac{5}{64} \nu^3
  \right) \gamma^3 \biggr\} \nonumber \\ && +\, {\cal O} \left(
  \frac{1}{c^8} \right).
  \label{130}
\end{eqnarray}%
This expression is that of a physical observable $E$; 
however, it depends on the choice of
a coordinate system, as it involves the post-Newtonian 
parameter $\gamma$ defined from the 
harmonic-coordinate separation $r_{12}$. But the \emph{numerical} 
value of $E$ should not depend on the choice of a coordinate system, 
so $E$ must admit a frame-invariant expression, the same in all 
coordinate systems. To find it we re-express $E$ 
with the help of a frequency-related parameter $x$ instead of 
the post-Newtonian parameter $\gamma$. Posing
\begin{equation}
  x \equiv \left(\frac{G \,m \,\omega}{c^3}\right)^{2/3} \!\! =
  {\cal O}\left(\frac{1}{c^2}\right),
  \label{131}
\end{equation}
we readily obtain from Equation~(\ref{128}) the expression 
of $\gamma$ in terms of $x$ at 3PN order, 
\begin{eqnarray}
  \gamma &=& x \biggl\{1+\left(1-{\nu\over3}\right)x +
  \left(1-{65\over 12} \nu \right) x^2
  \nonumber \\
  && \quad \;
  + \left( 1 + \left[-{2203\over 2520}-
  {41\over 192}\pi^2 - {22\over 3}\ln\left(r\over{r'}_0\right)
  \right] \nu +{229\over 36}\nu^2+
  {1\over 81}\nu^3\right)x^3
  \nonumber \\
  && \quad \; +\, {\cal O}\left(\frac{1}{c^8}\right) \biggr\},
  \label{132}
\end{eqnarray}%
that we substitute back into Equation~(\ref{130}), making all appropriate
post-Newtonian re-expansions. As a result, we gladly discover that the
logarithms together with their associated gauge constant $r'_0$ have
cancelled out. Therefore, our result is
\begin{eqnarray}
  E &=& -\frac{\mu c^2 x}{2} \biggl\{ 1 +\left(-\frac{3}{4} -
  \frac{1}{12}\nu\right) x + \left(-\frac{27}{8} +
  \frac{19}{8}\nu -\frac{1}{24}\nu^2\right) x^2
  \nonumber \\
  && \qquad \quad ~\;
  + \! \left( \! -\frac{675}{64} +
  \left[\frac{34445}{576} - \frac{205}{96}\pi^2 
  \right]\nu - \frac{155}{96}\nu^2 -
  \frac{35}{5184}\nu^3 \!\right) x^3 \!\biggr\}
  \nonumber \\
  && + {\cal O}\left(\frac{1}{c^8}\right).
  \label{133}
\end{eqnarray}%
For circular orbits one can check that there are no terms of order
$x^{7/2}$ in Equation~(\ref{133}), so our result for $E$ is actually valid up
to the 3.5PN order.


\subsection{The innermost circular orbit (ICO)}
\label{subsec:9.4}

Having in hand the circular-orbit energy, we define the innermost
circular orbit (ICO) as the minimum, when it exists, of the energy
function $E(x)$. Notice that we do not define the ICO as a point of
dynamical general-relativistic unstability. Hence, we prefer to call
this point the ICO rather than, strictly speaking, an innermost stable
circular orbit or ISCO. A study of the dynamical stability of circular
binary orbits in the post-Newtonian approximation of general relativity
can be found in Ref.~\cite{BI03CM}.

The previous definition of the ICO is motivated by our comparison with
the results of numerical relativity. Indeed we shall confront the
prediction of the standard (Taylor-based) post-Newtonian approach with a
recent result of numerical relativity by Gourgoulhon, Grandcl\'ement, and
Bonazzola~\cite{GGB1, GGB2}. These authors computed numerically the
energy of binary black holes under the assumptions of conformal flatness
for the spatial metric and of exactly circular orbits. The latter
restriction is implemented by requiring the existence of an ``helical''
Killing vector, which is time-like inside the light cylinder associated
with the circular motion, and space-like outside. In the numerical
approach~\cite{GGB1, GGB2} there are no gravitational waves, the field
is periodic in time, and the gravitational potentials tend to zero at
spatial infinity within a restricted model equivalent to solving five
out of the ten Einstein field equations (the so-called
Isenberg--Wilson--Mathews approximation; see Ref.~\cite{FUS02} for a
discussion). Considering an evolutionary sequence of equilibrium
configurations Refs.~\cite{GGB1, GGB2} obtained numerically the
circular-orbit energy $E(\omega)$ and looked for the ICO of binary black
holes (see also Refs.~\cite{BGM99, GGTMB01, LGG05} for related
calculations of binary neutron and strange quark stars).

Since the numerical calculation~\cite{GGB1, GGB2} has been performed in
the case of \emph{corotating} black holes, which are spinning with the
orbital angular velocity $\omega$, we must for the comparison include
within our post-Newtonian formalism the effects of spins appropriate to
two Kerr black holes rotating at the orbital rate. The total
relativistic mass of the Kerr black hole is given by\epubtkFootnote{In
this section we pose $G=1=c$, and the two individual black hole masses
are denoted $M_1$ and $M_2$.}
\begin{equation}
  M^2=M_\mathrm{irr}^2+\frac{S^2}{4M_\mathrm{irr}^2},
  \label{m2}
\end{equation}
where $S$ is the spin, related to the usual Kerr parameter by $S=M a$,
and $M_\mathrm{irr}$ is the irreducible mass given by
$M_\mathrm{irr}=\sqrt{A}/(4\pi)$ ($A$ is the hole's surface
area). The angular velocity of the corotating black hole is $\omega = \partial
M/\partial S$ hence, from Equation~(\ref{m2}),
\begin{equation}
  \omega = \frac{S}{2M^3\left[1+\sqrt{1-\frac{S^2}{M^4}}\right]}.
  \label{omS}
\end{equation}
Physically this angular velocity is the one of the outgoing photons that
remain for ever at the location of the light-like horizon. Combining
Equations~(\ref{m2}, \ref{omS}) we obtain $M$ and $S$ as functions of
$M_\mathrm{irr}$ and $\omega$,
\begin{equation}
  \begin{array}{rcl}
    M &=& \displaystyle
    \frac{M_\mathrm{irr}}{\sqrt{1-4M_\mathrm{irr}^2\,\omega^2}},
    \\ [1.5 em]
    S &=& \displaystyle
    \frac{4M_\mathrm{irr}^3\omega}{\sqrt{1-4M_\mathrm{irr}^2\,\omega^2}}.
  \end{array}
  \label{mS}
\end{equation}
This is the right thing to do since $\omega$ is the basic variable
describing each equilibrium configuration calculated numerically, and
because the irreducible masses are the ones which are held constant
along the numerical evolutionary sequences in Refs.~\cite{GGB1, GGB2}.
In the limit of slow rotation we get
\begin{equation}
  S = I\,\omega + {\cal O}\left(\omega^3\right),
  \label{SI}
\end{equation}
where $I=4 M_\mathrm{irr}^3$ is the moment of inertia of the black hole.
Next the total mass-energy is
\begin{equation}
  M = M_\mathrm{irr}+\frac{1}{2} I\,\omega^2 +
  {\cal O}\left(\omega^4\right),
  \label{mmirr}
\end{equation}
which involves, as we see, the usual kinetic energy of the spin.

To take into account the spin effects our first task is to replace all
the masses entering the energy function~(\ref{133}) by their equivalent
expressions in terms of $\omega$ and the two irreducible masses. It is
clear that the leading contribution is that of the spin kinetic energy
given by Equation~(\ref{mmirr}), and it comes from the replacement of the
rest mass-energy $m \,c^2$ (where $m=M_1+M_2$). From Equation~(\ref{mmirr})
this effect is of order $\omega^2$ in the case of corotating binaries,
which means by comparison with Equation~(\ref{133}) that it is equivalent to
an ``orbital'' effect at the 2PN order (i.e.\ $\propto x^2$).
Higher-order corrections in Equation~(\ref{mmirr}), which behave at least
like $\omega^4$, will correspond to the orbital 5PN order at least and
are negligible for the present purpose. In addition there will be a
subdominant contribution, of the order of $\omega^{8/3}$ equivalent to
3PN order, which comes from the replacement of the masses into the
``Newtonian'' part, proportional to $x\propto \omega^{2/3}$, of the
energy $E$ (see Equation~(\ref{133})). With the 3PN accuracy we do not need
to replace the masses that enter into the post-Newtonian corrections in
$E$, so in these terms the masses can be considered to be the
irreducible ones.

Our second task is to include the specific relativistic effects due to
the spins, namely the spin-orbit (SO) interaction and the spin-spin (SS)
one. In the case of spins $S_1$ and $S_2$ aligned parallel to the
orbital angular momentum (and right-handed with respect to the sense of
motion) the SO energy reads
\begin{equation}
  E_\mathrm{SO}= -\mu\, (m\omega)^{5/3}
  \left[ \left( \frac{4}{3}\frac{M_1^2}{m^2}+\nu \right)
  \frac{S_1}{M_1^2}+ \left(\frac{4}{3}\frac{M_2^2}{m^2}+\nu\right)
  \frac{S_2}{M_2^2} \right].
  \label{ESO}
\end{equation}
Here we are employing the formula given by Kidder et al.~\cite{KWW93,
  K95} (based on seminal works of Barker and
O'Connell~\cite{BOC75, BOC79}) who have computed the SO contribution and
expressed it by means of the orbital frequency $\omega$. The derivation
of Equation~(\ref{ESO}) in Ref.~\cite{KWW93, K95} takes into account the fact
that the relation between the orbital separation $r$ (in the harmonic
coordinate system) and the frequency $\omega$ depends on the spins. We
immediately infer from Equation~(\ref{ESO}) that in the case of corotating
black holes the SO effect is equivalent to a 3PN orbital effect and thus
must be retained with the present accuracy (with this approximation, the
masses in Equation~(\ref{ESO}) are the irreducible ones). As for the SS
interaction (still in the case of spins aligned with the orbital angular
momentum) it is given by
\begin{equation}
  E_\mathrm{SS} = \mu\,\nu\, (m\omega)^2 \frac{S_1\,S_2}{M_1^2\,M_2^2}.
  \label{ESS}
\end{equation}
The SS effect can be neglected here because it is of order 5PN for
corotating systems. Summing up all the spin contributions we find that
the suplementary energy due to the corotating spins is~\cite{B02ico}
\begin{equation}
  \Delta E^\mathrm{corot} = m \, c^2\, x
  \left\{ (2-6\nu)x^2+\left(-6\nu+13\nu^2\right)x^3+{\cal O}(x^4)\right\},
  \label{Ecorot}
\end{equation}
where $x=(m\omega)^{2/3}$. The complete 3PN energy of the corotating
binary is finally given by the sum of Equations~(\ref{133})
and~(\ref{Ecorot}), in which we must now understand all the masses as
being the irreducible ones (we no longer indicate the superscript
``irr''), which for the comparison with the numerical calculation must
be assumed to stay constant when the binary evolves.

\epubtkImage{fig1.png}{%
  \begin{figure}[htpb]
    \def\epsfsize#1#2{0.7#1}
    \centerline{\epsfbox{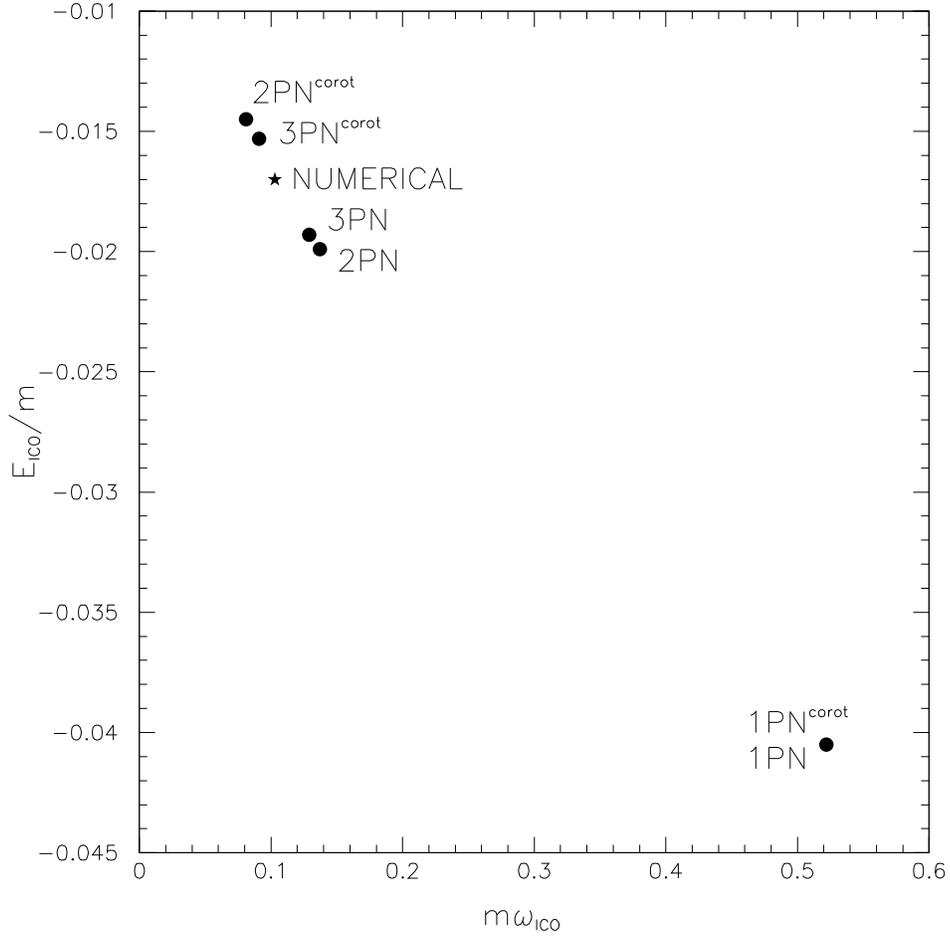}}
    \caption{\it Results for the binding energy $E_\mathrm{ICO}$
      versus $\omega_\mathrm{ICO}$ in the equal-mass case
      ($\nu=1/4$). The asterisk marks the result calculated by
      numerical relativity. The points indicated by $\mathrm{1PN}$,
      $\mathrm{2PN}$, and $\mathrm{3PN}$ are computed from the minimum
      of Equation~(\ref{133}), and correspond to irrotational
      binaries. The points denoted by $\mathrm{1PN}^\mathrm{corot}$,
      $\mathrm{2PN}^\mathrm{corot}$, and $\mathrm{3PN}^\mathrm{corot}$
      come from the minimum of the sum of Equations~(\ref{133}) and
      (\ref{Ecorot}), and describe corotational binaries.}
    \label{fig1}
  \end{figure}}

The Figure~\ref{fig1} (issued from Ref.~\cite{B02ico}) presents our
results for $E_\mathrm{ICO}$ in the case of irrotational and corotational
binaries. Since $\Delta E^\mathrm{corot}$, given by Equation~(\ref{Ecorot}), is
at least of order 2PN, the result for $\mathrm{1PN}^\mathrm{corot}$ is the same as
for 1PN in the irrotational case; then, obviously, $\mathrm{2PN}^\mathrm{corot}$
takes into account only the leading 2PN corotation effect (i.e.\ the spin
kinetic energy given by Equation~(\ref{mmirr})), while $\mathrm{3PN}^\mathrm{corot}$
involves also, in particular, the corotational SO coupling at the 3PN
order. In addition we present in Figure~\ref{fig1} the numerical point
obtained by numerical relativity under the assumptions of conformal
flatness and of helical symmetry~\cite{GGB1, GGB2}. As we can see the
3PN points, and even the 2PN ones, are rather close to the numerical
value. The fact that the 2PN and 3PN values are so close to each other
is a good sign of the convergence of the expansion; we shall further
comment this point in Section~\ref{subsec:9.5}. In fact one might say
that the role of the 3PN approximation is merely to ``confirm'' the
value already given by the 2PN one (but of course, had we not computed
the 3PN term, we would not be able to trust very much the 2PN value). As
expected, the best agreement we obtain is for the 3PN approximation and
in the case of corotation, i.e.\ the point $\mathrm{3PN}^\mathrm{corot}$.
However, the 1PN approximation is clearly not precise enough, but this
is not surprising in the highly relativistic regime of the ICO.

In conclusion, we find that the location of the ICO as computed by
numerical relativity, under the helical-symmetry and conformal-flatness
approximations, is in good agreement with the post-Newtonian prediction.
See also Ref.~\cite{DGG02} for the results calculated within the
effective-one-body approach method~\cite{BuonD98, BuonD00} at the 3PN
order, which are close to the ones reported in Figure~\ref{fig1}. This
agreement constitutes an appreciable improvement of the previous
situation, because the earlier estimates of the ICO in post-Newtonian
theory~\cite{KWWisco} and numerical relativity~\cite{PfTC00, Baum00}
strongly disagreed with each other, and do not match with the present
3PN results. The numerical calculation of quasi-equilibrium
configurations has been since then redone and refined by a number of
groups, for both corotational and irrotational binaries (see in
particular Ref.~\cite{CPf04}). These works confirm the previous
findings.


\subsection{Accuracy of the post-Newtonian approximation}
\label{subsec:9.5}

In this section we want to assess the validity of the post-Newtonian
approximation, and, more precisely, to address, and to some extent to
answer, the following questions: How accurate is the post-Newtonian
expansion for describing the dynamics of binary black hole systems? Is
the ICO of binary black holes, defined by the
minimum of the energy function $E(\omega)$, accurately determined at the
highest currently known post-Newtonian order? The latter question is
pertinent because the ICO represents a point in the late stage of
evolution of the binary which is very relativistic (orbital velocities
of the order of 50\% of the speed of light). How well does the 3PN
approximation as compared with the prediction provided by numerical
relativity (see Section~\ref{subsec:9.4})? What is the validity of the
various post-Newtonian resummation techniques~\cite{DIS98, DIS00,
BuonD98, BuonD00} which aim at ``boosting'' the convergence of the
standard post-Newtonian approximation?

The previous questions are interesting but difficult to settle down
rigorously. Indeed the very essence of an approximation is to cope with
our ignorance of the higher-order terms in some expansion, but the
higher-order terms are precisely the ones which would be needed for a
satisfying answer to these problems. So we shall be able to give only
some educated guesses and/or plausible answers, that we cannot justify
rigorously, but which seem very likely from the standard point of view
on the post-Newtonian theory, in particular that the successive orders
of approximation get smaller and smaller as they should (in average),
with maybe only few accidents occuring at high orders where a particular
approximation would be abnormally large with respect to the lower-order
ones. Admittedly, in addition, our faith in the estimation we shall give
regarding the accuracy of the 3PN order for instance, comes from the
historical perspective, thanks to the many successes achieved in the
past by the post-Newtonian approximation when confronting the theory and
observations. It is indeed beyond question, from our past experience,
that the post-Newtonian method does work.

Establishing the post-Newtonian expansion rigorously has been the
subject of numerous mathematical oriented works, see, e.g.,
\cite{Rend90, Rend92, Rend94}. In the present section we shall
simply look (much more modestly) at what can be said by inspection of
the explicit post-Newtonian coefficients which have been computed so
far. Basically, the point we would like to emphasize\epubtkFootnote{We
are following the discussion in Ref.~\cite{Baccuracy}. Note that the
arguments of this section are rather biased toward the author's own
work~\cite{B02ico, Baccuracy}.} is that the post-Newtonian
approximation, in standard form (without using the resummation
techniques advocated in Refs.~\cite{DIS98, BuonD98, BuonD00}), is able
to located the ICO of two black holes, in the case of \emph{comparable
masses} ($m_1\simeq m_2$), with a very good accuracy. At first sight
this statement is rather surprising, because the dynamics of two black
holes at the point of the ICO is so relativistic. Indeed one sometimes
hears about the ``bad convergence'', or the ``fundamental breakdown'',
of the post-Newtonian series in the regime of the ICO. However our
estimates do show that the 3PN approximation is good in this regime, for
comparable masses, and we have already confirmed this by the remarkable
agreement with the numerical calculations, as detailed in
Section~\ref{subsec:9.4}.

Let us center our discussion on the post-Newtonian expression of the
circular-orbit energy~(\ref{133}), developed to the 3PN order, which
is of the form
\begin{equation}
  E(x) = -\frac{\mu\,c^2 x}{2}
  \left\{ 1 + a_1(\nu)\,x + a_2(\nu)\,x^2 + a_3(\nu)\,x^3+{\cal O}(x^4) \right\}.
  \label{Ex}
\end{equation} 
The first term, proportional to $x$, is the Newtonian term, and then we
have many post-Newtonian corrections, the coefficients of which are
known up to 3PN order~\cite{JaraS98, JaraS99, DJSpoinc, DJSequiv, BF00,
BFeom, ABF01}:
\begin{equation}
  \begin{array}{rcl}
    a_1(\nu) &=& \displaystyle -\frac{3}{4}-\frac{\nu}{12},
    \\ [1.5 em]
    a_2(\nu) &=& \displaystyle -\frac{27}{8}+\frac{19}{8}\nu -
    \frac{\nu^2}{24},
    \\ [1.5 em]
    a_3(\nu) &=& \displaystyle -\frac{675}{64}+
    \left[\frac{209323}{4032}-\frac{205}{96}\pi^2 -
    \frac{110}{9}\lambda\right]\nu-\frac{155}{96}\nu^2 -
    \frac{35}{5184}\nu^3.
  \end{array}
  \label{aPN}
\end{equation} 
For the discussion it is helpful to keep the Hadamard regularization
ambiguity parameter $\lambda$ present in the 3PN coefficient $a_3(\nu)$.
Recall from Section~\ref{subsec:8.2} that this parameter was introduced
in Refs.~\cite{BF00, BFeom} and is equivalent to the parameter
$\omega_\mathrm{static}$ of Refs.~\cite{JaraS98, JaraS99}. We already gave
in Equation~(\ref{109b}) the relation linking them,
\begin{equation}
  \lambda = -\frac{3}{11} \omega_\mathrm{static}-\frac{1987}{3080}.
  \label{lambdaoms}
\end{equation}

Before its actual computation in general relativity, it has been argued
in Ref.~\cite{DJSisco} that the numerical value of $\omega_\mathrm{static}$
could be $\simeq -9$, because for such a value some different
resummation techniques, when they are implemented at the 3PN order, give
approximately the same result for the ICO. Even more, it was
suggested~\cite{DJSisco} that $\omega_\mathrm{static}$ might be precisely
equal to $\omega_\mathrm{static}^*$, with
\begin{equation}
  \omega_\mathrm{static}^*=
  -\frac{47}{3}+\frac{41}{64}\pi^2 =
  -9.34\cdots.
  \label{omegastar}
\end{equation}
However, as reviewed in Sections~\ref{subsec:8.2} and~\ref{subsec:8.3}, the
computations performed using dimensional regularization, within the
ADM-Hamiltonian formalism~\cite{DJSdim} and harmonic-coordinate
approaches~\cite{BDE04}, and the independent computation of
Refs.~\cite{itoh1, itoh2}, have settled the value of this parameter in
general relativity to be
\begin{equation}
  \omega_\mathrm{static}=0
  \qquad \Longleftrightarrow \qquad
  \lambda=-\frac{1987}{3080}.
  \label{omsres}
\end{equation}
We note that this result is quite different from
$\omega_\mathrm{static}^*$, Equation~(\ref{omegastar}). This already
suggests that different resummation techniques, namely Pad\'e
approximants~\cite{DIS98, DIS00,
DJSisco} and effective-one-body methods~\cite{BuonD98, BuonD00,
DJSisco}, which are designed to ``accelerate'' the convergence of the
post-Newtonian series, do \emph{not} in fact converge toward the same
exact solution (or, at least, not as fast as expected).

In the limiting case $\nu\to 0$, the expression~(\ref{Ex}, \ref{aPN})
reduces to the 3PN approximation of the energy for a test particle in
the Schwarzschild background,
\begin{equation}
  E^\mathrm{Sch}(x) =
  \mu\,c^2\,\left[\frac{1-2x}{\sqrt{1-3x}}-1\right].
  \label{Esch}
\end{equation}
The minimum of that function or Schwarzschild ICO occurs at
$x^\mathrm{Sch}_\mathrm{ICO}=1/6$, and we have $E^\mathrm{Sch}_\mathrm{ICO}=\mu
c^2\bigl(\sqrt{8/9}-1\bigr)$. We know that the Schwarzschild ICO is also
an innermost \emph{stable} circular orbit or ISCO, i.e.\ it
corresponds to a point of dynamical unstability. Another important
feature of Equation~(\ref{Esch}) is the singularity at the value
$x^\mathrm{Sch}_\mathrm{light\,ring}=1/3$ which corresponds to the famous circular
orbit of photons in the Schwarzschild metric (``light-ring''
singularity). This orbit can also be viewed as the last \emph{unstable}
circular orbit. We can check that the post-Newtonian coefficients
$a_n^\mathrm{Sch}\equiv a_n(0)$ corresponding to Equation~(\ref{Esch}) are given
by
\begin{equation}
  a_n^\mathrm{Sch} = - \frac{3^n(2n-1)!!(2n-1)}{2^n(n+1)!}.
  \label{aPNsch}
\end{equation}
They increase with $n$ by roughly a factor 3 at each order. This is
simply the consequence of the fact that the radius of convergence of the
post-Newtonian series is given by the Schwarzschild light-ring
singularity at the value $1/3$. We may therefore recover the light-ring
orbit by investigating the limit
\begin{equation}
  \lim_{n\to +\infty}\,\frac{a_{n-1}^\mathrm{Sch}}{a_n^\mathrm{Sch}} =
  \frac{1}{3} = x^\mathrm{Sch}_\mathrm{light\,ring}.
  \label{limit}
\end{equation}

Let us now discuss a few order-of-magnitude estimates. At the location
of the ICO we have found (see Figure~\ref{fig1} in
Section~\ref{subsec:9.4}) that the frequency-related parameter $x$
defined by Equation~(\ref{131}) is approximately of the order of $x\sim
(0.1)^{2/3}\sim 20\%$ for equal masses. Therefore, we might \emph{a
priori} expect that the contribution of the 1PN approximation to the
energy at the ICO should be of that order. For the present discussion we
take the pessimistic view that the order of magnitude of an
approximation represents also the order of magnitude of the higher-order
terms which are neglected. We see that the 1PN approximation should
yield a rather poor estimate of the ``exact'' result, but this is quite
normal at this very relativistic point where the orbital velocity is
$v/c \sim x^{1/2}\sim 50\%$. By the same argument we infer that the 2PN
approximation should do much better, with fractional errors of the order
of $x^2\sim 5\%$, while 3PN will be even better, with the accuracy
$x^3\sim 1\%$.

Now the previous estimate makes sense only if the numerical values of
the post-Newtonian coefficients in Equations~(\ref{aPN}) stay roughly of the
order of one. If this is not the case, and if the coefficients increase
dangerously with the post-Newtonian order $n$, one sees that the
post-Newtonian approximation might in fact be very bad. It has often
been emphasized in the litterature (see, e.g., Refs.~\cite{3mn, P95,
DIS98}) that in the test-mass limit $\nu\to 0$ the post-Newtonian series
converges slowly, so the post-Newtonian approximation is not very good
in the regime of the ICO. Indeed we have seen that when $\nu=0$ the
radius of convergence of the series is $1/3$ (not so far from
$x^\mathrm{Sch}_\mathrm{ICO}=1/6$), and that accordingly the post-Newtonian
coefficients increase by a factor $\sim 3$ at each order. So it is
perfectly correct to say that in the case of test particles in the
Schwarzschild background the post-Newtonian approximation is to be
carried out to a high order in order to locate the turning point of the
ICO.

What happens when the two masses are comparable ($\nu=\frac{1}{4}$)? It
is clear that the accuracy of the post-Newtonian approximation depends
crucially on how rapidly the post-Newtonian coefficients increase with
$n$. We have seen that in the case of the Schwarzschild metric the
latter increase is in turn related to the existence of a light-ring
orbit. For continuing the discussion we shall say that the relativistic
interaction between two bodies of comparable masses is
\emph{``Schwarzschild-like''} if the post-Newtonian coefficients
$a_n(\frac{1}{4})$ increase when $n\to +\infty$. If this is the case
this signals the existence of something like a light-ring singularity
which could be interpreted as the deformation, when the mass ratio $\nu$
is ``turned on'', of the Schwarzschild light-ring orbit. By analogy with
Equation~(\ref{limit}) we can estimate the location of this
``pseudo-light-ring'' orbit by
\begin{equation}
  \frac{a_{n-1}(\nu)}{a_n(\nu)} \sim \, x_\mathrm{light\,ring}(\nu)
  \qquad
  \mbox{with }n=3.
  \label{lightring}
\end{equation}
Here $n=3$ is the highest known post-Newtonian order. If the two-body
problem is ``Schwarzschild-like'' then the right-hand side of Equation
(\ref{lightring}) is small (say around $1/3$), the post-Newtonian
coefficients typically increase with $n$, and most likely it should be
difficult to get a reliable estimate by post-Newtonian methods of the
location of the ICO. So we ask: Is the gravitational interaction between
two comparable masses Schwarzschild-like?

\begin{table}[htbp]
  \renewcommand{\arraystretch}{1.2}
  \centering
  \begin{tabular}{l|crrr}
    \hline \hline
    & Newtonian &
    $a_1(\nu)$ &
    $a_2(\nu)$ &
    $a_3(\nu)$ \\
    \hline
    $\nu=0$ &
    $1$ &
    $-0.75$ &
    $-3.37$ &
    $-10.55$ \\
    $\nu=\frac{1}{4}$, \qquad $\omega^*_\mathrm{static}\simeq -9.34$ &
    $1$ &
    $-0.77$ &
    $-2.78$ &
    $-8.75$ \\
    $\nu=\frac{1}{4}$, \qquad $\omega_\mathrm{static}=0$ (GR) &
    $1$ &
    $-0.77$ &
    $-2.78$ &
    $-0.97$ \\
    \hline \hline
  \end{tabular}
  \caption{\it Numerical values of the sequence of coefficients of the
    post-Newtonian series composing the energy function $E(x)$ as
    given by Equations~(\ref{Ex}, \ref{aPN}).}
  \label{tab1}
  \renewcommand{\arraystretch}{1.0}
\end{table}

In Table~\ref{tab1} we present the values of the coefficients $a_n(\nu)$
in the test-mass limit $\nu=0$ (see Equation~(\ref{aPNsch}) for their
analytic expression), and in the equal-mass case $\nu=\frac{1}{4}$ when
the ambiguity parameter takes the ``uncorrect'' value
$\omega^*_\mathrm{static}$ defined by Equation~(\ref{omegastar}), and
the correct one $\omega_\mathrm{static}=0$ predicted by general
relativity. When $\nu=0$ we
clearly see the expected increase of the coefficients by roughly a
factor 3 at each step. Now, when $\nu=\frac{1}{4}$ and $\omega_\mathrm{static}=\omega^*_\mathrm{static}$ we notice that the coefficients increase
approximately in the same manner as in the test-mass case $\nu=0$. This
indicates that the gravitational interaction in the case of
$\omega^*_\mathrm{static}$ looks like that in a one-body problem. The
associated pseudo-light-ring singularity is estimated using
Equation~(\ref{lightring}) as
\begin{equation}
  x_\mathrm{light\,ring}
  \left( \frac{1}{4},{\omega^*_\mathrm{static}} \right) \sim 0.32.
  \label{xlr}
\end{equation}
The pseudo-light-ring orbit seems to be a very small deformation of the
Schwarzschild light-ring orbit given by Equation~(\ref{limit}). In this
Schwarzschild-like situation, we should not expect the post-Newtonian
series to be very accurate.

Now in the case $\nu=\frac{1}{4}$ but when the ambiguity parameter takes
the correct value $\omega_\mathrm{static}=0$, we see that the 3PN
coefficient $a_3(\frac{1}{4})$ is of the order of $ -1 $ instead of
being $\sim -10$. This suggests, unless 3PN happens to be quite
accidental, that the post-Newtonian coefficients in general relativity
do not increase very much with $n$. This is an interesting finding
because it indicates that the actual two-body interaction in general
relativity is \emph{not} Schwarzschild-like. There does not seem to exist
something like a light-ring orbit which would be a deformation of the
Schwarzschild one. Applying Equation~(\ref{lightring}) we obtain as an
estimate of the ``light ring'',
\begin{equation}
  x_\mathrm{light\,ring} \left( \frac{1}{4}, \mbox{GR} \right) \sim 2.88.
  \label{xlrGR}
\end{equation}
It is clear that if we believe the correctness of this estimate we must
conclude that there is in fact \emph{no} notion of a light-ring orbit in
the real two-body problem. Or, one might say (pictorially speaking) that
the light-ring orbit gets hidden inside the horizon of the final
black hole formed by coalescence. Furthermore, if we apply
Equation~(\ref{lightring}) using the 2PN approximation $n=2$ instead of the
3PN one $n=3$, we get the value $\sim 0.28$ instead of
Equation~(\ref{xlrGR}). So at the 2PN order the metric seems to admit a light
ring, while at the 3PN order it apparently does not admit any. This
erratic behaviour reinforces our idea that it is meaningless (with our
present 3PN-based knowledge, and until fuller information is available)
to assume the existence of a light-ring singularity when the masses are
equal.

It is impossible of course to be thoroughly confident about the validity
of the previous statement because we know only the coefficients up to
3PN order. Any tentative conclusion based on 3PN can be ``falsified''
when we obtain the next 4PN order. Nevertheless, we feel that the mere
fact that $a_3(\frac{1}{4})=-0.97$ in Table~\ref{tab1} is sufficient to
motivate our conclusion that the gravitational field generated by two
bodies is more complicated than the Schwarzschild space-time. This
appraisal should look cogent to relativists and is in accordance with
the author's respectfulness of the complexity of the Einstein field
equations.

We want next to comment on a possible implication of our conclusion as
regards the so-called post-Newtonian resummation techniques, i.e.\
Pad\'e approximants~\cite{DIS98, DIS00, DJSisco}, which aim at
``boosting'' the convergence of the post-Newtonian series in the
pre-coalescence stage, and the effective-one-body (EOB)
method~\cite{BuonD98, BuonD00, DJSisco}, which attempts at describing
the late stage of the coalescence of two black holes. These techniques
are based on the idea that the gravitational two-body interaction is a
``deformation'' -- with $\nu\leq\frac{1}{4}$ being the deformation
parameter -- of the Schwarzschild space-time. The Pad\'e approximants
are valuable tools for giving accurate representations of functions
having some singularities. In the problem at hands they would be
justified if the ``exact'' expression of the energy (whose 3PN expansion
is given by Equations~(\ref{Ex}, \ref{aPN})) would admit a singularity at
some reasonable value of $x$ (e.g., $\leq 0.5$). In the
Schwarzschild case, for which Equation~(\ref{limit}) holds, the Pad\'e series
converges rapidly~\cite{DIS98}: The Pad\'e constructed only from the 2PN
approximation of the energy -- keeping only $a_1^\mathrm{Sch}$ and
$a_2^\mathrm{Sch}$ -- already coincide with the exact result given by
Equation~(\ref{Esch}). On the other hand, the EOB method maps the
post-Newtonian two-body dynamics (at the 2PN or 3PN orders) on the
geodesic motion on some effective metric which happens to be a
$\nu$-deformation of the Schwarzschild space-time. In the EOB method the
effective metric looks like Schwarzschild \emph{by definition}, and we
might of course expect the two-body interaction to own the main
Schwarzschild-like features.

Our comment is that the validity of these post-Newtonian resummation
techniques does not seem to be compatible with the value
$\omega_\mathrm{static}=0$, which suggests that the two-body
interaction in general
relativity is not Schwarzschild-like. This doubt is confirmed by the
finding of Ref.~\cite{DJSisco} (already alluded to above) that in the
case of the wrong ambiguity parameter $\omega^*_\mathrm{static}\simeq
-9.34$ the Pad\'e approximants and the EOB method at the 3PN order give
the same result for the ICO. From the previous discussion we see that
this agreement is to be expected because a deformed light-ring
singularity seems to exist with that value $\omega^*_\mathrm{static}$. By
contrast, in the case of general relativity, $\omega_\mathrm{static}=0$,
the Pad\'e and EOB methods give quite different results (cf.\ the
Figure 2 in~\cite{DJSisco}). Another confirmation comes from the
light-ring singularity which is determined from the Pad\'e approximants
at the 2PN order (see Equation~(3.22) in~\cite{DIS98}) as
\begin{equation}
  x_\mathrm{light\,ring}
  \left( \frac{1}{4}, \mbox{Pad\'e} \right) \sim 0.44.
  \label{pade}
\end{equation}
This value is rather close to Equation~(\ref{xlr}) but strongly disagrees
with Equation~(\ref{xlrGR}). Our explanation is that the Pad\'e series
converges toward a theory having
$\omega_\mathrm{static}\simeq\omega^*_\mathrm{static}$; such a theory is
different from general relativity.

Finally we come to the good news that, if really the post-Newtonian
coefficients when $\nu=\frac{1}{4}$ stay of the order of one (or minus
one) as it seems to, this means that the \emph{standard} post-Newtonian
approach, based on the standard Taylor approximants, is probably very
accurate. The post-Newtonian series is likely to ``converge well'', with
a ``convergence radius'' of the order of one\epubtkFootnote{Actually,
the post-Newtonian series could be only asymptotic (hence divergent),
but nevertheless it should give excellent results provided that the
series is truncated near some optimal order of approximation. In this
discussion we assume that the 3PN order is not too far from that
optimum.}. Hence the order-of-magnitude estimate we proposed at the
beginning of this section is probably correct. In particular the 3PN
order should be close to the ``exact'' solution for comparable masses
even in the regime of the ICO.

\newpage


\section{Gravitational Waves from Compact Binaries}
\label{sec:10}

We pointed out that the 3PN equations of motion,
Equations~(\ref{127}, \ref{128}), are merely Newtonian as regards the
radiative aspects of the problem, because with that precision the
radiation reaction force is at the lowest 2.5PN order. A solution
would be to extend the precision of the equations of motion so as to
include the full relative 3PN or 3.5PN precision into the radiation
reaction force, but, needless to say, the equations of motion up to
the 5.5PN or 6PN order are quite impossible to derive with the present
technology. The much better alternative solution is to apply
the wave-generation formalism described in Part~\ref{part:a}, and to determine by
its means the work done by the radiation reaction force directly as a
total energy flux at future null infinity. In this approach, we replace
the knowledge of the higher-order radiation reaction force by the
computation of the total flux ${\cal L}$, and we apply the energy
balance equation as in the test of the $\dot P$ of the binary pulsar
(see Equations~(\ref{4}, \ref{5})):
\begin{equation}
  \frac{d E}{dt}=-{\cal L}.
  \label{134}
\end{equation}
Therefore, the result~(\ref{133}) that we found for the 3.5PN binary's
center-of-mass energy $E$ constitutes only ``half'' of the solution of
the problem. The second ``half'' consists of finding the rate of
decrease $dE/dt$, which by the balance equation is nothing but finding
the total gravitational-wave flux ${\cal L}$ at the 3.5PN order.
Because the orbit of inspiralling binaries is circular, the balance
equation for the energy is sufficient (no need of a balance equation
for the angular momentum). This all sounds perfect, but it is important to
realize that we shall use Equation~(\ref{134}) at the very high
3.5PN order, at which order there are no proofs (following from first
principles in general relativity) that the equation is correct,
despite its physically obvious character. Nevertheless, Equation~(\ref{134})
has been checked to be valid, both in the cases of point-particle
binaries~\cite{IW93, IW95} and extended weakly self-gravitating
fluids~\cite{B93, B97}, at the 1PN order and even at 1.5PN (the 1.5PN
approximation is especially important for this check because it
contains the first wave tails).

Obtaining ${\cal L}$ can be divided into two equally important
steps: (1) the computation of the \emph{source} multipole moments
$\mathrm{I}_L$ and $\mathrm{J}_L$ of the compact binary and (2) the control and
determination of the tails and related non-linear effects occuring in
the relation between the binary's source moments and the
\emph{radiative} ones $\mathrm{U}_L$ and $\mathrm{V}_L$ (cf.\ the general
formalism of Part~\ref{part:a}).


\subsection{The binary's multipole moments}
\label{subsec:10.1}

The general expressions of the source multipole moments given by
Theorem~\ref{th6}, Equations~(\ref{71}), are first to be worked out
explicitly for general fluid systems at the 3PN order. For this
computation one uses the formula~(\ref{75}), and we insert the 3PN
metric coefficients (in harmonic coordinates) expressed in
Equations~(\ref{93}) by means of the retarded-type elementary
potentials~(\ref{95}, \ref{96}, \ref{97}). Then we specialize each of
the (quite numerous) terms to the case of point-particle binaries by
inserting, for the matter stress-energy tensor $T^{\alpha\beta}$, the
standard expression made out of Dirac delta-functions. The infinite
self-field of point-particles is removed by means of the Hadamard
regularization; and dimensional regularization is used to compute the
few ambiguity parameters (see Section~\ref{sec:8}). This computation has
been performed in~\cite{BS89} at the 1PN order, and in~\cite{BDI95} at
the 2PN order; we report below the most accurate 3PN results obtained in
Refs.~\cite{BIJ02, BI04mult, BDEI04, BDEI05dr}.

The difficult part of the analysis is to find the closed-form
expressions, fully explicit in terms of the particle's positions and
velocities, of many non-linear integrals. We refer to~\cite{BIJ02}
for full details; nevertheless, let us give a few examples of the type
of technical formulas that are employed in this calculation. Typically
we have to compute some integrals like
\begin{equation}
  \stackrel{(n, p)}{Y_L}({\bf y}_1,{\bf y}_2) =
  -\frac{1}{2\pi}{\cal FP} \int d^3{\bf x} \, \hat{x}_L \,  r_1^n r_2^p,
  \label{135}
\end{equation}
where $r_1=|{\bf x}-{\bf y}_1|$ and $r_2=|{\bf x}-{\bf y}_2|$. When $n >
-3$ and $p > -3$, this integral is perfectly well-defined (recall that
the finite part ${\cal FP}$ deals with the bound at infinity). When
$n\leq -3$ or $p\leq -3$, our basic ansatz is that we apply the
definition of the Hadamard \emph{partie finie} provided by
Equation~(\ref{101}). Two examples of closed-form formulas that we get, which
do not necessitate the Hadamard \emph{partie finie}, are (quadrupole case
$l=2$)
\begin{equation}
  \begin{array}{rcl}
    \stackrel{(-1,-1)}{Y_{ij}} &=& \displaystyle
    \frac{r_{12}}{3} \left[
    y_1^{\langle ij \rangle}+y_1^{\langle i} y_2^{j \rangle} +
    y_2^{\langle ij \rangle}\right],
    \\ [2.0 em]
    \stackrel{(-2,-1)}{Y_{ij}} &=& \displaystyle
    y_1^{\langle ij \rangle}
    \left[\frac{16}{15}\ln\left(\frac{r_{12}}{r_0}\right) -
    \frac{188}{225} \right] + y_1^{\langle i}y_2^{j \rangle}
    \left[\frac{8}{15} \ln\left(\frac{r_{12}}{r_0}\right) -
    \frac{4}{225} \right] + y_2^{\langle ij \rangle}
    \left[\frac{2}{5} \ln \left( \frac{r_{12}}{r_0}\right) -
    \frac{2}{25} \right]. \hspace{-2.5 em}
  \end{array}
\end{equation}
We denote for example $y_1^{\langle ij \rangle}=y_1^{\langle i}y_1^{j
\rangle}$ (and $r_{12}=r\vert {\bf y}_1-{\bf y}_2\vert$); the constant
$r_0$ is the one pertaining to the finite-part process (see
Equation~(\ref{29_1})). One example where the integral diverges at the
location of the particle 1 is
\begin{equation}
  \stackrel{(-3, 0)}{Y_{ij}} =
  \left[2\ln\left(\frac{s_1}{r_0}\right)+\frac{16}{15} \right]
  y_1^{\langle ij \rangle},
  \label{137}
\end{equation}
where $s_1$ is the Hadamard-regularization constant introduced in
Equation~(\ref{101})\epubtkFootnote{When computing the gravitational-wave
flux in Ref.~\cite{BIJ02} we preferred to call the
Hadamard-regularization constants $u_1$ and $u_2$, in order to
distinguish them from the constants $s_1$ and $s_2$ that were used in
our previous computation of the equations of motion in
Ref.~\cite{BFeom}. Indeed these regularization constants need not
neccessarily be the same when employed in different contexts.}.

The crucial input of the computation of the flux at the 3PN order is
the mass quadrupole moment $\mathrm{I}_{ij}$, since this moment
necessitates the full 3PN precision. The result of Ref.~\cite{BIJ02}
for this moment (in the case of circular orbits) is
\begin{equation}
  \mathrm{I}_{ij} = \mu \left(A \, x_{\langle ij \rangle}+B \, 
  \frac{r_{12}^3}{G m}v_{\langle ij \rangle} + \frac{48}{7} \,
  \frac{G^2 m^2\nu}{c^5r_{12}}x_{\langle i}v_{j \rangle}\right)+
  {\cal O}\left(\frac{1}{c^7}\right),
  \label{138}
\end{equation}
where we pose $x_i = x^i \equiv y_{12}^i$ and $v_i = v^i \equiv v_{12}^i$. 
The third term is the 2.5PN radiation-reaction term, which does not
contribute to the energy flux for circular orbits. The two important coefficients 
are $A$ and $B$, whose expressions through 3PN order are
\begin{equation}
  \begin{array}{rcl}
    A &=& \displaystyle
    1 + \gamma \left(-\frac{1}{42} - \frac{13}{14}\nu \right) +
    \gamma^2 \left(-\frac{461}{1512} - \frac{18395}{1512}\nu -
    \frac{241}{1512} \nu^2\right)
    \\ [1.5 em]
    && \displaystyle
    + \gamma^3 \left\{\! \frac{395899}{13200} - \frac{428}{105} \ln
    \left( \frac{r_{12}}{r_0}\right) + \left[ \frac{139675}{33264}
    - \frac{44}{3} \xi - \frac{88}{3} \kappa -
    \frac{44}{3} \ln \left(\frac{r_{12}}{{r'}_0} \right) \right]
    \nu \right.
    \\ [1.5 em]
    && \qquad ~ \displaystyle
    \left. + \frac{162539}{16632} \nu^2 +
    \frac{2351}{33264}\nu^3 \right\},
    \\ [1.5 em]
    B &=& \displaystyle
    \gamma \left(\frac{11}{21} - \frac{11}{7} \nu \right) +
    \gamma^2 \left( \frac{1607}{378} - \frac{1681}{378} \nu +
    \frac{229}{378} \nu^2 \right)
    \\ [1.5 em]
    && \displaystyle
    + \gamma^3 \left(\!\! - \frac{357761}{19800} +
    \frac{428}{105} \ln \left( \frac{r_{12}}{r_0} \right) +
    \left[ \! - \frac{75091}{5544} + \frac{44}{3} \zeta \right]
    \nu + \frac{35759}{924} \nu^2 + \frac{457}{5544} \nu^3
    \right).
  \end{array}
  \label{139}
\end{equation}
These expressions are valid in harmonic coordinates \emph{via} the
post-Newtonian parameter $\gamma$ given by Equation~(\ref{126}). As we see,
there are two types of logarithms in the moment: One type involves the
length scale $r'_0$ related by Equation~(\ref{lnr0}) to the two gauge
constants $r'_1$ and $r'_2$ present in the 3PN equations of motion; the
other type contains the different length scale $r_0$ coming from the
general formalism of Part~\ref{part:a} -- indeed, recall that there is a
${\cal FP}$ operator in front of the source multipole moments in
Theorem~\ref{th6}. As we know, that $r_0'$ is pure gauge; it will
disappear from our physical results at the end. On the other hand, we
have remarked that the multipole expansion outside a general
post-Newtonian source is actually free of $r_0$, since the $r_0$'s
present in the two terms of Equation~(\ref{58}) cancel out. We shall indeed
find that the constants $r_0$ present in Equations~(\ref{139}) are
compensated by similar constants coming from the non-linear wave ``tails
of tails''. Finally, the constants $\xi$, $\kappa$, and $\zeta$ are the
Hadamard-regularization ambiguity parameters which take the
values~(\ref{xikappazeta}).

Besides the 3PN mass quadrupole~(\ref{138}, \ref{139}), we need also the
mass octupole moment $\mathrm{I}_{ijk}$ and current quadrupole moment
$\mathrm{J}_{ij}$, both of them at the 2PN order; these are given by~\cite{BIJ02}
\begin{equation}
  \begin{array}{rcl}
    \mathrm{I}_{ijk} &=& \displaystyle
    \mu \frac{\delta m}{m} \hat{x}_{ijk} \left[
    - 1 + \gamma \nu + \gamma^2 \left(\frac{139}{330} +
    \frac{11923}{660}\nu + \frac{29}{110} \nu^2 \right) \right]
    \\ [1.5 em]
    && \displaystyle
    +\, \mu \frac{\delta m}{m} x_{\langle i} v_{jk \rangle}
    \frac{r_{12}^2}{c^2} \left[ - 1 + 2 \nu + \gamma \left(
    - \frac{1066}{165} + \frac{1433}{330} \nu - \frac{21}{55}
    \nu^2 \right) \right] + {\cal O}\left(\frac{1}{c^5}\right),
    \\ [1.5 em]
    \mathrm{J}_{ij} &=& \displaystyle
    \mu \frac{\delta m}{m}
    \varepsilon_{ab \langle i} x_{j \rangle a}v_b \left[ - 1 +
    \gamma \left( - \frac{67}{28} + \frac{2}{7} \nu \right) +
    \gamma^2 \left( - \frac{13}{9} + \frac{4651}{252} \nu +
    \frac{1}{168} \nu^2 \right) \right] +
    {\cal O} \left( \frac{1}{c^5} \right). \hspace{-1 em}
  \end{array}
  \label{140}
\end{equation}
Also needed are the 1PN mass $2^4$-pole, 1PN current $2^3$-pole
(octupole), Newtonian mass $2^5$-pole and Newtonian current
$2^4$-pole:
\begin{equation}
  \begin{array}{rcl}
    \mathrm{I}_{ijkl} &=& \displaystyle
    \mu \, {\hat x}_{ijkl} \left[1 - 3\nu + \gamma
    \left(\frac{3}{110} - \frac{25}{22}\nu + \frac{69}{22} \nu^2
    \right) \right] + \frac{78}{55} \mu \, x_{\langle ij} v_{kl \rangle}
    \frac{r_{12}^2}{c^2} (1 - 5\nu + 5\nu^2) +
    {\cal O} \left( \frac{1}{c^3} \right), \hspace{-3.5 em}
    \\ [1.5 em]
    \mathrm{J}_{ijk} &=& \displaystyle
    \mu \, \varepsilon_{ab \langle i}
    x_{jk \rangle a} v_b \left[ 1 - 3 \nu + \gamma \left(
    \frac{181}{90} - \frac{109}{18} \nu + \frac{13}{18} \nu^2 \right)
    \right] + \frac{7}{45} \mu \, (1 - 5\nu + 5 \nu^2)
    \varepsilon_{ab\langle i} v_{jk \rangle b} x_a
    \frac{r_{12}^2}{c^2} \hspace{-3.5 em}
    \\ [1.5 em]
    && \displaystyle
    +\, {\cal O} \left( \frac{1}{c^3} \right),
    \\ [1.5 em]
    \mathrm{I}_{ijklm} &=& \displaystyle
    \mu \frac{\delta m}{m} (- 1 + 2\nu)
    {\hat x}_{ijklm} + {\cal O} \left( \frac{1}{c} \right),
    \\ [1.5 em]
    \mathrm{J}_{ijkl} &=& \displaystyle
    \mu \frac{\delta m}{m} (- 1 + 2 \nu)
    \varepsilon_{ab \langle i} x_{jkl \rangle a} v_b +
    {\cal O} \left( \frac{1}{c} \right).
  \end{array}
  \label{141}
\end{equation}

These results permit one to control what can be called the
``instantaneous'' part, say ${\cal L}_\mathrm{inst}$, of the total energy
flux, by which we mean that part of the flux that is generated solely by
the \emph{source} multipole moments, i.e.\ not counting the
``non-instantaneous'' tail integrals. The instantaneous flux is defined
by the replacement into the general expression of ${\cal L}$ given by
Equation~(\ref{51}) of all the radiative moments $\mathrm{U}_L$ and $\mathrm{V}_L$
by the corresponding ($l$th time derivatives of the) source moments
$\mathrm{I}_L$ and $\mathrm{J}_L$. Actually, we prefer to define ${\cal
L}_\mathrm{inst}$ by means of the intermediate moments $\mathrm{M}_L$ and
$\mathrm{S}_L$. Up to the 3.5PN order we have
\begin{eqnarray}
  {\cal L}_\mathrm{inst} &=& \frac{G}{c^5} \left\{ \frac{1}{5}
  \mathrm{M}^{(3)}_{ij} \mathrm{M}^{(3)}_{ij} + \frac{1}{c^2}
  \left[ \frac{1}{189} \mathrm{M}^{(4)}_{ijk} \mathrm{M}^{(4)}_{ijk} +
  \frac{16}{45} \mathrm{S}^{(3)}_{ij} \mathrm{S}^{(3)}_{ij}\right] +
  \frac{1}{c^4} \left[ \frac{1}{9072}
  \mathrm{M}^{(5)}_{ijkm} \mathrm{M}^{(5)}_{ijkm} +
  \frac{1}{84} \mathrm{S}^{(4)}_{ijk} \mathrm{S}^{(4)}_{ijk}\right] \right.
  \nonumber \\
  && \qquad
  \left. + \frac{1}{c^6} \left[ \frac{1}{594000}
  \mathrm{M}^{(6)}_{ijkmn} \mathrm{M}^{(6)}_{ijkmn} + \frac{4}{14175}
  \mathrm{S}^{(5)}_{ijkm} \mathrm{S}^{(5)}_{ijkm}\right] +
  {\cal O}\left(\frac{1}{c^8}\right) \right\}.
  \label{142}
\end{eqnarray}%
The time derivatives of the source moments~(\ref{138}, \ref{139},
\ref{140}, \ref{141}) are computed by means of the circular-orbit
equations of motion given by Equation~(\ref{127}, \ref{128}); then we
substitute them into Equation~(\ref{142})\epubtkFootnote{For circular orbits
there is no difference at this order between $\mathrm{I}_L$, $\mathrm{J}_L$
and $\mathrm{M}_L$, $\mathrm{S}_L$.}. The net result is
\begin{eqnarray}
  {\cal L}_\mathrm{inst} &=& \frac{32c^5}{5G}\nu^2 \gamma^5
  \biggl\{ 1 + \left(-\frac{2927}{336}-\frac{5}{4}\nu \right)
  \gamma + \left(\frac{293383}{9072}+\frac{380}{9}\nu\right) \gamma^2
  \nonumber \\
  && \qquad \qquad \quad
  + \left[\frac{53712289}{1108800} -
  \frac{1712}{105} \ln \left(\frac{r_{12}}{r_0} \right) \right.
  \nonumber \\
  && \qquad \qquad \qquad ~
  \left. + \left(-\frac{50625}{112} +
  \frac{123}{64}\pi^2+\frac{110}{3}
  \ln \left(\frac{r_{12}}{r_0'}\right) 
  \right)\nu - \frac{383}{9}\nu^2\right] \gamma^3
  \nonumber \\
  && \qquad \qquad \quad
  + {\cal O}\left(\frac{1}{c^8}\right)\biggr\}.
  \label{143}
\end{eqnarray}%
The Newtonian approximation, ${\cal L}_\mathrm{N}=(32c^5/5G) \,\nu^2
\gamma^5$, is the prediction of the Einstein quadrupole
formula~(\ref{4}), as computed by Landau and Lifchitz~\cite{LL}. In
Equation~(\ref{143}), we have replaced the Hadamard regularization ambiguity
parameters $\lambda$ and $\theta$ arising at the 3PN order by their
values~(\ref{lambda}) and~(\ref{theta}).


\subsection{Contribution of wave tails}
\label{subsec:10.2}

To the ``instantaneous'' part of the flux, we must add the
contribution of non-linear multipole interactions contained in the
relationship between the source and radiative moments. The needed
material has already been provided in Equations~(\ref{78}, \ref{79}). Up
to the 3.5PN level we have the dominant quadratic-order tails, the
cubic-order tails or tails of tails, and the non-linear memory
integral. We shall see that the tails play a crucial role in the
predicted signal of compact binaries. By contrast, the non-linear
memory effect, given by the integral inside the 2.5PN term in
Equation~(\ref{78}), does not contribute to the gravitational-wave energy
flux before the 4PN order in the case of circular-orbit binaries
(essentially because the memory integral is actually ``instantaneous''
in the flux), and therefore has rather poor observational consequences
for future detections of inspiralling compact binaries. We split
the energy flux into the different terms
\begin{equation}
  {\cal L} = {\cal L}_\mathrm{inst}+{\cal L}_\mathrm{tail}+
  {\cal L}_\mathrm{(tail)^2}+{\cal L}_\mathrm{tail(tail)},
  \label{144}
\end{equation}
where ${\cal L}_\mathrm{inst}$ has just been found in Equation~(\ref{143});
${\cal L}_\mathrm{tail}$ is made of the quadratic (multipolar) tail
integrals in Equation~(\ref{79}); ${\cal L}_\mathrm{(tail)^2}$ is
the square of the quadrupole tail in Equation~(\ref{78}); and ${\cal
L}_\mathrm{tail(tail)}$ is the quadrupole tail of tail in Equation~(\ref{78}).
We find that ${\cal L}_\mathrm{tail}$ contributes at the half-integer
1.5PN, 2.5PN, and 3.5PN orders, while both ${\cal L}_\mathrm{(tail)^2}$
and ${\cal L}_\mathrm{tail(tail)}$ appear only at the 3PN order. It is
quite remarkable that so small an effect as a ``tail of tail'' should
be relevant to the present computation, which is aimed at preparing
the ground for forthcoming experiments.

The results follow from the reduction to the case of circular compact
binaries of the general formulas~(\ref{78}, \ref{79}), in which we make
use of the explicit expressions for the source moments of compact
binaries as found in Section~\ref{subsec:10.1}. Without going into
accessory details (see Ref.~\cite{B98tail}), let us point out that following
the general formalism of Part~\ref{part:a}, the total mass $\mathrm{M}$ in
front of the tail integrals is the ADM mass of the binary which is given
by the sum of the rest masses, $m=m_1+m_2$ (which is the one appearing
in the $\gamma$-parameter, Equation~(\ref{126})), plus some relativistic
corrections. At the 2PN relative order needed here to compute the tail
integrals we have
\begin{equation}
  \mathrm{M} =
  m \left[ 1 - \frac{\nu}{2}\gamma+\frac{\nu}{8}
  \left(7-\nu\right)\gamma^2 + {\cal O}\left(\frac{1}{c^6}\right)\right].
  \label{calM}
\end{equation}
Let us give the two basic technical formulas needed when carrying out
this reduction:
\begin{equation}
  \begin{array}{rcl}
    \displaystyle \int^{+\infty}_0 \!\!\!\!\! d\tau\ln \tau \,
    e^{-\sigma \tau} &=& \displaystyle - \frac{1}{\sigma} (C + \ln \sigma),
    \\ [1.5 em]
    \displaystyle \int^{+\infty}_0 \!\!\!\!\! d\tau \ln^2\tau \,
    e^{-\sigma \tau} &=& \displaystyle \frac{1}{\sigma}
    \left[ \frac{\pi^2}{6} + (C + \ln \sigma)^2\right],
  \end{array}
\end{equation}
where $\sigma\in {\mathbb C}$ and $C=0.577\cdots$ denotes the Euler
constant~\cite{GZ}. The tail integrals are evaluated thanks to these
formulas for a \emph{fixed} (non-decaying) circular orbit. Indeed it can
be shown~\cite{BS93} that the ``remote-past'' contribution to the tail
integrals is negligible; the errors due to the fact that the orbit
actually spirals in by gravitational radiation do not affect the signal
before the 4PN order. We then find, for the quadratic tail term
\emph{stricto sensu}, the 1.5PN, 2.5PN, and 3.5PN contributions\epubtkFootnote{All
formulas incorporate the changes in some equations following the
published Errata (2005) to the works~\cite{B96, B98tail, BIJ02, BFIJ02,
ABIQ04}.}
\begin{eqnarray}
  {\cal L}_\mathrm{tail} &=& \frac{32c^5}{5G} \gamma^5 \nu^2
  \left\{ 4\pi \gamma^{3/2} + \left(-\frac{25663}{672} -
  \frac{125}{8}\nu\right) \pi \gamma^{5/2} + \left(\frac{90205}{576} +
  \frac{505747}{1512}\nu + \frac{12809}{756}\nu^2\right) \right.
  \pi\gamma^{7/2}
  \nonumber \\
  && \qquad \qquad \quad
  + \left. {\cal O} \left(\frac{1}{c^8}\right)\!\right\}.
  \label{146}
\end{eqnarray}%
For the sum of squared tails and cubic tails of tails at 3PN, we get
\begin{eqnarray}
  {\cal L}_{(\mathrm{tail})^2+\mathrm{tail}(\mathrm{tail})} &=&
  \frac{32c^5}{5G} \gamma^5 \nu^2
  \left\{ \left( \!\! -\frac{116761}{3675} + \frac{16}{3}\pi^2 -
  \frac{1712}{105}C + \frac{1712}{105}\ln
  \left(\frac{r_{12}}{r_0}\right) - \frac{856}{105}\ln
  \left(16\gamma\right) \right)\gamma^3 \right.
  \nonumber \\
  && \qquad \qquad \quad ~
  \left. + {\cal O}\left(\frac{1}{c^8}\right)\right\}.
  \label{147}
\end{eqnarray}%
By comparing Equations~(\ref{143}) and~(\ref{147}) 
we observe that the constants $r_0$ cleanly cancel out. 
Adding together all these contributions we obtain
\begin{eqnarray}
  {\cal L} &=& \frac{32c^5}{5G} \gamma^5 \nu^2
  \left\{ 1+\left(\!-\frac{2927}{336} -
  \frac{5}{4}\nu \!\right) \gamma + 4\pi \gamma^{3/2} +
  \left(\!\frac{293383}{9072} +
  \frac{380}{9}\nu \!\right) \gamma^2  +\left(\! -\frac{25663}{672} -
  \frac{125}{8}\nu \!\right)\pi \gamma^{5/2} \right.
  \nonumber \\
  && \qquad \qquad \quad
  + \left[ \frac{129386791}{7761600} +
  \frac{16\pi^2}{3}-\frac{1712}{105}C -
  \frac{856}{105} \ln (16\gamma) \right.
  \nonumber \\
  && \qquad \qquad \qquad ~
  + \left. \left( -\frac{50625}{112} +
  \frac{110}{3} \ln \left( \frac{r_{12}}{{r'}_0} \right) +
  \frac{123\pi^2}{64} \right) \nu -
  \frac{383}{9}\nu^2 \right] \gamma^3
  \nonumber \\
  && \qquad \qquad \quad
  + \left. \left(\frac{90205}{576} +
  \frac{505747}{1512}\nu + \frac{12809}{756} \nu^2\right)
  \pi \gamma^{7/2} + {\cal O}\left(\frac{1}{c^8}\right)\!\right\}.
  \label{148}
\end{eqnarray}%
The gauge constant $r'_0$ has not yet disappeared because 
the post-Newtonian expansion is still parametrized by $\gamma$ 
instead of the frequency-related parameter $x$ defined by Equation~(\ref{131}) -- 
just as for $E$ 
when it was given by Equation~(\ref{130}). After substituting 
the expression $\gamma (x)$ given by Equation~(\ref{132}), 
we find that $r'_0$ does cancel as well. Because the 
relation $\gamma (x)$ is issued from the
equations of motion, the latter cancellation represents an 
interesting test of the consistency of the two 
computations, in harmonic coordinates, of the 
3PN multipole moments and the 3PN equations of motion. 
At long last we obtain our end result:
\begin{eqnarray}
  {\cal L} &=& \frac{32c^5}{5G}\nu^2 x^5 \biggl\{ 1 +
  \left(-\frac{1247}{336} - \frac{35}{12}\nu \right) x + 4\pi x^{3/2} +
  \left(-\frac{44711}{9072} + \frac{9271}{504}\nu +
  \frac{65}{18} \nu^2\right) x^2 
  \nonumber \\
  && \qquad \qquad \quad
  + \left(-\frac{8191}{672}-\frac{583}{24}\nu\right)\pi x^{5/2}
  \nonumber \\
  && \qquad \qquad \quad
  + \left[\frac{6643739519}{69854400}+
  \frac{16}{3}\pi^2-\frac{1712}{105}C -
  \frac{856}{105} \ln (16\,x) \right.
  \nonumber \\
  && \qquad \qquad \qquad ~
  + \left. \left(-\frac{134543}{7776} +
  \frac{41}{48}\pi^2 
  \right)\nu - \frac{94403}{3024}\nu^2 -
  \frac{775}{324}\nu^3 \right] x^3
  \nonumber \\
  && \qquad \qquad \quad
  + \left(-\frac{16285}{504} +
  \frac{214745}{1728}\nu + \frac{193385}{3024}\nu^2\right)\pi x^{7/2} +
  {\cal O}\left(\frac{1}{c^8}\right) \biggr\}.
  \label{149}
\end{eqnarray}%
In the test-mass limit $\nu\to 0$ for one of the bodies, we recover
exactly the result following from linear black-hole perturbations
obtained by Tagoshi and Sasaki~\cite{TSasa94}. In particular, the
rational fraction $6643739519/69854400$ comes out exactly the same as in
black-hole perturbations. On the other hand, the ambiguity parameters
$\lambda$ and $\theta$ are part of the rational fraction $-134543/7776$,
belonging to the coefficient of the term at 3PN order proportional to
$\nu$ (hence this coefficient cannot be computed by linear black hole
perturbations)\epubtkFootnote{Generalizing the flux formula~(\ref{149})
to point masses moving on \emph{quasi elliptic} orbits dates back to
the work of Peters and Mathews~\cite{PM63} at Newtonian order. The
result was obtained in~\cite{WagW76, BS89} at 1PN order, and then
further extended by Gopakumar and Iyer~\cite{GopuI97} up to 2PN order
using an explicit quasi-Keplerian representation of the
motion~\cite{DS88, SW93}. No complete result at 3PN order is yet
available.}.


\subsection{Orbital phase evolution}
\label{subsec:10.3}

We shall now deduce the laws of variation with time of the orbital
frequency and phase of an inspiralling compact binary from the energy
balance equation~(\ref{134}). The center-of-mass energy $E$ is given by
Equation~(\ref{133}) and the total flux ${\cal L}$ by Equation~(\ref{149}). For
convenience we adopt the dimensionless time
variable\epubtkFootnote{Notice the ``strange'' post-Newtonian order of
this time variable: $\Theta={\cal O}(c^{+8})$.}
\begin{equation}
  \Theta \equiv \frac{\nu c^3}{5Gm}(t_\mathrm{c}-t),
  \label{150}
\end{equation}
where $t_\mathrm{c}$ denotes the instant of coalescence, at which the
frequency tends to infinity (evidently, the post-Newtonian method breaks
down well before this point). We transform the balance equation into an
ordinary differential equation for the parameter $x$, which is
immediately integrated with the result
\begin{eqnarray}
  x &=& \frac{1}{4}\Theta^{-1/4}\biggl\{ 1 +
  \left( \frac{743}{4032} + \frac{11}{48}\nu\right)\Theta^{-1/4} -
  \frac{1}{5}\pi\Theta^{-3/8} + \left( \frac{19583}{254016} +
  \frac{24401}{193536} \nu + \frac{31}{288} \nu^2 \right) \Theta^{-1/2}
  \nonumber \\
  && \qquad \qquad \:
  + \left(-\frac{11891}{53760} +
  \frac{109}{1920}\nu\right) \pi \Theta^{-5/8}
  \nonumber \\
  && \qquad \qquad \:
  + \left[-\frac{10052469856691}{6008596070400} +
  \frac{1}{6}\pi^2 + \frac{107}{420}C - \frac{107}{3360}
  \ln\left(\frac{\Theta}{256}\right)\right.
  \nonumber \\
  && \qquad \qquad \quad ~\:
  + \left(\frac{3147553127}{780337152} -
  \frac{451}{3072}\pi^2 \right) \nu
  - \left.\frac{15211}{442368}\nu^2 +
  \frac{25565}{331776}\nu^3\right] \Theta^{-3/4}
  \nonumber \\
  && \qquad \qquad \:
  + \left(-\frac{113868647}{433520640} -
  \frac{31821}{143360}\nu + \frac{294941}{3870720}\nu^2\right) \pi
  \Theta^{-7/8} + {\cal O}\left(\frac{1}{c^8}\right)\biggr\}.
  \label{151}
\end{eqnarray}%
The orbital phase is defined as the angle $\phi$, oriented in the
sense of the motion, between the separation of the two bodies and the
direction of the ascending node ${\cal N}$ within the plane of the
sky, namely the point on the orbit at which the bodies cross the plane
of the sky moving toward the detector. We have
$d\phi/dt=\omega$, which translates, with our notation, into
$d\phi/d\Theta = -5/\nu \cdot x^{3/2}$, from which we
determine
\begin{eqnarray}
  \phi &=& - \frac{1}{\nu}\Theta^{5/8}\biggl\{ 1 +
  \left( \frac{3715}{8064} + \frac{55}{96} \nu \right)
  \Theta^{-1/4} - \frac{3}{4}\pi\Theta^{-3/8}
  + \left( \frac{9275495}{14450688} +
  \frac{284875}{258048} \nu + \frac{1855}{2048} \nu^2 \right)
  \Theta^{-1/2}
  \nonumber \\
  && \qquad \qquad \;
  + \left( -\frac{38645}{172032} +
  \frac{65}{2048} \nu \right) \pi \Theta^{-5/8}
  \ln \left(\frac{\Theta}{\Theta_0} \right)
  \nonumber \\
  && \qquad \qquad \;
  + \left[\frac{831032450749357}{57682522275840} -
  \frac{53}{40}\pi^2 - \frac{107}{56}C + \frac{107}{448}
  \ln\left(\frac{\Theta}{256}\right)\right.
  \nonumber \\
  && \qquad \qquad \quad ~\;
  + \left.\left(-\frac{126510089885}{4161798144} +
  \frac{2255}{2048}\pi^2 \right)\nu\right.
  \nonumber \\
  && \qquad \qquad \quad ~\;
  + \left.\frac{154565}{1835008}\nu^2 -
  \frac{1179625}{1769472}\nu^3\right] \Theta^{-3/4}
  \nonumber \\
  && \qquad \qquad \;
  + \left(\frac{188516689}{173408256} +
  \frac{488825}{516096}\nu - \frac{141769}{516096}\nu^2\right)\pi
  \Theta^{-7/8} + {\cal O}\left(\frac{1}{c^8}\right)\biggr\},
  \label{152}
\end{eqnarray}%
where $\Theta_0$ is a constant of integration that can be fixed 
by the initial conditions when the wave frequency enters
the detector's bandwidth. Finally we want also to dispose of the
important expression of the phase in terms of the frequency $x$. For this we
get
\begin{eqnarray}
  \phi &=& - \frac{x^{-5/2}}{32\nu}\biggl\{ 1 +
  \left( \frac{3715}{1008} + \frac{55}{12}\nu\right)x -
  10\pi x^{3/2} 
  + \left( \frac{15293365}{1016064} +
  \frac{27145}{1008} \nu + \frac{3085}{144} \nu^2 \right) x^2
  \nonumber \\
  && \qquad \qquad \;
  + \left(\frac{38645}{1344} -
  \frac{65}{16}\nu\right) \pi x^{5/2}
  \ln \left(\frac{x}{x_0}\right)
  \nonumber \\
  && \qquad \qquad \;
  + \left[\frac{12348611926451}{18776862720} -
  \frac{160}{3}\pi^2 - \frac{1712}{21}C -
  \frac{856}{21} \ln (16\,x)\right.
  \nonumber \\
  && \qquad \qquad \quad ~~
  + \left.\left(-\frac{15737765635}{12192768} +
  \frac{2255}{48}\pi^2 \right)\nu
  + \frac{76055}{6912}\nu^2 -
  \frac{127825}{5184}\nu^3\right] x^3
  \nonumber \\
  && \qquad \qquad \;
  + \left(\frac{77096675}{2032128} +
  \frac{378515}{12096}\nu - \frac{74045}{6048}\nu^2\right)
  \pi x^{7/2} + {\cal O}\left(\frac{1}{c^8}\right)\biggr\},
  \label{153}
\end{eqnarray}%
where $x_0$ is another constant of integration. With the
formula~(\ref{153}) the orbital phase is complete up to the 3.5PN order.
The effects due to the spins of the particles, i.e.\ the spin-orbit
(SO) coupling arising at the 1.5PN order for maximally rotating compact
bodies and the spin-spin (SS) coupling at the 2PN order, can be added if
necessary; they are known up to the 2.5PN order included~\cite{KWW93,
K95, OTO98, TOO01, FBB06spin, BBF06spin}. On the other hand, the
contribution of the quadrupole moments of the compact objects, which are
induced by tidal effects, is expected to come only at the 5PN order (see
Equation~(\ref{8})).

As a rough estimate of the relative importance of the various
post-Newtonian terms, let us give in Table~\ref{tab2} their
contributions to the accumulated number of gravitational-wave cycles
${\cal N}$ in the bandwidth of the LIGO and VIRGO detectors (see also
Table~I in Ref.~\cite{BDIWW95} for the contributions of the SO and SS
effects). Note that such an estimate is only indicative, because a full
treatment would require the knowledge of the detector's power spectral
density of noise, and a complete simulation of the parameter estimation
using matched filtering~\cite{CF94, PW95, KKS95}. We define ${\cal N}$
by
\begin{equation}
  {\cal N}={1\over \pi}\bigl[\phi_\mathrm{ISCO}-\phi_\mathrm{seismic}\bigr].
  \label{GWcycle}
\end{equation}
The frequency of the signal at the entrance of the bandwidth is the
seismic cut-off frequency $f_\mathrm{seismic}$ of ground-based detectors;
the terminal frequency $f_\mathrm{ISCO}$ is assumed for simplicity's sake
to be given by the Schwarzschild innermost stable circular orbit. Here
$f = \omega/\pi = 2/P$ is the signal frequency at the
dominant harmonics (twice the orbital frequency). As we see in
Table~\ref{tab2}, with the 3PN or 3.5PN approximations we reach an
acceptable level of, say, a few cycles, that roughly corresponds to the
demand which was made by data-analysists in the case of neutron-star
binaries~\cite{3mn, CFPS93, CF94, P95, BCV03a, BCV03b}. Indeed, the
above estimation suggests that the neglected 4PN terms will yield some
systematic errors that are, at most, of the same order of magnitude,
i.e.\ a few cycles, and perhaps much less (see also the discussion
in Section~\ref{subsec:9.5}).

\begin{table}[htbp]
  \renewcommand{\arraystretch}{1.2}
  \centering
  \begin{tabular}{l|rrr}
    \hline \hline
    & $2 \times 1.4 \, M_\odot$ &
    $10 \, M_\odot+1.4 \, M_\odot$ &
    $2 \times 10 \, M_\odot$ \\
    \hline
    Newtonian order &
    $16031\pz$ &
    $3576\pz$ &
    $602\pz$ \\
    1PN &
    $441\pz$ &
    $213\pz$ &
    $59\pz$ \\
    1.5PN (dominant tail) &
    $-211\pz$ &
    $-181\pz$ &
    $-51\pz$ \\
    2PN &
    $9.9\ws$ &
    $9.8\ws$ &
    $4.1\ws$ \\
    2.5PN &
    $-11.7\ws$ &
    $-20.0\ws$ &
    $-7.1\ws$ \\
    3PN &
    $2.6\ws$ &
    $2.3\ws$ &
    $2.2\ws$ \\
    3.5PN &
    $-0.9\ws$ &
    $-1.8\ws$ &
    $-0.8\ws$ \\
    \hline \hline
  \end{tabular}
  \caption{\it Contributions of post-Newtonian orders to the
    accumulated number of gravitational-wave cycles ${\cal N}$
    (defined by Equation~(\ref{GWcycle})) in the bandwidth of VIRGO
    and LIGO detectors. Neutron stars have mass $1.4 \, M_\odot$, and
    black holes $10 \, M_\odot$. The entry frequency is
    $f_\mathrm{seismic}= 10 \mathrm{\ Hz}$, and the terminal frequency
    is $f_\mathrm{ISCO}=c^3/(6^{3/2}\pi G m)$.}
  \label{tab2}
  \renewcommand{\arraystretch}{1.0}
\end{table}


\subsection{The two polarization waveforms}
\label{subsec:10.4}

The theoretical templates of the compact binary inspiral follow from
insertion of the previous solutions for the 3.5PN-accurate orbital
frequency and phase into the binary's two polarization waveforms $h_+$
and $h_\times$. We shall include in $h_+$ and $h_\times$ all the
harmonics, besides the dominant one at twice the orbital frequency, up
to the 2.5PN order, as they have been calculated in Refs.~\cite{BIWW96,
ABIQ04}. The polarization waveforms are defined with respect to two
polarization vectors ${\bf p}=(p_i)$ and ${\bf q}=(q_i)$,
\begin{equation}
  \begin{array}{rcl}
    h_+ &=& \displaystyle
    \frac{1}{2}(p_ip_j-q_iq_j)h_{ij}^\mathrm{TT},
    \\ [1.5 em]
    h_\times &=& \displaystyle
    \frac{1}{2}(p_iq_j+p_jq_i)h_{ij}^\mathrm{TT},
  \end{array}
\end{equation}
where ${\bf p}$ and ${\bf q}$ are chosen to lie
along the major and minor axis, respectively, of the projection onto
the plane of the sky of the circular orbit, with ${\bf p}$ oriented
toward the ascending node ${\cal N}$.
To the 2PN order we have
\begin{eqnarray}
  h_{+,\times} &=& \frac{2G\mu x}{c^2 R} \left\{ H^{(0)}_{+,\times} +
  x^{1/2} H^{(1/2)}_{+,\times} + x H^{(1)}_{+,\times} + x^{3/2}
  H^{(3/2)}_{+,\times} + x^2 H^{(2)}_{+,\times} + x^{5/2}
  H^{(5/2)}_{+,\times} + {\cal O}\left(\frac{1}{c^6}\right)
  \right\}.
  \nonumber\\
  \label{154}
\end{eqnarray}%
The post-Newtonian terms are ordered by means of the frequency-related variable
$x$. They depend on the binary's 3.5PN-accurate phase $\phi$ through the auxiliary
phase variable
\begin{equation}
  \psi =\phi - \frac{2G \mathrm{M} \omega}{c^3}
  \ln \left(\frac{\omega}{\omega_0}\right),
  \label{155}
\end{equation}
where $\mathrm{M} = m \left[ 1 -\nu\gamma/2+ {\cal
O}\left(1/c^4\right)\right]$ is the ADM mass (cf.\
Equation~(\ref{calM})), and where $\omega_0$ is a constant frequency that can
conveniently be chosen to be the entry frequency of a
laser-interferometric detector (say $\omega_0/\pi=10 \mathrm{\ Hz}$). For
the plus polarization we have\epubtkFootnote{We neglect the non-linear
memory (DC) term present in the Newtonian plus polarization $H^{(0)}_+$.
See Wiseman and Will~\cite{WW91} and Arun et al.~\cite{ABIQ04} for
the computation of this term.}
\begin{eqnarray}
  H^{(0)}_+ &=&
  -(1+c_i^2) \cos 2 \psi,
  \nonumber \\
  H^{(1/2)}_+ &=&
  -\frac{s_i}{8} \frac{\delta m}{m}
  \left[ (5+c_i^2) \cos \psi - 9 (1+c_i^2) \cos 3 \psi \right],
  \nonumber \\
  H^{(1)}_+ &=&
  \frac{1}{6} \left[ 19 + 9 c_i^2 -
  2 c_i^4 - \nu ( 19 - 11 c_i^2 - 6 c_i^4 ) \right] \cos 2 \psi -
  \frac{4}{3} s_i^2 (1+c_i^2) (1-3\nu) \cos 4 \psi,
  \nonumber \\
  H^{(3/2)}_+ &=&
  \frac{s_i}{192} \frac{\delta m}{m}
  \biggl\{ \! \left[ 57 + 60 c_i^2 - c_i^4 - 2 \nu
  (49 - 12 c_i^2 - c_i^4) \right] \cos \psi
  \nonumber \\
  && \qquad \qquad
  - \frac{27}{2} \left[ 73 + 40 c_i^2 -
  9 c_i^4 - 2 \nu (25 - 8 c_i^2 - 9 c_i^4) \right] \cos 3\psi
  \nonumber \\
  && \qquad \qquad
  + \frac{625}{2} (1-2\nu) s_i^2 (1 + c_i^2) \cos 5 \psi \biggr\} -
  2 \pi (1 + c_i^2) \cos 2 \psi,
  \nonumber \\
  H^{(2)}_+ &=&
  \frac{1}{120} \biggl[ 22 + 396 c_i^2 +
  145 c_i^4 - 5 c_i^6 + \frac{5}{3} \nu
  (706 - 216 c_i^2 - 251 c_i^4 + 15 c_i^6)
  \nonumber \\
  && \qquad \, 
  - 5 \nu^2 (98 - 108 c_i^2 + 7 c_i^4 + 5 c_i^6) \biggr] \cos 2 \psi
  \nonumber \\
  &&
  + \frac{2}{15} s_i^2 \left[ 59 + 35 c_i^2 - 8 c_i^4 -
  \frac{5}{3} \nu ( 131 + 59 c_i^2 - 24 c_i^4) + 5 \nu^2 (21 - 3 c_i^2 -
  8 c_i^4) \right] \cos 4 \psi
  \nonumber \\
  &&
  - \frac{81}{40} (1-5\nu +5\nu^2) s_i^4 (1+ c_i^2) \cos 6\psi
  \nonumber \\
  &&
  + \frac{s_i}{40} \frac{\delta m}{m} \biggl\{ \!
  \left[ 11 + 7 c_i^2 + 10 (5+c_i^2) \ln 2 \right] \sin \psi -
  {5\pi} (5+c_i^2) \cos \psi
  \nonumber \\
  && \qquad \qquad \;
  - 27 \left[ 7 - 10 \ln (3/2) \right]
  (1 + c_i^2) \sin 3\psi + 135 \pi (1 + c_i^2) \cos 3 \psi
  \biggr\}.
  \nonumber
\end{eqnarray}%
For the cross polarization, we have
\begin{eqnarray}
  H^{(0)}_\times &=&
  - 2c_i \sin 2\psi,
  \nonumber \\
  H^{(1/2)}_\times &=&
  - \frac{3}{4} s_i c_i \frac{\delta m}{m}
  \left[ \sin \psi - 3 \sin 3 \psi \right],
  \nonumber \\
  H^{(1)}_\times &=&
  \frac{c_i}{3} \left[ 17 - 4 c_i^2 -
  \nu (13 - 12 c_i^2) \right] \sin 2\psi -
  \frac{8}{3} (1-3\nu) c_i s_i^2 \sin 4 \psi,
  \nonumber \\
  H^{(3/2)}_\times &=&
  \frac{s_i c_i}{96} \frac{\delta m}{m}
  \biggl\{ \! \left[ 63 - 5 c_i^2 - 2 \nu ( 23 - 5 c_i^2) \right]
  \sin \psi
  - \frac{27}{2} \left[ 67 - 15 c_i^2 - 2 \nu
  (19 - 15 c_i^2) \right] \sin 3 \psi \hspace{-1.0 em}
  \nonumber \\
  && \qquad \qquad 
  + \frac{625}{2} (1 - 2\nu) s_i^2 \sin 5 \psi \biggr\} -
  4 \pi c_i \sin 2\psi,
  \nonumber \\
  H^{(2)}_\times &=&
  \frac{c_i}{60} \left[ 68 + 226 c_i^2 -
  15 c_i^4 + \frac{5}{3} \nu (572 - 490 c_i^2 + 45 c_i^4) -
  5 \nu^2 (56 - 70 c_i^2 + 15 c_i^4 ) \right] \sin 2 \psi \hspace{-1.0 em}
  \nonumber \\
  &&
  + \frac{4}{15} c_is_i^2 \left[ 55 - 12 c_i^2 - \frac{5}{3} \nu
  (119 - 36 c_i^2) + 5 \nu^2 (17 - 12 c_i^2) \right] \sin 4 \psi
  \nonumber \\
  &&
  - \frac{81}{20} (1 - 5 \nu + 5 \nu^2) c_i s_i^4 \sin 6 \psi
  \nonumber \\
  &&
  - \frac{3}{20} s_i c_i \frac{\delta m}{m} \left\{ \! \left[
  3 + 10 \ln 2 \right] \cos \psi + 5 \pi \sin \psi -
  9 \left[ 7 - 10 \ln (3/2) \right] \cos 3\psi -
  45 \pi \sin 3 \psi \right\}.
  \nonumber
\end{eqnarray}%
We use the shorthands $c_i=\cos i$ and $s_i=\sin i$ for the cosine and
sine of the inclination angle $i$ between the direction of the detector
as seen from the binary's center-of-mass, and the normal to the orbital
plane (we always suppose that the normal is right-handed with respect to
the sense of motion, so that $0\leq i\leq \pi$). Finally, the more
recent calculation of the 2.5PN order in Ref.~\cite{ABIQ04} is reported
here:
\begin{eqnarray}
  H^{(5/2)}_+ &=& s_i{\delta m\over m}\cos\psi
  \bigg[ {1771\over 5120}-{1667\over5120} c_i^2+{217\over 9216} c_i^4-
  {1\over 9216} c_i^6
  \nonumber \\
  && \qquad \qquad \quad \,\,
  +\nu \left( {681\over256}+{13\over768} c_i^2-
  {35\over 768} c_i^4+{1\over2304} c_i^6 \right)
  \nonumber \\
  && \qquad \qquad \quad \,\,
  + \nu^2 \left( \! -{3451\over9216}+{673\over3072} c_i^2-
  {5\over9216} c_i^4-{1\over3072} c_i^6 \right) \bigg]
  \nonumber \\
  &&
  +\pi\cos2\psi \left[ {19\over3}+3 c_i^2-{2\over3} c_i^4+
  \nu \left( \! -{16\over3}+{14\over3} c_i^2+2 c_i^4 \right) \right]
  \nonumber \\
  &&
  +s_i{\delta m\over m}\cos3\psi \bigg[ {3537\over1024}-
  {22977\over5120} c_i^2-{15309\over5120} c_i^4+
  {729\over5120} c_i^6
  \nonumber \\
  && \qquad \qquad \qquad ~\,
  +\nu \left( \! -{23829\over1280}+{5529\over1280} c_i^2+
  {7749\over1280} c_i^4-{729\over1280} c_i^6 \right)
  \nonumber \\
  && \qquad \qquad \qquad ~\,
  +\nu^2 \left( {29127\over5120}-{27267\over5120} c_i^2-
  {1647\over5120} c_i^4+{2187\over 5120} c_i^6 \right) \bigg]
  \nonumber \\
  &&
  +\cos4\psi \left[ -{16\pi\over3} (1+ c_i^2) s_i^2(1-3\nu) \right]
  \nonumber \\
  &&
  +s_i {\delta m\over m}\cos5\psi \bigg[ \! -{108125\over9216}+
  {40625\over9216} c_i^2+{83125\over9216} c_i^4-{15625\over 9216} c_i^6
  \nonumber \\
  && \qquad \qquad \qquad ~\,
  +\nu \left( {8125\over 256}-{40625\over2304} c_i^2-
  {48125\over2304} c_i^4+{15625\over 2304} c_i^6 \right)
  \nonumber \\
  && \qquad \qquad \qquad ~\,
  +\nu^2 \left( \! -{119375\over9216}+{40625\over3072} c_i^2+
  {44375\over 9216} c_i^4-{15625\over3072} c_i^6 \right) \bigg]
  \nonumber \\
  &&
  +{\delta m\over m}\cos7\psi \left[ {117649\over46080}
  s_i^5(1+ c_i^2)(1-4\nu+3\nu^2) \right]
  \nonumber \\
  &&
  +\sin2\psi \left[ -{9\over5}+{14\over5} c_i^2+{7\over5} c_i^4+
  \nu \left( {96\over5}-{8\over5} c_i^2-{28\over5} c_i^4 \right) \right]
  \nonumber \\
  &&
  +s_i^2(1+ c_i^2)\sin4\psi \left[ {56\over5}-{32\ln2 \over3}-
  \nu \left( {1193\over30}-32\ln2 \right) \right],
  \label{H25+}
  \\ [1 em]
  H^{(5/2)}_\times &=&{6\over5} s_i^2 c_i \nu
  \nonumber \\
  &&
  + c_i\cos2\psi \left[ 2-{22\over5} c_i^2+\nu\left( \! -{154\over5}+
  {94\over5} c_i^2 \right) \right]
  \nonumber \\
  &&
  + c_i s_i^2 \cos4\psi \left[ -{112\over5}+{64\over3}\ln2+
  \nu \left( {1193\over15}-64\ln2 \right) \right]
  \nonumber \\
  &&
  + s_i c_i{\delta m \over m} \sin\psi \bigg[ \! -{913\over 7680}+
  {1891\over11520} c_i^2-{7\over4608} c_i^4
  \nonumber \\
  && \qquad \qquad \qquad ~\:
  +\nu \left( {1165\over384}-{235\over576} c_i^2+{7\over1152} c_i^4 \right)
  \nonumber \\
  && \qquad \qquad \qquad ~\:
  +\nu^2 \left( \!-{1301\over4608}+{301\over2304} c_i^2-
  {7\over1536} c_i^4 \right) \bigg]
  \nonumber \\
  &&
  +\pi c_i\sin2\psi \left[ {34\over3}-{8\over3} c_i^2-
  \nu \left( {20\over3}-8 c_i^2 \right) \right]
  \nonumber \\
  &&
  + s_i c_i{\delta m \over m}\sin3\psi \bigg[ {12501\over2560}-
  {12069\over1280} c_i^2+{1701\over2560} c_i^4
  \nonumber \\
  && \qquad \qquad \qquad \quad \;
  +\nu \left( \!-{19581\over640}+{7821\over320} c_i^2-
  {1701\over640} c_i^4 \right)
  \nonumber \\
  && \qquad \qquad \qquad \quad \;
  +\nu^2\left({18903\over2560}-{11403\over 1280} c_i^2+
  {5103\over2560} c_i^4 \right) \bigg]
  \nonumber \\
  &&
  +s_i^2 c_i\sin4\psi \left[ -{32\pi \over3}(1-3\nu) \right]
  \nonumber \\
  &&
  +{\delta m \over m} s_i c_i\sin5\psi \bigg[ \! -{101875\over4608}+
  {6875\over256} c_i^2-{21875\over4608} c_i^4
  \nonumber \\
  && \qquad \qquad \qquad \quad \;
  +\nu \left( {66875\over1152}-{44375\over576} c_i^2+
  {21875\over1152}c_i^4 \right)
  \nonumber \\
  && \qquad \qquad \qquad \quad \;
  +\nu^2\left( \! -{100625\over4608}+{83125\over2304} c_i^2-
  {21875\over1536} c_i^4 \right) \bigg]
  \nonumber \\
  &&
  +{\delta m\over m} s_i^5 c_i\sin7\psi \left[ {117649\over23040}
  \left( 1-4\nu+3\nu^2 \right) \right].
  \label{H25x}
\end{eqnarray}%

The practical implementation of the theoretical templates in the data
analysis of detectors follows the standard matched filtering technique.
The raw output of the detector $o(t)$ consists of the superposition of
the real gravitational wave signal $h_\mathrm{real}(t)$ and of noise
$n(t)$. The noise is assumed to be a stationary Gaussian random
variable, with zero expectation value, and with (supposedly known)
frequency-dependent power spectral density $S_n(\omega)$. The
experimenters construct the correlation between $o(t)$ and a filter
$q(t)$, i.e.
\begin{equation}
  c(t) = \int^{+\infty}_{-\infty} \!\!\!\! dt' \, o (t') q(t+t'),
  \label{ct}
\end{equation}
and divide $c(t)$ by the square root of its variance, or correlation
noise. The expectation value of this ratio defines the filtered
signal-to-noise ratio (SNR). Looking for the useful signal
$h_\mathrm{real}(t)$ in the detector's output $o(t)$, the experimenters adopt for
the filter
\begin{equation}
  \tilde q (\omega) = {\tilde h (\omega)\over S_n (\omega)},
  \label{qtilde}
\end{equation}
where ${\tilde q} (\omega)$ and ${\tilde h} (\omega)$ are the Fourier
transforms of $q(t)$ and of the \emph{theoretically computed} template
$h(t)$. By the matched filtering theorem, the filter (\ref{qtilde})
maximizes the SNR if $h(t)=h_\mathrm{real}(t)$. The maximum SNR is then the
best achievable with a linear filter. In practice, because of systematic
errors in the theoretical modelling, the template $h(t)$ will not
exactly match the real signal $h_\mathrm{real} (t)$, but if the template is
to constitute a realistic representation of nature the errors will be
small. This is of course the motivation for computing high order
post-Newtonian templates, in order to reduce as much as possible the
systematic errors due to the unknown post-Newtonian remainder.

To conclude, the use of theoretical templates based on the preceding
2.5PN wave forms, and having their frequency evolution built in
\emph{via} the 3.5PN phase evolution~(\ref{152}, \ref{153}), should yield
some accurate detection and measurement of the binary signals.
Interestingly, it should also permit some new tests of general
relativity, because we have the possibility of checking that the
observed signals do obey each of the terms of the phasing
formulas~(\ref{152}, \ref{153}), e.g., those associated with the
specific non-linear tails, exactly as they are predicted by Einstein's
theory~\cite{BSat94, BSat95, AIQS06}. Indeed, we don't know of any other
physical systems for which it would be possible to perform such tests.

\newpage


\section{Acknowledgments}
\label{sec:11}

It is a great pleasure to thank Silvano Bonazzola, Alessandra Buonanno,
Thibault Damour, J\"urgen Ehlers, Gilles Esposito-Far\`ese, Guillaume
Faye, Eric Gourgoulhon, Bala Iyer, Sergei Kopeikin, Misao Sasaki,
Gerhard Sch\"afer, Bernd Schmidt, Kip Thorne, and Clifford Will for
interesting discussions and/or collaborations.

\newpage


\bibliography{refs}

\begin{thebibliography}{100}

\bibitem{AIRS05}
Ajith, P., Iyer, B.R., Robinson, C.A.K., and Sathyaprakash, B.S., ``New class
  of post-Newtonian approximants to the waveform templates of inspiralling
  compact binaries: Test mass in the Schwarzschild spacetime'', {\em Phys. Rev.
  D}, {\bf 71}, 044029--1--21, (2005). Related online version (cited on 26
  April 2006): \newline\url{http://arXiv.org/abs/gr-qc/0412033}.
  \epubtkKeywords{Post-Newtonian approximations, Data analysis}

\bibitem{ADec75}
Anderson, J.L., and DeCanio, T.C., ``Equations of hydrodynamics in general
  relativity in the slow motion approximation'', {\em Gen. Relativ. Gravit.},
  {\bf 6}, 197--238, (1975). \epubtkKeywords{Post-Newtonian approximations,
  Radiation reaction}

\bibitem{ACST94}
Apostolatos, T.A., Cutler, C., Sussman, G.J., and Thorne, K.S., ``Spin-induced
  orbital precession and its modulation of the gravitational waveforms from
  merging binaries'', {\em Phys. Rev. D}, {\bf 49}, 6274--6297, (1994).
  \epubtkKeywords{Gravitational radiation, Compact binaries, Spin, Data
  analysis}

\bibitem{ABIQ04}
Arun, K.G., Blanchet, L., Iyer, B.R., and Qusailah, M.S., ``The 2.5PN
  gravitational wave polarisations from inspiralling compact binaries in
  circular orbits'', {\em Class. Quantum Grav.}, {\bf 21}, 3771, (2004).
  Related online version (cited on 26 April 2006):
  \newline\url{http://arXiv.org/abs/gr-qc/0404185}. Erratum {\it Class.\
  Quantum Grav.}, {\bf 22}, 3115, (2005). \epubtkKeywords{Gravitational
  radiation, Compact binaries, Post-Newtonian approximations}

\bibitem{AIQS06}
Arun, K.G., Iyer, B.R., Qusailah, M.S., and Sathyaprakash, B.S., ``Probing the
  non-linear structure of general relativity with black hole mergers'', (2006).
  URL (cited on 26 April 2006):
  \newline\url{http://arXiv.org/abs/gr-qc/0604067}. \epubtkKeywords{Alternative
  theories, Data analysis, Post-Newtonian approximations}

\bibitem{AISS05}
Arun, K.G., Iyer, B.R., Sathyaprakash, B.S., and Sundararajan, P.A.,
  ``Parameter estimation of inspiralling compact binaries using 3.5
  post-Newtonian gravitational wave phasing: The nonspinning case'', {\em Phys.
  Rev. D}, {\bf 71}, 084008--1--16, (2005). Related online version (cited on 26
  April 2006): \newline\url{http://arXiv.org/abs/gr-qc/0411146}.
  \epubtkKeywords{Post-Newtonian approximations, Data analysis}

\bibitem{BOC75}
Barker, B.M., and O'Connell, R.F., ``Gravitational two-body problem with
  arbitrary masses, spins, and quadrupole moments'', {\em Phys. Rev. D}, {\bf
  12}, 329--335, (1975). \epubtkKeywords{Spin, ADM formalism}

\bibitem{BOC79}
Barker, B.M., and O'Connell, R.F., ``The gravitational interaction: Spin,
  rotation, and quantum effects - A review'', {\em Gen. Relativ. Gravit.}, {\bf
  11}, 149--175, (1979). \epubtkKeywords{Spin, ADM formalism}

\bibitem{Baum00}
Baumgarte, T.W., ``Innermost stable circular orbit of binary black holes'',
  {\em Phys. Rev. D}, {\bf 62}, 024018--1--8, (2000). \epubtkKeywords{Compact
  binaries, Innermost stable circular orbits, Numerical relativity}

\bibitem{Bek73}
Bekenstein, J.D., ``Gravitational Radiation Recoil and Runaway Black Holes'',
  {\em Astrophys. J.}, {\bf 183}, 657--664, (1973).
  \epubtkKeywords{Relativistic astrophysics, Gravitational radiation}

\bibitem{BeDD81}
Bel, L., Damour, T., Deruelle, N., Iba{\~{n}}ez, J., and Martin, J.,
  ``Poincar{\'e}-invariant gravitational-field and equations of motion of 2
  point-like objects -- The post-linear approximtion of general-relativity'',
  {\em Gen. Relativ. Gravit.}, {\bf 13}, 963--1004, (1981).
  \epubtkKeywords{Equations of motion, Post-Minkowskian approximations}

\bibitem{B87}
Blanchet, L., ``Radiative gravitational fields in general-relativity. II.
  Asymptotic-behaviour at future null infinity'', {\em Proc. R. Soc. London,
  Ser. A}, {\bf 409}, 383--399, (1987). \epubtkKeywords{Post-Minkowskian
  approximations, Asymptotics}

\bibitem{B90}
Blanchet, L., {\em Contribution \`{a} l'\'{e}tude du rayonnement gravitationnel
  \'{e}mis par un syst\`{e}me isol\'{e}}, Habilitation, (Universit\'{e} Paris
  VI, Paris, France, 1990). \epubtkKeywords{Gravitational radiation}

\bibitem{B93}
Blanchet, L., ``Time-asymmetric structure of gravitational radiation'', {\em
  Phys. Rev. D}, {\bf 47}, 4392--4420, (1993). \epubtkKeywords{Radiation
  reaction, Post-Newtonian approximations}

\bibitem{B95}
Blanchet, L., ``Second-post-Newtonian generation of gravitational radiation'',
  {\em Phys. Rev. D}, {\bf 51}, 2559--2583, (1995). Related online version
  (cited on 24 January 1995): \newline\url{http://arXiv.org/abs/gr-qc/9501030}.
  \epubtkKeywords{Gravitational radiation, Post-Newtonian approximations}

\bibitem{B96}
Blanchet, L., ``Energy losses by gravitational radiation in inspiralling
  compact binaries to 5/2 post-Newtonian order'', {\em Phys. Rev. D}, {\bf 54},
  1417--1438, (1996). \epubtkKeywords{Gravitational radiation, Binary systems,
  Post-Newtonian approximations}

\bibitem{Bhouches}
Blanchet, L., ``Gravitational Radiation from Relativistic Sources'', in Marck,
  J.A., and Lasota, J.P., eds., {\em Relativistic Gravitation and Gravitational
  Radiation}, Proceedings of the Les Houches School of Physics, held in Les
  Houches, Haute Savoie, 26 September -- 6 October, 1995,  33--66, (Cambridge
  University Press, Cambridge, U.K., 1997). Related online version (cited on 11
  July 1996): \newline\url{http://arXiv.org/abs/gr-qc/9607025}.
  \epubtkKeywords{Gravitational radiation, Binary systems, Post-Newtonian
  approximations}

\bibitem{B97}
Blanchet, L., ``Gravitational radiation reaction and balance equations to
  post-Newtonian order'', {\em Phys. Rev. D}, {\bf 55}, 714--732, (1997).
  Related online version (cited on 20 September 1996):
  \newline\url{http://arXiv.org/abs/gr-qc/9609049}. \epubtkKeywords{Radiation
  reaction, Post-Newtonian approximations}

\bibitem{B98tail}
Blanchet, L., ``Gravitational-wave tails of tails'', {\em Class. Quantum
  Grav.}, {\bf 15}, 113--141, (1998). Related online version (cited on 7
  October 1997): \newline\url{http://arXiv.org/abs/gr-qc/9710038}.
  \epubtkKeywords{Gravitational radiation, Multipole expansions}

\bibitem{B98mult}
Blanchet, L., ``On the multipole expansion of the gravitational field'', {\em
  Class. Quantum Grav.}, {\bf 15}, 1971--1999, (1998). Related online version
  (cited on 29 January 1998): \newline\url{http://arXiv.org/abs/gr-qc/9710038}.
  \epubtkKeywords{Multipole expansions}

\bibitem{B98quad}
Blanchet, L., ``Quadrupole-quadrupole gravitational waves'', {\em Class.
  Quantum Grav.}, {\bf 15}, 89--111, (1998). Related online version (cited on 7
  October 1997): \newline\url{http://arXiv.org/abs/gr-qc/9710037}.
  \epubtkKeywords{Gravitational radiation, Multipole expansions}

\bibitem{Behlers}
Blanchet, L., ``Post-Newtonian Gravitational Radiation'', in Schmidt, B.G.,
  ed., {\em Einstein's Field Equations and Their Physical Implications:
  Selected Essays in Honour of J\"{u}rgen Ehlers}, vol. 540 of Lecture Notes in
  Physics,  225--271, (Springer, Berlin, Germany; New York, U.S.A., 2000).
  \epubtkKeywords{Gravitational radiation, Binary systems, Equations of motion,
  Approximation methods}

\bibitem{B02ico}
Blanchet, L., ``Innermost circular orbit of binary black holes at the third
  post-Newtonian approximation'', {\em Phys. Rev. D}, {\bf 65}, 124009, (2002).
  Related online version (cited on 26 April 2006):
  \newline\url{http://arXiv.org/abs/gr-qc/0112056}. \epubtkKeywords{Equations
  of motion, Innermost stable circular orbits, Post-Newtonian approximations}

\bibitem{Baccuracy}
Blanchet, L., ``On the accuracy of the post-Newtonian approximation'', in
  Ciufolini, I., Dominici, D., and Lusanna, L., eds., {\em 2001: A Relativistic
  Spacetime Odyssey}, Proceedings of the Johns Hopkins Workshop on Current
  Problems in Particle Theory 25, Firenze, 2001 (September 3--5),  411, (World
  Scientific, River Edge, U.S.A., 2003). Related online version (cited on 26
  April 2006): \newline\url{http://arXiv.org/abs/gr-qc/0207037}.
  \epubtkKeywords{Equations of motion, Innermost stable circular orbits,
  Post-Newtonian approximations}

\bibitem{BBF06spin}
Blanchet, L., Buonanno, A., and Faye, G., ``Higher-order spin effects in the
  dynamics of compact binaries II. Radiation field'', in preparation, (2006).
  \epubtkKeywords{Gravitational radiation, Compact binaries, Spin,
  Post-Newtonian approximations}

\bibitem{BD86}
Blanchet, L., and Damour, T., ``Radiative gravitational fields in general
  relativity. I. General structure of the field outside the source'', {\em
  Philos. Trans. R. Soc. London, Ser. A}, {\bf 320}, 379--430, (1986).
  \epubtkKeywords{Post-Minkowskian approximations}

\bibitem{BD88}
Blanchet, L., and Damour, T., ``Tail-transported temporal correlations in the
  dynamics of a gravitating system'', {\em Phys. Rev. D}, {\bf 37}, 1410--1435,
  (1988). \epubtkKeywords{Radiation reaction, Post-Newtonian approximations}

\bibitem{BD89}
Blanchet, L., and Damour, T., ``Post-Newtonian generation of gravitational
  waves'', {\em Ann. Inst. Henri Poincare A}, {\bf 50}, 377--408, (1989).
  \epubtkKeywords{Gravitational radiation, Post-Newtonian approximations}

\bibitem{BD92}
Blanchet, L., and Damour, T., ``Hereditary effects in gravitational
  radiation'', {\em Phys. Rev. D}, {\bf 46}, 4304--4319, (1992).
  \epubtkKeywords{Gravitational radiation, Post-Newtonian approximations}

\bibitem{BDE04}
Blanchet, L., Damour, T., and Esposito-Far{\`e}se, G., ``Dimensional
  regularization of the third post-Newtonian dynamics of point particles in
  harmonic coordinates'', {\em Phys. Rev. D}, {\bf 69}, 124007, (2004). Related
  online version (cited on 26 April 2006):
  \newline\url{http://arXiv.org/abs/gr-qc/0311052}. \epubtkKeywords{Equations
  of motion, Post-Newtonian approximations}

\bibitem{BDEI04}
Blanchet, L., Damour, T., Esposito-Far{\`e}se, G., and Iyer, B.R.,
  ``Gravitational radiation from inspiralling compact binaries completed at the
  third post-Newtonian order'', {\em Phys. Rev. Lett.}, {\bf 93}, 091101,
  (2004). Related online version (cited on 26 April 2006):
  \newline\url{http://arXiv.org/abs/gr-qc/0406012}.
  \epubtkKeywords{Gravitational radiation, Equations of motion, Post-Newtonian
  approximations, ADM formalism}

\bibitem{BDEI05dr}
Blanchet, L., Damour, T., Esposito-Far{\`e}se, G., and Iyer, B.R.,
  ``Dimensional regularization of the third post-Newtonian gravitational wave
  generation of two point masses'', {\em Phys. Rev. D}, {\bf 71},
  124004--1--36, (2005). \epubtkKeywords{Gravitational radiation, Compact
  binaries, Post-Newtonian approximations}

\bibitem{BDI95}
Blanchet, L., Damour, T., and Iyer, B.R., ``Gravitational waves from
  inspiralling compact binaries: Energy loss and waveform to
  second-post-Newtonian order'', {\em Phys. Rev. D}, {\bf 51}, 5360--5386,
  (1995). Related online version (cited on 24 January 1995):
  \newline\url{http://arXiv.org/abs/gr-qc/9501029}. Erratum {\it Phys.\ Rev.\
  D}, {\bf 54}, 1860, (1996). \epubtkKeywords{Gravitational radiation, Binary
  systems, Post-Newtonian approximations}

\bibitem{BDI04zeta}
Blanchet, L., Damour, T., and Iyer, B.R., ``Surface-integral expressions for
  the multipole moments of post-Newtonian sources and the boosted Schwarzschild
  solution'', {\em Class. Quantum Grav.}, {\bf 22}, 155, (2005). Related online
  version (cited on 26 April 2006):
  \newline\url{http://arXiv.org/abs/gr-qc/0410021}.
  \epubtkKeywords{Gravitational radiation, Multipole moments, Post-Newtonian
  approximations}

\bibitem{BDIWW95}
Blanchet, L., Damour, T., Iyer, B.R., Will, C.M., and Wiseman, A.G.,
  ``Gravitational-Radiation Damping of Compact Binary Systems to Second
  Post-Newtonian Order'', {\em Phys. Rev. Lett.}, {\bf 74}, 3515--3518, (1995).
  Related online version (cited on 23 January 1995):
  \newline\url{http://arXiv.org/abs/gr-qc/9501027}.
  \epubtkKeywords{Gravitational radiation, Binary systems, Post-Newtonian
  approximations}

\bibitem{BFreg}
Blanchet, L., and Faye, G., ``Hadamard regularization'', {\em J. Math. Phys.},
  {\bf 41}, 7675--7714, (2000). Related online version (cited on 28 July 2000):
  \newline\url{http://arXiv.org/abs/gr-qc/0004008}.
  \epubtkKeywords{Regularization methods}

\bibitem{BF00}
Blanchet, L., and Faye, G., ``On the equations of motion of point-particle
  binaries at the third post-Newtonian order'', {\em Phys. Lett. A}, {\bf 271},
  58--64, (2000). Related online version (cited on 22 May 2000):
  \newline\url{http://arXiv.org/abs/gr-qc/0004009}. \epubtkKeywords{Equations
  of motion, Post-Newtonian approximations}

\bibitem{BFeom}
Blanchet, L., and Faye, G., ``General relativistic dynamics of compact binaries
  at the third post-Newtonian order'', {\em Phys. Rev. D}, {\bf 63},
  062005--1--43, (2001). Related online version (cited on 18 November 2000):
  \newline\url{http://arXiv.org/abs/gr-qc/0007051}. \epubtkKeywords{Equations
  of motion, Post-Newtonian approximations}

\bibitem{BFregM}
Blanchet, L., and Faye, G., ``Lorentzian regularization and the problem of
  point-like particles in general relativity'', {\em J. Math. Phys.}, {\bf 42},
  4391--4418, (2001). Related online version (cited on 4 April 2001):
  \newline\url{http://arXiv.org/abs/gr-qc/0006100}.
  \epubtkKeywords{Regularization methods}

\bibitem{BFIJ02}
Blanchet, L., Faye, G., Iyer, B.R., and Joguet, B., ``Gravitational-wave
  inspiral of compact binary systems to 7/2 post-Newtonian order'', {\em Phys.
  Rev. D}, {\bf 65}, 061501--1--5, (2002). Related online version (cited on 26
  May 2001): \newline\url{http://arXiv.org/abs/gr-qc/0105099}.
  \epubtkKeywords{Gravitational radiation, Binary systems, Post-Newtonian
  approximations}

\bibitem{BFN05}
Blanchet, L., Faye, G., and Nissanke, S., ``Structure of the post-Newtonian
  expansion in general relativity'', {\em Phys. Rev. D}, {\bf 72}, 044024,
  (2005). \epubtkKeywords{Post-Newtonian approximations, Radiation reaction}

\bibitem{BFP98}
Blanchet, L., Faye, G., and Ponsot, B., ``Gravitational field and equations of
  motion of compact binaries to 5/2 post-Newtonian order'', {\em Phys. Rev. D},
  {\bf 58}, 124002--1--20, (1998). Related online version (cited on 11 August
  1998): \newline\url{http://arXiv.org/abs/gr-qc/9804079}.
  \epubtkKeywords{Equations of motion, Post-Newtonian approximations}

\bibitem{BI03CM}
Blanchet, L., and Iyer, B.R., ``Third post-Newtonian dynamics of compact
  binaries: Equations of motion in the center-of-mass frame'', {\em Class.
  Quantum Grav.}, {\bf 20}, 755, (2003). Related online version (cited on 26
  April 2006): \newline\url{http://arXiv.org/abs/gr-qc/0209089}.
  \epubtkKeywords{Equations of motion, Compact binaries, Post-Newtonian
  approximations}

\bibitem{BI04mult}
Blanchet, L., and Iyer, B.R., ``Hadamard regularization of the third
  post-Newtonian gravitational wave generation of two point masses'', {\em
  Phys. Rev. D}, {\bf 71}, 024004, (2004). Related online version (cited on 26
  April 2006): \newline\url{http://arXiv.org/abs/gr-qc/0409094}.
  \epubtkKeywords{Gravitational radiation, Multipole expansions, Post-Newtonian
  approximations}

\bibitem{BIJ02}
Blanchet, L., Iyer, B.R., and Joguet, B., ``Gravitational waves from
  inspiralling compact binaries: Energy flux to third post-Newtonian order'',
  {\em Phys. Rev. D}, {\bf 65}, 064005--1--41, (2002). Related online version
  (cited on 26 May 2001): \newline\url{http://arXiv.org/abs/gr-qc/0105098}.
  \epubtkKeywords{Gravitational radiation, Binary systems, Post-Newtonian
  approximations}

\bibitem{BIWW96}
Blanchet, L., Iyer, B.R., Will, C.M., and Wiseman, A.G., ``Gravitational
  waveforms from inspiralling compact binaries to second-post-Newtonian
  order'', {\em Class. Quantum Grav.}, {\bf 13}, 575--584, (1996). Related
  online version (cited on 13 February 1996):
  \newline\url{http://arXiv.org/abs/gr-qc/9602024}.
  \epubtkKeywords{Gravitational radiation, Binary systems, Post-Newtonian
  approximations}

\bibitem{BSat94}
Blanchet, L., and Sathyaprakash, B.S., ``Signal analysis of gravitational wave
  tails'', {\em Class. Quantum Grav.}, {\bf 11}, 2807--2831, (1994).
  \epubtkKeywords{Binary systems, Gravitational wave detectors}

\bibitem{BSat95}
Blanchet, L., and Sathyaprakash, B.S., ``Detecting a tail effect in
  gravitational-wave experiments'', {\em Phys. Rev. Lett.}, {\bf 74},
  1067--1070, (1995). \epubtkKeywords{Binary systems, Gravitational wave
  detectors}

\bibitem{BS89}
Blanchet, L., and Sch{\"{a}}fer, G., ``Higher-order gravitational-radiation
  losses in binary systems'', {\em Mon. Not. R. Astron. Soc.}, {\bf 239},
  845--867, (1989). \epubtkKeywords{Gravitational radiation, Post-Newtonian
  approximations}

\bibitem{BS93}
Blanchet, L., and Sch{\"{a}}fer, G., ``Gravitational wave tails and binary star
  systems'', {\em Class. Quantum Grav.}, {\bf 10}, 2699--2721, (1993).
  \epubtkKeywords{Gravitational radiation, Binary systems, Post-Newtonian
  approximations}

\bibitem{Bollini}
Bollini, C.G., and Giambiagi, J.J., ``Lowest order ``divergent'' graphs in
  v-dimensional space'', {\em Phys. Lett. B}, {\bf 40}, 566--568, (1972).
  \epubtkKeywords{Quantum field theory, Perturbation theory, Gauge theories}

\bibitem{BGM99}
Bonazzola, S., Gourgoulhon, E., and Marck, J.-A., ``Numerical models of
  irrotational binary neutron stars in general relativity'', {\em Phys. Rev.
  Lett.}, {\bf 82}, 892, (1999). Related online version (cited on 26 April
  2006): \newline\url{http://arXiv.org/abs/gr-qc/9810072}.
  \epubtkKeywords{Numerical relativity, Compact binaries, Neutron stars}

\bibitem{BBM62}
Bondi, H., van~der Burg, M.G.J., and Metzner, A.W.K., ``Gravitational waves in
  general relativity VII. Waves from axi-symmetric isolated systems'', {\em
  Proc. R. Soc. London, Ser. A}, {\bf 269}, 21--52, (1962).
  \epubtkKeywords{Asymptotics}

\bibitem{Bo59}
Bonnor, W.B., ``Spherical gravitational waves'', {\em Philos. Trans. R. Soc.
  London, Ser. A}, {\bf 251}, 233--271, (1959). \epubtkKeywords{Multipole
  expansions}

\bibitem{BoR61}
Bonnor, W.B., and Rotenberg, M.A., ``Transport of momentum by gravitational
  waves -- Linear approximation'', {\em Proc. R. Soc. London, Ser. A}, {\bf
  265}, 109, (1961). \epubtkKeywords{Multipole expansions}

\bibitem{BoR66}
Bonnor, W.B., and Rotenberg, M.A., ``Gravitational waves from isolated
  sources'', {\em Proc. R. Soc. London, Ser. A}, {\bf 289}, 247--274, (1966).
  \epubtkKeywords{Multipole expansions}

\bibitem{Breitenlohner}
Breitenlohner, P., and Maison, D., ``Dimensional renormalization and the action
  principle'', {\em Commun. Math. Phys.}, {\bf 52}, 11--38, (1977).
  \epubtkKeywords{Quantum field theory, Perturbation theory, Gauge theories}

\bibitem{BCV03b}
Buonanno, A., Chen, Y., and Vallisneri, M., ``Detecting gravitational waves
  from precessing binaries of spinning compact objects: Adiabatic limit'', {\em
  Phys. Rev. D}, {\bf 67}, 104025--1--31, (2003). Related online version (cited
  on 26 April 2006): \newline\url{http://arXiv.org/abs/gr-qc/0211087}.
  \epubtkKeywords{Gravitational radiation, Data analysis, Post-Newtonian
  approximations}

\bibitem{BCV03a}
Buonanno, A., Chen, Y., and Vallisneri, M., ``Detection template families for
  gravitational waves from the final stages of binary black-holes binaries:
  Nonspinning case'', {\em Phys. Rev. D}, {\bf 67}, 024016, (2003). Related
  online version (cited on 26 April 2006):
  \newline\url{http://arXiv.org/abs/gr-qc/0205122}.
  \epubtkKeywords{Gravitational radiation, Data analysis, Post-Newtonian
  approximations}

\bibitem{BuonD98}
Buonanno, A., and Damour, T., ``Effective one-body approach to general
  relativistic two-body dynamics, ADM formalism'', {\em Phys. Rev. D}, {\bf
  59}, 084006, (1999). Related online version (cited on 26 April 2006):
  \newline\url{http://arXiv.org/abs/gr-qc/9811091}. \epubtkKeywords{Compact
  binaries, Post-Newtonian approximations}

\bibitem{BuonD00}
Buonanno, A., and Damour, T., ``Transition from inspiral to plunge in binary
  black hole coalescences'', {\em Phys. Rev. D}, {\bf 62}, 064015, (2000).
  Related online version (cited on 26 April 2006):
  \newline\url{http://arXiv.org/abs/gr-qc/0001013}. \epubtkKeywords{Compact
  binaries, Post-Newtonian approximations, ADM formalism}

\bibitem{Bu71}
Burke, W.L., ``Gravitational radiation damping of slowly moving systems
  calculated using matched asymptotic expansions'', {\em J. Math. Phys.}, {\bf
  12}(3), 401--418, (1971). \epubtkKeywords{Radiation reaction}

\bibitem{BuTh70}
Burke, W.L., and Thorne, K.S., ``Gravitational Radiation Damping'', in Carmeli,
  M., Fickler, S.I., and Witten, L., eds., {\em Relativity}, Proceedings of the
  Relativity Conference in the Midwest, held at Cincinnati, Ohio, June 2--6,
  1969,  209--228, (Plenum Press, New York, U.S.A.; London, U.K., 1970).
  \epubtkKeywords{Radiation reaction}

\bibitem{CMM77}
Campbell, W.B., Macek, J., and Morgan, T.A., ``Relativistic time-dependent
  multipole analysis for scalar, electromagnetic, and gravitational fields'',
  {\em Phys. Rev. D}, {\bf 15}, 2156--2164, (1977). \epubtkKeywords{Multipole
  expansions}

\bibitem{CM71}
Campbell, W.B., and Morgan, T.A., ``Debye Potentials For Gravitational Field'',
  {\em Physica}, {\bf 53}(2), 264, (1971). \epubtkKeywords{Multipole
  expansions}

\bibitem{C65}
Chandrasekhar, S., ``The Post-Newtonian Equations of Hydrodynamics in General
  Relativity'', {\em Astrophys. J.}, {\bf 142}, 1488--1540, (1965).
  \epubtkKeywords{Relativistic hydrodynamics, Post-Newtonian approximations}

\bibitem{CE70}
Chandrasekhar, S., and Esposito, F.P., ``The 5/2-Post-Newtonian Equations of
  Hydrodynamics and Radiation Reaction in General Relativity'', {\em Astrophys.
  J.}, {\bf 160}, 153--179, (1970). \epubtkKeywords{Relativistic hydrodynamics,
  Radiation reaction, Post-Newtonian approximations}

\bibitem{CN69}
Chandrasekhar, S., and Nutku, Y., ``The Second Post-Newtonian Equations of
  Hydrodynamics in General Relativity'', {\em Astrophys. J.}, {\bf 158},
  55--79, (1969). \epubtkKeywords{Relativistic hydrodynamics, Post-Newtonian
  approximations}

\bibitem{CKMR01}
Chicone, C., Kopeikin, S.M., Mashhoon, B., and Retzloff, D.G., ``Delay
  equations and radiation damping'', {\em Phys. Lett. A}, {\bf 285}, 17--26,
  (2001). Related online version (cited on 2 May 2001):
  \newline\url{http://arXiv.org/abs/gr-qc/0101122}. \epubtkKeywords{Radiation
  reaction}

\bibitem{cho98}
Cho, H.T., ``Post-Newtonian approximation for spinning particles'', {\em Class.
  Quantum Grav.}, {\bf 15}, 2465, (1998). Related online version (cited on 26
  April 2006): \newline\url{http://arXiv.org/abs/gr-qc/9703071}.
  \epubtkKeywords{Gravitational radiation, Spin, Post-Newtonian approximations}

\bibitem{Chr91}
Christodoulou, D., ``Nonlinear Nature of Gravitation and Gravitational-Wave
  Experiments'', {\em Phys. Rev. Lett.}, {\bf 67}, 1486--1489, (1991).
  \epubtkKeywords{Asymptotics}

\bibitem{CS79}
Christodoulou, D., and Schmidt, B.G., ``Convergent and asymptotic iteration
  methods in general-relativity'', {\em Commun. Math. Phys.}, {\bf 68},
  275--289, (1979). \epubtkKeywords{Mathematical relativity, Perturbation
  methods}

\bibitem{Collins}
Collins, J.C., {\em Renormalization: An introduction to renormalization, the
  renormalization group, and the operator-product expansion}, (Cambridge
  University Press, Cambridge, U.K.; New York, U.S.A., 1984).
  \epubtkKeywords{Quantum field theory, Perturbation theory, Gauge theories}

\bibitem{CPf04}
Cook, G.B., and Pfeiffer, H.P., ``Excision boundary conditions for black-hole
  initial data'', {\em Phys. Rev. D}, {\bf 70}, 104016--1--24, (2004).
  \epubtkKeywords{Compact binaries, Innermost stable circular orbits, Numerical
  relativity}

\bibitem{CB69}
Cooperstock, F.I., and Booth, D.J., ``Angular-Momentum Flux For Gravitational
  Radiation To Octupole Order'', {\em Nuovo Cimento}, {\bf 62}(1), 163, (1969).
  \epubtkKeywords{Gravitational radiation, Multipole expansions}

\bibitem{CTh77}
Crowley, R.J., and Thorne, K.S., ``Generation of gravitational waves. II.
  Post-linear formalism revisited'', {\em Astrophys. J.}, {\bf 215}, 624--635,
  (1977). \epubtkKeywords{Post-Minkowskian approximations}

\bibitem{3mn}
Cutler, C., Apostolatos, T.A., Bildsten, L., Finn, L.S., Flanagan,
  {\'{E}}.{\'{E}}., Kennefick, D., Markovi{\'{c}}, D.M., Ori, A., Poisson, E.,
  Sussman, G.J., and Thorne, K.S., ``The last three minutes: Issues in
  gravitational wave measurements of coalescing compact binaries'', {\em Phys.
  Rev. Lett.}, {\bf 70}, 2984--2987, (1993). \epubtkKeywords{Relativistic
  astrophysics, Binary systems}

\bibitem{CFPS93}
Cutler, C., Finn, L.S., Poisson, E., and Sussman, G.J., ``Gravitational
  radiation from a particle in circular orbit around a black hole. II.
  Numerical results for the nonrotating case'', {\em Phys. Rev. D}, {\bf 47},
  1511--1518, (1993). \epubtkKeywords{Binary systems, Gravitational wave
  detectors}

\bibitem{CF94}
Cutler, C., and Flanagan, {\'{E}}.{\'{E}}., ``Gravitational waves from merging
  compact binaries: How accurately can one extract the binary's parameters from
  the inspiral waveform?'', {\em Phys. Rev. D}, {\bf 49}, 2658--2697, (1994).
  \epubtkKeywords{Binary systems, Gravitational wave detectors}

\bibitem{D82}
Damour, T., ``The two-body problem and radiation damping in
  general-relativity'', {\em C. R. Acad. Sci. Ser. II}, {\bf 294}, 1355--1357,
  (1982). \epubtkKeywords{Equations of motion, Post-Newtonian approximations}

\bibitem{D83houches}
Damour, T., ``Gravitational radiation and the motion of compact bodies'', in
  Deruelle, N., and Piran, T., eds., {\em Gravitational Radiation}, NATO
  Advanced Study Institute, Centre de physique des Houches, 2--21 June 1982,
  59--144, (North-Holland; Elsevier, Amsterdam, Netherlands; New York, U.S.A.,
  1983). \epubtkKeywords{Equations of motion, Post-Newtonian approximations}

\bibitem{D83}
Damour, T., ``Gravitational Radiation Reaction in the Binary Pulsar and the
  Quadrupole-Formula Controversy'', {\em Phys. Rev. Lett.}, {\bf 51},
  1019--1021, (1983). \epubtkKeywords{Relativistic astrophysics, Binary
  systems, Pulsars, Radiation reaction}

\bibitem{Dcargese}
Damour, T., ``An Introduction to the Theory of Gravitational Radiation'', in
  Carter, B., and Hartle, J.B., eds., {\em Gravitation in Astrophysics:
  Carg\`{e}se 1986}, Proceedings of a NATO Advanced Study Institute on
  Gravitation in Astrophysics, held July 15--31, 1986 in Carg\'{e}se, France,
  vol. 156 of NATO ASI Series B,  3--62, (Plenum Press, New York, U.S.A.,
  1987). \epubtkKeywords{Gravitational radiation, Equations of motion}

\bibitem{D300}
Damour, T., ``The problem of motion in Newtonian and Einsteinian gravity'', in
  Hawking, S.W., and Israel, W., eds., {\em Three Hundred Years of
  Gravitation},  128--198, (Cambridge University Press, Cambridge, U.K.; New
  York, U.S.A., 1987). \epubtkKeywords{Equations of motion, Approximation
  methods}

\bibitem{DD81b}
Damour, T., and Deruelle, N., ``Generalized lagrangian of two point masses in
  the post-post-Newtonian approximation of general-relativity'', {\em C. R.
  Acad. Sci. Ser. II}, {\bf 293}, 537--540, (1981). \epubtkKeywords{Equations
  of motion, Post-Newtonian approximations}

\bibitem{DD81a}
Damour, T., and Deruelle, N., ``Radiation reaction and angular momentum loss in
  small angle gravitational scattering'', {\em Phys. Lett. A}, {\bf 87},
  81--84, (1981). \epubtkKeywords{Radiation reaction, Equations of motion}

\bibitem{Dgef96}
Damour, T., and Esposito-Far{\`e}se, G., ``Testing gravity to second
  post-Newtonian order: A Field theory approach'', {\em Phys. Rev. D}, {\bf
  53}, 5541--5578, (1996). Related online version (cited on 26 April 2006):
  \newline\url{http://arXiv.org/abs/gr-qc/9506063}. \epubtkKeywords{Alternative
  theories, Post-Newtonian approximations}

\bibitem{DGG02}
Damour, T., Gourgoulhon, E., and Grandcl{\'e}ment, P., ``Circular orbits of
  corotating binary black holes: Comparison between analytical and numerical
  results'', {\em Phys. Rev. D}, {\bf 66}, 024007--1--15, (2002). Related
  online version (cited on 26 April 2006):
  \newline\url{http://arXiv.org/abs/gr-qc/0204011}. \epubtkKeywords{Equations
  of motion, Innermost stable circular orbits, Post-Newtonian approximations,
  Numerical relativity}

\bibitem{DI91b}
Damour, T., and Iyer, B.R., ``Multipole analysis for electromagnetism and
  linearized gravity with irreducible Cartesian tensors'', {\em Phys. Rev. D},
  {\bf 43}, 3259--3272, (1991). \epubtkKeywords{Linearized gravity, Multipole
  expansions}

\bibitem{DI91a}
Damour, T., and Iyer, B.R., ``Post-Newtonian generation of gravitational waves.
  II. The spin moments'', {\em Ann. Inst. Henri Poincare A}, {\bf 54},
  115--164, (1991). \epubtkKeywords{Gravitational radiation, Post-Newtonian
  approximations}

\bibitem{DIJS03}
Damour, T., Iyer, B.R., Jaranowski, P., and Sathyaprakash, B.S.,
  ``Gravitational waves from black hole binary inspiral and merger: The span of
  third post-Newtonian effective-one-body templates'', {\em Phys. Rev. D}, {\bf
  67}, 064028, (2003). Related online version (cited on 26 April 2006):
  \newline\url{http://arXiv.org/abs/gr-qc/0211041}.
  \epubtkKeywords{Post-Newtonian approximations, Data analysis}

\bibitem{DIS98}
Damour, T., Iyer, B.R., and Sathyaprakash, B.S., ``Improved filters for
  gravitational waves from inspiraling compact binaries'', {\em Phys. Rev. D},
  {\bf 57}, 885--907, (1998). Related online version (cited on 18 August 1997):
  \newline\url{http://arXiv.org/abs/gr-qc/9708034}. \epubtkKeywords{Binary
  systems, Gravitational wave detectors}

\bibitem{DIS00}
Damour, T., Iyer, B.R., and Sathyaprakash, B.S., ``Frequency-domain
  P-approximant filters for time-truncated inspiral gravitational wave signals
  from compact binaries'', {\em Phys. Rev. D}, {\bf 62}, 084036, (2000).
  Related online version (cited on 26 April 2006):
  \newline\url{http://arXiv.org/abs/gr-qc/0001023}. \epubtkKeywords{Binary
  systems, Gravitational wave detectors}

\bibitem{DJSisco}
Damour, T., Jaranowski, P., and Sch{\"a}fer, G., ``On the determination of the
  last stable orbit for circular general relativistic binaries at the third
  post-Newtonian approximation'', {\em Phys. Rev. D}, {\bf 62}, 084011--1--21,
  (2000). Related online version (cited on 26 April 2006):
  \newline\url{http://arXiv.org/abs/gr-qc/0005034}. \epubtkKeywords{Equations
  of motion, Post-Newtonian approximations, ADM formalism}

\bibitem{DJSpoinc}
Damour, T., Jaranowski, P., and Sch{\"{a}}fer, G., ``Poincar{\'e} invariance in
  the ADM Hamiltonian approach to the general relativistic two-body problem'',
  {\em Phys. Rev. D}, {\bf 62}, 021501--1--5, (2000). Related online version
  (cited on 21 October 2000): \newline\url{http://arXiv.org/abs/gr-qc/0003051}.
  Erratum {\it Phys.\ Rev.\ D}, {\bf 63}, 029903, (2001).
  \epubtkKeywords{Equations of motion, Post-Newtonian approximations, ADM
  formalism}

\bibitem{DJSdim}
Damour, T., Jaranowski, P., and Sch{\"{a}}fer, G., ``Dimensional regularization
  of the gravitational interaction of point masses'', {\em Phys. Lett. B}, {\bf
  513}, 147--155, (2001). Related online version (cited on 11 May 2001):
  \newline\url{http://arXiv.org/abs/gr-qc/0105038}. \epubtkKeywords{Equations
  of motion, Post-Newtonian approximations, ADM formalism}

\bibitem{DJSequiv}
Damour, T., Jaranowski, P., and Sch{\"{a}}fer, G., ``Equivalence between the
  ADM-Hamiltonian and the harmonic-coordinates approaches to the third
  post-Newtonian dynamics of compact binaries'', {\em Phys. Rev. D}, {\bf 63},
  044021, (2001). Related online version (cited on 10 November 2000):
  \newline\url{http://arXiv.org/abs/gr-qc/0010040}. Erratum {\it Phys.\ Rev.\
  D}, {\bf 66}, 029901, (2002). \epubtkKeywords{Equations of motion,
  Post-Newtonian approximations, ADM formalism}

\bibitem{DS85}
Damour, T., and Sch{\"{a}}fer, G., ``Lagrangians for n point masses at the
  second post-Newtonian approximation of general-relativity'', {\em Gen.
  Relativ. Gravit.}, {\bf 17}, 879--905, (1985). \epubtkKeywords{Equations of
  motion, ADM formalism}

\bibitem{DS88}
Damour, T., and Sch{\"a}fer, G., ``Higher order relativistic periastron
  advances in binary pulsars'', {\em Nuovo Cimento B}, {\bf 101}, 127, (1988).
  \epubtkKeywords{Binary pulsars, Post-Newtonian approximations}

\bibitem{DS90}
Damour, T., and Schmidt, B., ``Reliability of perturbation theory in general
  relativity'', {\em J. Math. Phys.}, {\bf 31}, 2441--2458, (1990).
  \epubtkKeywords{Mathematical relativity, Perturbation methods}

\bibitem{DSX91}
Damour, T., Soffel, M., and Xu, C., ``General-relativistic celestial mechanics.
  I. Method and definition of reference systems'', {\em Phys. Rev. D}, {\bf
  43}, 3273--3307, (1991). \epubtkKeywords{Equations of motion}

\bibitem{DT91}
Damour, T., and Taylor, J.H., ``On the orbital period change of the Binary
  Pulsar PSR 1913+16'', {\em Astrophys. J.}, {\bf 366}, 501--511, (1991).
  \epubtkKeywords{Relativistic astrophysics, Binary systems, Pulsars}

\bibitem{ABF01}
de~Andrade, V.C., Blanchet, L., and Faye, G., ``Third post-Newtonian dynamics
  of compact binaries: Noetherian conserved quantities and equivalence between
  the harmonic-coordinate and ADM-Hamiltonian formalisms'', {\em Class. Quantum
  Grav.}, {\bf 18}, 753--778, (2001). Related online version (cited on 19
  December 2000): \newline\url{http://arXiv.org/abs/gr-qc/0011063}.
  \epubtkKeywords{Equations of motion, Post-Newtonian approximations}

\bibitem{Dthese}
Deruelle, N., {\em Sur les \'{e}quations du mouvement et le rayonnement
  gravitationnel d'un syst\`{e}me binaire en Relativit\'{e} G\'{e}n\'{e}rale},
  Ph.D. Thesis, (Universit\'{e} Pierre et Marie Curie, Paris, 1982).
  \epubtkKeywords{Equations of motion, Post-Newtonian approximations}

\bibitem{E18}
Einstein, A., ``\"{U}ber Gravitationswellen'', {\em Sitzungsber. K. Preuss.
  Akad. Wiss.}, {\bf 1918}, 154--167, (1918). \epubtkKeywords{Gravitational
  waves}

\bibitem{EIH}
Einstein, A., Infeld, L., and Hoffmann, B., ``The Gravitational Equations and
  the Problem of Motion'', {\em Ann. Math.}, {\bf 39}, 65--100, (1938).
  \epubtkKeywords{Equations of motion, Post-Newtonian approximations}

\bibitem{EW75}
Epstein, R., and Wagoner, R.V., ``Post-Newtonian generation of gravitational
  waves'', {\em Astrophys. J.}, {\bf 197}, 717--723, (1975).
  \epubtkKeywords{Gravitational radiation, Post-Newtonian approximations}

\bibitem{EH75}
Esposito, L.W., and Harrison, E.R., ``Properties of the Hulse-Taylor binary
  pulsar system'', {\em Astrophys. J. Lett.}, {\bf 196}, L1--L2, (1975).
  \epubtkKeywords{Relativistic astrophysics, Binary systems, Pulsars}

\bibitem{FayeThesis}
Faye, G., {\em Equations du mouvement d'un syst\`eme binaire d'objets compact
  \`a l'approximation post-newtonienne}, Ph.D. Thesis, (Universit{\'e} Paris
  VI, Paris, France, 1999). \epubtkKeywords{Equations of motion, Post-Newtonian
  approximations}

\bibitem{FBB06spin}
Faye, G., Blanchet, L., and Buonanno, A., ``Higher-order spin effects in the
  dynamics of compact binaries I. Equations of motion'', in preparation,
  (2006). \epubtkKeywords{Equations of motion, Compact binaries, Spin,
  Post-Newtonian approximations}

\bibitem{FCh93}
Finn, L.S., and Chernoff, D.F., ``Observing binary inspiral in gravitational
  radiation: One interferometer'', {\em Phys. Rev. D}, {\bf 47}, 2198--2219,
  (1993). \epubtkKeywords{Binary systems, Gravitational wave detectors}

\bibitem{Fock39}
Fock, V.A., ``On motion of finite masses in general relativity'', {\em J. Phys.
  (Moscow)}, {\bf 1}(2), 81--116, (1939). \epubtkKeywords{Equations of motion}

\bibitem{Fock}
Fock, V.A., {\em Theory of space, time and gravitation}, (Pergamon, London,
  U.K., 1959). \epubtkKeywords{General relativity}

\bibitem{FUS02}
Friedman, J.L., Ury{\={u}}, K., and Shibata, M., ``Thermodynamics of binary
  black holes and neutron stars'', {\em Phys. Rev. D}, {\bf 65}, 064035--1--20,
  (2002). \epubtkKeywords{Black hole thermodynamics, Numerical relativity}

\bibitem{F87}
Futamase, T., ``Strong-field point-particle limit and the equations of motion
  in the binary pulsar'', {\em Phys. Rev. D}, {\bf 36}, 321--329, (1987).
  \epubtkKeywords{Equations of motion, Compact binaries, Post-Newtonian
  approximations}

\bibitem{Galtsov}
Gal'tsov, D.V., Matiukhin, A.A., and Petukhov, V.I., ``Relativistic corrections
  to the gravitational radiation of a binary system and the fine structure of
  the spectrum'', {\em Phys. Lett. A}, {\bf 77}, 387--390, (1980).
  \epubtkKeywords{Binary systems, Perturbation methods}

\bibitem{G70}
Geroch, R., ``Multipole Moments. II. Curved Space'', {\em J. Math. Phys.}, {\bf
  11}, 2580--2588, (1970). \epubtkKeywords{Asymptotics}

\bibitem{GH78}
Geroch, R., and Horowitz, G.T., ``Asymptotically simple does not imply
  asymptotically Minkowskian'', {\em Phys. Rev. Lett.}, {\bf 40}, 203--206,
  (1978). \epubtkKeywords{Asymptotics}

\bibitem{GopuI97}
Gopakumar, A., and Iyer, B.R., ``Gravitational waves from inspiraling compact
  binaries: Angular momentum flux, evolution of the orbital elements and the
  waveform to the second post-Newtonian order'', {\em Phys. Rev. D}, {\bf 56},
  7708--7731, (1997). Related online version (cited on 15 October 1997):
  \newline\url{http://arXiv.org/abs/gr-qc/9710075}.
  \epubtkKeywords{Gravitational radiation, Binary systems, Post-Newtonian
  approximations}

\bibitem{GGB1}
Gourgoulhon, E., Grandcl{\'e}ment, P., and Bonazzola, S., ``Binary black holes
  in circular orbits. I. A global spacetime approach'', {\em Phys. Rev. D},
  {\bf 65}, 044020--1--19, (2002). Related online version (cited on 26 April
  2006): \newline\url{http://arXiv.org/abs/gr-qc/0106015}.
  \epubtkKeywords{Compact binaries, Numerical relativity}

\bibitem{GGTMB01}
Gourgoulhon, E., Grandcl{\'e}ment, P., Taniguchi, K., Marck, J.-A., and
  Bonazzola, S., ``Quasi-equilibrium sequences of synchronized and irrotational
  binary neutron stars in general relativity'', {\em Phys. Rev. D}, {\bf 63},
  064029, (2001). Related online version (cited on 26 April 2006):
  \newline\url{http://arXiv.org/abs/gr-qc/0007028}. \epubtkKeywords{Numerical
  relativity, Compact binaries, Neutron stars}

\bibitem{GZ}
Gradshteyn, I.S., and Ryzhik, I.M., {\em Table of Integrals, Series and
  Products}, (Academic Press, San Diego, U.S.A.; London, U.K., 1980).
  \epubtkKeywords{Mathematical methods}

\bibitem{GGB2}
Grandcl{\'e}ment, P., Gourgoulhon, E., and Bonazzola, S., ``Binary black holes
  in circular orbits. II. Numerical methods and first results'', {\em Phys.
  Rev. D}, {\bf 65}, 044021--1--18, (2002). Related online version (cited on 26
  April 2006): \newline\url{http://arXiv.org/abs/gr-qc/0106015}.
  \epubtkKeywords{Compact binaries, Numerical relativity}

\bibitem{GKop86}
Grishchuk, L.P., and Kopeikin, S.M., ``Equations of motion for isolated bodies
  with relativistic corrections including the radiation-reaction force'', in
  Kovalevsky, J., and Brumberg, V.A., eds., {\em Relativity in Celestial
  Mechanics and Astrometry: High Precision Dynamical Theories and Observational
  Verifications}, Proceedings of the 114th Symposium of the International
  Astronomical Union, held in Leningrad, USSR, May 28--31, 1985,  19--34,
  (Reidel, Dordrecht, Netherlands; Boston, U.S.A., 1986).
  \epubtkKeywords{Equations of motion, Post-Newtonian approximations}

\bibitem{Hadamard}
Hadamard, J., {\em Le probl\`{e}me de Cauchy et les \'{e}quations aux
  d\'{e}riv\'{e}es partielles lin\'{e}aires hyperboliques}, (Hermann, Paris,
  France, 1932). \epubtkKeywords{Mathematical methods}

\bibitem{H74}
Hansen, R.O., ``Multipole moments of stationary space-times'', {\em J. Math.
  Phys.}, {\bf 15}, 46--52, (1974). \epubtkKeywords{Asymptotics}

\bibitem{HR69}
Hunter, A.J., and Rotenberg, M.A., ``The double-series approximation method in
  general relativity. I. Exact solution of the (24) approximation. II.
  Discussion of 'wave tails' in the (2s) approximation'', {\em J. Phys. A},
  {\bf 2}, 34--49, (1969). \epubtkKeywords{Multipole expansions}

\bibitem{IW68}
Isaacson, R.A., and Winicour, J., ``Harmonic and Null Descriptions of
  Gravitational Radiation'', {\em Phys. Rev.}, {\bf 168}, 1451--1456, (1968).
  \epubtkKeywords{Asymptotics}

\bibitem{itoh2}
Itoh, Y., ``Equation of motion for relativistic compact binaries with the
  strong field point particle limit: Third post-Newtonian order'', {\em Phys.
  Rev. D}, {\bf 69}, 064018--1--43, (2004). \epubtkKeywords{Equations of
  motion, Post-Newtonian approximations}

\bibitem{itoh1}
Itoh, Y., and Futamase, T., ``New derivation of a third post-Newtonian equation
  of motion for relativistic compact binaries without ambiguity'', {\em Phys.
  Rev. D}, {\bf 68}, 121501(R), (2003). \epubtkKeywords{Equations of motion,
  Post-Newtonian approximations}

\bibitem{IFA00}
Itoh, Y., Futamase, T., and Asada, H., ``Equation of motion for relativistic
  compact binaries with the strong field point particle limit: Formulation, the
  first post-Newtonian order, and multipole terms'', {\em Phys. Rev. D}, {\bf
  62}, 064002--1--12, (2000). Related online version (cited on 17 May 2000):
  \newline\url{http://arXiv.org/abs/gr-qc/9910052}. \epubtkKeywords{Equations
  of motion, Post-Newtonian approximations}

\bibitem{IFA01}
Itoh, Y., Futamase, T., and Asada, H., ``Equation of motion for relativistic
  compact binaries with the strong field point particle limit: The second and
  half post-Newtonian order'', {\em Phys. Rev. D}, {\bf 63}, 064038--1--21,
  (2001). Related online version (cited on 30 January 2001):
  \newline\url{http://arXiv.org/abs/gr-qc/0101114}. \epubtkKeywords{Equations
  of motion, Post-Newtonian approximations}

\bibitem{IW93}
Iyer, B.R., and Will, C.M., ``Post-Newtonian gravitational radiation reaction
  for two-body systems'', {\em Phys. Rev. Lett.}, {\bf 70}, 113--116, (1993).
  \epubtkKeywords{Radiation reaction, Post-Newtonian approximations}

\bibitem{IW95}
Iyer, B.R., and Will, C.M., ``Post-Newtonian gravitational radiation reaction
  for two-body systems: Nonspinning bodies'', {\em Phys. Rev. D}, {\bf 52},
  6882--6893, (1995). \epubtkKeywords{Radiation reaction, Post-Newtonian
  approximations}

\bibitem{JaraS97}
Jaranowski, P., and Sch{\"a}fer, G., ``Radiative 3.5 post-Newtonian ADM
  Hamiltonian for many-body point-mass systems'', {\em Phys. Rev. D}, {\bf 55},
  4712--4722, (1997). \epubtkKeywords{Equations of motion, Post-Newtonian
  approximations, ADM formalism, Radiation reaction}

\bibitem{JaraS98}
Jaranowski, P., and Sch{\"{a}}fer, G., ``Third post-Newtonian higher order ADM
  Hamilton dynamics for two-body point-mass systems'', {\em Phys. Rev. D}, {\bf
  57}, 7274--7291, (1998). Related online version (cited on 17 December 1997):
  \newline\url{http://arXiv.org/abs/gr-qc/9712075}. Erratum {\it Phys.\ Rev.\
  D}, {\bf 63}, 029902, (2001). \epubtkKeywords{Equations of motion,
  Post-Newtonian approximations, ADM formalism}

\bibitem{JaraS99}
Jaranowski, P., and Sch{\"{a}}fer, G., ``The binary black-hole problem at the
  third post-Newtonian approximation in the orbital motion: Static part'', {\em
  Phys. Rev. D}, {\bf 60}, 124003--1--7, (1999). Related online version (cited
  on 23 June 1999): \newline\url{http://arXiv.org/abs/gr-qc/9906092}.
  \epubtkKeywords{Equations of motion, Post-Newtonian approximations, ADM
  formalism}

\bibitem{JaraS00}
Jaranowski, P., and Sch{\"{a}}fer, G., ``The binary black-hole dynamics at the
  third post-Newtonian order in the orbital motion'', {\em Ann. Phys.
  (Berlin)}, {\bf 9}, 378--383, (2000). Related online version (cited on 14
  March 2000): \newline\url{http://arXiv.org/abs/gr-qc/0003054}.
  \epubtkKeywords{Equations of motion, Post-Newtonian approximations, ADM
  formalism}

\bibitem{K80a}
Kerlick, G.D., ``Finite reduced hydrodynamic equations in the slow-motion
  approximation to general relativity. Part I. First post-Newtonian
  equations'', {\em Gen. Relativ. Gravit.}, {\bf 12}, 467--482, (1980).
  \epubtkKeywords{Post-Newtonian approximations, Radiation reaction}

\bibitem{K80b}
Kerlick, G.D., ``Finite reduced hydrodynamic equations in the slow-motion
  approximation to general relativity. Part II. Radiation reaction and
  higher-order divergent terms'', {\em Gen. Relativ. Gravit.}, {\bf 12},
  521--543, (1980). \epubtkKeywords{Radiation reaction}

\bibitem{K95}
Kidder, L.E., ``Coalescing binary systems of compact objects to
  (post)$^{5/2}$-Newtonian order. V. Spin effects'', {\em Phys. Rev. D}, {\bf
  52}, 821--847, (1995). Related online version (cited on 8 June 1995):
  \newline\url{http://arXiv.org/abs/gr-qc/9506022}.
  \epubtkKeywords{Gravitational radiation, Equations of motion}

\bibitem{KWWisco}
Kidder, L.E., Will, C.M., and Wiseman, A.G., ``Coalescing binary systems of
  compact objects to (post)$^5/2$-Newtonian order. III. Transition from
  inspiral to plunge'', {\em Phys. Rev. D}, {\bf 47}, 3281--3291, (1993).
  \epubtkKeywords{Gravitational radiation, Equations of motion}

\bibitem{KWW93}
Kidder, L.E., Will, C.M., and Wiseman, A.G., ``Spin effects in the inspiral of
  coalescing compact binaries'', {\em Phys. Rev. D}, {\bf 47}, R4183--R4187,
  (1993). \epubtkKeywords{Gravitational radiation, Equations of motion}

\bibitem{Kochanek}
Kochanek, C.S., ``Coalescing Binary Neutron Stars'', {\em Astrophys. J.}, {\bf
  398}(1), 234--247, (1992). \epubtkKeywords{Relativistic astrophysics, Neutron
  stars}

\bibitem{KFS03}
K{\"o}nigsd{\"o}rffer, C., Faye, G., and Sch{\"a}fer, G., ``Binary black-hole
  dynamics at the third-and-a-half post-Newtonian order in the ADM formalism'',
  {\em Phys. Rev. D}, {\bf 68}, 044004--1--19, (2003). Related online version
  (cited on 26 April 2006): \newline\url{http://arXiv.org/abs/gr-qc/0305048}.
  \epubtkKeywords{Equations of motion, Post-Newtonian approximations, ADM
  formalism, Radiation reaction}

\bibitem{Kop85}
Kopeikin, S.M., ``The equations of motion of extended bodies in
  general-relativity with conservative corrections and radiation damping taken
  into account'', {\em Astron. Zh.}, {\bf 62}, 889--904, (1985).
  \epubtkKeywords{Equations of motion, Post-Newtonian approximations}

\bibitem{Kop88}
Kopeikin, S.M., ``Celestial Coordinate Reference Systems in Curved Spacetime'',
  {\em Celest. Mech.}, {\bf 44}, 87, (1988). \epubtkKeywords{Equations of
  motion}

\bibitem{KSGE}
Kopeikin, S.M., Sch{\"{a}}fer, G., Gwinn, C.R., and Eubanks, T.M.,
  ``Astrometric and timing effects of gravitational waves from localized
  sources'', {\em Phys. Rev. D}, {\bf 59}, 084023--1--29, (1999). Related
  online version (cited on 17 February 1999):
  \newline\url{http://arXiv.org/abs/gr-qc/9811003}.
  \epubtkKeywords{Gravitational radiation}

\bibitem{KKS95}
Kr{\'{o}}lak, A., Kokkotas, K.D., and Sch{\"{a}}fer, G., ``Estimation of the
  post-Newtonian parameters in the gravitational-wave emission of a coalescing
  binary'', {\em Phys. Rev. D}, {\bf 52}, 2089--2111, (1995). Related online
  version (cited on 7 March 1995):
  \newline\url{http://arXiv.org/abs/gr-qc/9503013}. \epubtkKeywords{Binary
  systems, Gravitational wave detectors}

\bibitem{LL}
Landau, L.D., and Lifshitz, E.M., {\em The classical theory of fields},
  (Pergamon Press, Oxford, U.K.; New York, U.S.A., 1971), 3rd edition.
  \epubtkKeywords{General relativity, Field theory}

\bibitem{LGG05}
Limousin, F., Gondek-Rosi\'nska, D., and Gourgoulhon, E., ``Last orbits of
  binary strange quark stars'', {\em Phys. Rev. D}, {\bf 71}, 064012--1--11,
  (2005). Related online version (cited on 26 April 2006):
  \newline\url{http://arXiv.org/abs/gr-qc/0411127}. \epubtkKeywords{Numerical
  relativity, Innermost stable circular orbits, Strange stars}

\bibitem{LW90}
Lincoln, C.W., and Will, C.M., ``Coalescing binary systems of compact objects
  to (post)$^{5/2}$-Newtonian order: Late time evolution and gravitational
  radiation emission'', {\em Phys. Rev. D}, {\bf 42}, 1123--1143, (1990).
  \epubtkKeywords{Equations of motion, Post-Newtonian approximations}

\bibitem{LD17}
Lorentz, H.A., and Droste, J., in {\em The Collected Papers of H.A. Lorentz,
  Vol. 5}, (Nijhoff, The Hague, Netherlands, 1937), {\it Versl.\ K.\ Akad.\
  Wet.\ Amsterdam}, {\bf 26}, 392 and 649, (1917). \epubtkKeywords{Equations of
  motion}

\bibitem{Madore}
Madore, J., ``Gravitational radiation from a bounded source. I'', {\em Ann.
  Inst. Henri Poincare}, {\bf 12}, 285--305, (1970). Related online version
  (cited on 02 May 2006):
  \newline\url{http://www.numdam.org/item?id=AIHPA_1970__12_3_285_0}.
  \epubtkKeywords{Asymptotics}

\bibitem{MS}
Martin, J., and Sanz, J.L., ``Slow motion approximation in predictive
  relativistic mechanics. II. Non-interaction theorem for interactions derived
  from the classical field-theory'', {\em J. Math. Phys.}, {\bf 20}, 25--34,
  (1979). \epubtkKeywords{Field theory, Equations of motion}

\bibitem{M62}
Mathews, J., ``Gravitational multipole radiation'', {\em J. Soc. Ind. Appl.
  Math.}, {\bf 10}, 768--780, (1962). \epubtkKeywords{Multipole expansions}

\bibitem{MSSTT}
Mino, Y., Sasaki, M., Shibata, M., Tagoshi, H., and Tanaka, T., ``Black Hole
  Perturbation'', {\em Prog. Theor. Phys. Suppl.}, {\bf 128}, 1--121, (1997).
  Related online version (cited on 12 December 1997):
  \newline\url{http://arXiv.org/abs/gr-qc/9712057}. \epubtkKeywords{Binary
  systems, Perturbation methods}

\bibitem{MW03}
Mora, T., and Will, C.M., ``A post-Newtonian diagnostic of quasi-equilibrium
  binary configurations of compact objects'', {\em Phys. Rev. D}, {\bf 69},
  104021, (2004). Related online version (cited on 26 April 2006):
  \newline\url{http://arXiv.org/abs/gr-qc/0312082}. \epubtkKeywords{Equations
  of motion, Post-Newtonian approximations}

\bibitem{Moritz}
Moritz, H., {\em Advanced Physical Geodesy}, (H. Wichmann, Karlsruhe, Germany,
  1980). \epubtkKeywords{Geodesy}

\bibitem{NSW}
Newhall, X.X., Standish, E.M., and Williams, J.G., ``DE-102 -- A Numerically
  Integrated Ephemeris of the Moon and Planets Spanning 44 Centuries'', {\em
  Astron. Astrophys.}, {\bf 125}, 150--167, (1983). \epubtkKeywords{Celestial
  mechanics}

\bibitem{NB05}
Nissanke, S., and Blanchet, L., ``Gravitational radiation reaction in the
  equations of motion of compact binaries to 3.5 post-Newtonian order'', {\em
  Class. Quantum Grav.}, {\bf 22}, 1007, (2005). Related online version (cited
  on 26 April 2006): \newline\url{http://arXiv.org/abs/gr-qc/0412018}.
  \epubtkKeywords{Equations of motion, Post-Newtonian approximations, Radiation
  reaction}

\bibitem{OO73}
Ohta, T., Okamura, H., Kimura, T., and Hiida, K., ``Physically acceptable
  solution of Eintein's equation for many-body system'', {\em Prog. Theor.
  Phys.}, {\bf 50}, 492--514, (1973). \epubtkKeywords{Equations of motion}

\bibitem{OO74b}
Ohta, T., Okamura, H., Kimura, T., and Hiida, K., ``Coordinate condition and
  higher-order gravitational potential in canonical formalism'', {\em Prog.
  Theor. Phys.}, {\bf 51}, 1598--1612, (1974). \epubtkKeywords{Equations of
  motion}

\bibitem{OO74a}
Ohta, T., Okamura, H., Kimura, T., and Hiida, K., ``Higher-order gravitational
  potential for many-body system'', {\em Prog. Theor. Phys.}, {\bf 51},
  1220--1238, (1974). \epubtkKeywords{Equations of motion}

\bibitem{OTO98}
Owen, B.J., Tagoshi, H., and Ohashi, A., ``Nonprecessional spin-orbit effects
  on gravitational waves from inspiraling compact binaries to second
  post-Newtonian order'', {\em Phys. Rev. D}, {\bf 57}, 6168--6175, (1998).
  Related online version (cited on 31 October 1997):
  \newline\url{http://arXiv.org/abs/gr-qc/9710134}.
  \epubtkKeywords{Gravitational radiation, Equations of motion}

\bibitem{Papa51}
Papapetrou, A., ``Equations of motion in general relativity'', {\em Proc. Phys.
  Soc. London, Sect. B}, {\bf 64}, 57--75, (1951). \epubtkKeywords{Equations of
  motion}

\bibitem{Papa71}
Papapetrou, A., {\em Ann. Inst. Henri Poincare}, {\bf XIV}, 79, (1962).
  \epubtkKeywords{Gravitational radiation, Multipole expansions}

\bibitem{Papa62}
Papapetrou, A., ``Relativit\'{e} -- une formule pour le rayonnement
  gravitationnel en premi\`{e}re approximation'', {\em C. R. Acad. Sci. Ser.
  II}, {\bf 255}, 1578, (1962). \epubtkKeywords{Gravitational radiation,
  Multipole expansions}

\bibitem{PapaL81}
Papapetrou, A., and Linet, B., ``Equation of motion including the reaction of
  gravitational radiation'', {\em Gen. Relativ. Gravit.}, {\bf 13}, 335,
  (1981). \epubtkKeywords{Radiation reaction}

\bibitem{PW00}
Pati, M.E., and Will, C.M., ``Post-Newtonian gravitational radiation and
  equations of motion via direct integration of the relaxed Einstein equations:
  Foundations'', {\em Phys. Rev. D}, {\bf 62}, 124015--1--28, (2000). Related
  online version (cited on 31 July 2000):
  \newline\url{http://arXiv.org/abs/gr-qc/0007087}. \epubtkKeywords{Equations
  of motion, Post-Newtonian approximations}

\bibitem{PW02}
Pati, M.E., and Will, C.M., ``Post-Newtonian gravitational radiation and
  equations of motion via direct integration of the relaxed Einstein equations.
  II. Two-body equations of motion to second post-Newtonian order, and
  radiation-reaction to 3.5 post-Newtonian order'', {\em Phys. Rev. D}, {\bf
  65}, 104008--1--21, (2001). Related online version (cited on 31 December
  2001): \newline\url{http://arXiv.org/abs/gr-qc/0201001}.
  \epubtkKeywords{Gravitational radiation, Post-Newtonian approximations}

\bibitem{P63}
Penrose, R., ``Asymptotic Properties of Fields and Space-Times'', {\em Phys.
  Rev. Lett.}, {\bf 10}, 66--68, (1963). \epubtkKeywords{Asymptotics}

\bibitem{P65}
Penrose, R., ``Zero rest-mass fields including gravitation: asymptotic
  behaviour'', {\em Proc. R. Soc. London, Ser. A}, {\bf 284}, 159--203, (1965).
  \epubtkKeywords{Asymptotics}

\bibitem{Pe64}
Peters, P.C., ``Gravitational Radiation and the Motion of Two Point Masses'',
  {\em Phys. Rev.}, {\bf 136}, B1224--B1232, (1964).
  \epubtkKeywords{Gravitational radiation, Binary systems}

\bibitem{PM63}
Peters, P.C., and Mathews, J., ``Gravitational Radiation from Point Masses in a
  Keplerian Orbit'', {\em Phys. Rev.}, {\bf 131}, 435--440, (1963).
  \epubtkKeywords{Gravitational radiation, Binary systems}

\bibitem{Petrova}
Petrova, N.M., ``Ob Uravnenii Dvizheniya i Tenzore Materii dlya Sistemy
  Konechnykh Mass v Obshchei Teorii Otnositielnosti'', {\em J. Exp. Theor.
  Phys.}, {\bf 19}(11), 989--999, (1949). \epubtkKeywords{Equations of motion}

\bibitem{PfTC00}
Pfeiffer, H.P., Teukolsky, S.A., and Cook, G.B., ``Quasicircular orbits for
  spinning binary black holes'', {\em Phys. Rev. D}, {\bf 62}, 104018--1--11,
  (2000). \epubtkKeywords{Compact binaries, Innermost stable circular orbits,
  Numerical relativity}

\bibitem{Pi64}
Pirani, F.A.E., ``Introduction to Gravitational Radiation Theory'', in
  Trautman, A., Pirani, F.A.E., and Bondi, H., eds., {\em Lectures on General
  Relativity, Vol. 1}, Brandeis Summer Institute in Theoretical Physics,
  249--373, (Prentice-Hall, Englewood Cliffs, U.S.A., 1964).
  \epubtkKeywords{Gravitational radiation, Asymptotics}

\bibitem{P93}
Poisson, E., ``Gravitational radiation from a particle in circular orbit around
  a black hole. I. Analytic results for the nonrotating case'', {\em Phys. Rev.
  D}, {\bf 47}, 1497--1510, (1993). \epubtkKeywords{Binary systems,
  Perturbation methods}

\bibitem{P95}
Poisson, E., ``Gravitational radiation from a particle in circular orbit around
  a black-hole. VI. Accuracy of the post-Newtonian expansion'', {\em Phys. Rev.
  D}, {\bf 52}, 5719--5723, (1995). Related online version (cited on 11
  February 1997): \newline\url{http://arXiv.org/abs/gr-qc/9505030}. Addendum
  Phys. Rev. D 55 (1997) 7980--7981. \epubtkKeywords{Binary systems,
  Perturbation theory}

\bibitem{PW95}
Poisson, E., and Will, C.M., ``Gravitational waves from inspiralling compact
  binaries: Parameter estimation using second-post-Newtonian waveforms'', {\em
  Phys. Rev. D}, {\bf 52}, 848--855, (1995). Related online version (cited on
  24 February 1995): \newline\url{http://arXiv.org/abs/gr-qc/9502040}.
  \epubtkKeywords{Binary systems, Gravitational wave detectors}

\bibitem{PB02}
Poujade, O., and Blanchet, L., ``Post-Newtonian approximation for isolated
  systems calculated by matched asymptotic expansions'', {\em Phys. Rev. D},
  {\bf 65}, 124020--1--25, (2002). Related online version (cited on 21 December
  2001): \newline\url{http://arXiv.org/abs/gr-qc/0112057}.
  \epubtkKeywords{Post-Newtonian approximations}

\bibitem{Press77}
Press, W.H., ``Gravitational Radiation from Sources Which Extend Into Their Own
  Wave Zone'', {\em Phys. Rev. D}, {\bf 15}, 965--968, (1977).
  \epubtkKeywords{Gravitational radiation, Quadrupole formula}

\bibitem{Rend90}
Rendall, A.D., ``Convergent and divergent perturbation series and the
  post-Minkowskian scheme'', {\em Class. Quantum Grav.}, {\bf 7}, 803, (1990).
  \epubtkKeywords{Mathematical relativity, Approximation methods,
  Post-Minkowskian approximations}

\bibitem{Rend92}
Rendall, A.D., ``On the definition of post-Newtonian approximations'', {\em
  Proc. R. Soc. London, Ser. A}, {\bf 438}, 341--360, (1992).
  \epubtkKeywords{Mathematical relativity, Post-Newtonian approximations,
  Newtonian limit}

\bibitem{Rend94}
Rendall, A.D., ``The Newtonian limit for asymptotically flat solutions of the
  Vlasov--Einstein system'', {\em Commun. Math. Phys.}, {\bf 163}, 89, (1994).
  Related online version (cited on 26 April 2006):
  \newline\url{http://arXiv.org/abs/gr-qc/9303027}.
  \epubtkKeywords{Mathematical relativity, Approximation methods,
  Post-Newtonian approximations}

\bibitem{Riesz}
Riesz, M., ``L'int\'{e}grale de Riemann--Liouville et le probl\`{e}me de
  Cauchy'', {\em Acta Math.}, {\bf 81}, 1--218, (1949).
  \epubtkKeywords{Mathematical methods}

\bibitem{SB58}
Sachs, R., and Bergmann, P.G., ``Structure of Particles in Linearized
  Gravitational Theory'', {\em Phys. Rev.}, {\bf 112}, 674--680, (1958).
  \epubtkKeywords{Linearized gravity, Asymptotics}

\bibitem{S61}
Sachs, R.K., ``Gravitational waves in general relativity VI. The outgoing
  radiation condition'', {\em Proc. R. Soc. London, Ser. A}, {\bf 264},
  309--338, (1961). \epubtkKeywords{Asymptotics}

\bibitem{Sachs62}
Sachs, R.K., ``Gravitational waves in general relativity VIII. Waves in
  asymptotically flat space-time'', {\em Proc. R. Soc. London, Ser. A}, {\bf
  270}, 103--126, (1962). \epubtkKeywords{Asymptotics}

\bibitem{Sasa94}
Sasaki, M., ``Post-Newtonian Expansion of the Ingoing-Wave Regge--Wheeler
  Function'', {\em Prog. Theor. Phys.}, {\bf 92}, 17--36, (1994).
  \epubtkKeywords{Binary systems, Perturbation methods}

\bibitem{S85}
Sch{\"{a}}fer, G., ``The Gravitational Quadrupole Radiation-Reaction Force and
  the Canonical Formalism of ADM'', {\em Ann. Phys. (N.Y.)}, {\bf 161},
  81--100, (1985). \epubtkKeywords{Equations of motion, ADM formalism}

\bibitem{S86}
Sch{\"{a}}fer, G., ``The ADM Hamiltonian at the Postlinear Approximation'',
  {\em Gen. Relativ. Gravit.}, {\bf 18}, 255--270, (1986).
  \epubtkKeywords{Equations of motion, ADM formalism}

\bibitem{SW93}
Sch{\"a}fer, G., and Wex, N., ``Second post-Newtonian motion of compact
  binaries'', {\em Phys. Lett. A}, {\bf 174}, 196--205, (1993). Erratum {\it
  Phys.\ Lett.\ A}, {\bf 177}, 461, (1993). \epubtkKeywords{Binary pulsars,
  Post-Newtonian approximations}

\bibitem{Schwartz54}
Schwartz, L., ``Sur l'impossibilit\'e de la multiplication des distributions'',
  {\em C. R. Acad. Sci. Ser. II}, {\bf 239}, 847--848, (1954).
  \epubtkKeywords{Mathematical methods}

\bibitem{Schwartz}
Schwartz, L., {\em Th\'{e}orie des distributions}, (Hermann, Paris, France,
  1978). \epubtkKeywords{Mathematical methods}

\bibitem{Sellier}
Sellier, A., ``Hadamard's finite part concept in dimension $n \ge 2$,
  distributional definition, regularization forms and distributional
  derivatives'', {\em Proc. R. Soc. London, Ser. A}, {\bf 445}, 69--98, (1994).
  \epubtkKeywords{Mathematical methods}

\bibitem{SB83}
Simon, W., and Beig, R., ``The multipole structure of stationary space-times'',
  {\em J. Math. Phys.}, {\bf 24}, 1163--1171, (1983).
  \epubtkKeywords{Asymptotics}

\bibitem{tHooft}
't~Hooft, G., and Veltman, M.J.G., ``Regularization and renormalization of
  gauge fields'', {\em Nucl. Phys. B}, {\bf 44}, 189--213, (1972).
  \epubtkKeywords{Quantum field theory, Perturbation theory, Gauge theories}

\bibitem{TNaka94}
Tagoshi, H., and Nakamura, T., ``Gravitational waves from a point particle in
  circular orbit around a black hole: Logarithmic terms in the post-Newtonian
  expansion'', {\em Phys. Rev. D}, {\bf 49}, 4016--4022, (1994).
  \epubtkKeywords{Binary systems, Perturbation theory}

\bibitem{TOO01}
Tagoshi, H., Ohashi, A., and Owen, B.J., ``Gravitational field and equations of
  motion of spinning compact binaries to 2.5-post-Newtonian order'', {\em Phys.
  Rev. D}, {\bf 63}, 044006--1--14, (2001). Related online version (cited on 4
  October 2000): \newline\url{http://arXiv.org/abs/gr-qc/0010014}.
  \epubtkKeywords{Gravitational radiation, Equations of motion}

\bibitem{TSasa94}
Tagoshi, H., and Sasaki, M., ``Post-Newtonian Expansion of Gravitational Waves
  from a Particle in Circular Orbit around a Schwarzschild Black Hole'', {\em
  Prog. Theor. Phys.}, {\bf 92}, 745--771, (1994). \epubtkKeywords{Binary
  systems, Perturbation methods}

\bibitem{TTS96}
Tanaka, T., Tagoshi, H., and Sasaki, M., ``Gravitational Waves by a Particle in
  Circular Orbit around a Schwarzschild Black Hole'', {\em Prog. Theor. Phys.},
  {\bf 96}, 1087--1101, (1996). \epubtkKeywords{Binary systems, Perturbation
  methods}

\bibitem{T93}
Taylor, J.H., ``Pulsar timing and relativistic gravity'', {\em Class. Quantum
  Grav.}, {\bf 10}, 167--174, (1993). \epubtkKeywords{Relativistic
  astrophysics, Binary systems, Pulsars}

\bibitem{TFMc79}
Taylor, J.H., Fowler, L.A., and McCulloch, P.M., ``Measurements of general
  relativistic effects in the binary pulsar PSR 1913+16'', {\em Nature}, {\bf
  277}, 437--440, (1979). \epubtkKeywords{Relativistic astrophysics, Binary
  systems, Pulsars}

\bibitem{TW82}
Taylor, J.H., and Weisberg, J.M., ``A New Test of General Relativity:
  Gravitational Radiation and the Binary Pulsar PSR~1913+16'', {\em Astrophys.
  J.}, {\bf 253}, 908--920, (1982). \epubtkKeywords{Relativistic astrophysics,
  Binary systems, Pulsars}

\bibitem{Th80}
Thorne, K.S., ``Multipole expansions of gravitational radiation'', {\em Rev.
  Mod. Phys.}, {\bf 52}, 299--340, (1980). \epubtkKeywords{Multipole
  expansions}

\bibitem{Thhouches}
Thorne, K.S., ``The theory of gravitational radiation: An introductory
  review'', in Deruelle, N., and Piran, T., eds., {\em Gravitational
  Radiation}, NATO Advanced Study Institute, Centre de physique des Houches,
  2--21 June 1982,  1--57, (North-Holland; Elsevier, Amsterdam, Netherlands;
  New York, U.S.A., 1983). \epubtkKeywords{Gravitational radiation}

\bibitem{Th300}
Thorne, K.S., ``Gravitational radiation'', in Hawking, S.W., and Israel, W.,
  eds., {\em Three Hundred Years of Gravitation},  330--458, (Cambridge
  University Press, Cambridge, U.K.; New York, U.S.A., 1987).
  \epubtkKeywords{Gravitational radiation}

\bibitem{Th92}
Thorne, K.S., ``Gravitational-wave bursts with memory: The Christodoulou
  effect'', {\em Phys. Rev. D}, {\bf 45}, 520, (1992).
  \epubtkKeywords{Gravitational radiation, Multipole expansions}

\bibitem{ThH85}
Thorne, K.S., and Hartle, J.B., ``Laws of motion and precession for black holes
  and other bodies'', {\em Phys. Rev. D}, {\bf 31}, 1815--1837, (1985).
  \epubtkKeywords{Equations of motion}

\bibitem{ThK75}
Thorne, K.S., and Kov{\`{a}}cs, S.J., ``Generation of gravitational waves. I.
  Weak-field sources'', {\em Astrophys. J.}, {\bf 200}, 245--262, (1975).
  \epubtkKeywords{Post-Minkowskian approximations}

\bibitem{Wag75}
Wagoner, R.V., ``Test for Existence of Gravitational Radiation'', {\em
  Astrophys. J. Lett.}, {\bf 196}, L63--L65, (1975).
  \epubtkKeywords{Relativistic astrophysics, Binary systems, Pulsars}

\bibitem{WagW76}
Wagoner, R.V., and Will, C.M., ``Post-Newtonian gravitational radiation from
  orbiting point masses'', {\em Astrophys. J.}, {\bf 210}, 764--775, (1976).
  \epubtkKeywords{Gravitational radiation, Post-Newtonian approximations}

\bibitem{W94}
Will, C.M., ``Gravitational Waves from Inspiralling Compact Binaries: A
  Post-Newtonian Approach'', in Sasaki, M., ed., {\em Relativistic Cosmology},
  Proceedings of the 8th Nishinomiya-Yukawa Memorial Symposium, on October
  28--29, 1993, Shukugawa City Hall, Nishinomiya, Hyogo, Japan, vol.~8 of
  NYMSS,  83--98, (Universal Academy Press, Tokyo, Japan, 1994).
  \epubtkKeywords{Gravitational radiation, Post-Newtonian approximations}

\bibitem{W99}
Will, C.M., ``Generation of Post-Newtonian Gravitational Radiation via Direct
  Integration of the Relaxed Einstein Equations'', {\em Prog. Theor. Phys.
  Suppl.}, {\bf 136}, 158--167, (1999). Related online version (cited on 15
  October 1999): \newline\url{http://arXiv.org/abs/gr-qc/9910057}.
  \epubtkKeywords{Gravitational radiation, Post-Newtonian approximations}

\bibitem{WW96}
Will, C.M., and Wiseman, A.G., ``Gravitational radiation from compact binary
  systems: Gravitational waveforms and energy loss to second post-Newtonian
  order'', {\em Phys. Rev. D}, {\bf 54}, 4813--4848, (1996). Related online
  version (cited on 5 August 1996):
  \newline\url{http://arXiv.org/abs/gr-qc/9608012}.
  \epubtkKeywords{Gravitational radiation, Post-Newtonian approximations}

\bibitem{Wiseman93}
Wiseman, A.G., ``Coalescing binary-systems of compact objects to
  5/2-post-Newtonian order. IV. The gravitational-wave tail'', {\em Phys. Rev.
  D}, {\bf 48}, 4757--4770, (1993). \epubtkKeywords{Gravitational radiation,
  Binary systems, Post-Newtonian approximations}

\bibitem{WW91}
Wiseman, A.G., and Will, C.M., ``Christodoulou's nonlinear gravitational-wave
  memory: Evaluation in the quadrupole approximation'', {\em Phys. Rev. D},
  {\bf 44}, R2945--R2949, (1991). \epubtkKeywords{Gravitational radiation,
  Multipole expansions}

\end{thebibliography}

\end{document}